\documentclass[11pt]{article}
\usepackage{amssymb,amsmath,epsfig}
\usepackage[colorlinks=false,breaklinks=true,linkcolor=blue]{hyperref}
\usepackage{graphics,psfrag,graphicx,color}
\usepackage{float,hhline}
\usepackage{multicol}
\usepackage{comment}
\usepackage[dvipsnames]{xcolor}

\numberwithin{equation}{section}
\allowdisplaybreaks

\newcommand{\ds}{\displaystyle}

\newcommand{\bbeta}{{\boldsymbol\eta}}

\newcommand{\bsi}{{\boldsymbol\sigma}}

\newcommand{\btau}{{\boldsymbol\tau}}
\newcommand{\bzeta}{{\boldsymbol\zeta}}

\newcommand{\bv}{{\mathbf{v}}}
\newcommand{\bw}{{\mathbf{w}}}

\newcommand{\bu}{\mathbf{u}}

\newcommand{\bt}{{\mathbf{t}}}
\newcommand{\bn}{{\mathbf{n}}}
\newcommand{\be}{{\mathbf{e}}}

\newcommand{\0}{{\mathbf{0}}}

\def\bK{\mathbf{K}}
\def\bI{\mathbf{I}}

\def\bM{\mathbf{M}}

\def\bx{\mathbf{x}}

\newcommand\bbM{\mathbb{M}}

\newcommand{\cO}{\mathcal{O}}

\def\R{\mathrm{R}}

\def\M{\mathrm{M}}

\def\rt{\mathrm{t}}

\def\BJS{\mathtt{BJS}}

\def\bdiv{\mathbf{div}}

\def\div{\mathrm{div}}

\def\qin{{\quad\hbox{in}\quad}}
\def\qon{{\quad\hbox{on}\quad}}
\def\qan{{\quad\hbox{and}\quad}}

\numberwithin{equation}{section}
\numberwithin{figure}{section}
\numberwithin{table}{section}
\allowdisplaybreaks

\allowdisplaybreaks

\title{Non-Newtonian and poroelastic effects in simulations of arterial flows}

\author{
{\sc Tongtong Li}\thanks{Department of Mathematics, University of Pittsburgh, Pittsburgh, PA 15260, USA, email: {\tt tol24@pitt.edu}. Supported in part by NSF grant DMS 1818775}	
\quad
{\sc Xing Wang}\thanks{Department of Mathematics, University of Pittsburgh, Pittsburgh, PA 15260, USA, email: {\tt xiw117@pitt.edu}. Supported in part by NSF grant DMS 1818775}	
\quad
{\sc Ivan Yotov}\thanks{Department of Mathematics, University of Pittsburgh, Pittsburgh, PA 15260, USA, email: {\tt yotov@math.pitt.edu}. Supported in part by NSF grant DMS 1818775}}

\date{\today}

\begin{document}

\maketitle

\begin{abstract}
\noindent In this paper, we focus on investigating the influence on hydrodynamic factors of different coupled computational models describing the interaction between an incompressible fluid and two symmetric elastic or poroelastic structures. The fluid region is governed by time dependent Navier-Stokes equations; while for the structure region, we employ two different types of fully dynamic models to study the effects of elasticity and poroelasticity. It is known that blood flow shows a non-Newtonian property in small vessels and in situations of complex geometries. On one hand, we perform numerical experiments for blood flow using the Carreau-Yasuda model to simulate the viscosity and study the influence of non-Newtonian blood rheology as well as the poroelasticity on a benchmark vessel, by means of comparing computational results with models with Newtonian fluids or elastic structures. On the other hand, we present a two-dimensional simulation of blood flow in an axisymmetric stenosis artery, considering not only the non-Newtonian fluids properties but also the fluid-structure interaction. The results of this study demonstrate that the flow characteristics, including velocity and pressure fields, wall shear stress, relative residence time, displacement and filtration velocity, are affected by different models, geometries and parameters, such as permeability and Lam\'{e} coefficients.
\end{abstract}

\maketitle


\section{Introduction}
 Due to changes of lifestyles such as bad dietary habits, smoking, staying up late and sedentary, cardiovascular diseases have become a major concern of modern society. There has been considerable evidence that hydrodynamic factors would play an important role in identifying, diagnosising and understanding the development and progression of arterial lesions. Intrigued by this purpose, we want to focus on investigating the prototype problem arising in blood flow in this paper. We study fully dynamic blood flow models for the interaction of an incompressible Newtonian or non-Newtonian fluid and a fluid within a poroelastic or an elastic vessel. This is a challenging problem with predicting, modeling or controling process in blood rheology. Arterial flow is not only affected by the poroelastic nature of the arterial wall \cite{badia2009coupling, bodnar2014fluid, bukavc2015operator, bungartz2006fluid}, but also the geometrical complexity of the vessel \cite{guerciotti2018computational, rabby2014pulsatile}. Therefore, it is important to build a mathematical model, which simulates properly the interaction of a free viscous fluid with a porous material and also accounts for the elasticity of the medium in various computational domains.
 
 Related literature on the blood flow problems is rich \cite{bukavc2015partitioning, bukavc2015operator, guerciotti2018computational, rabby2014pulsatile, badea2010numerical, discacciati2002mathematical}. In \cite{guerciotti2018computational}, based on three-dimensional patient-specific stenotic vessels, the influences of the degrees of stenosis on Newtonian and non-Newtonian behavior of blood have been studied. The limitation of this work are the absence of turbulence in the stenotic models and also of a fluid-structure interaction model. In \cite{deyranlou2016non}, non-Newtonian blood behavior on LDL (low-density lipoprotein) accumulation is analyzed and fluid-multilayer arteries are adopted for healthy and stenotic vessel models. Numerical investigation of non-Newtonian modeling effects on unsteading periodic flows in a two-dimensional vessel with multiple idealized stenoses of different degrees has been learned in \cite{rabby2014pulsatile}. However, to our knowledge, few works have considered that the arterial wall is porous and deformable and the non-Newtonian property of blood flow in different vessel geometries at the same time. 
 
 In this work, we use time-dependent Navier-Stokes equations to model free fluids in the fluid region and fully dynamic Biot system to govern the poroelastic structure. We focus on differences between Newtonian and non-Newtonian fluids using Carreau-Yasuda model. In addition to the prototype benchmark problem, we also conduct simulation for stenosis cases. This work considers four different parts essential to blood flow problem: (1) the difference between elasticity and poroelasticity models; (2) how the parameter permeability affects the dynamic characteristics; (3) differences of Newtonian and non-Newtonian blood rheology in numerical simulations; (4) practicability of coupled elastic and poroelasitc models in stenotic regions. 

Earlier works \cite{deyranlou2016non, bazilevs2013computational} about the fluid-structure interaction (FSI) models focused on establishing and exploring rheological phenomema through the porous arterial walls. It has been categorized that there are three distinct wall layer groups: wall-free model, fluid-wall single layer model and fluid-wall multilayer model \cite{prosi2005mathematical, deyranlou2015low}. However, they don't account for the natural deformation of arterial walls. Instead of simple porous vessel, we will employ a coupled Biot system which includes the second derivative of displacement to govern the fluids in deformable and porous structures. The fluids-structure regions are coupled through dynamic and kinematic interface conditions and also the Beavers-Joseph-Saffman slip with friction conditions.  In addition, we study how the structure parameter permeability and L\'{a}me coefficients affect the blood flow. 

On the other aspect, blood is comprised of different types of elements such as red blood cells, platelets, proteins, water etc., and will exhibit complex rheological properties \cite{deyranlou2016non, guerciotti2018computational, koshiba2007multiphysics}. And a lot biophysical research has confirmed that blood shows a shear-thinning behaviour: the fluids viscosity decreases with increasing shear rates, and it will reach approximately a constant value. However, the assumption of Newtonian fluids is generally used and accepted for blood flow studies in large-sized vessels \cite{gijsen1999influence, guerciotti2016computational}. But so far, no universal agreement is reached about the proper model to describe the viscous properties of blood especially for medium-sized or small-sized arteries. In addition to comparing the difference between poroelastic and elastic models, we also want to study how non-Newtonian properties affect on blood flow characteristics.

Last but not the least, we want to apply our simulation to a relatively complicated geometry. Due to the deposition of lipid, cholesterol, some other substances, there are high risks that stenosis would be initiated preferentially in arteries and regions with high curvature or bifurcations \cite{guerciotti2018computational,rabby2014pulsatile}. A few studies have even been carried out on multiple stenoses \cite{rabby2014pulsatile}. Few of these studies worked on the coupled fluids-structure model using non-Newtonian viscosity. We will enhance our numerical simulations through considering the poroelastic nature of arterial wall and also the non-Newtonian behavior of blood flow in a stenotic geometry. 

The outline of the remaining paper is presented as follow: in section \ref{sec:model-problem}, we describe two mathematical models: Navier-Stokes/Elasticity (NSE/E) model and Navier-Stokes/Poroelasticity(NSE/P) model together with appropriate interface, boundary and initial conditions. The next section is devoted to the numerical simulations of blood flow models with real world parameters. We will mainly focus on the differences between the four types of  models: Newtonian NSE/E, Newtonian NSE/P, non-Newtonian NSE/E, non-Newtonian NSE/P models and several effects on the blood rheology.

We end up this section by introducing some definitions and fixing some notations.
Let $\cO\subseteq \R^2$, denote a domain with Lipschitz boundary $\Gamma$.
By $\bM$ and $\bbM$ we will denote the corresponding vectorial and tensorial counterparts of the generic scalar functional space $\M$.
In turn, for any vector field $\bv=(v_i)_{i=1,n}$, we set the gradient and divergence operators, as
\begin{equation*}
\nabla\bv:=\left(\frac{\partial v_i}{\partial x_j}\right)_{i,j=1,n} \qan 
\div(\bv):=\sum_{j=1}^n \frac{\partial v_j}{\partial x_j}.
\end{equation*}
Furthermore, for any tensor field $\btau:=(\tau_{ij})_{i,j=1,n}$ and $\bzeta:=(\zeta_{ij})_{i,j=1,n}$, we let $\bdiv(\btau)$ be the divergence operator $\div$ acting along the rows of $\btau$, and define the transpose and the deviatoric tensor as
\begin{equation*}
\btau^\rt := (\tau_{ji})_{i,j=1,n},\quad \btau:\bzeta:=\sum_{i,j=1}^n \tau_{ij}\zeta_{ij}.
\end{equation*}

\section{Simulation domains and methods}\label{sec:model-problem}

\subsection{Simulation domains}
\begin{figure}[ht]
	\includegraphics[trim=0 10 0 30,scale=0.37]{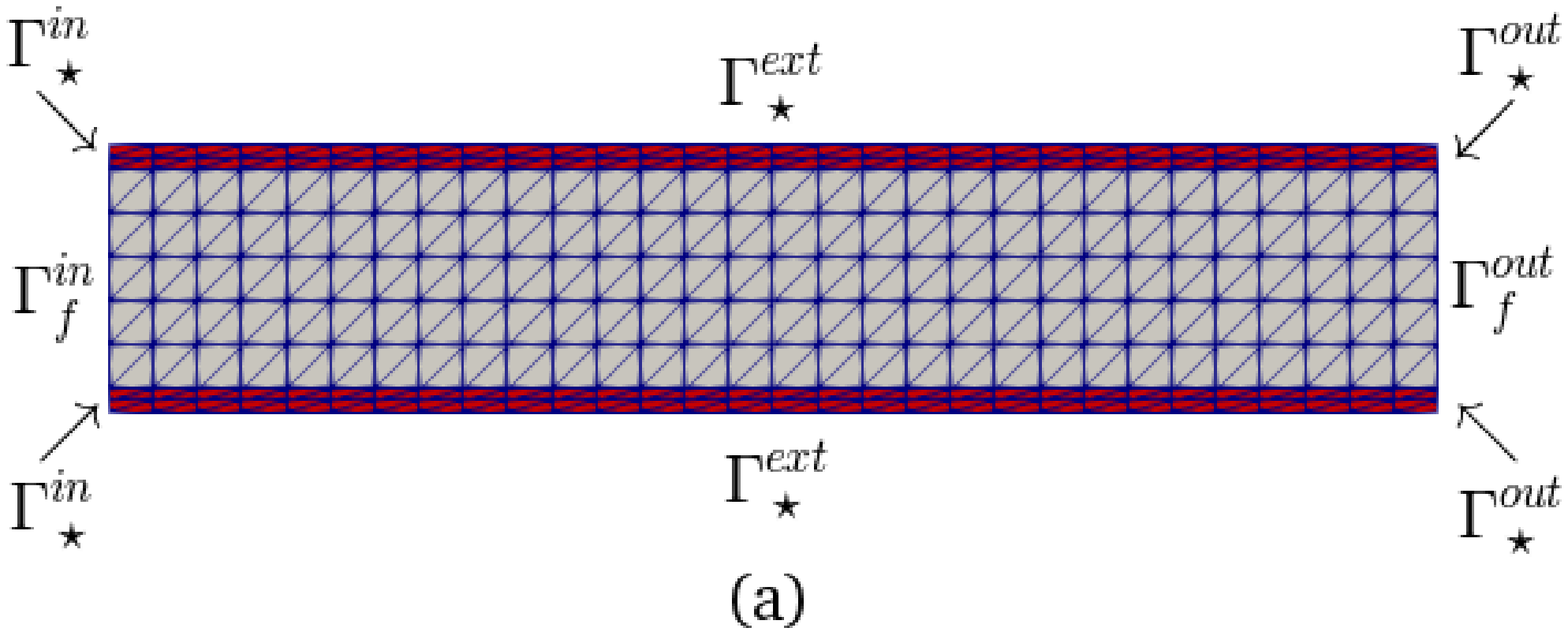}
	\includegraphics[trim=0 10 0 30,scale=0.37]{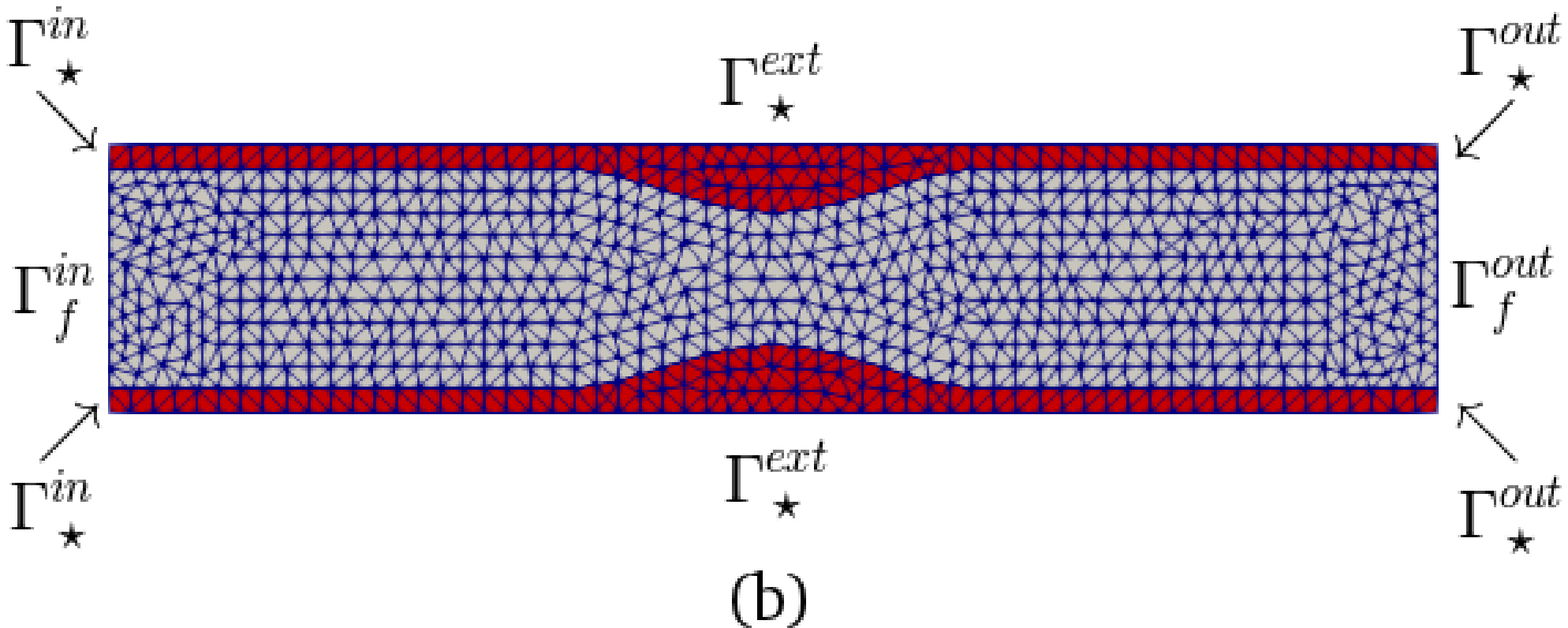}
	\centering
	\caption{Simulation domains. (a) is the computational domain $\Omega$ using in section \ref{Newtonian} and section \ref{Nonnewtonian}, where $\star\in \{p,e\}$. (b) is the computational domain $\Omega$ using in section \ref{stenosis}. The red areas are structure regions $\Omega_{\star}$ and the grey areas are fluid region $\Omega_{f}$. The mesh we are using in the simulation is non-matching and much finer than what is showed in (a) and (b). }
		\label{mesh11}
\end{figure}

We will first focus on the prototype benchmark problem arising from FSI modeling of blood flows \cite{bukavc2015partitioning, bukavc2015operator, formaggia2001coupling} in section \ref{Newtonian} and section \ref{Nonnewtonian}, and then consider an ideal stenotic model in section \ref{stenosis}. We show the two simulation domains in Fig.\ref{mesh11}. 
\begin{figure}[ht]
	\includegraphics[trim=0 10 0 13,scale=0.07]{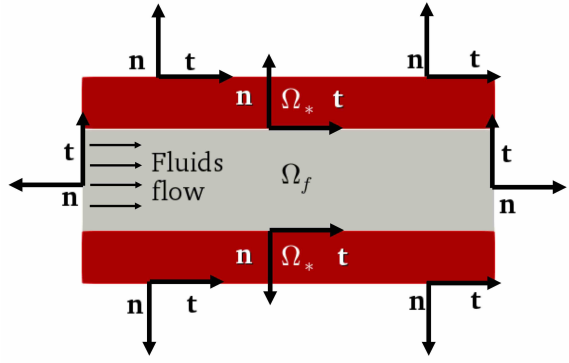}
	\centering
	\caption{Normal vectors $\bn$ and tangential vector $\bt$ on each boundary or interface. Note that we exaggerate the thickness of the structure region $\Omega_{\star}$ to have a better view and understanding. }
	\label{mesh12}
\end{figure}

As shown in Fig.\ref{mesh11}, we consider a Lipschitz rectangular domain $\Omega\subset\R^2$, which is subdivided into three non-overlapping regions: fluid region $\Omega_f$ in the middle and we will use $\Omega_{e}$ to represent the strcuture regions for elastic model, and correspondly, $\Omega_p$ stands for the poroelastic model. For notation purpose, we will use $\Omega_{\star}$, $\star\in \{p,e\}$ to represent the structure regions. Let $\Gamma_{f \star} = \partial\Omega_f\cap\partial\Omega_\star$ denote the nonempty interfaces between these regions. For the fluid region, we denote the inlet and outlet boundaries by $\Gamma_f^{in}=\{(0,y)|-R<y<R\}$ and $\Gamma_f^{out}=\{(L,y)|-R<y<R\}$. And for the structure region with $\star\in\{p,e\}$, we denote the inlet and outlet poroelastic/elastic structure boundaries, respectively, by $\Gamma_\star^{in}=\{(0,y)|-R-r_\star<y<R\,\ \text{or} \,\ R<y<R+r_\star \}$ and $\Gamma_\star^{in}=\{(L,y)|-R-r_\star<y<R\,\ \text{or} \,\ R<y<R+r_\star \}$, where $r_\star$ is the poroelastic/elastic wall thickness. In addition, we let $\Gamma_\star^{ext}=\{(x,y)|0<x<L,  y=R+r_\star \,\ \text{or}\,\ y=-R-r_\star \}$ be the external structure boundaries. Finally, we denote by $\bn$ the unit outward normal vector, which points outward from fluid domain on $\partial \Omega_{f}$. And we denote by $\bt$ the unit tangential vector, which points toward the direction of the blood flow on the interface $\Gamma_{f \star}$ and other horizontal exterior boundaries $\Gamma_\star^{ext}$ and upward on the vertical boundaries $\Gamma_f^{in}$,  $\Gamma_f^{out}$, $\Gamma_\star^{in}$ and $\Gamma_\star^{out}$.

\subsection {Mathematical models and numerical methods}
We focus on studying the propagation of a single pressure wave whose amplitude is comparable to the pressure difference between systolic and diatolic phase of the heartbeat. A time-dependent pressure as follow will drive the blood flow:
\begin{eqnarray}\label{pressurefunc}
p_{in}(t)=\begin{cases}
\frac{P_{max}}{2}(1-\cos(\frac{2\pi t}{T_{max}})), & \text{if} \,\ t\leq T_{max}; \\
0, & \text{if}\,\  t>T_{max},
\end{cases}
\end{eqnarray}\label{inflow}
where $P_{max}=13,334$ dyn/cm$^2$ and $T_{max}=0.003$ s.

\subsubsection {Navier-Stokes/Elasticity model problem}
We assume that the flow in $\Omega_f$ is governed by the time-dependent Navier-Stokes equations with density $\rho_{f}$ and viscosity $\mu_f$, which are written in the following stress-velocity-pressure formulation:
\begin{equation}\label{eq:Stokes-1} 
\ds 
\rho_f\,\left(\frac{\partial \bu_f}{\partial t} + (\nabla \, \bu_f )\bu_f\right) -\,\bdiv(\bsi_f) \,=\, \0,\quad 
\div(\bu_f) \,=\, 0, \qin \Omega_f\times (0,T],
\end{equation}
where $\bu_f$ is the fluid velocity, $\bsi_f\,=\, -p_f\,\bI + 2\,\mu_f\,\be(\bu_f)$ is the stress tensor, and $\be(\bu_f) := \dfrac{1}{2}\,\left( \nabla\bu_f + (\nabla\bu_f)^\rt \right)$. We adopt commonly used boundary conditions in blood flow models and prescribe the normal stress at the inlet and outlet boundaries \cite{bukavc2015operator, bukavc2015partitioning} as follows:
\begin{equation}
\begin{array}{c}
\ds \bsi_f\bn=-p_{in}(t)\bn, \quad \text{on} \quad \Gamma_f^{in}\times (0,T],\\[1ex]
\ds \bsi_f\bn=\0, \quad \text{on} \quad \Gamma_f^{out}\times (0,T].
\end{array}
\end{equation}

We then state the elastic model which is used to govern the structure region. Let $\bbeta_e$ be the displacement in $\Omega_e$, and let $\bsi_e$ be the elastic stress tensor defined as follows:
\begin{equation}\label{eq:bsie-definitions}
\bsi_e \,:=\, \lambda_e\,\div(\bbeta_e)\,\bI + 2\,\mu_e\,\be(\bbeta_e), \qin \Omega_e\times (0,T],
\end{equation}
where $0<\lambda_{\min}\leq \lambda_e(\bx)\leq \lambda_{\max}$ and $0<\mu_{\min} \leq\mu_e(\bx) \leq\mu_{\max}$ are the Lam\'e parameters and are determined by Young's modulus which we will discuss later \cite{MR4022710}.
We use the following governing equation to better represent the behavior of an artery \cite{bukavc2015operator}:
\begin{equation}\label{eq: extra xi term}
\rho_{e}\frac{\partial^2 {\bbeta}_e}{\partial{t^2}}+\xi \bbeta_e-\bdiv(\bsi_e)=\0, \quad \mbox{in}\quad \Omega_e\times(0,T].
\end{equation}
where $\rho_{e}>0$ is the wall density, $\xi>0$ is the spring coefficient. The term $\xi \bbeta_{e}$ comes from the axially symmetric formulation, accounting for the recoil due to the circumferential strain \cite{bukavc2015partitioning}. In other words, it acts like a spring term to keep the top and bottom structure displacements connected. For boundary conditions, we assume that elastic structure is fixed at the inlet and outlet boundaries, namely:
\begin{equation}\label{eq: bc1}
\bbeta_{e}=\0, \quad \text{on} \quad \Gamma_e^{in}\cup\Gamma_e^{out}\times (0,T].
\end{equation}
For the external structure boundary $\Gamma_e^{ext}$, we assume that the external ambient pressure and the displacement in tangential direction of the exterior boundary $\Gamma_e^{ext}$ are zero:
\begin{equation}\label{eq: bc2}
\ds \bn \cdot \bsi_e \bn=0, \quad \text{and} \quad \bbeta_{e}\cdot \bt=0, \quad \text{on} \quad \Gamma_e^{ext}\times (0,T].
\end{equation}

Next, we introduce the transmission conditions on the top interface as well as the bottom interface $\Gamma_{fe}$ \cite{beavers1967boundary}:
\begin{equation}\label{eq:interface-conditions}
\ds \bu_f = \frac{\partial\,\bbeta_e}{\partial t}, \quad  \bsi_f\bn \,=\, \bsi_e\bn , \qon \Gamma_{fe}\times (0,T].
\end{equation}
The first equation in \eqref{eq:interface-conditions} corresponds to the continuity of the velocity vector, and the second one represents the continuity of the normal stress vector on $\Gamma_{fe}$.

Finally, the above coupling system is complemented by a set of initial conditions:
\begin{equation*}
\begin{array}{c}
\ds \bu_f (\bx, 0) = \0 \qan p_f(\bx,0)=p_{in}(0), \qin \Omega_f, \\[2ex]
\ds \bbeta_e(\bx,0) = \0 \qan
\frac{\partial \bbeta_e}{\partial t}(\bx,0) = \0, \qin \Omega_e.
\end{array}
\end{equation*}

\subsubsection {Navier-Stokes/Poroelasticity model problem}

The Navier-Stokes equations are exactly the same as (\ref{eq:Stokes-1}). As for the poroelastic region, let $\bsi_e^p$ and $\bsi_p$ be the elastic and poroelastic stress tensors, respectively,
\begin{equation}\label{eq:bsie-bsip-definitions}
\bsi_e^p \,:=\, \lambda_p\,\div(\bbeta_p)\,\bI + 2\,\mu_p\,\be(\bbeta_p) \qan
\bsi_p \,:=\, \bsi_e^p - \alpha_p\,p_p\,\bI \qin \Omega_p\times (0,T],
\end{equation}
where $0<\lambda_{\min}\leq \lambda_p(\bx)\leq \lambda_{\max}$ and $0<\mu_{\min} \leq\mu_p(\bx) \leq\mu_{\max}$ are the Lam\'e parameters and $0\leq \alpha_p \leq 1$ is the Biot--Willis constant.
The poroelasticity region $\Omega_p$ is governed by the fully dynamic Biot system \cite{MR3851065, biot1941general}:
\begin{equation}\label{eq:Biot-model}
\begin{array}{c}
\ds \rho_p \frac{\partial^2 \bbeta_p}{\partial t^2}+\xi\bbeta_{p}-\,\bdiv(\bsi_p) = \0,\quad 
\mu_f\,\bK^{-1}\bu_p + \nabla\,p_p = \0, \qin \Omega_p\times(0,T], \\[2ex]
\ds \frac{\partial}{\partial t}\left( s_0\,p_p + \alpha_p\,\div(\bbeta_p) \right) + \div(\bu_p) = 0, \qin \Omega_p\times(0,T], 
\end{array}
\end{equation}
where $(\bu_p, p_p)$ is the velocity-pressure pair in $\Omega_p$, $s_0\geq 0$ is a storage coefficient and $\bK$ the symmetric and uniformly positive definite permeability tensor, satisfying, for some constants $0< k_{\min}\leq k_{\max}$,
\begin{equation}\label{eq:K-uniform-bound}
\forall\, \bw\in\R^n \quad k_{\min}\,|\bw|^2 
\,\leq\, \bw^\rt\,\bK^{-1}(\bx)\bw 
\,\leq\, k_{\max}\,|\bw|^2 \quad \forall\, \bx\in\Omega_p.
\end{equation}
And we complement the boundary conditions for $\bbeta_p$ and $\bsi_p$:
\begin{equation}\label{eq: bc33}
\bbeta_{p}=\0, \quad \text{on} \quad \Gamma_p^{in}\cup\Gamma_p^{out}\times (0,T].
\end{equation}
\begin{equation}\label{eq: bc44}
\ds \bn \cdot \bsi_p \bn=0, \quad \text{and} \quad \bbeta_{p}\cdot \bt=0, \quad \text{on} \quad \Gamma_p^{ext}\times (0,T].
\end{equation}
Additionally, for the fluids in the poroelastic medium, we impose the following boundary conditions:
\begin{equation}\label{darcyinflow}
\begin{array}{c}
\bu_{p}\cdot \bn=0, \quad \text{on}\quad \Gamma_p^{in}\cup\Gamma_p^{out}\times (0,T],\\[1ex]
p_p=0, \quad \text{on}\quad \Gamma_p^{ext}\times (0,T].
\end{array}
\end{equation}

Next, we introduce the transmission conditions on the interface $\Gamma_{fp}$ \cite{MR3851065}:
\begin{equation}\label{eq:interface-conditions2}
\begin{array}{c}
\ds \bu_f\cdot\bn \,=\, \left(\frac{\partial\,\bbeta_p}{\partial t} + \bu_p\right)\cdot\bn , \quad
\bsi_f\bn \,=\, \bsi_p\bn, \qon \Gamma_{fp}\times (0,T], \\ [2ex]
 \left(\ds \bsi_f\bn\right)\cdot \bn=-p_p,\quad \mu_f\,\alpha_{\BJS}\sqrt{K^{-1}}\left(\bu_f - \frac{\partial\,\bbeta_p}{\partial t}\right)\cdot\bt \,=\, -\left(\ds \bsi_f\bn\right)\cdot \bt, \qon \Gamma_{fp}\times (0,T],
\end{array}
\end{equation}
where $K= (\bK\,\bt)\cdot\bt$, and $\alpha_{\BJS} \geq 0$ is an experimentally determined friction coefficient.
The first and second equations in \eqref{eq:interface-conditions2} correspond to conservation of mass and balance of momentum on $\Gamma_{fp}$, respectively, whereas the third one 
represents the balance of normal stress and the last one represents Beaver--Joseph--Saffman (BJS) slip with friction condition, respectively.

Finally, the above coupling system is complemented by a set of initial conditions,
\begin{equation*}
\begin{array}{c}
\ds \bu_f (\bx, 0) = \0, \quad p_f(\bx,0)=p_{in}(0), \qin \Omega_f, \\[2ex]
\ds p_p(\bx,0) = p_{in}(0), \quad \bbeta_p(\bx,0) = \0 \qan
\frac{\partial \bbeta_p}{\partial t}(\bx,0) = \0, \qin \Omega_p.
\end{array}
\end{equation*}

\subsubsection{Discretized models}
For the time discretization, we consider the backward Euler method with a semi-implicit way. We indicate with $z^n$ the approximation of a generic function $z(t)$ evaluated at $t^n=n\Delta t$, $n=1, 2, \cdots N$. At each time step $t^n$, we have the following discretized in time Navier-Stokes equation in the fluid region $\Omega_{f}$:
\begin{equation}
\begin{array}{c}
\ds \rho_{f}\frac{\bu_{f}^{n+1}-\bu_{f}^{n}}{\Delta t}+\rho_{f}(\nabla\bu_{f}^{n+1})\bu_{f}^{n}-\bdiv(\bsi_f^{n+1})=\0,\\[2ex]
\ds \div(\bu_{f}^{n+1})=0.
\end{array}
\end{equation}

For the elastic model, we have the following for the governing equation in the structure region $\Omega_{e}$:
\begin{equation}
\ds \rho_e\frac{\bbeta_{e}^{n+1}-2\bbeta_{e}^n+\bbeta_{e}^{n-1}}{\Delta t^2}+\xi \bbeta_e^{n+1}-\bdiv(\bsi_e^{n+1})=\0.
\end{equation}
While for the poroelastic model, we have the following for the Biot system in the structure region $\Omega_{p}$:
\begin{equation}
\begin{array}{c}
\ds \rho_p\frac{\bbeta_{p}^{n+1}-2\bbeta_{p}^n+\bbeta_{p}^{n-1}}{\Delta t^2}+\xi \bbeta_p^{n+1}-\bdiv(\bsi_p^{n+1})=\0,\\[2ex]
\ds \rho_{f}\bK^{-1}\bu_{p}^{n+1}+\nabla p_p^{n+1}=\0,\\[2ex]
\ds s_0\frac{p_p^{n+1}-p_p^{n}}{\Delta t}+\alpha_p\frac{\div(\bbeta_{p}^{n+1})-\div(\bbeta_{p}^n)}{\Delta t}+\div(\bu_{p}^{n+1})=0.
\end{array}
\end{equation}
All the numerical tests in the following sections are implemented by finite element library Freefem++\cite{freefem}. For the discretization of space, we will choose the Taylor-Hood $\mathcal{P}_2 - \mathcal{P}_1$ finite elements for variables $(\bu_{f},p_f)$ in $\Omega_{f}$. For elastic model, we use continuous Lagrangian  $\mathcal{P}_1$ for variable $\bbeta_{e}$ in $\Omega_{e}$, while for the poroelastic model, we utilize Raviart-Thomas $\mathcal{RT}_0-\mathcal{P}_0$ for $(\bu_{p},p_p)$ and
$\mathcal{P}_2$ for $p_p$ in $\Omega_{p}$. In NSE/E and NSE/P models, a Lagrange multiplier method is employed to impose the normal stress on the interface and continuity of flux condition respectively \cite{MR3851065}. We have finished the convergent test for our numerical methods in \cite{MR3851065}. We set the time discrezation parameter $\Delta t=0.00006 s$ in section \ref{Newtonian} and \ref{Nonnewtonian}; and $\Delta t=0.00003s$ in section \ref{stenosis}.

\section{Numerical applications and discussions}\label{inflow/outflow}
\begin{figure}[ht!]
	\centering
	\includegraphics[trim=0 0 0 0,scale=0.74]{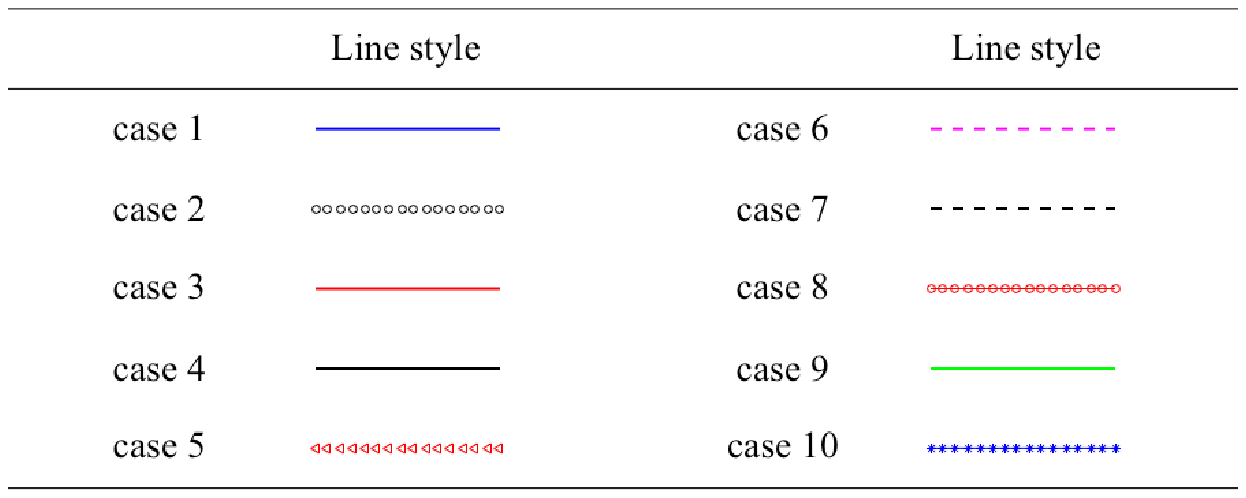}
	\caption{Line styles for different cases.}
	\label{LL}
\end{figure}

In this section, in order to study non-Newtonian and poroelastic effects in blood flow simulations, we present numerical results for ten different cases. We fix the line styles for these cases in the following sections as shown in table \ref{LL}. 

Also note that for the velocity and pressure waves, and viscosity plots in all the following sections, we use solid dark blue to simply present the structure region $\Omega_{e}$ for elastic models. In the structure region $\Omega_{e}$, elastic models don't have velocity, pressure and viscosity as variables.  
\subsection{Newtonian models}\label{Newtonian}
In this section, we focus on the Newtonian fluids in the computational domain shown in Fig.\ref{mesh11} part (a). With Newtonian fluids assumption, parameter viscosity, a propotional constant between shear stress and shear rate, is enough to describe the blood rheological behavior. In this case, it is important to fully understand the effects of poroelasticity and elasticity, and the effects of different permeability $\bK$ on NSE/P models. The reference values of the parameters used in this study fall within the range of physical values for blood flow and are reported in Table \ref{T2}. The Lam\'{e} coefficients are determined from the Young's modulus $E$ and the Poisson's ratio $\tilde{\nu}$ via the the following relationship \cite{MR3851065}:
\begin{equation}
	\lambda_\star=\frac{E\tilde{\nu}}{(1+\tilde{\nu})(1-2\tilde{\nu})},  \quad \mu_\star=\frac{E}{2(1+\tilde{\nu})}.
\end{equation}
The propagation of the pressure wave is analyzed over the time zone $[0, 0.006]$ s. The final time is selected so that the pressure wave can reach the outflow section. We introduce the following cases:
\begin{itemize}
\item case 1: Newtonian NSE/E model;
\item case 2: Newtonian NSE/P model, with $\bK=diag(1,1)\times10^{-9}$;
\item case 3: Newtonian NSE/P model, with $\bK=diag(1,1)\times10^{-7}$.
\end{itemize}

\begingroup
\def\arraystretch{1.1}
\begin{table}[ht!]
	\begin{center}
		\begin{tabular}{l c l l c l c}
			\hline
	Parameter(Units)                                           & Symbol                                                 & Values                      &Parameter(Units)                             &Symbol                        &Values         \\ \hline
	Radius(cm)                                                      & $R$                                                       &$0.5$                        &Lam\'{e} coeff.(dyn/cm$^2$)    & $\mu_\star$       &$4.28\times10^{6}$                        \\
	Length(cm)                                                      & $L$                                                      &$6$                           &Lam\'{e} coeff.(dyn/cm$^2$)    & $\lambda_\star$    & $1.07\times 10^{6}$                   \\
			wall thickness(cm)                     &$r_\star$                      &$0.1$                   &Total time(s)         &$T$                                                        &0.006    \\
		 wall density(g/cm$^3$)            &$\rho_{\star}$             &$1.1$                       &Spring coeff.(dyn/cm$^4$)                                        & $\xi$                                                     & $5\times 10^{7}$                 \\    
			Fluid density (g/cm$^3$)                                              &$\rho_{f}$                 &$1$   &BJS coeff.                      & $\alpha_{BJS}$                                 & 1.0 \\                     
			Dyn. viscosity(g/cm-s)                                                          &$\mu_f$                     &$0.035$           &Young's modulus(dyn/cm$^2$)    &$E$       &$2.996\times 10^{6}$  \\                                                                         
			Mass storativity (cm$^{2}$/dyn)                             & $s_0$                                                    & $5\times 10^{-6}$      &Poisson's ratio &$\tilde{\nu}$    & 0.4\\
		Permeability(cm$^2$) &$\bK$ 	& &Biot-Willis constant                                                        & $\alpha$                                              & 1.0                             \\
 \hline
		\end{tabular}
	\end{center}
	\caption{Geometry, elasticity, poroelasticity and fluid parameters. Note that $\star\in \{p,e\}$.}
	\label{T2}
\end{table}
\endgroup
\noindent Note that the permeability valus are within the physical range for arterial walls estabilished in the literature \cite{chung2012effect, deyranlou2016non, discacciati2002mathematical}.

\begin{figure}[ht!]
	\includegraphics[trim=0 0 0 30,scale=0.28]{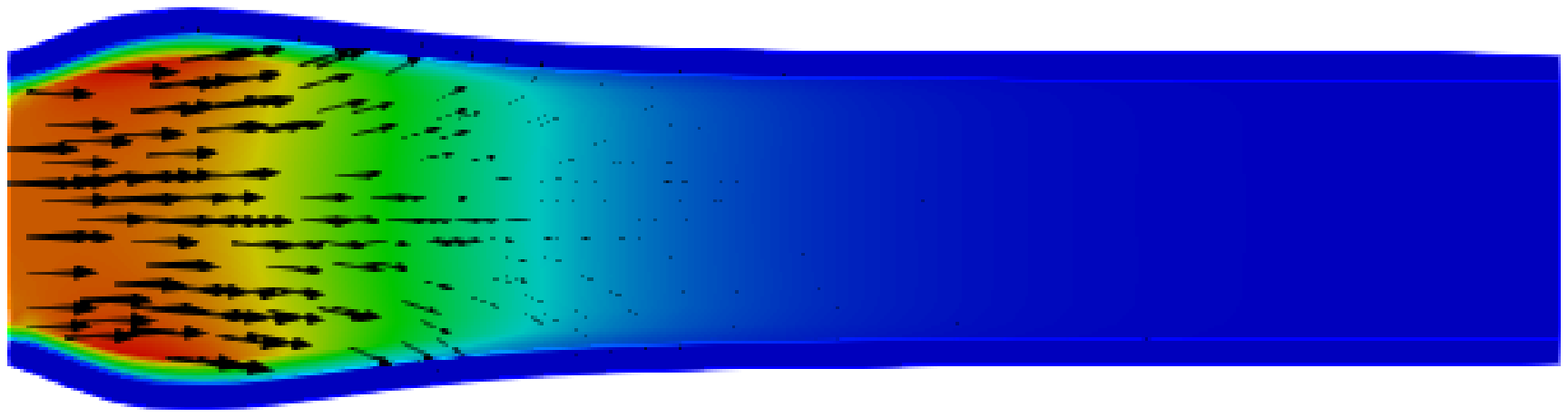}
	\includegraphics[trim=0 0 0 30,scale=0.28]{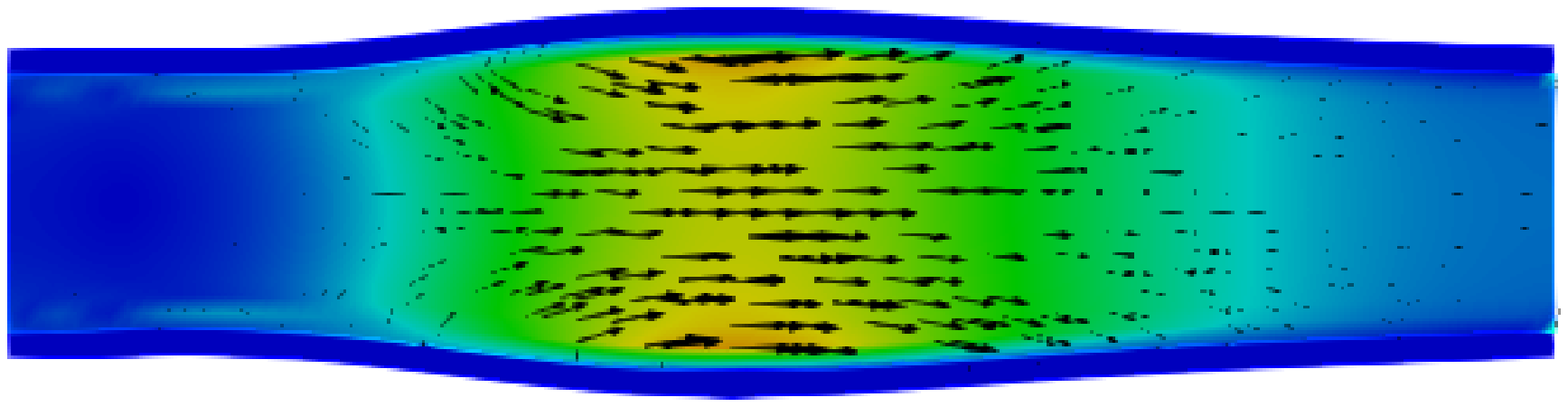}
	\includegraphics[trim=0 0 0 30,scale=0.28]{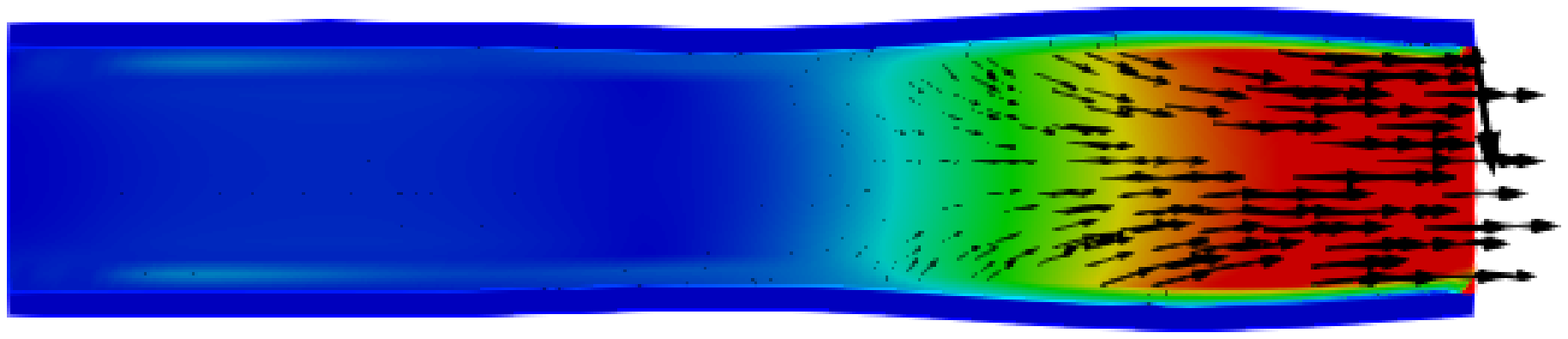}
	\includegraphics[trim=0 0 0 0,scale=0.28]{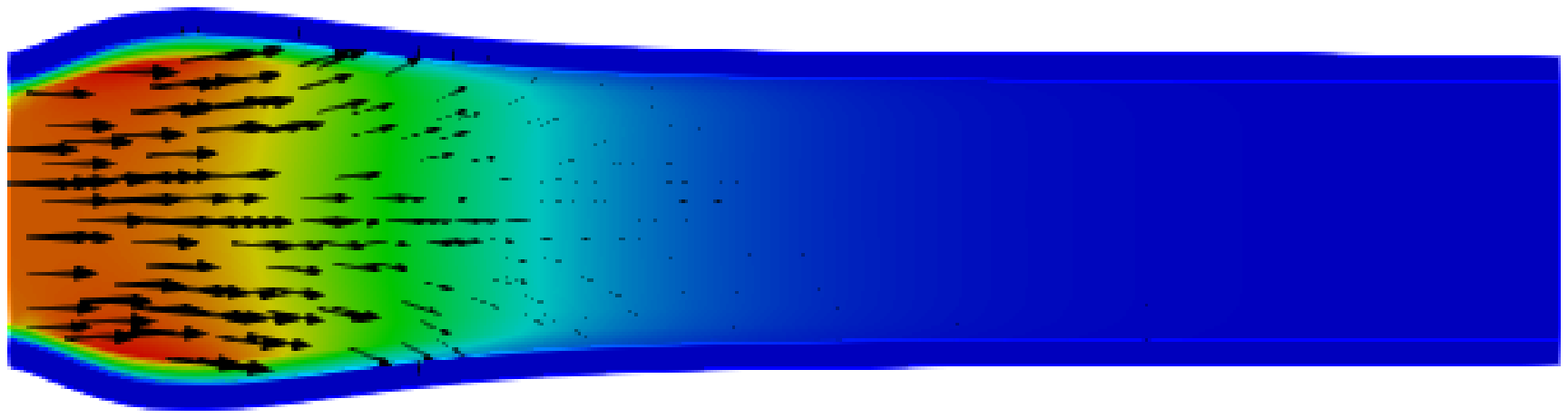}
	\includegraphics[trim=0 0 0 0,scale=0.28]{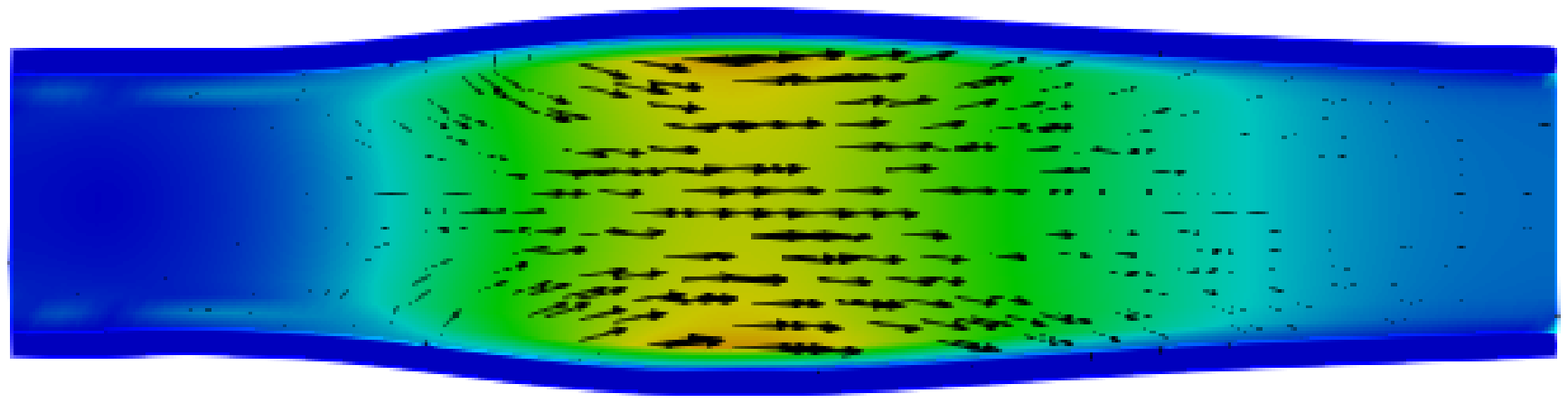}
	\includegraphics[trim=0 0 0 0,scale=0.28]{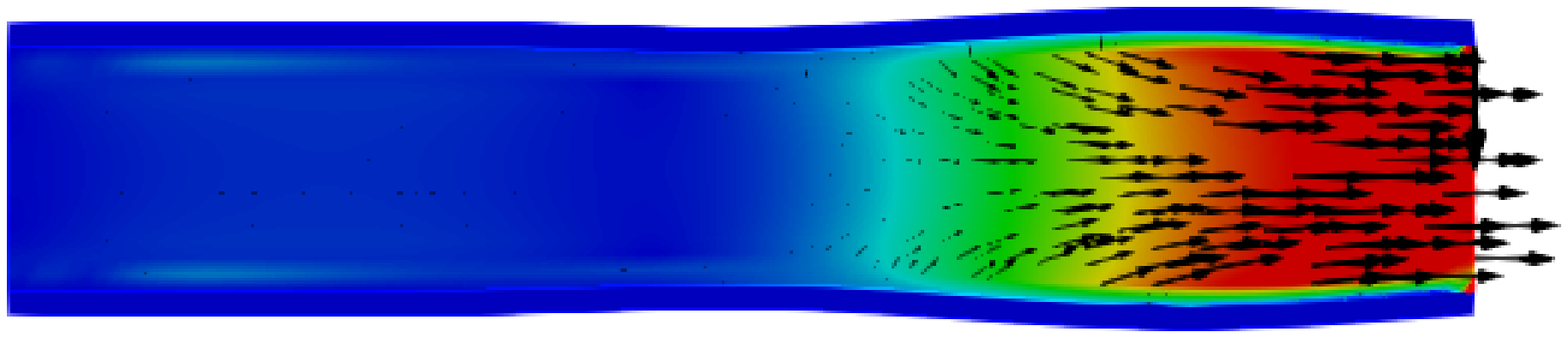}
	\includegraphics[trim=0 10 0 0,scale=0.28]{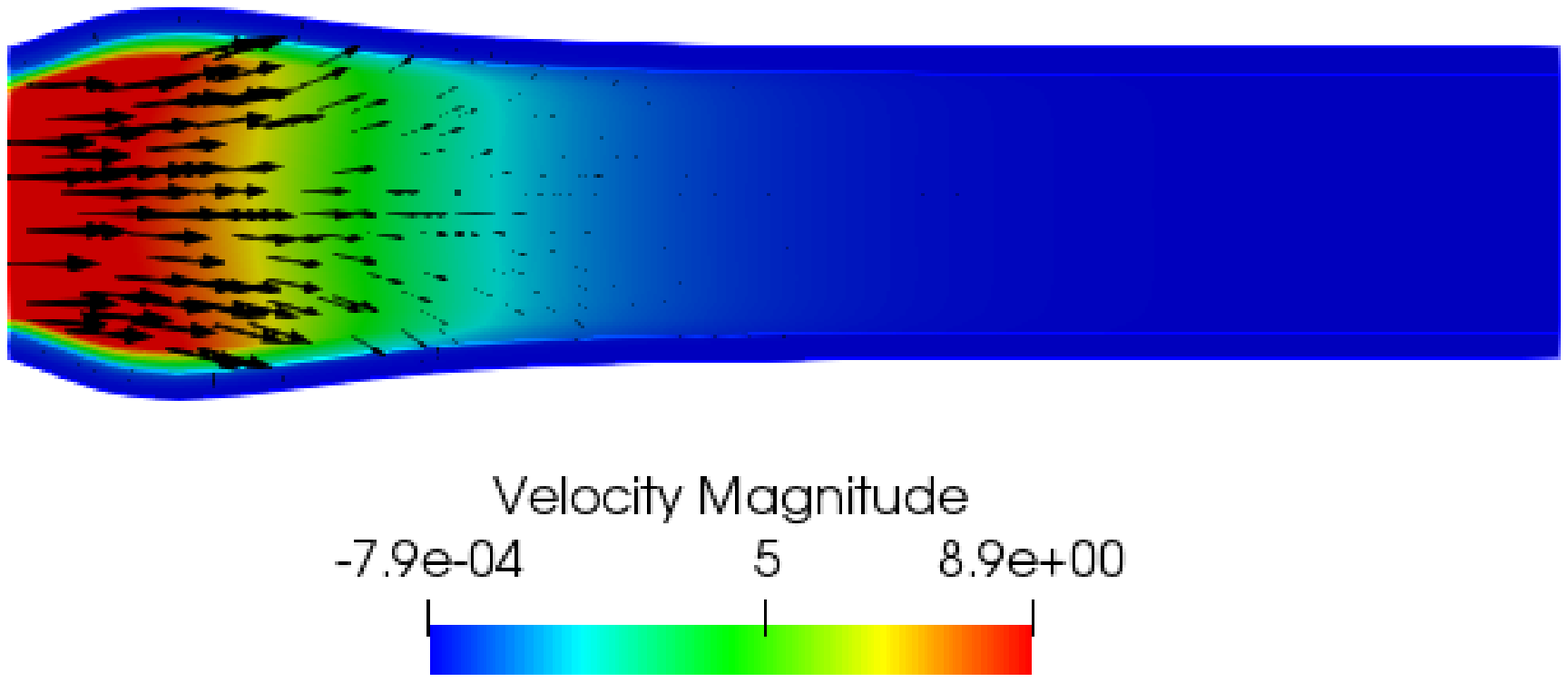}
	\includegraphics[trim=0 10 0 0,scale=0.28]{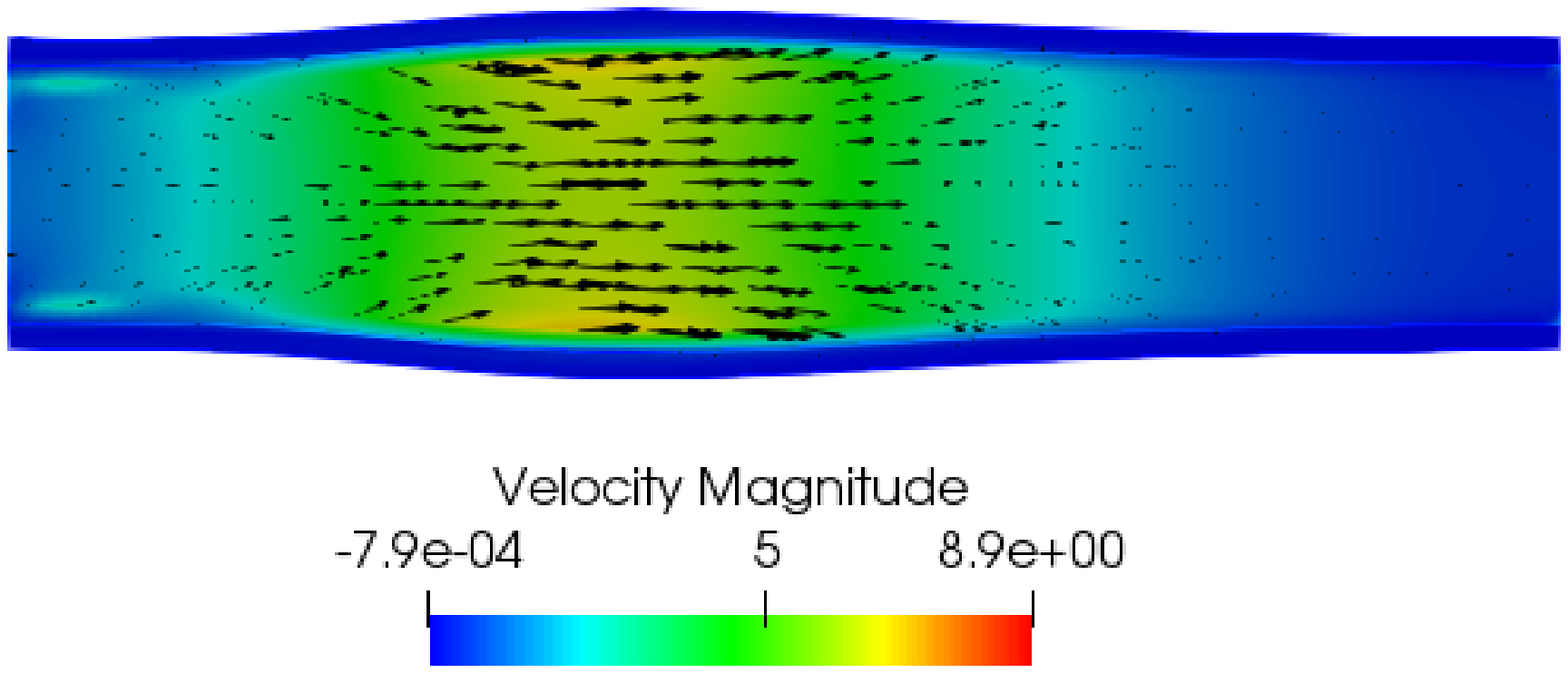}
	\includegraphics[trim=0 10 0 0,scale=0.28]{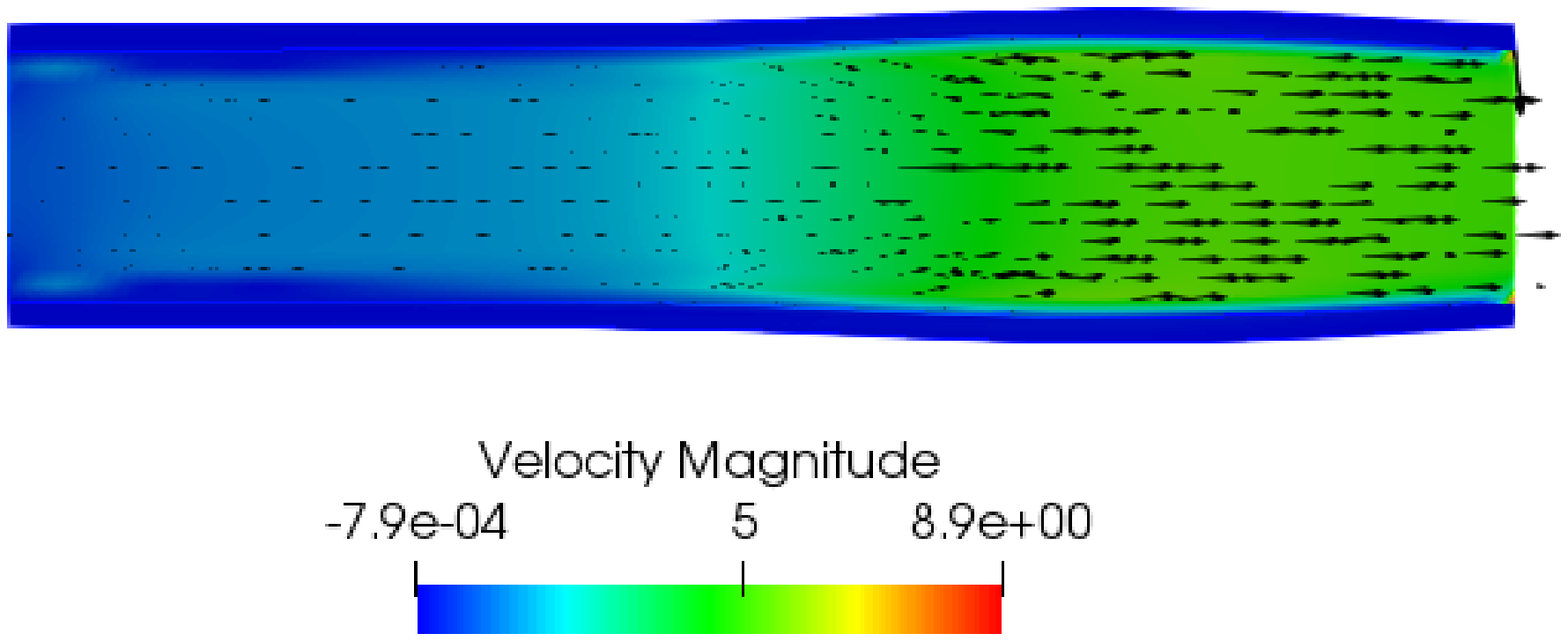}
	\caption{Fluid velocity magnitude together with velocity arrows scaled with the magnitude at time t=1.8 ms, t=3.6 ms, t=5.4 ms for case 1, case 2 and case 3. Note that for case 1, there is no Darcy velocity. We are using solid blue to present structure $\Omega_{e}$. This pattern follows in all the following plots of elastic models. }
	\label{velocitycase1}
\end{figure}

\begin{figure}[ht!]
	\includegraphics[trim=0 -10 0 30,scale=0.28]{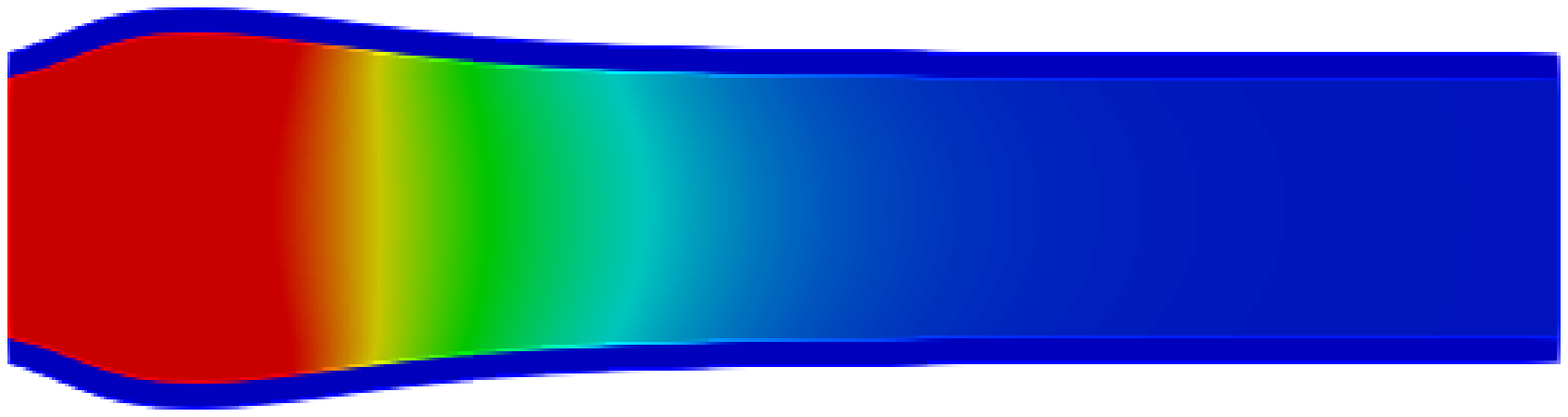}
	\includegraphics[trim=0 -10 0 30,scale=0.28]{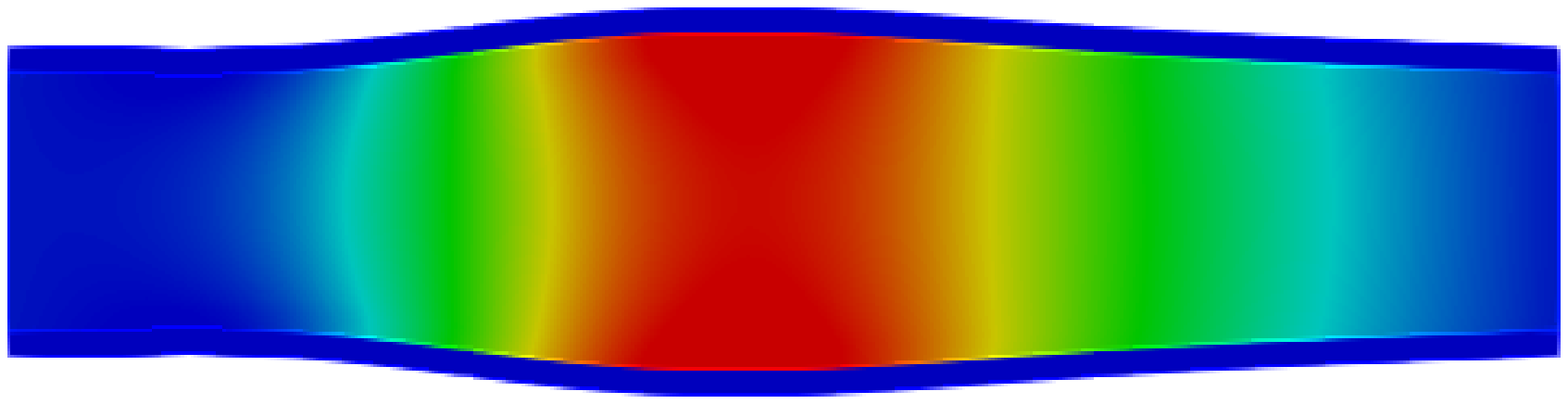}
	\includegraphics[trim=0 -10 0 30,scale=0.28]{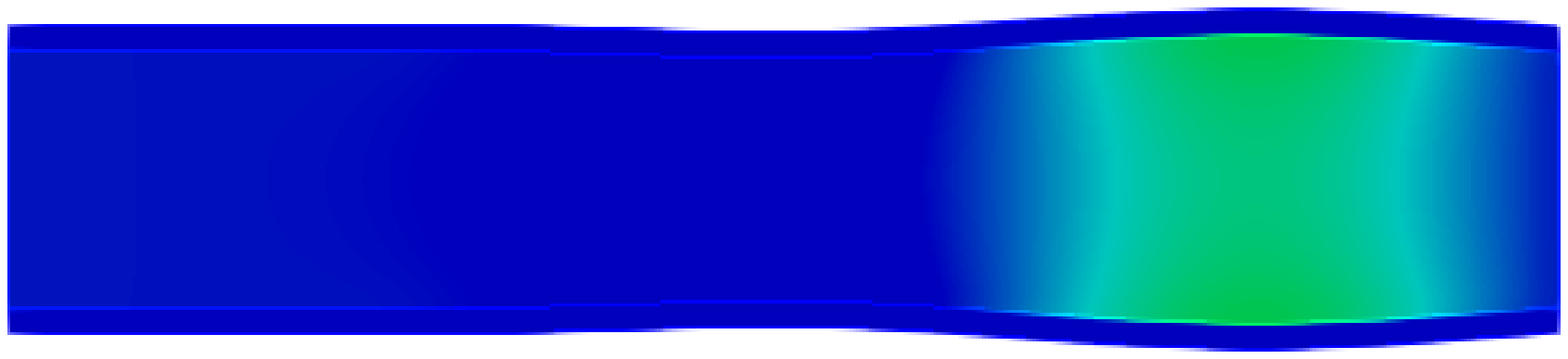}
	\includegraphics[trim=0 -15 0 0,scale=0.28]{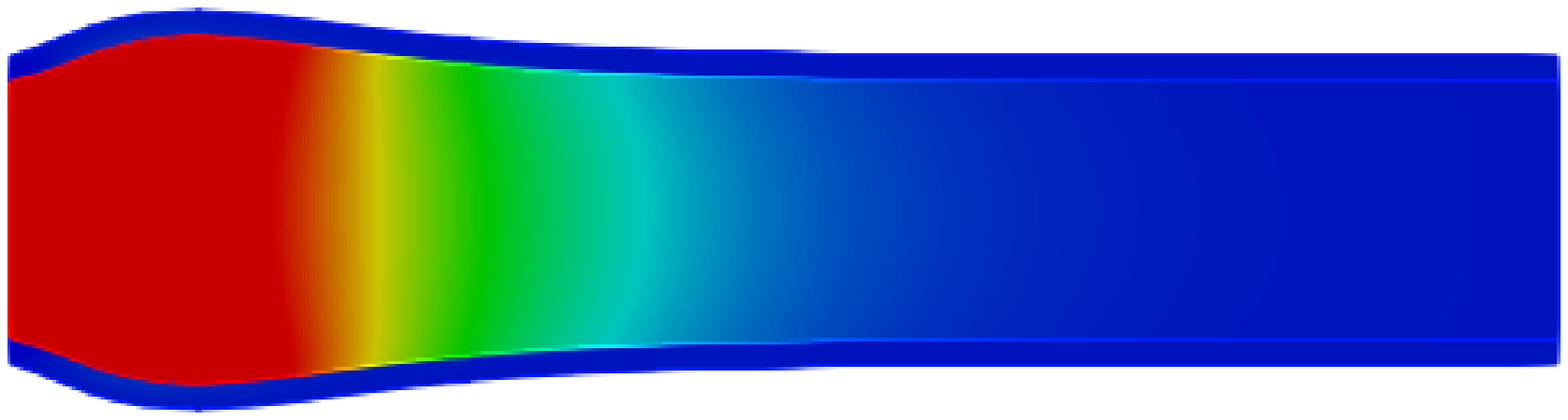}
	\includegraphics[trim=0 -15 0 0,scale=0.28]{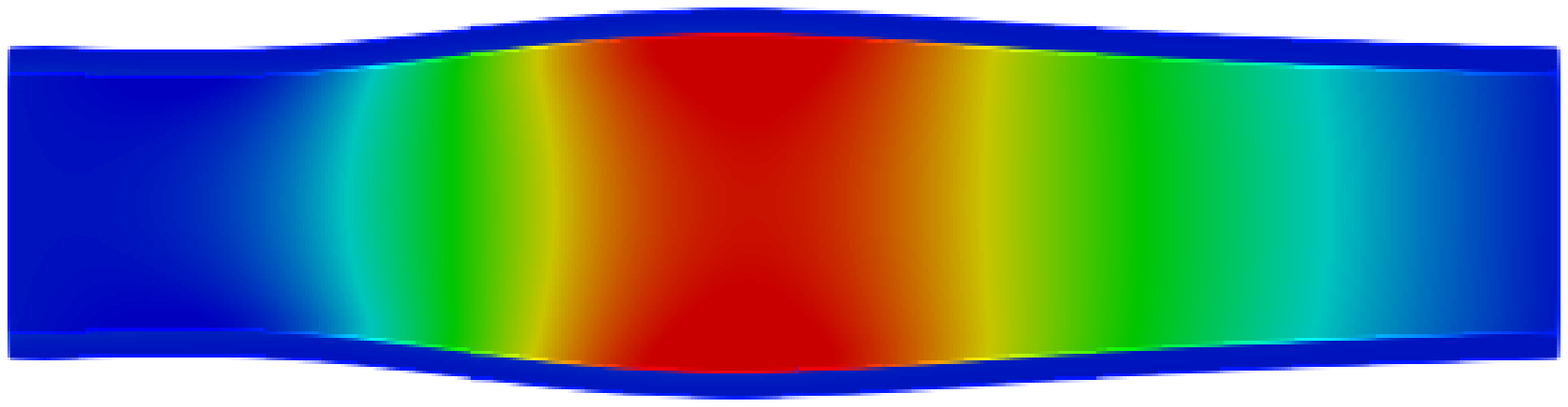}
	\includegraphics[trim=0 -15 0 0,scale=0.28]{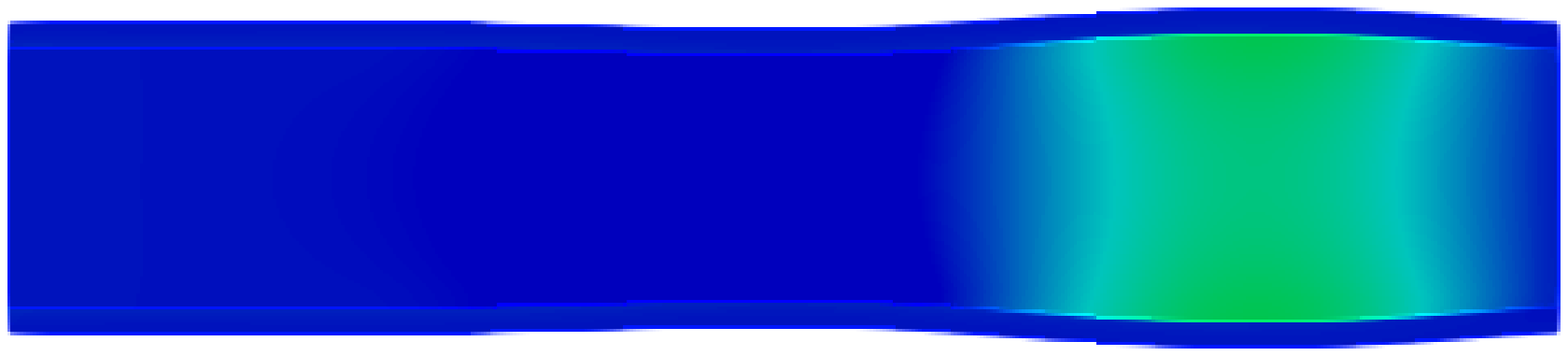}
	\includegraphics[trim=0 10 0 0,scale=0.28]{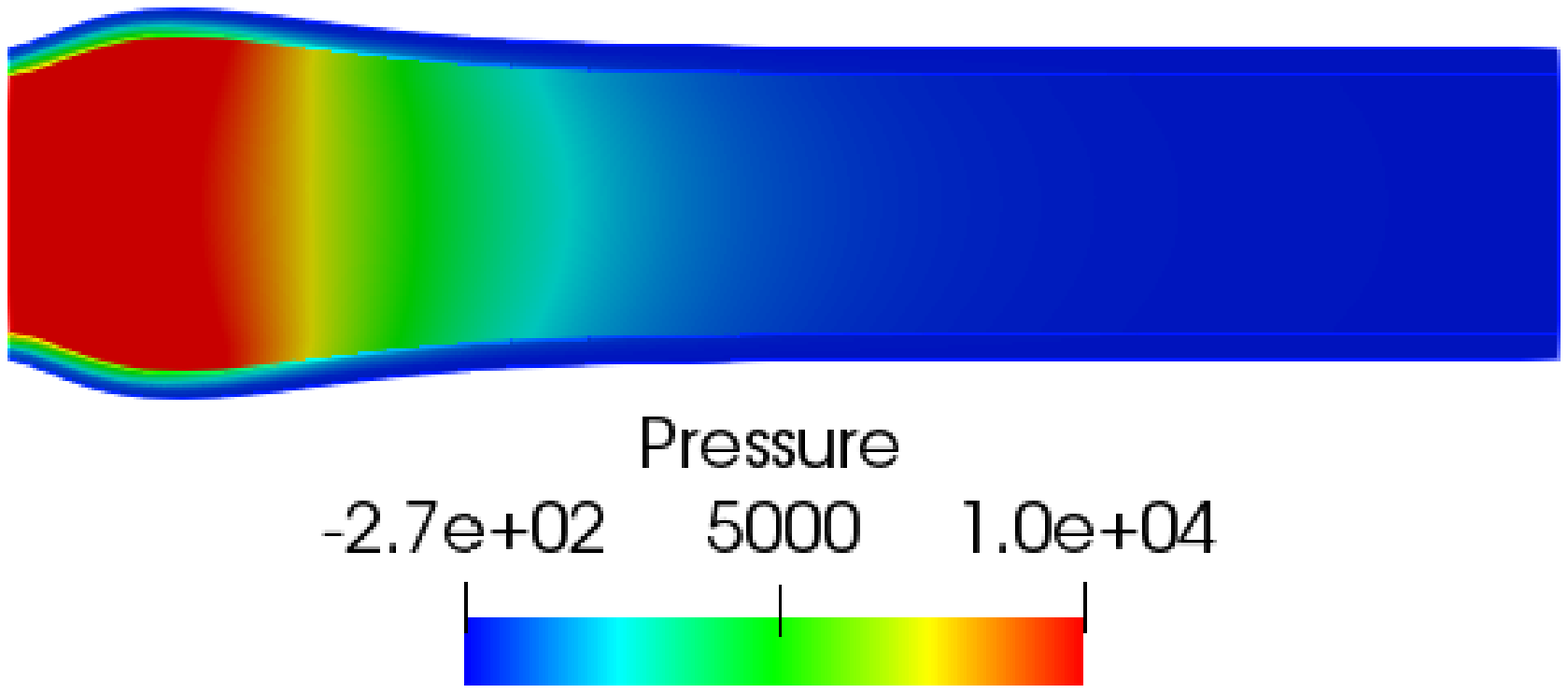}
	\includegraphics[trim=0 10 0 0,scale=0.28]{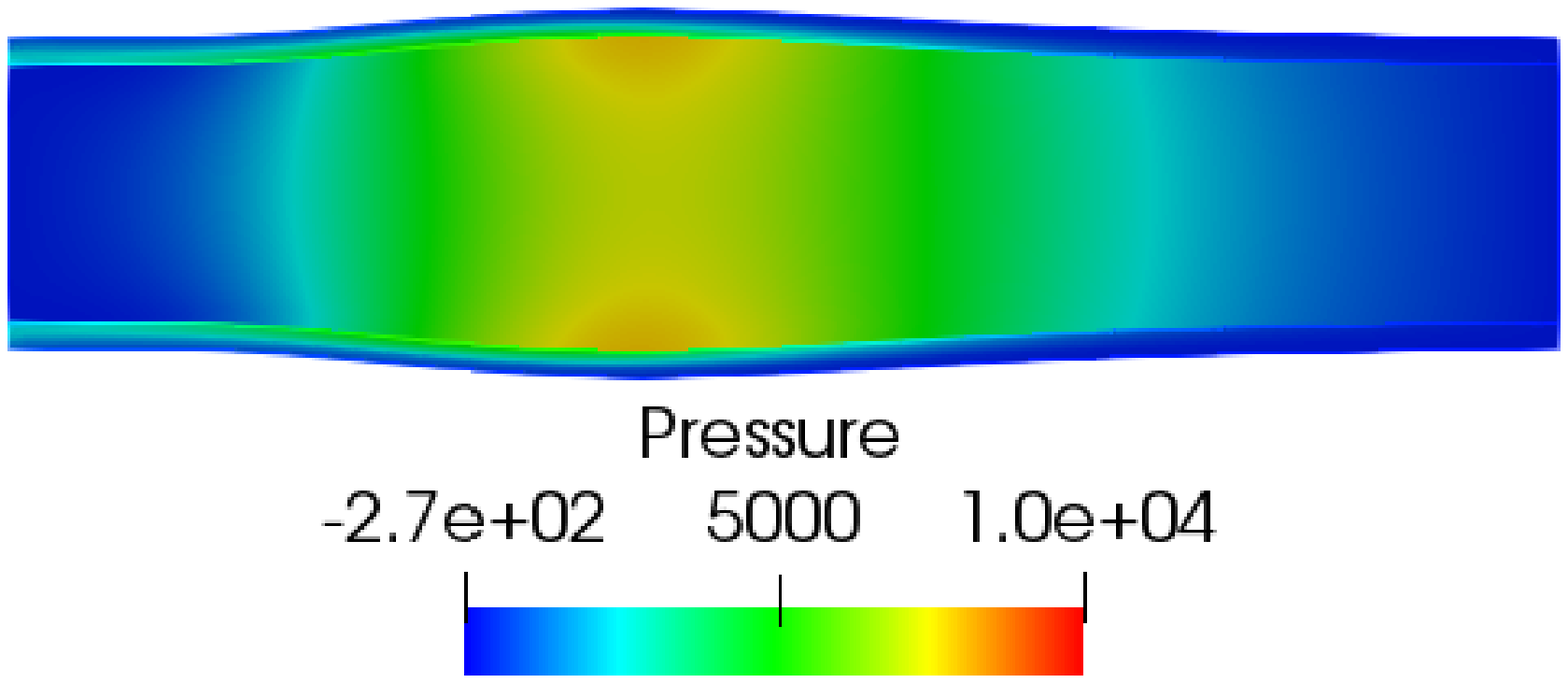}
	\includegraphics[trim=0 10 0 0,scale=0.28]{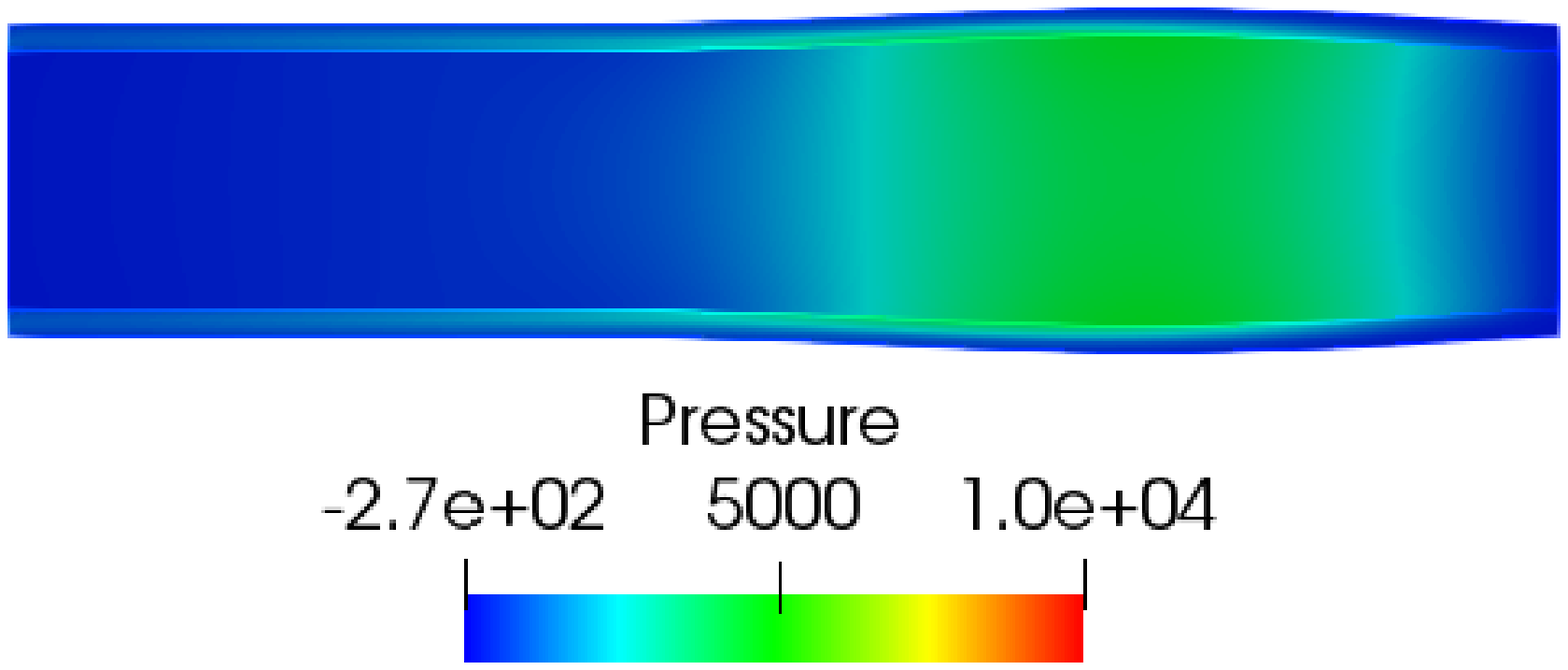}
	\caption{Pressure waves at time t=1.8 ms, t=3.6 ms, t=5.4 ms for case 1, case 2 and case 3.  }
	\label{pressurecom1}
\end{figure}

Some visualization of the solutions are reported in this section. Velocity and pressure waves along the channel are presented in Fig.\ref{velocitycase1} and Fig.\ref{pressurecom1}, together with the corresponding deformation at time $t=1.8, 3.6, 5.4$ ms. For visualization purpose, deformations of 2D plots in this and the next section are magnified 40 times. Fig.\ref{velocitycase1} shows that the variable inflow velocity together with the fluid-structure iteration generates a wave from left to right. More differences are noticeable at time $t=3.6$ and $t=5.4$ ms. In particular, for case 3, the flow has smaller velocity magnitude and takes more time to reach the outflow region. Correspondingly in Fig.\ref{pressurecom1}, we can observe a clear increase of pressure along the interface for case 3. We will look into more quantities closely on the top interfaces. 

\begin{figure}[ht!]
	\centering
	\includegraphics[trim=25 60 25 0,scale=0.149]{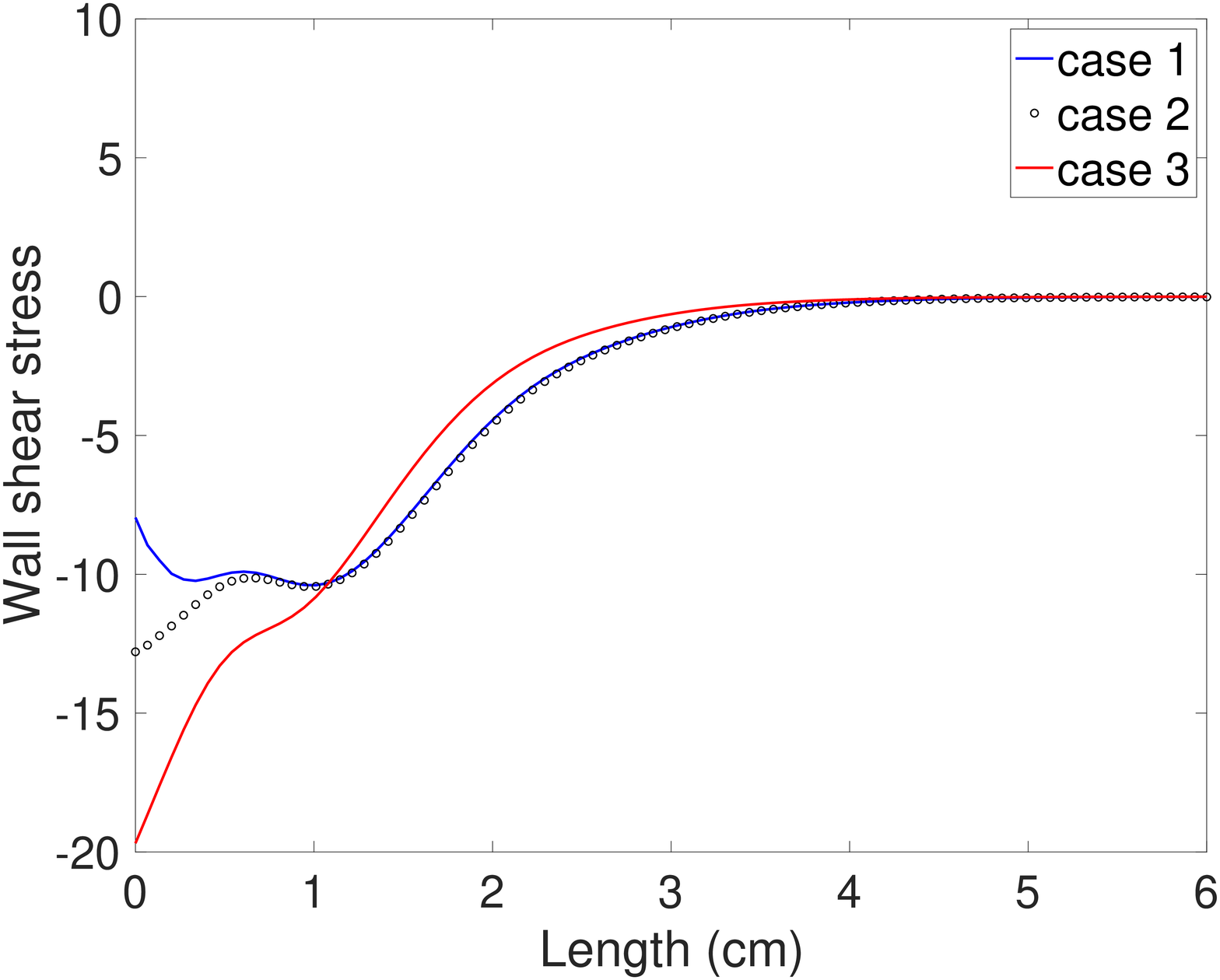}
	\includegraphics[trim=25 60 25 0,scale=0.149]{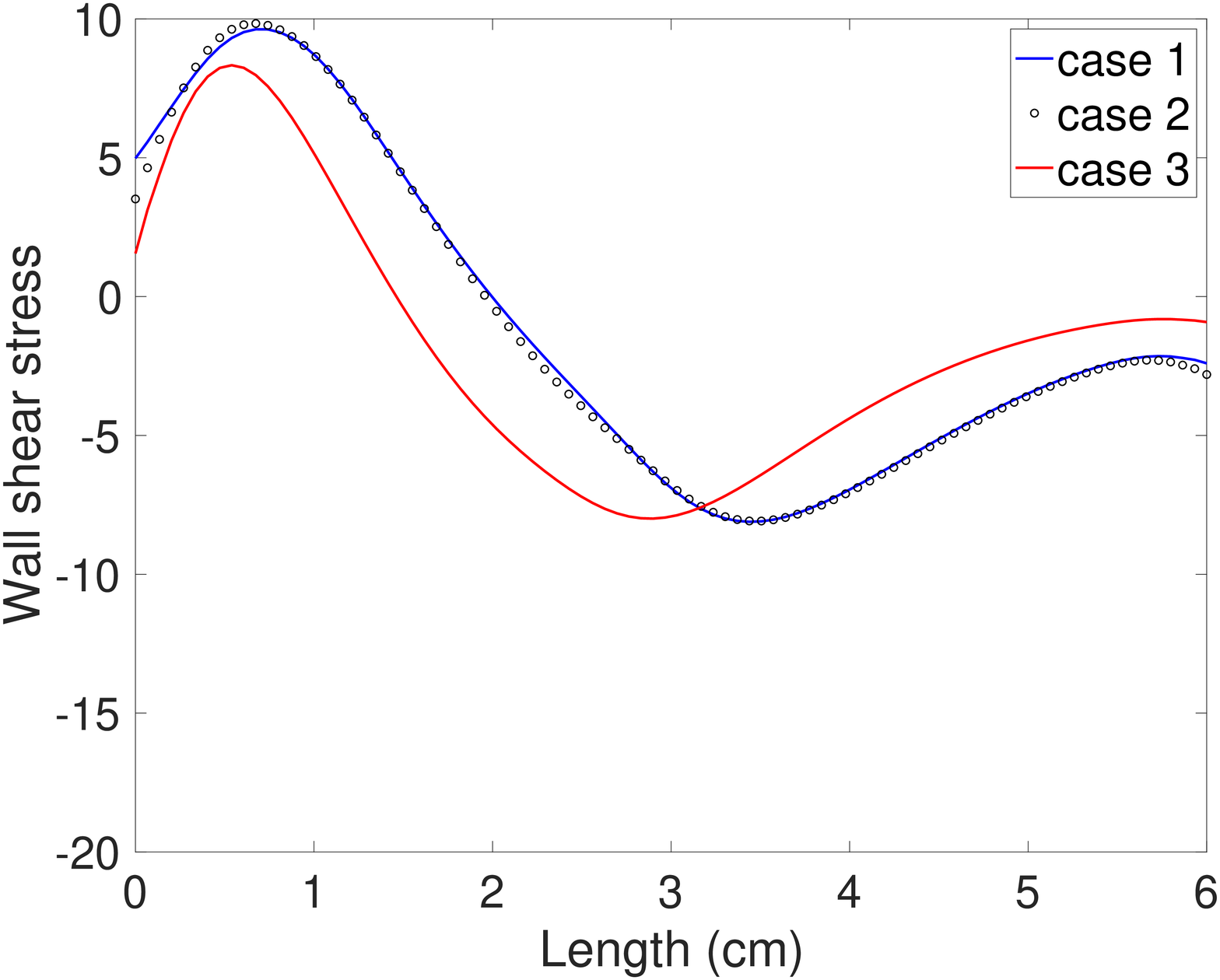}
	\includegraphics[trim=0 60 50 0,scale=0.149]{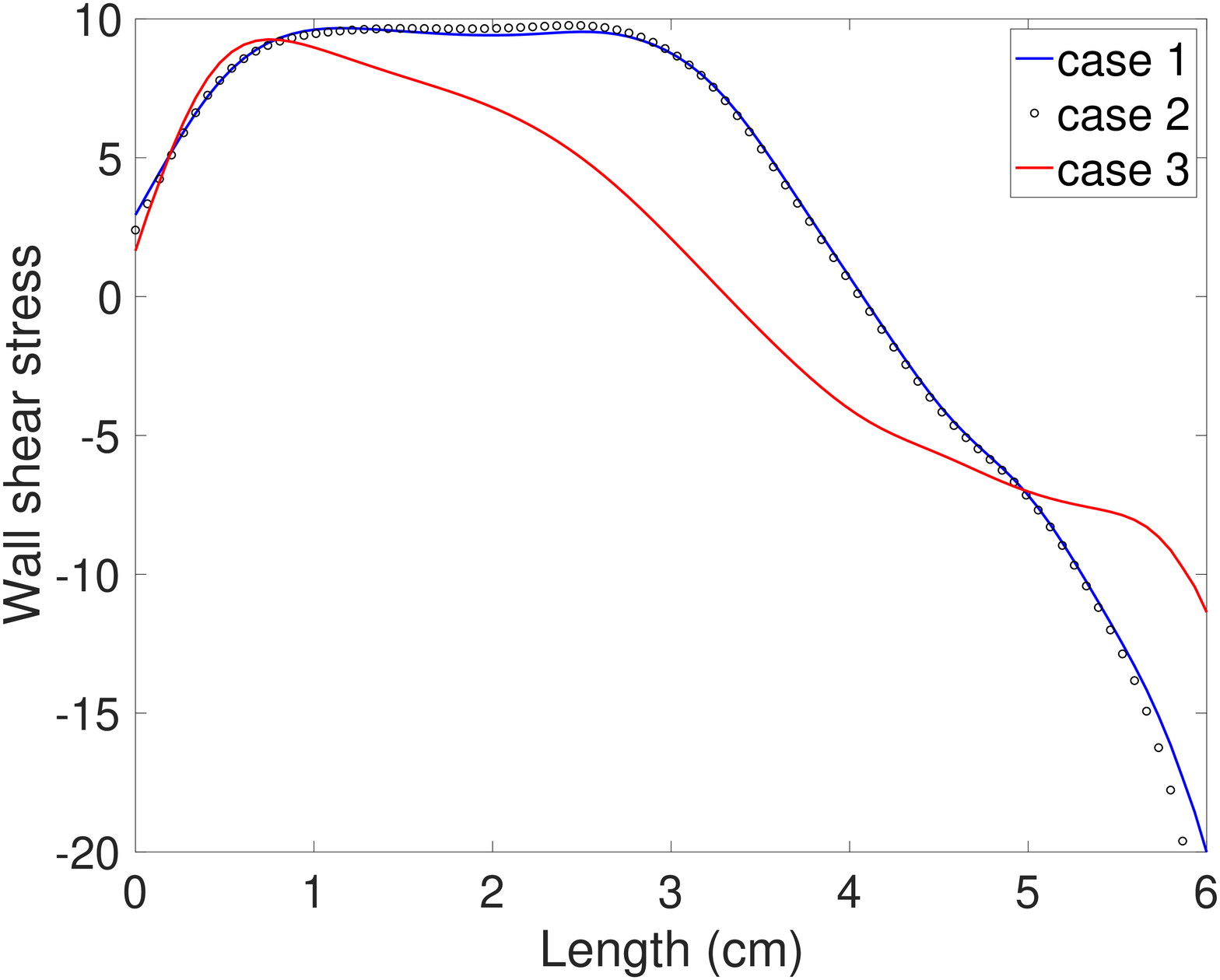}
	\caption{Wall shear stress $\bsi_f\bn\cdot \bt$ along the top arterial wall at time t=1.8 ms, t=3.6 ms, t=5.4 ms  for case 1, case 2, and case 3.}
	\label{shearcompare1}
\end{figure}

Wall shear stress (WSS) is an important index for the risk of plaque rupture in blood flow dynamic model \cite{guerciotti2018computational, chen2006non} and it is interesting to evaluate the possible effects of different models and different permeability values on this index. We will continue to pick up time $t=1.8, 3.6, 5.4$ ms for comparison, as shown in Fig.\ref{shearcompare1}. The wall shear stress is defined as follow \cite{john2017influence}:
\begin{equation}\label{shearStress}
WSS=\bsi_f\bn\cdot \bt.
\end{equation}

From the definition in \eqref{shearStress}, WSS is only a local physical quantity at a specific time. Therefore, to consider WSS over a period of time, we are going to introduce three more quantities: time averaged wall shear stress (TAWSS), oscillatory shear index (OSI) and relative residence time (RRT). For TAWSS, OSI and RRT, functions of space on the lumen boundary \cite{guerciotti2018computational, john2017influence}, the definitions are as follows:
\begin{equation}\label{TAWSS}
TAWSS(\bold{x})=\frac{1}{T}\int_{0}^{T}|\bsi_f \bn\cdot \bt(t,\bold{x})|dt;
\end{equation}
\begin{equation}\label{OSI}
OSI(\bold{x})=\frac{1}{2}\left(1-\frac{|\int_{0}^{T}\bsi_f \bn\cdot \bt (t,\bold{x})dt|}{\int_{0}^{T}|\bsi_f \bn\cdot \bt (t,\bold{x})|dt}\right);
\end{equation}
\begin{equation}\label{RRT}
RRT(\bold{x})=\frac{1}{(1-2OSI(\bold{x}))TAWSS(\bold{x})}=\frac{T}{|\int_{0}^{T}\bsi_f \bn\cdot \bt (t,\bold{x})dt|}.
\end{equation} 
We will focus on RRT, since high RRT distribution is emerging as an appropriate tool for identifying the possible regions of atheromatous concentrations and potential arterial lesions \cite{guerciotti2018computational, soulis2011relative}.

\begin{figure}[ht!]
	\centering
	\includegraphics[trim=25 60 25 0,scale=0.149]{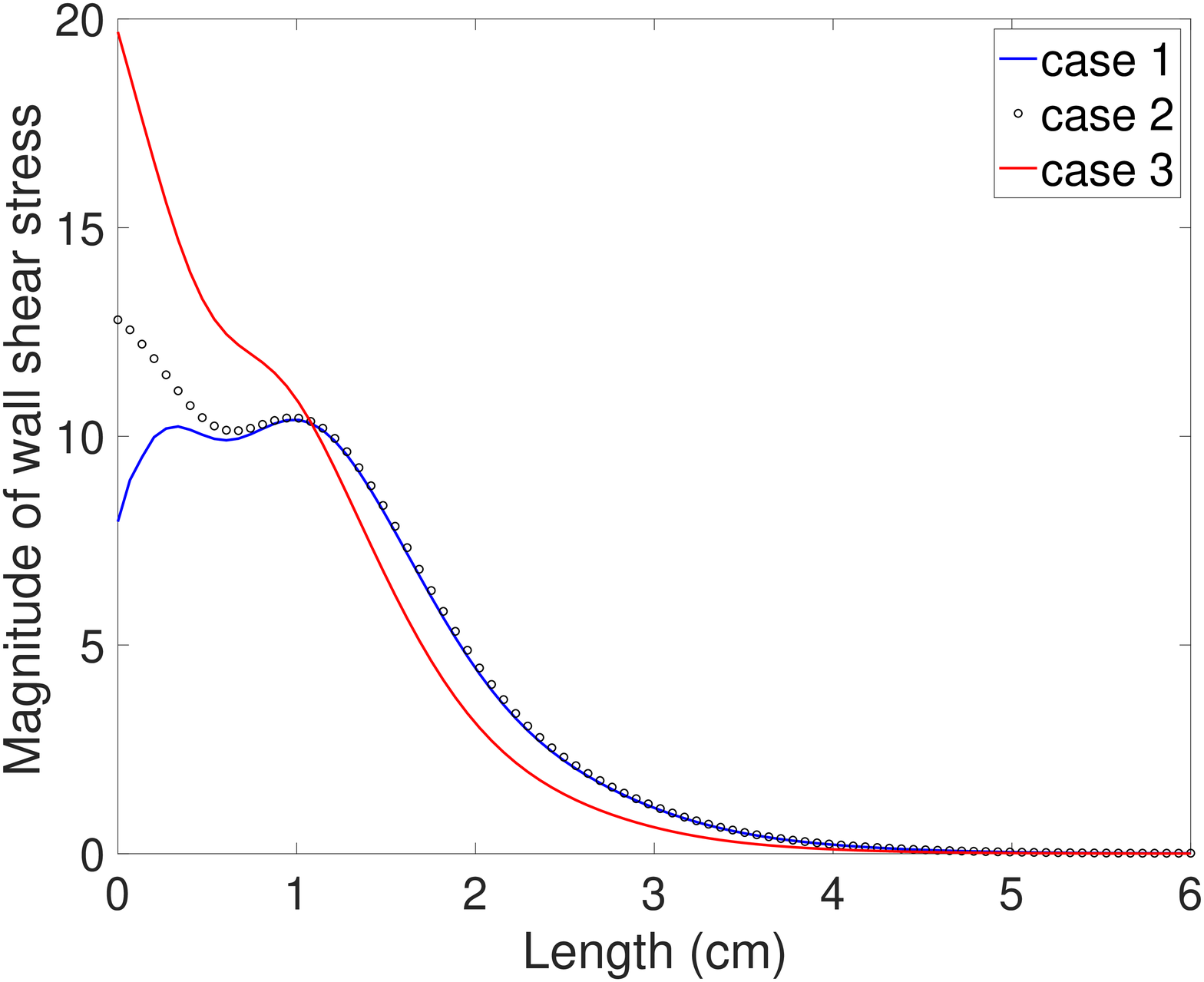}
	\includegraphics[trim=25 60 25 0,scale=0.149]{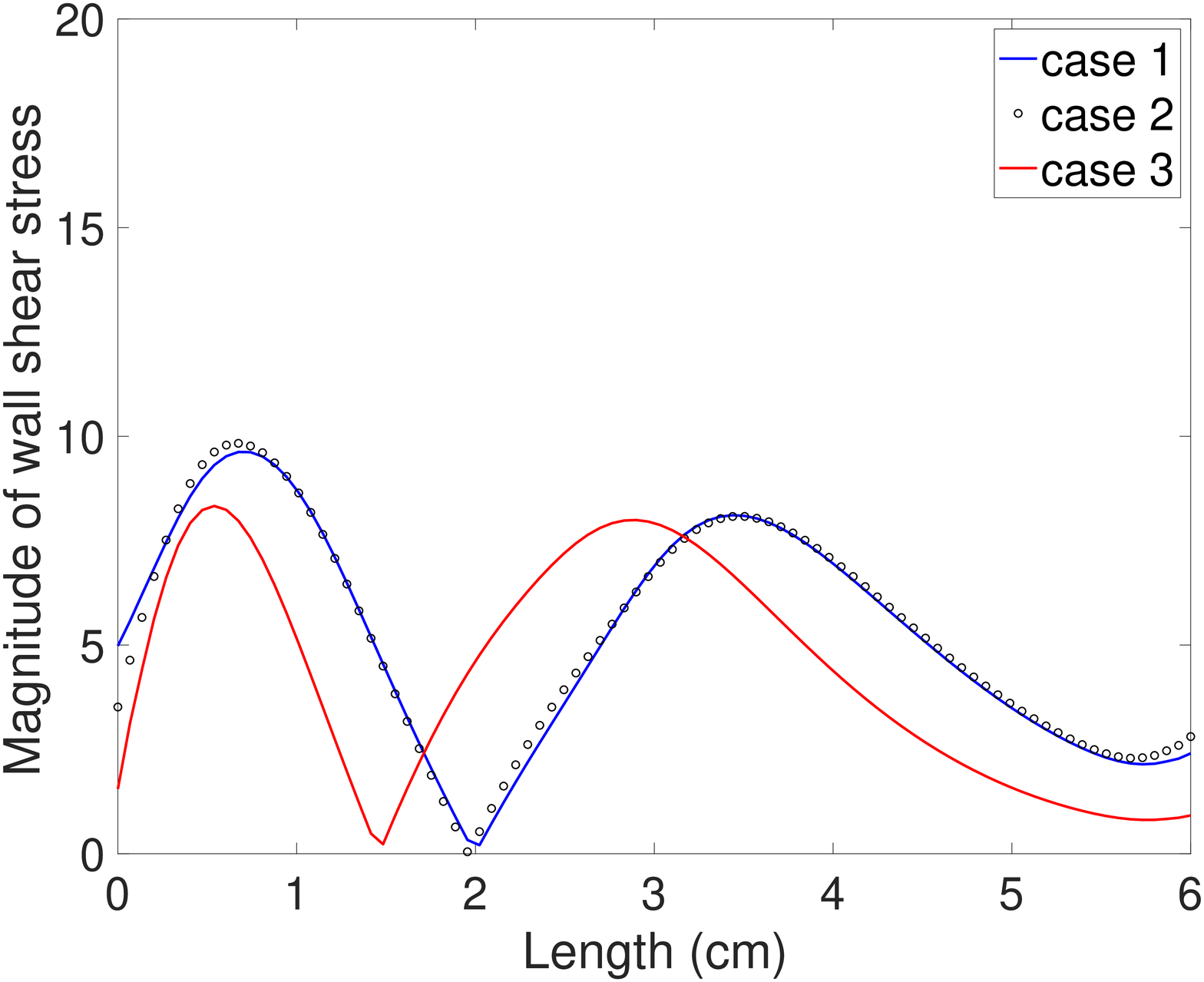}
	\includegraphics[trim=0 60 50 0,scale=0.149]{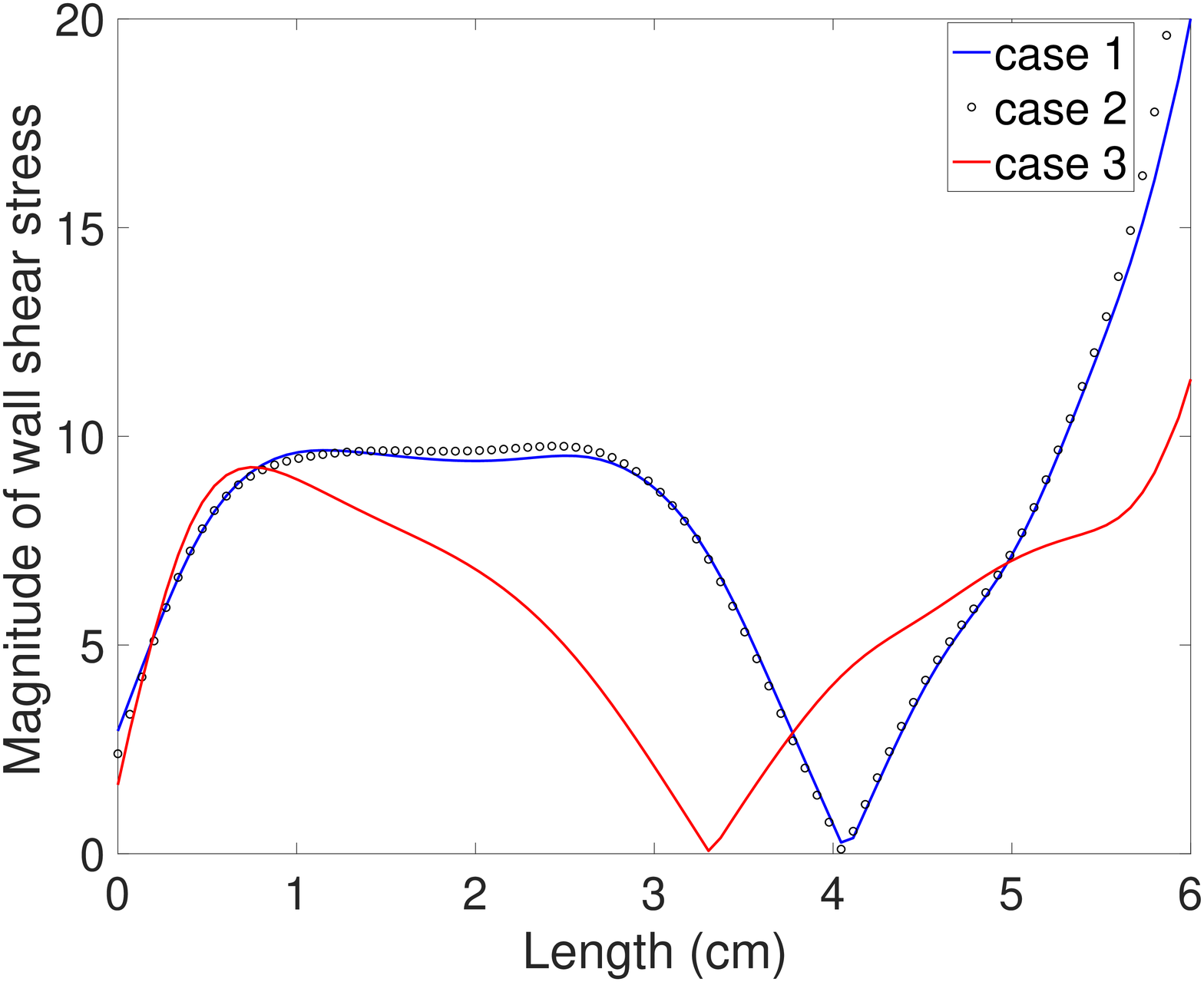}
	\caption{The magnitude of wall shear stress $|\bsi_f\bn\cdot \bt|$ along the top arterial wall at time t=1.8 ms, t=3.6 ms, t=5.4 ms  for case 1, case 2, and case 3.}
	\label{shearcompare11}
\end{figure}

\begin{figure}[ht!]
	\centering
	\includegraphics[trim=0 60 0 0,scale=0.149]{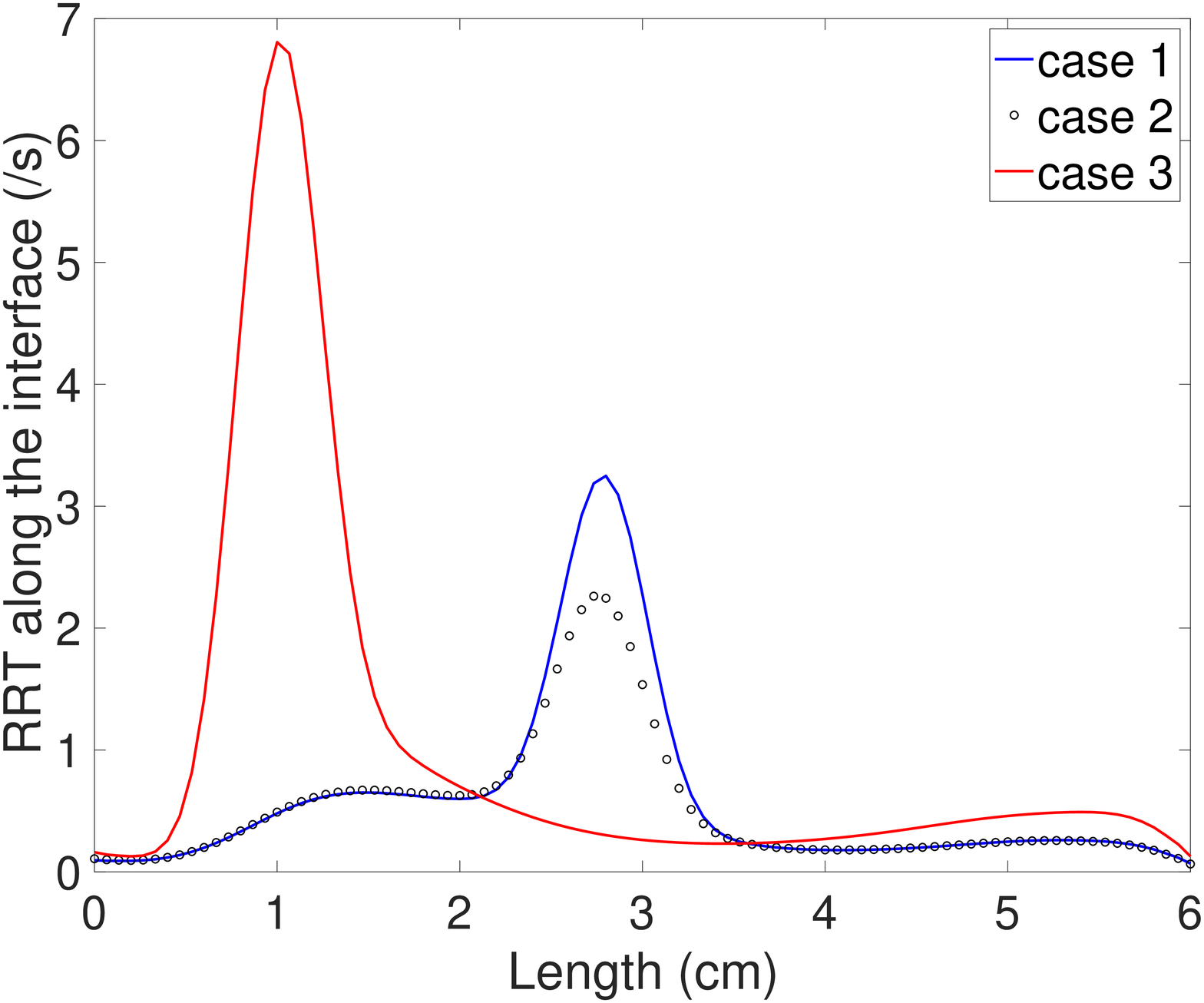}
	\caption{RRT along the top arterial wall for case 1, case 2, and case 3.}
	\label{RRT1}
\end{figure}

We can see from Fig.\ref{RRT1} that RRT is greatly influenced by permeability. Even though the behaviors of case 1 and case 2 are very similar at different time spots, we observe differences in values between case 1 and case 2. 

\begin{figure}[ht!]
	\includegraphics[trim=0 60 0 0,scale=0.149]{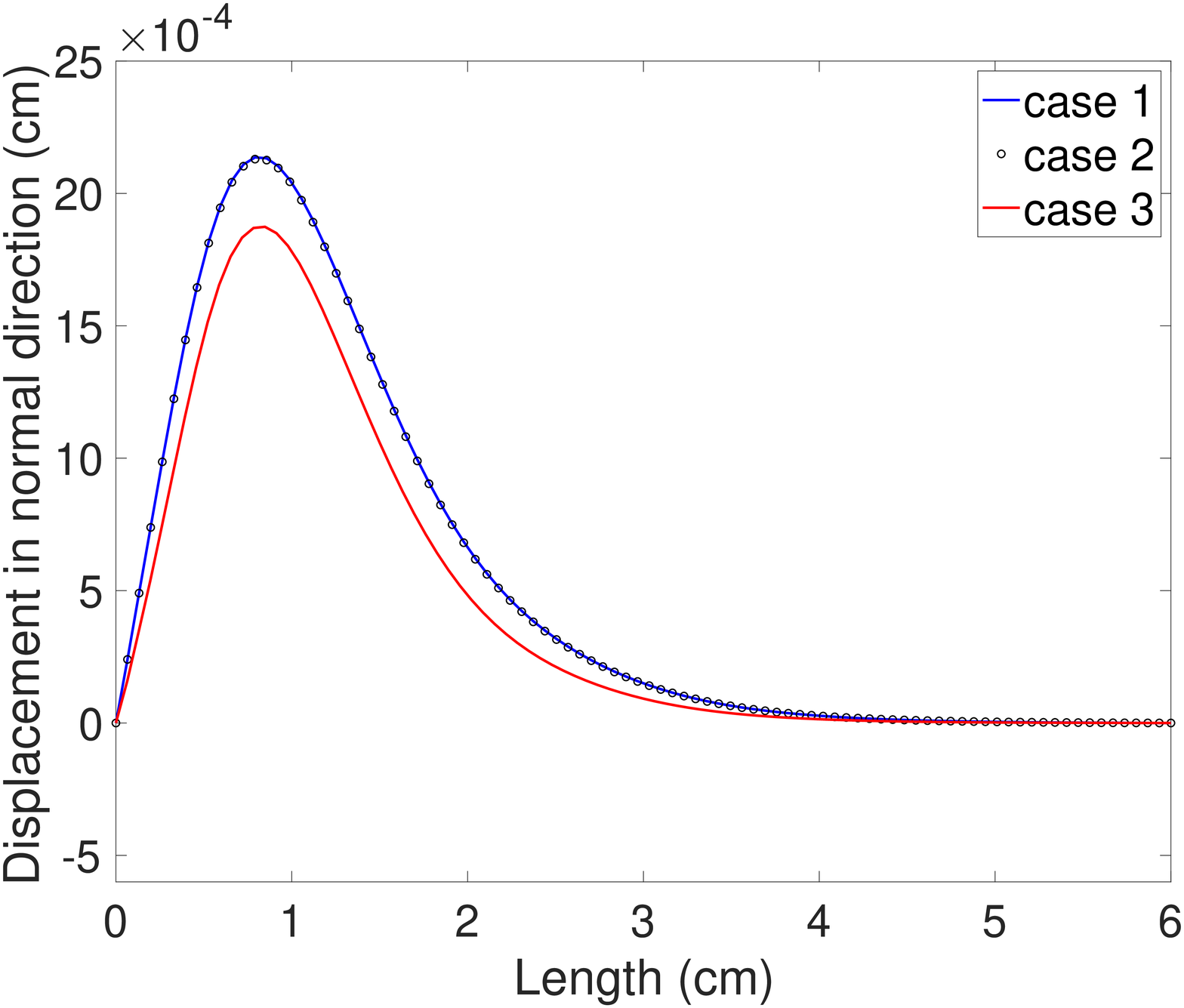}
	\includegraphics[trim=0 60 0 0,scale=0.149]{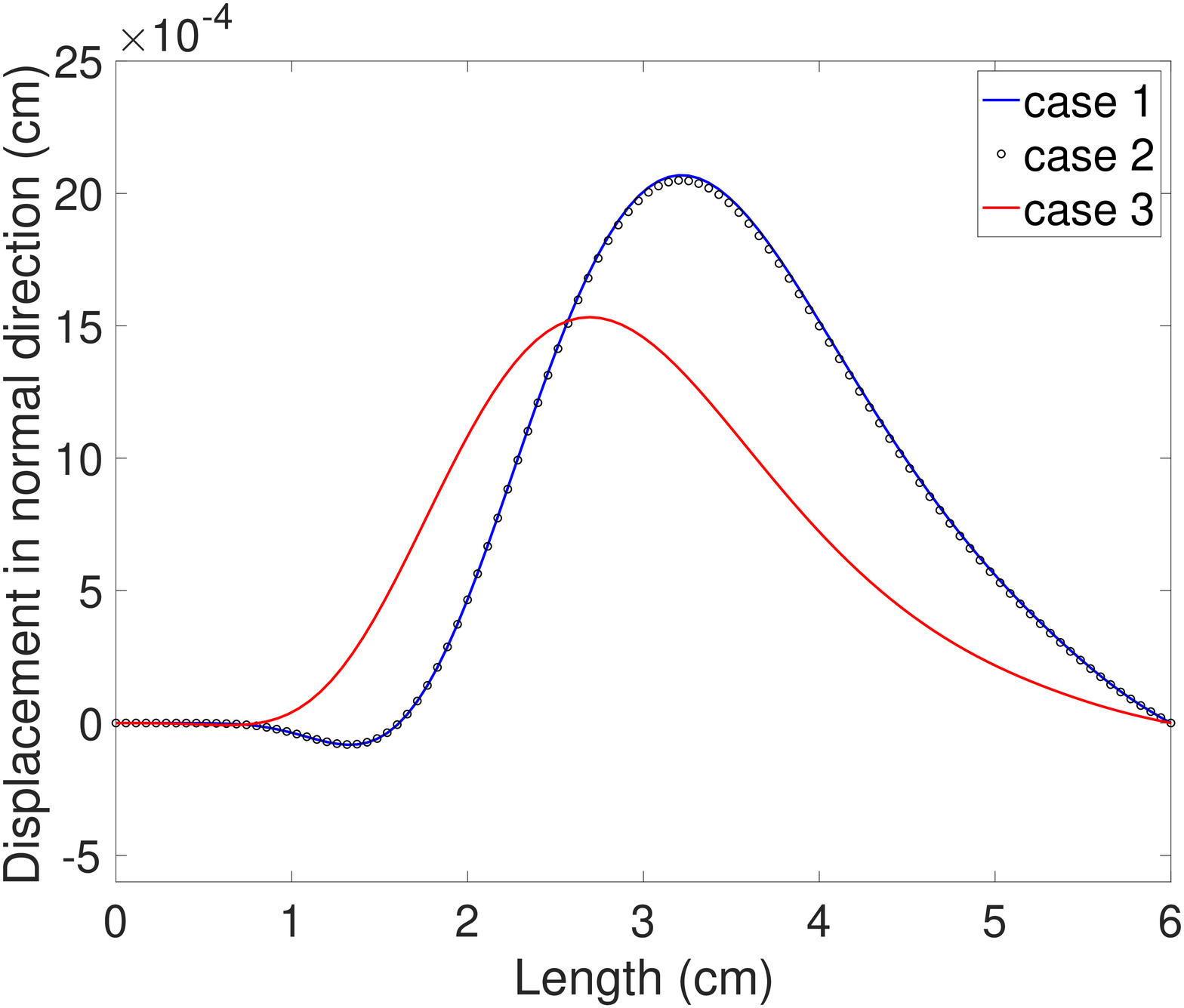}
	\includegraphics[trim=0 60 0 0,scale=0.149]{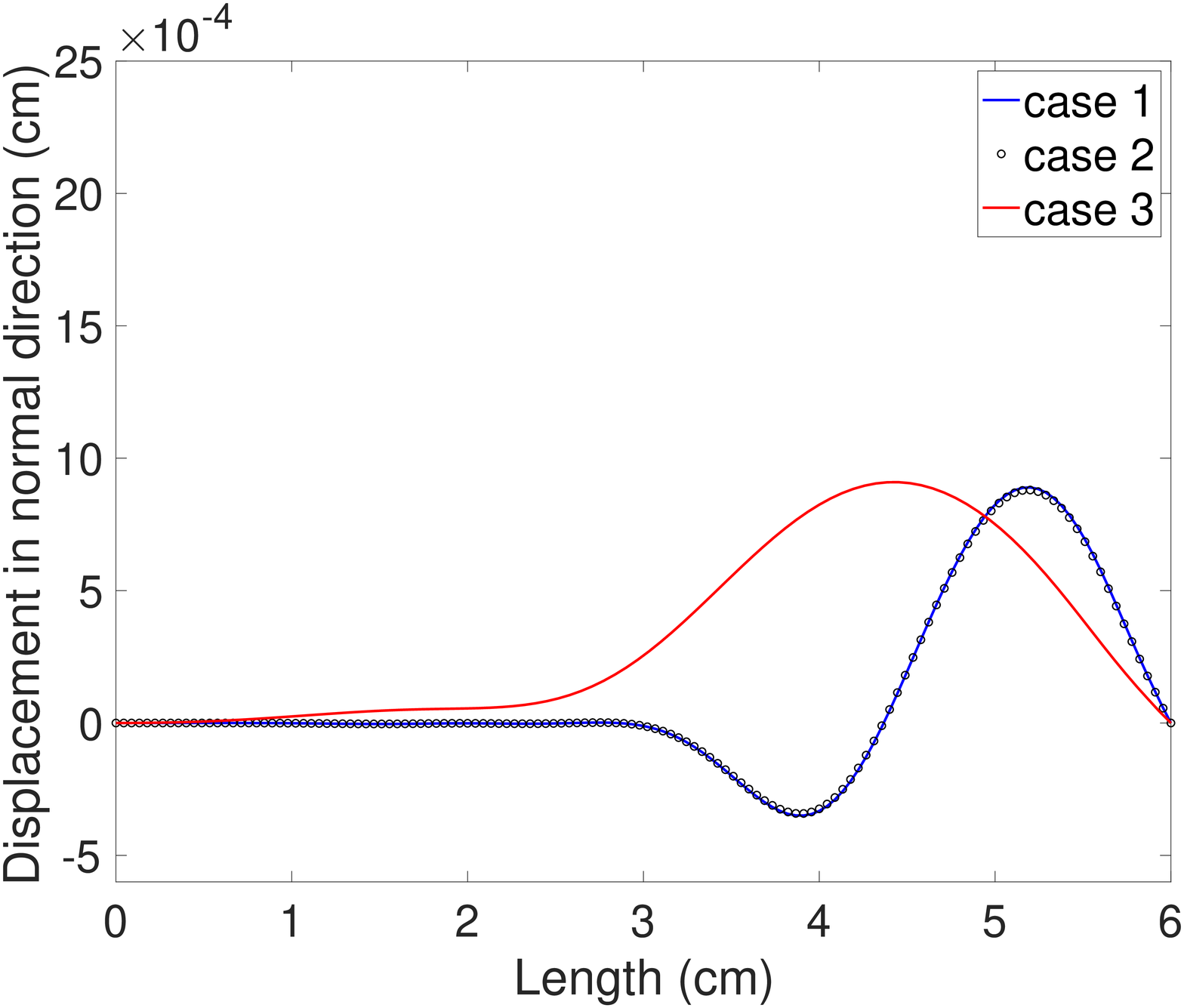}
	\caption{Displacement in the normal direction $\bbeta_{\star}\cdot \bn$ along the top arterial wall at time t=1.8 ms, t=3.6 ms, t=5.4 ms for case 1, case 2, and case 3. }
	\label{discompare1}
\end{figure}
In Fig. \ref{discompare1}, we present the displacement in the normal direction $\bbeta_\star \cdot \bn$ along the top arterial wall. We observe that the peaks of it coincide with the peaks of velocity magnitude and pressure. Notice that the peaks of $\bbeta_{p}\cdot \bn$ in case 3 is overall smaller than case 1 and case 2. For larger permeability $\bK$ in case 3, the peaks of normal displacement would be smaller. 

\begin{figure}[ht!]
	\includegraphics[trim=0 60 0 0,scale=0.149]{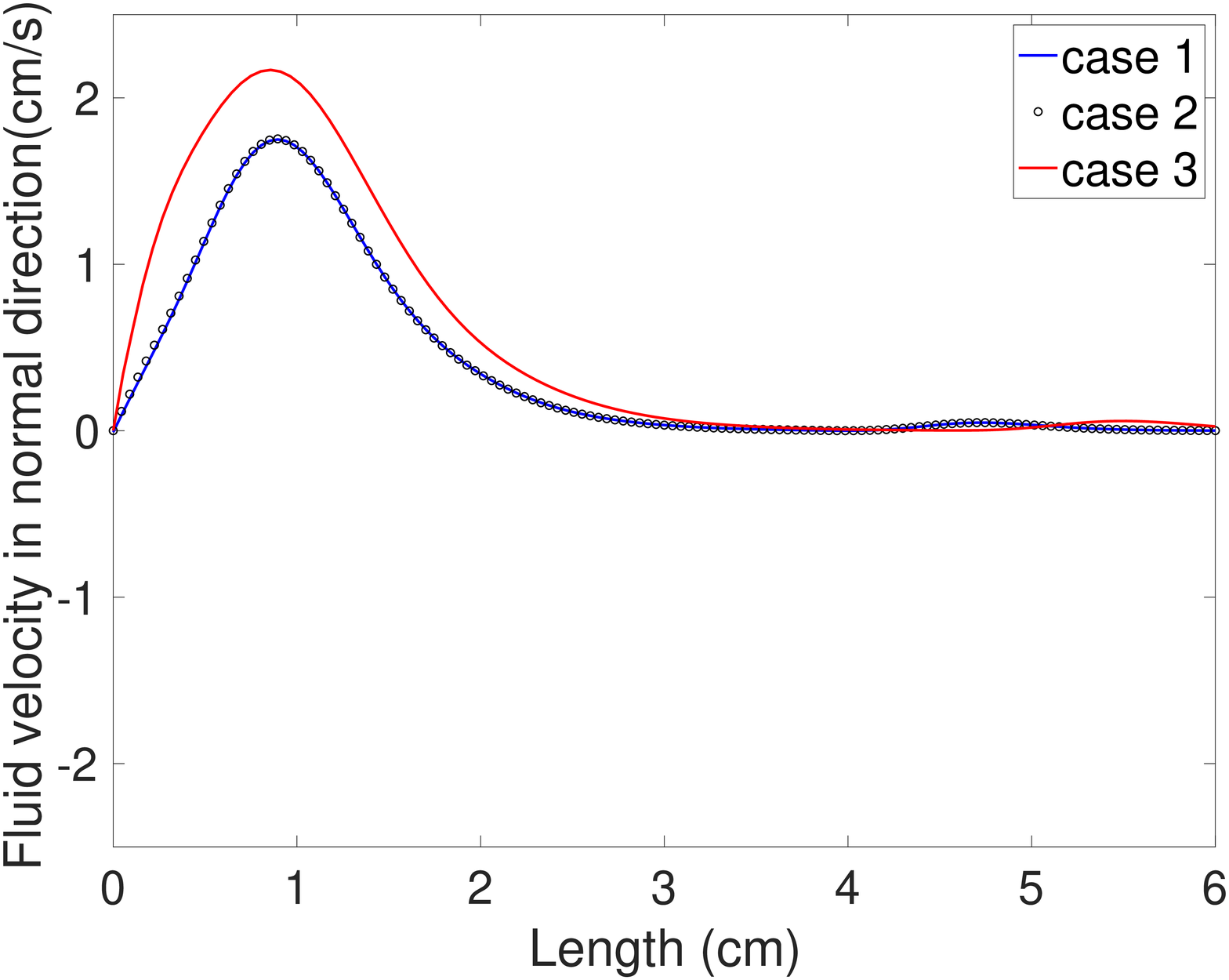}
	\includegraphics[trim=0 60 0 0,scale=0.149]{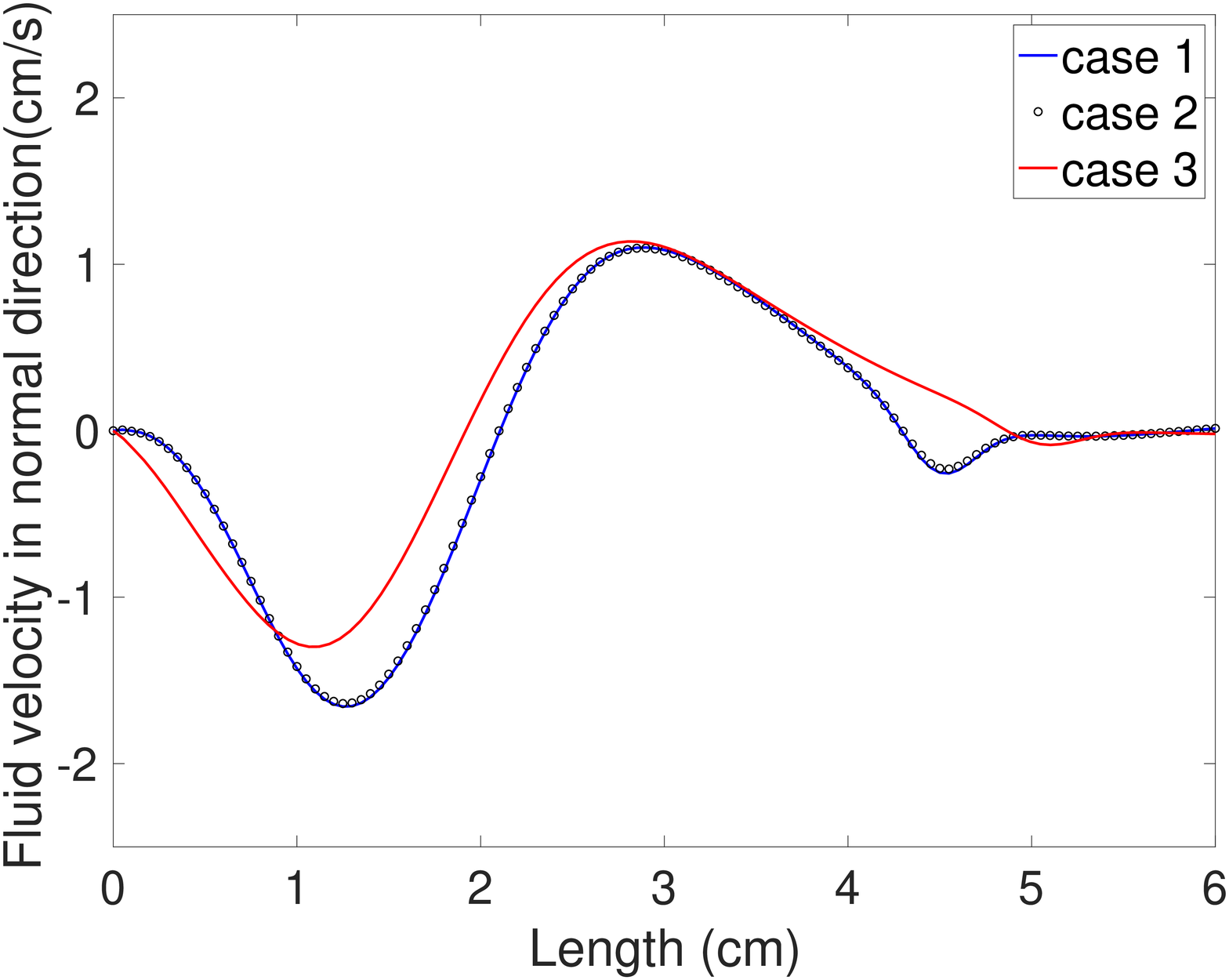}
	\includegraphics[trim=0 60 0 0,scale=0.149]{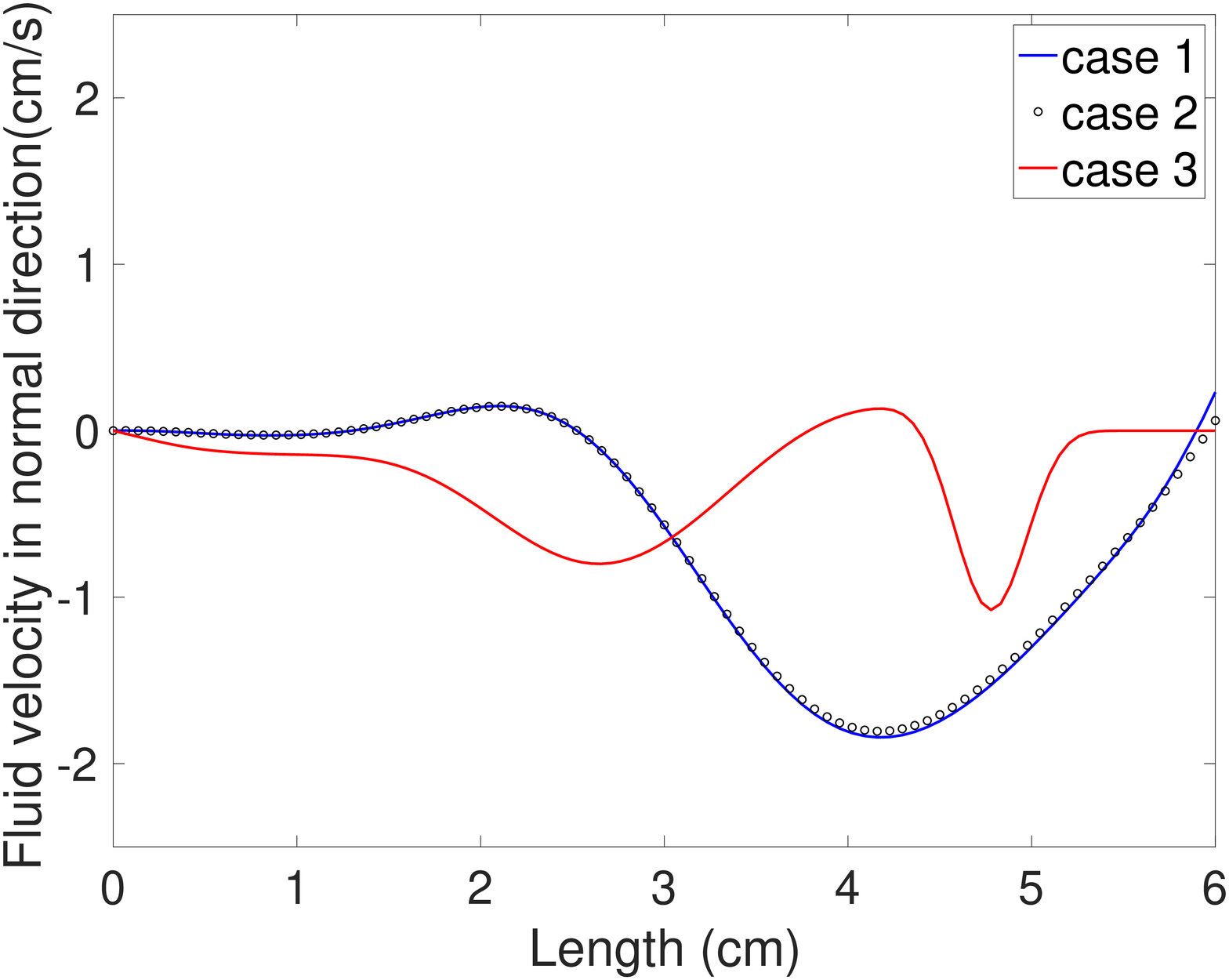}
	\caption{Velocity in the normal direction along the top arterial wall $\bu_f\cdot \bn$ at time t=1.8 ms, t=3.6 ms, t=5.4 ms for case 1, case 2, and case 3.}
	\label{utncompare1}
\end{figure}
\begin{figure}[ht!]
	\includegraphics[trim=0 60 0 0,scale=0.149]{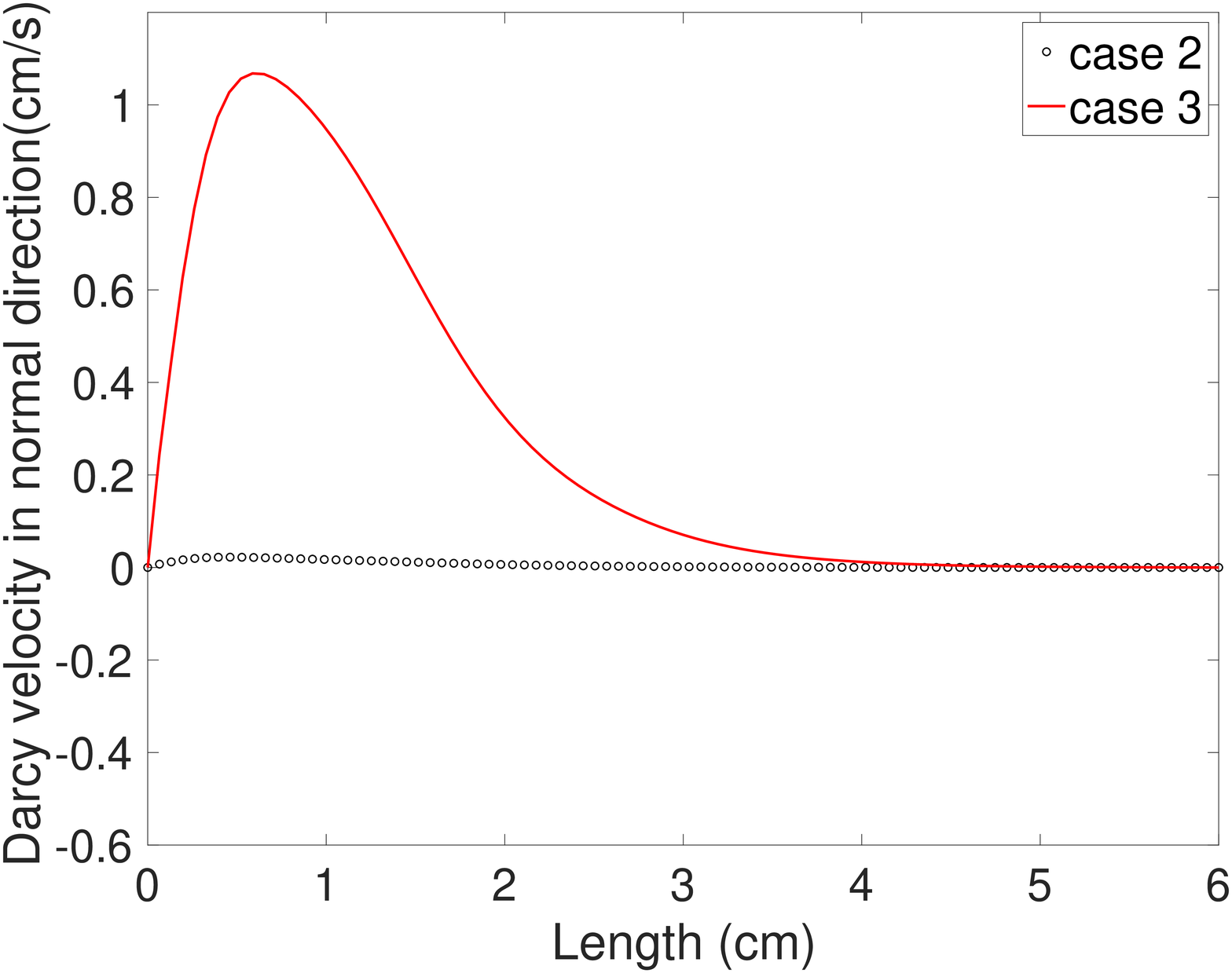}
	\includegraphics[trim=0 60 0 0,scale=0.149]{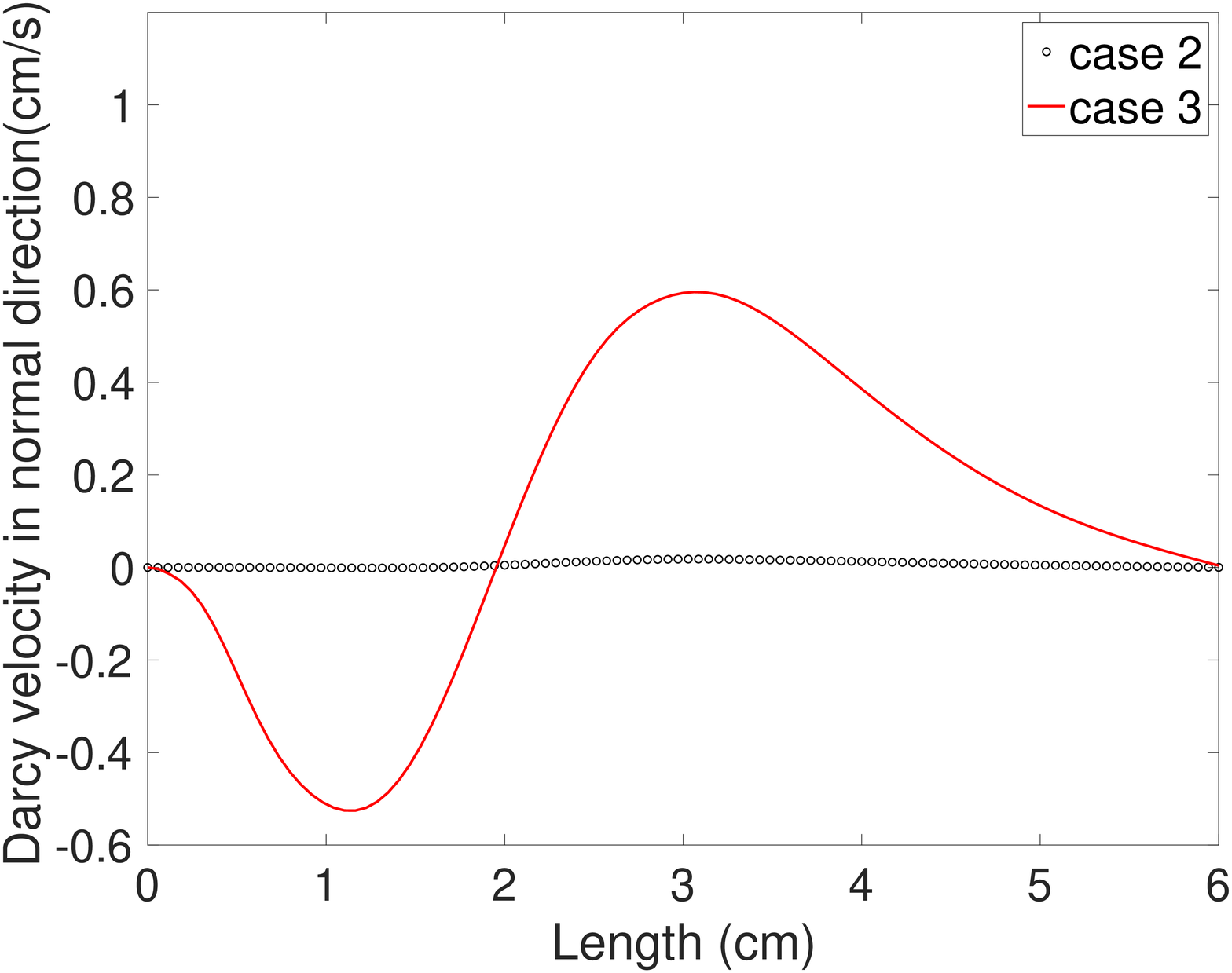}
	\includegraphics[trim=0 60 0 0,scale=0.149]{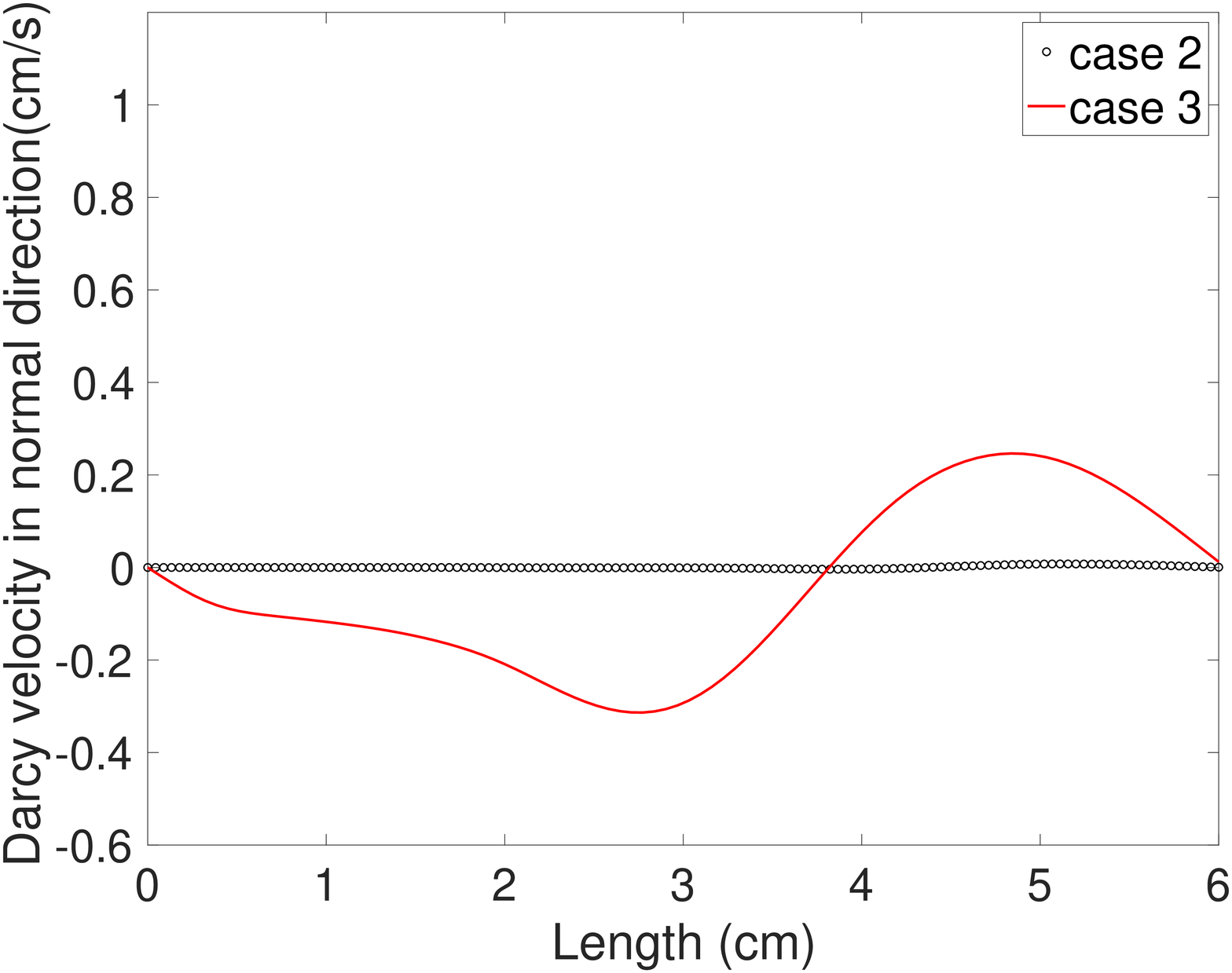}
	\caption{Darcy velocity in the normal direction $\bu_p\cdot \bn$ along the top arterial wall at time t=1.8 ms, t=3.6 ms, t=5.4 ms for case 2 and case 3. Note that the normal velocity of case 2 is not zero. It's small comparing with case 3 since a relatively small permeability. }
	\label{upncompare1}
\end{figure}
We present fluid velocity in normal $\bu_{f}\cdot \bn$ along the interface in Fig.\ref{utncompare1} and the Darcy velocity in normal direction $\bu_{p}\cdot \bn$ along the interface, which is known as filtration velocity, for case 2 and case 3 in Fig.\ref{upncompare1}. Note that the absence of filtration velocity in case 1 is due to NSE/E model. For both cases, the peaks of filtration velocity coincide with the ones of structure displacement in the normal direction $\bbeta_p \cdot \bn$ in Fig. \ref{discompare1}. We also notice that the Darcy velocity $\bu_{p}\cdot \bn$ in both cases decrease when the fluid penetrates into the arterial wall gradually, which is a consequence of the third equation in (\ref{eq:Biot-model}). This equation describes that $\nabla \cdot \bu_{p}$ is not locally preserved and it depends on the rate of change in pressure and displacement. At time $t=1.8, 3.6, 5.4$ ms, the value of $\bu_{p}\cdot \bn$ in case 3 is much bigger than the one of case 2. As a conclusion, larger permeability would result in larger filtration velocity.

\subsection{Non-Newtonian models }\label{Nonnewtonian}
In this section, we will compare the effects of non-Newtonian property as well as the effects of permeability on NSE/P models. We use the same computational domain as in the Newtonian models shown in Fig.\ref{mesh11} part (a). In non-Newtonian fluids, viscosity can change with force, velocity or temperature to either more liquid or more solid. Most commonly, the viscosity of non-Newtonian fluids depends on shear rate or shear rate history, showing a shear-thinning property. Even though in computational fluid dynamic, it is widely accepted to assume the blood flow is Newtonian, the non-Newtonian behavior of blood needs to be taken into consideration, especially in small vessels, like the capillaries \cite{guerciotti2018computational, mandal2005unsteady}. Moreover, for the middle sized vessels, for example, the carotid or coronary vessels, we are not completely clear whether it is validated to assume the Newtonian property of blood. For these reasons, we want to investigate how the non-Newtonian behavior affects the blood flow characteristics. There are different non-Newtonian models, such as Power Law \cite{johnston2006non}, Casson \cite{perktold1991pulsatile} and Carreau-Yasuda model \cite{guerciotti2018computational}. In this paper, we will perform simulation using the Carreau-Yasuda model to describe the non-Newtonian blood rheology. For the non-Newtonian fluid, instead of constant viscosity $\mu_f$, we will using the following nonlinear viscosity in the fluid regions for both NSE/E and NSE/P models \cite{guerciotti2018computational, boyd2007comparison, cho1991effects}, 
\begin{equation}\label{Non-New Viscosity}
\nu(x,y,t)=\nu_{\text{inf}}+(\nu_0-\nu_{\text{inf}})(1+(\delta\dot{\gamma}(x,y,t)^a)^{\frac{n-1}{a}}),
\end{equation}
where $\dot{\gamma}(x,y,t)=\sqrt{\frac{1}{2}\be(\bu_f):\be(\bu_f)}$ is the shear rate. The values of parameters defined in (\ref{Non-New Viscosity}) are chosen as $\delta=1.902 s$, $n=0.22$, $a=1.25$, $\nu_0=0.56$ Poi, $\nu_{\text{inf}}=0.035$ Poi. For NSE/P model, we use the above shear rate in the fluid region while we will use $\dot{\gamma}(x,y,t)=\sqrt{\bu_p\cdot \bu_p}$ as  shear rate for structure region.

Let's consider the following cases:
\begin{itemize}
\item case 4: non-Newtonian NSE/E model;
\item case 5: non-Newtonian NSE/P model, with $\bK=diag(5,5)\times10^{-11}$;
\item case 6: non-Newtonian NSE/P model, with $\bK=diag(1,1)\times10^{-9}$.
\end{itemize}
We don't pick $\bK=diag(1,1)\times 10^{-7}$ as in case 3 since the flow rate of non-Newtonian fluids with this permeability is too small. When $t=0.006$ s, this non-Newtonian flow would barely reach the middle position of the vessel, therefore it will lose the necessity in our comparison models. 

\begin{figure}[ht!]
	\includegraphics[trim=0 0 0 30,scale=0.28]{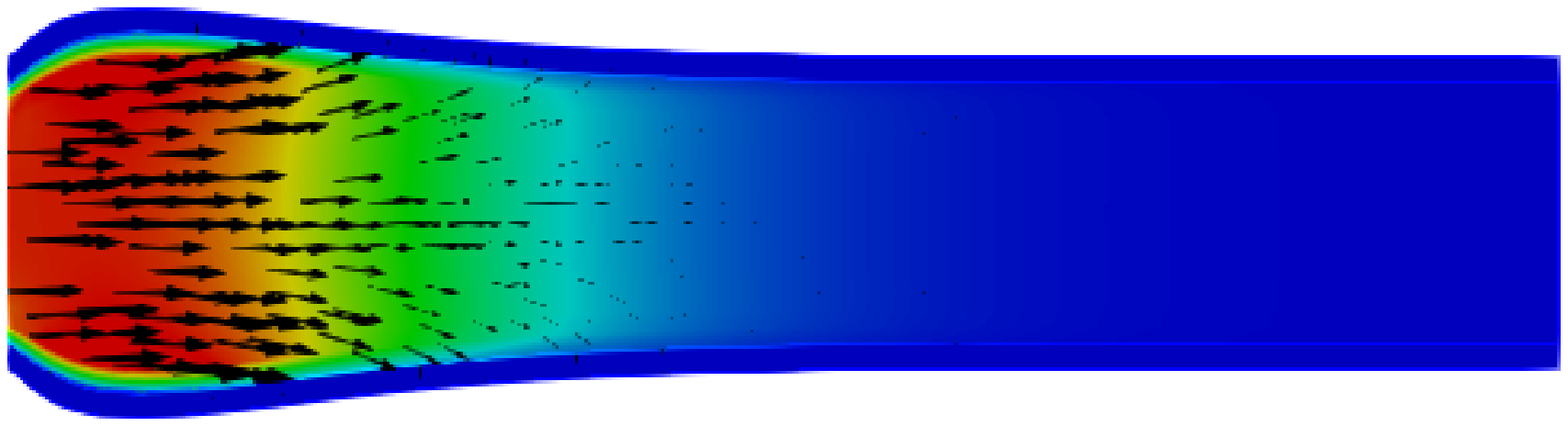}
	\includegraphics[trim=0 0 0 30,scale=0.28]{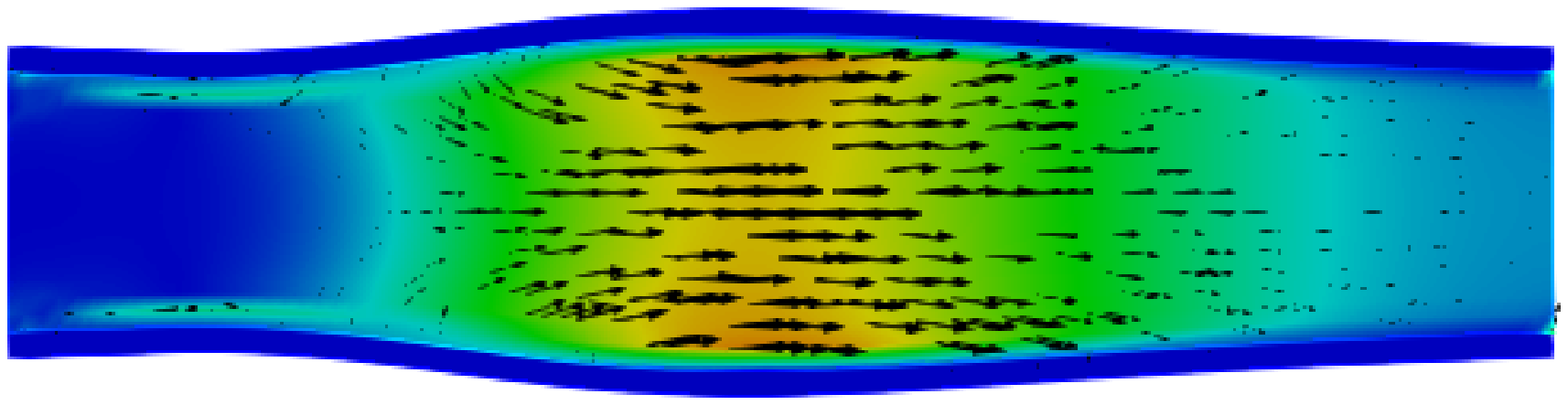}
	\includegraphics[trim=0 0 0 30,scale=0.28]{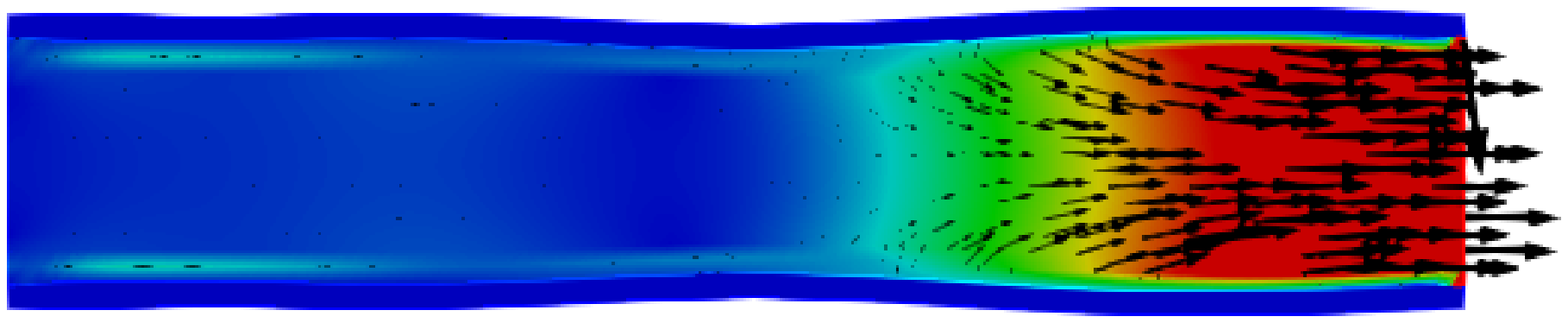}
	\includegraphics[trim=0 0 0 0,scale=0.28]{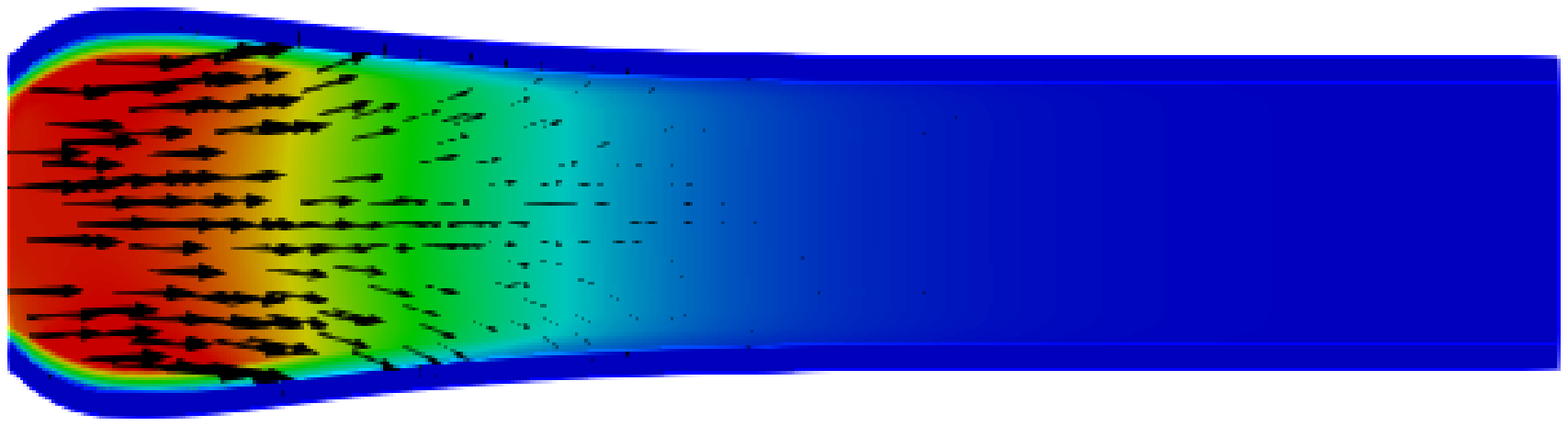}
	\includegraphics[trim=0 0 0 0,scale=0.28]{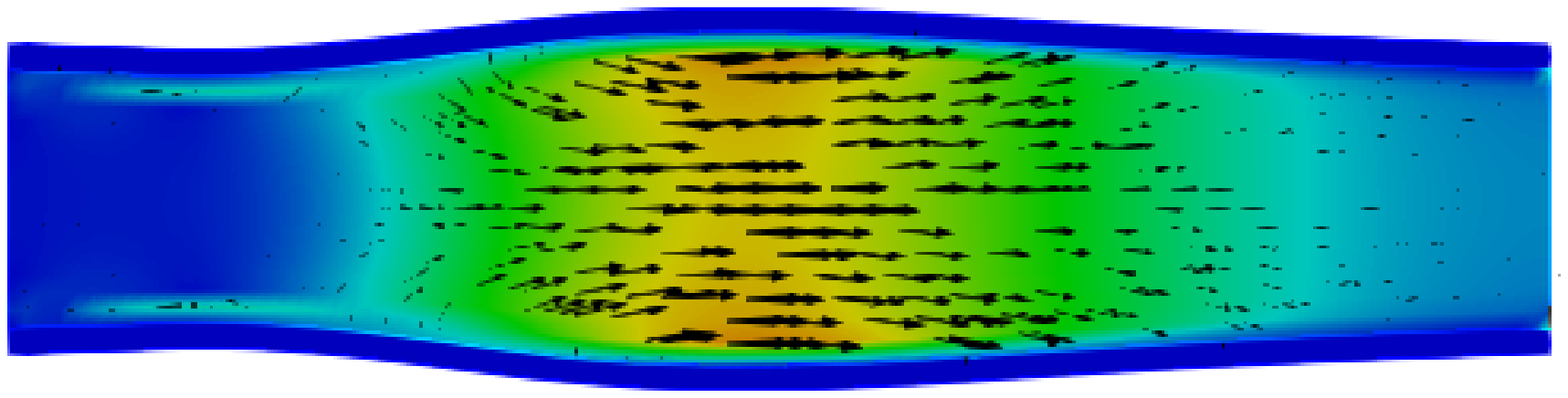}
	\includegraphics[trim=0 0 0 0,scale=0.28]{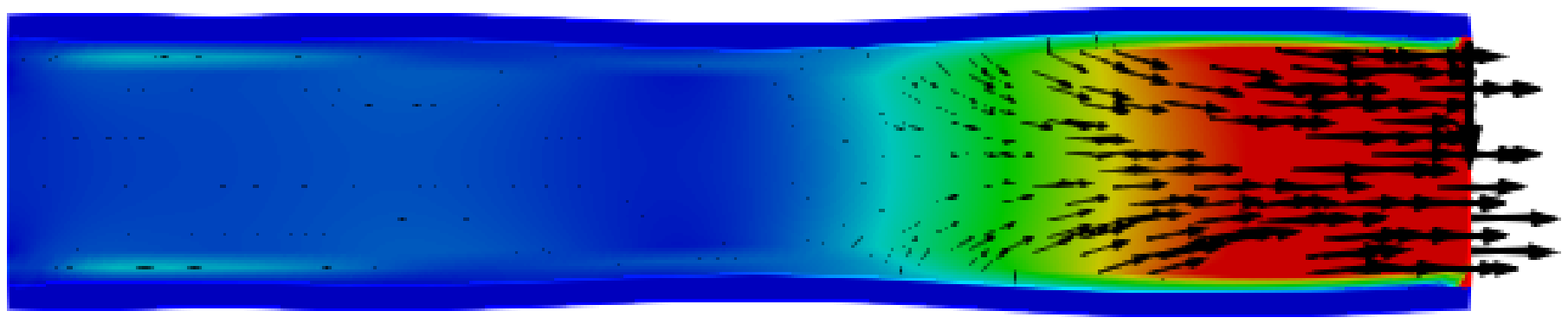}
	\includegraphics[trim=0 10 0 0,scale=0.28]{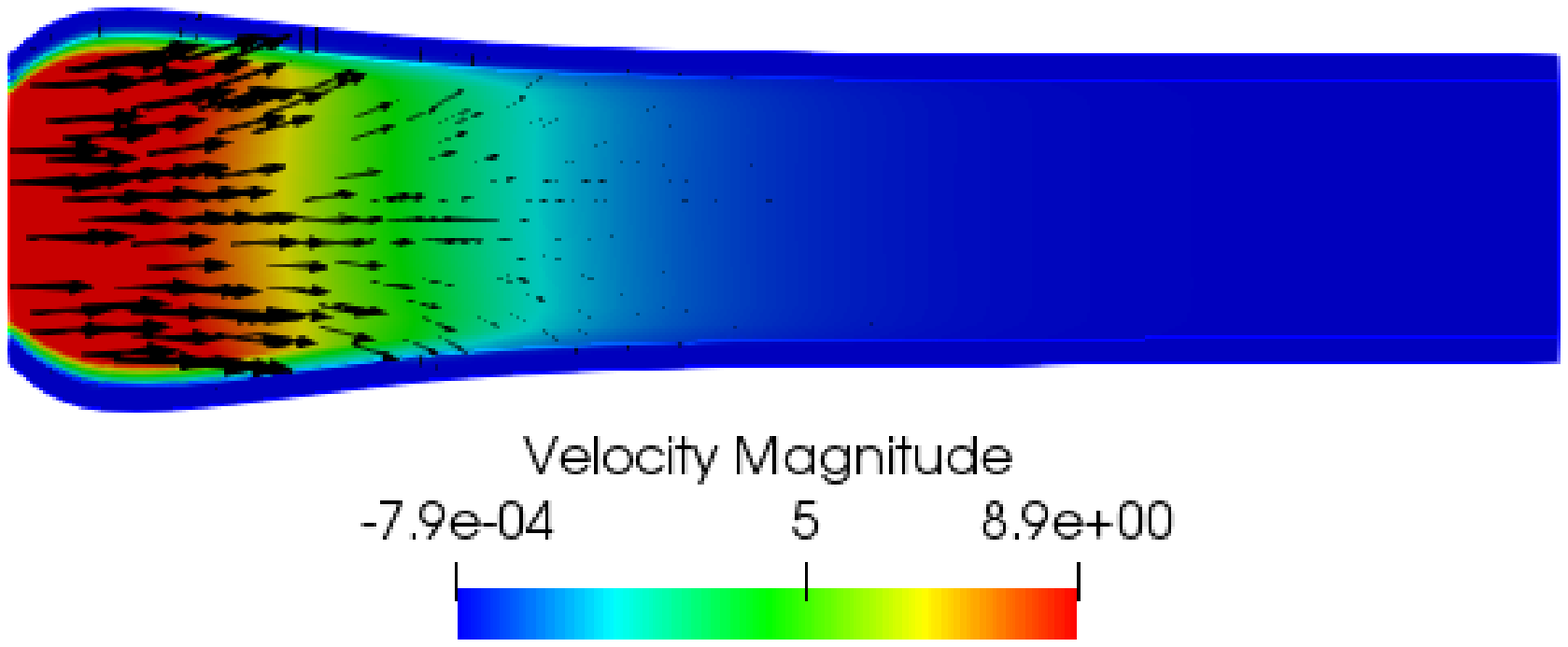}
	\includegraphics[trim=0 10 0 0,scale=0.28]{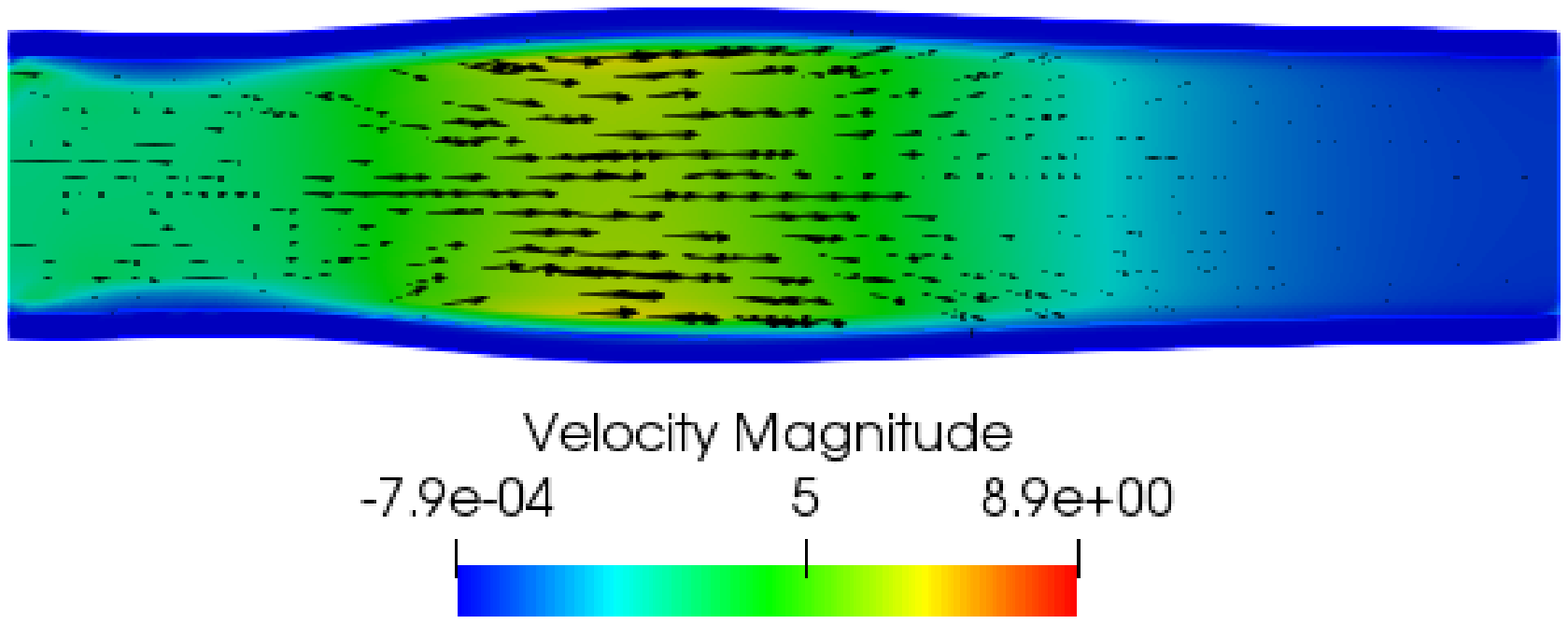}
	\includegraphics[trim=0 10 0 0,scale=0.28]{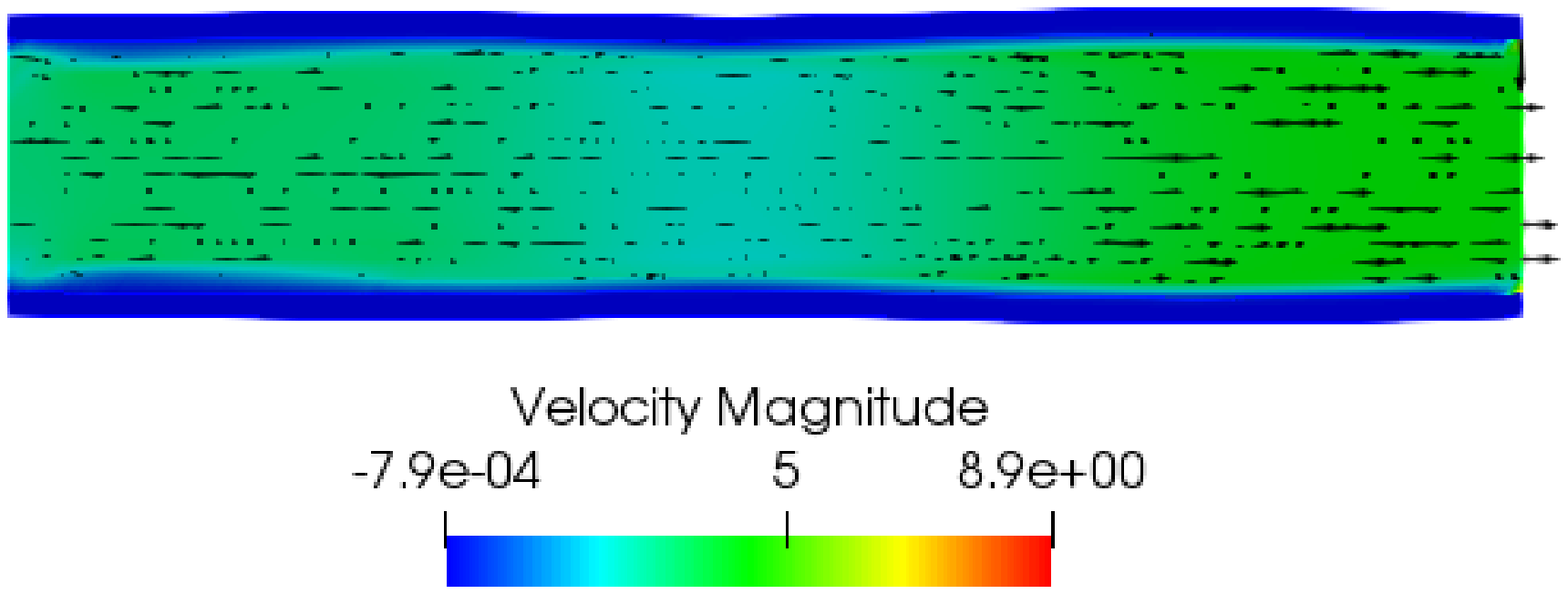}
	\caption{Fluid velocity magnitude together with scaled velocity arrows at time t=1.8 ms, t=3.6 ms, t=5.4 ms for case 4, case 5 and case 6. }
	\label{velocitycase2}
\end{figure}

\begin{figure}[ht!]
	\includegraphics[trim=0 -10 0 30,scale=0.28]{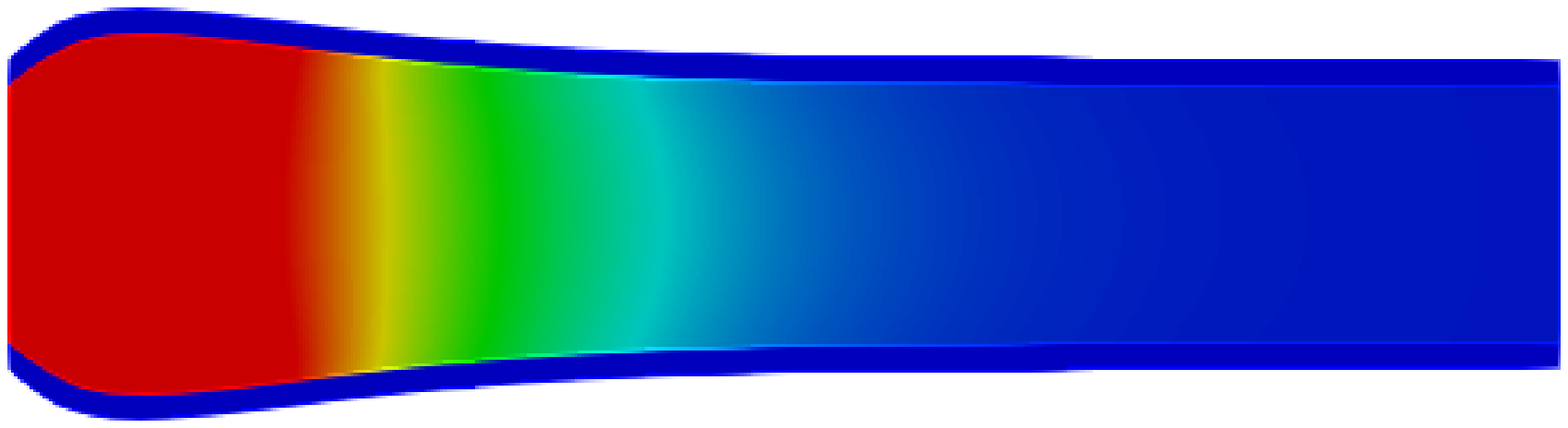}
	\includegraphics[trim=0 -10 0 30,scale=0.28]{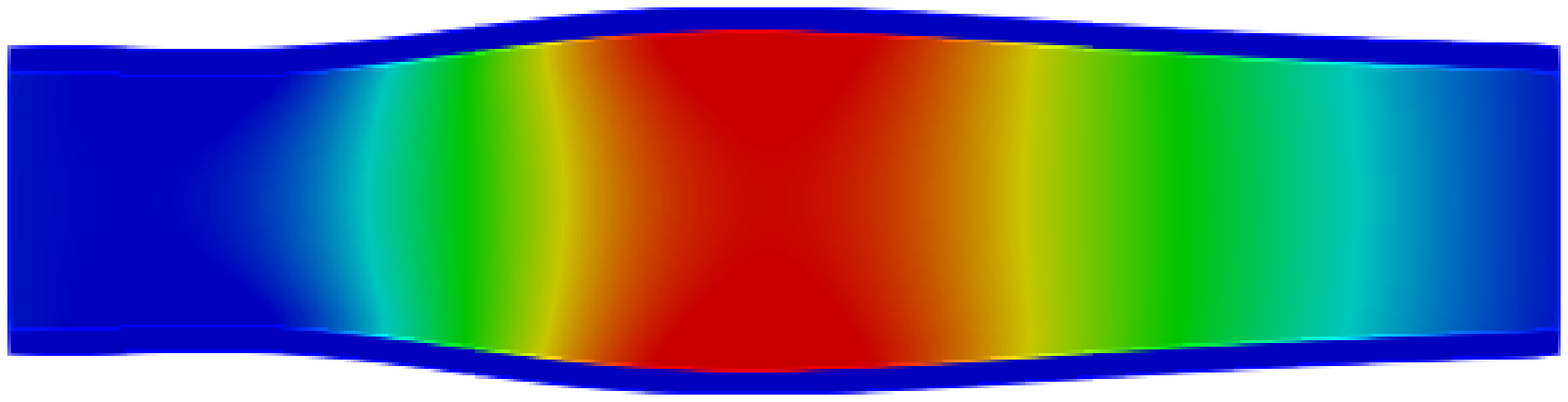}
	\includegraphics[trim=0 -10 0 30,scale=0.28]{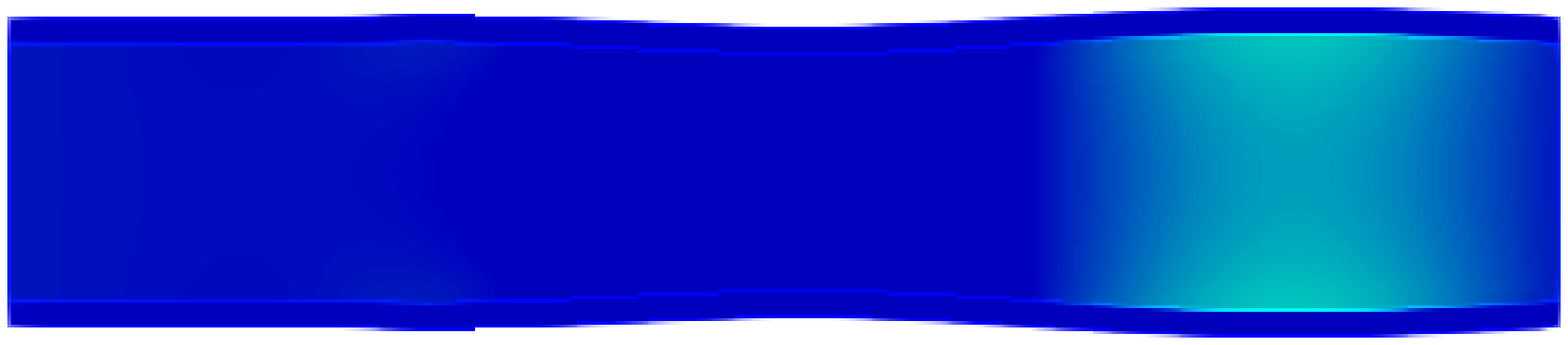}
	\includegraphics[trim=0 0 0 0,scale=0.28]{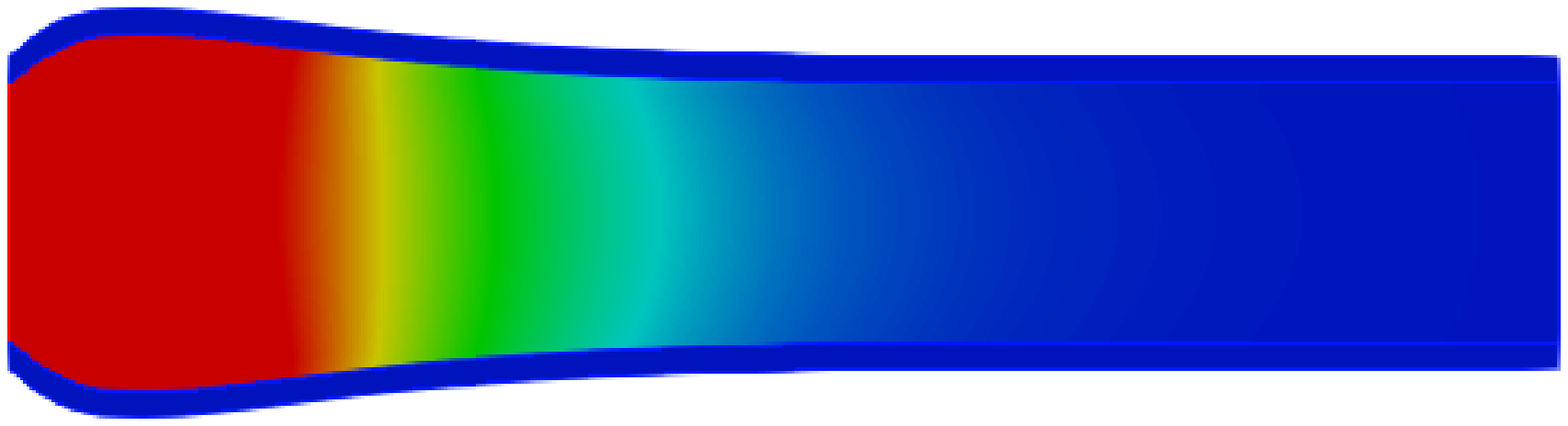}
	\includegraphics[trim=0 0 0 0,scale=0.28]{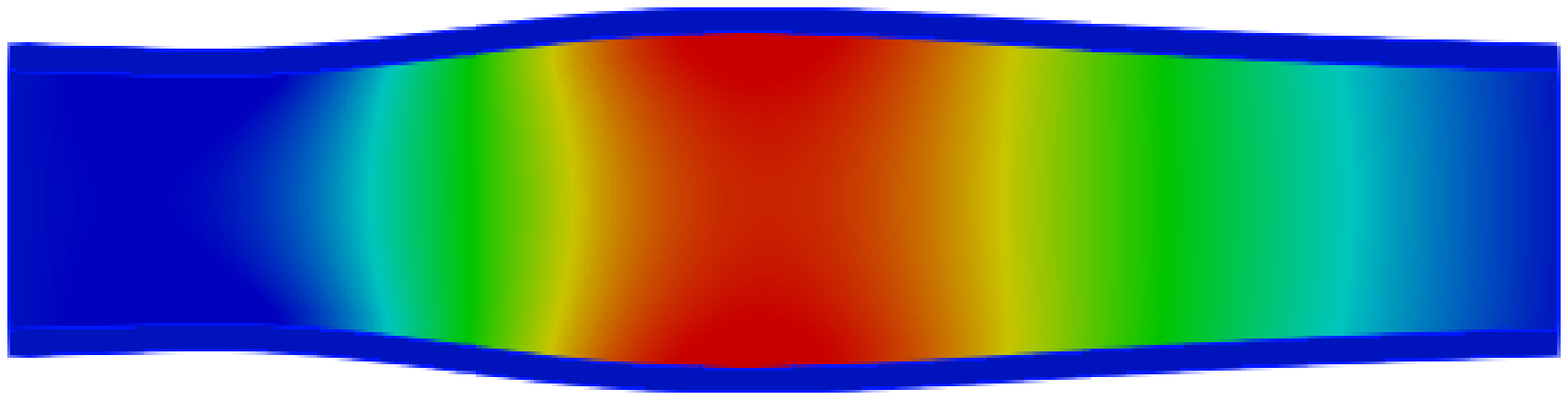}
	\includegraphics[trim=0 0 0 0,scale=0.28]{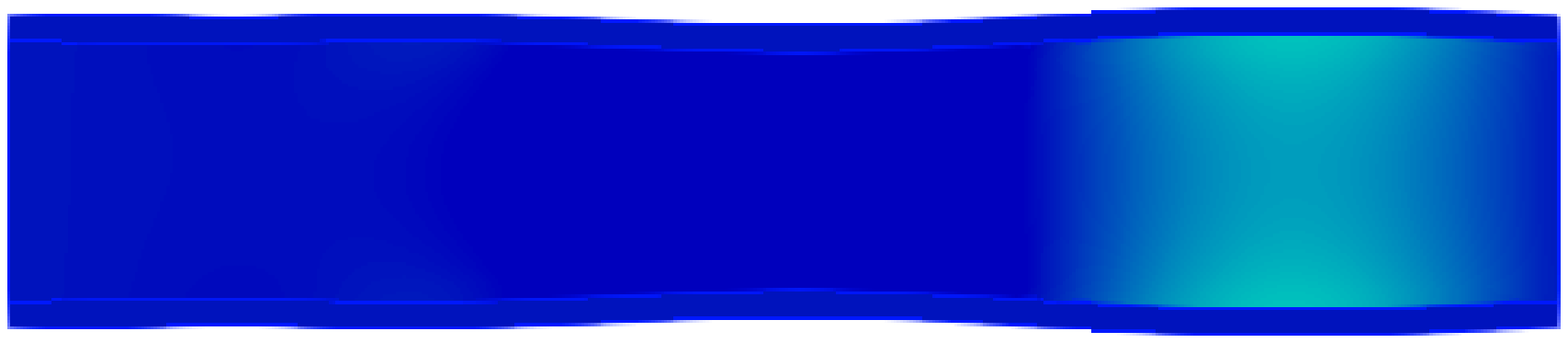}
	\includegraphics[trim=0 10 0 0,scale=0.28]{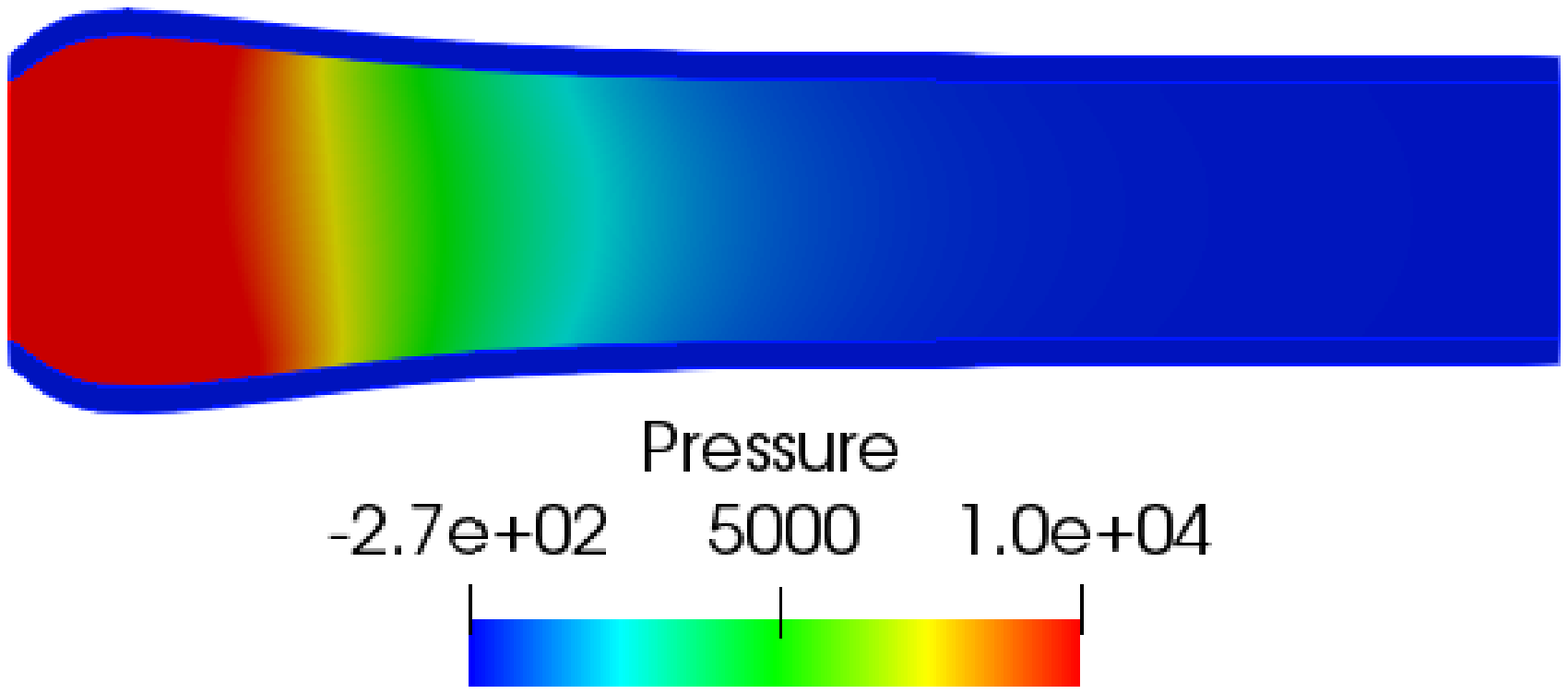}
	\includegraphics[trim=0 10 0 0,scale=0.28]{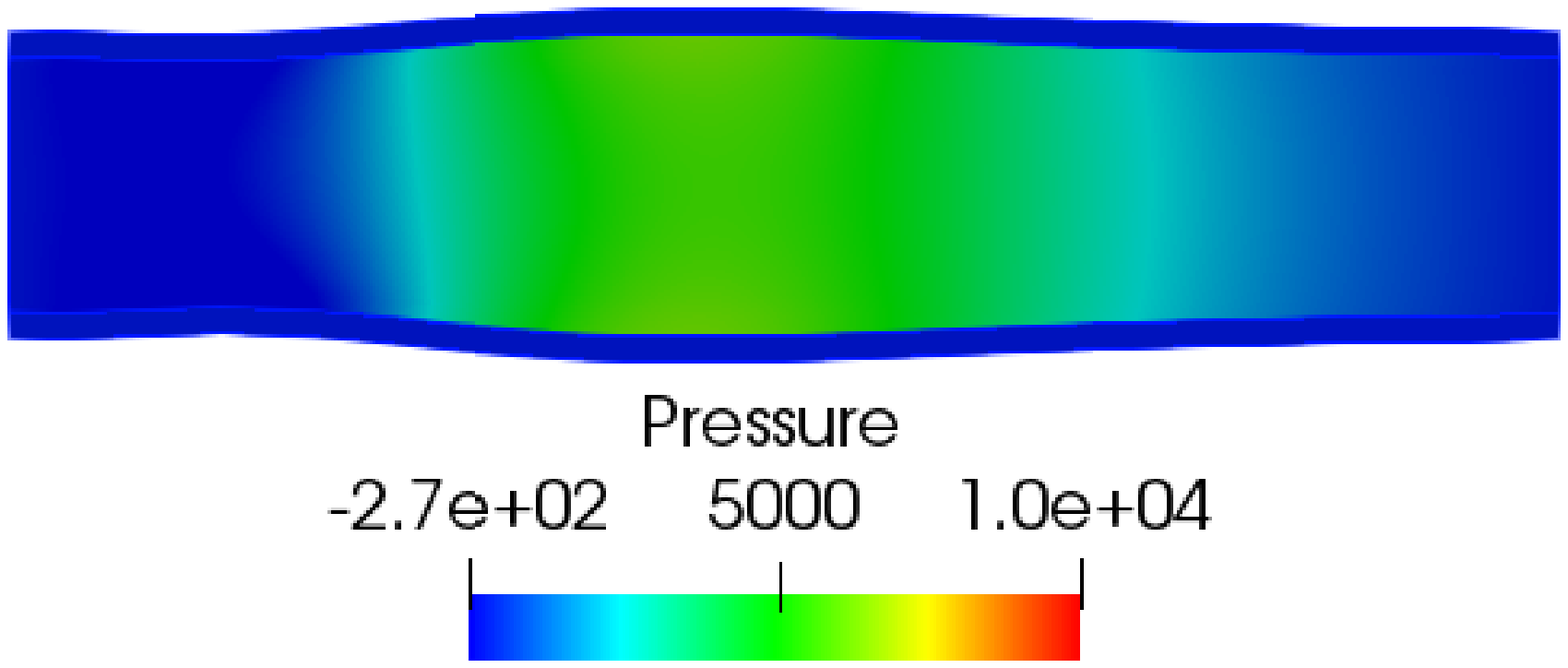}
	\includegraphics[trim=0 10 0 0,scale=0.28]{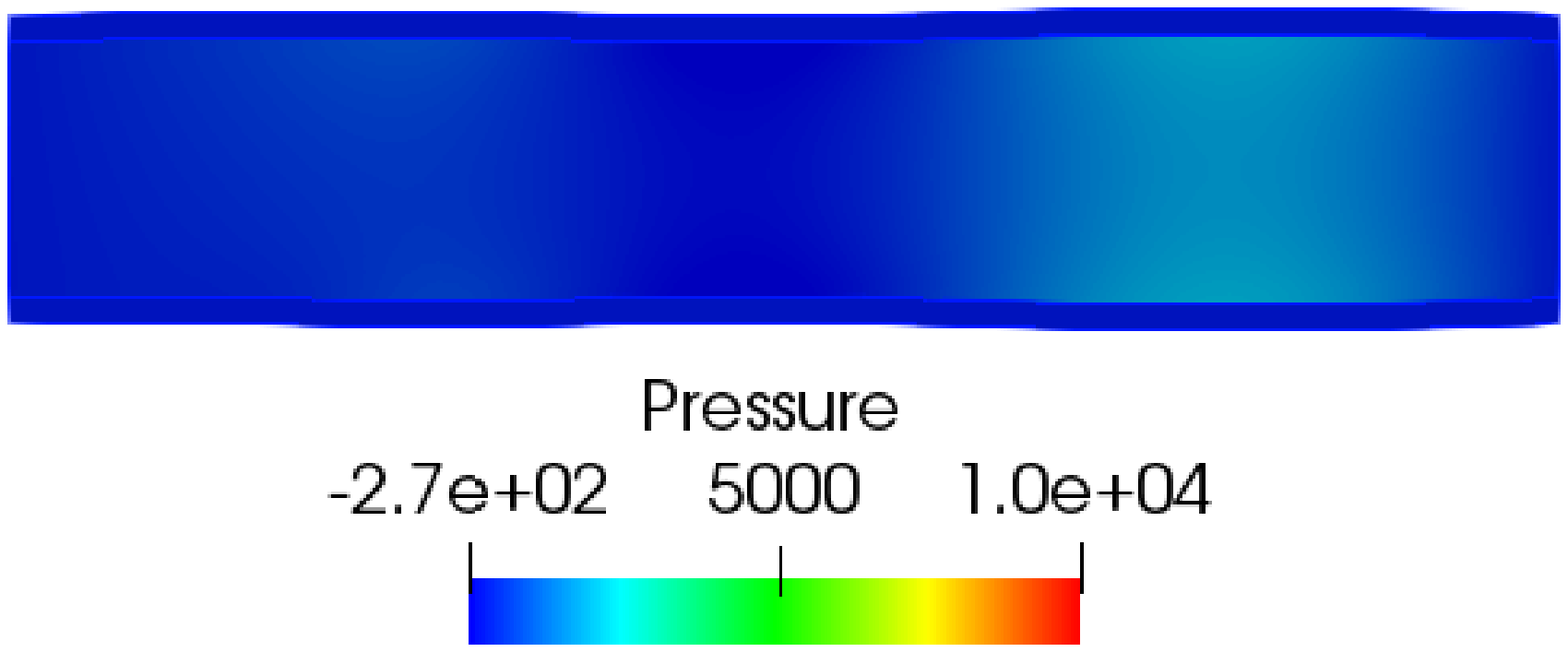}
	\caption{Pressure waves at time t=1.8 ms, t=3.6 ms, t=5.4 ms for case 4, case 5 and case 6. Note that for case 4, there is no pressure variable in the structure region $\Omega_{e}$. }
	\label{pressurecom2}
\end{figure}

\begin{figure}[ht!]
	\includegraphics[trim=0 0 0 30,scale=0.28]{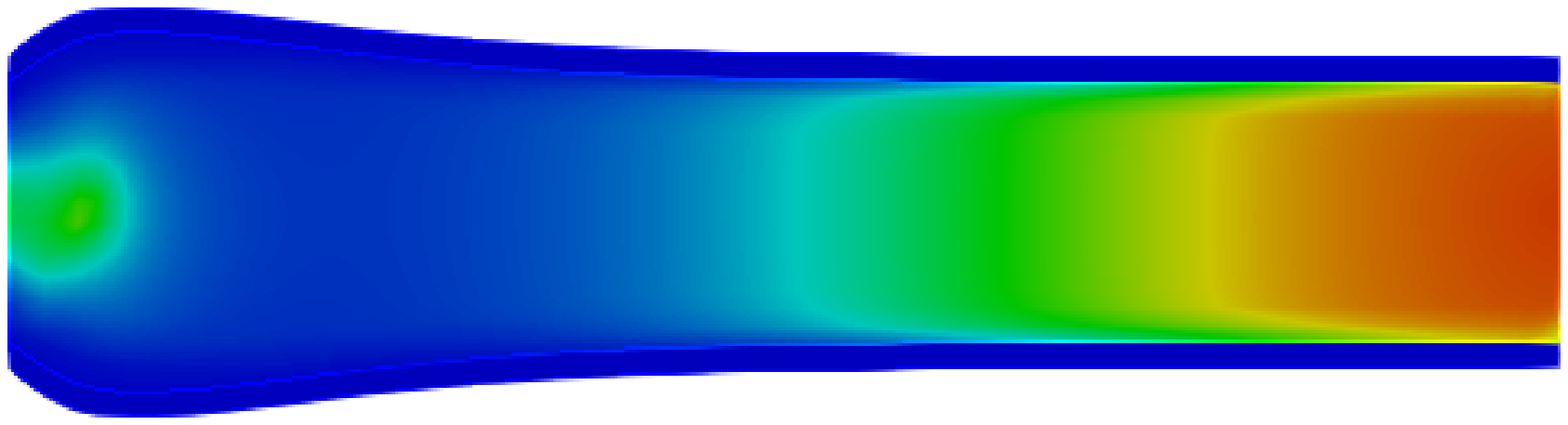}
	\includegraphics[trim=0 0 0 30,scale=0.28]{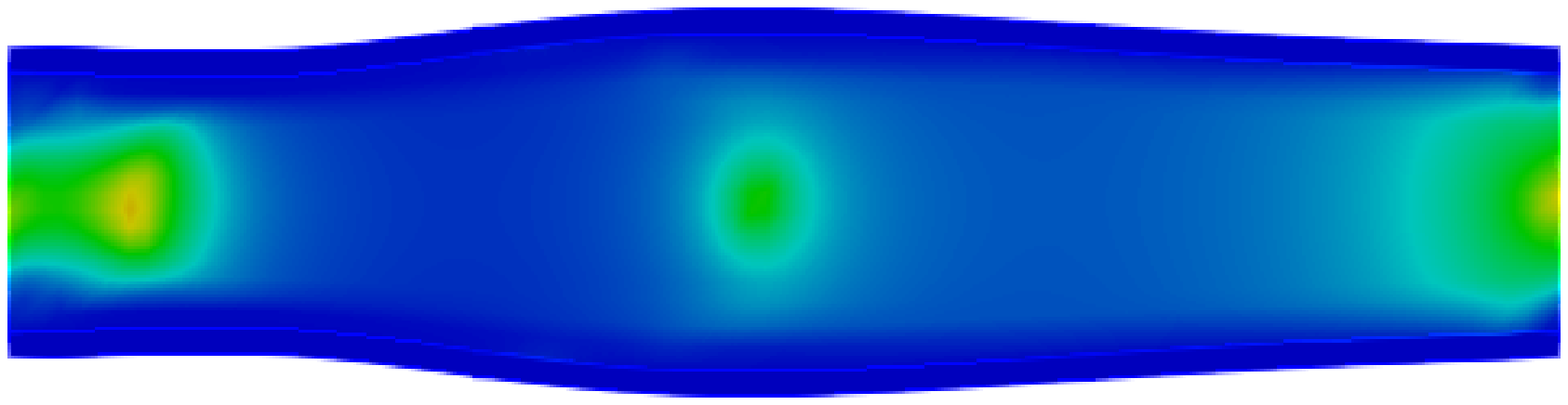}
	\includegraphics[trim=0 0 0 0,scale=0.28]{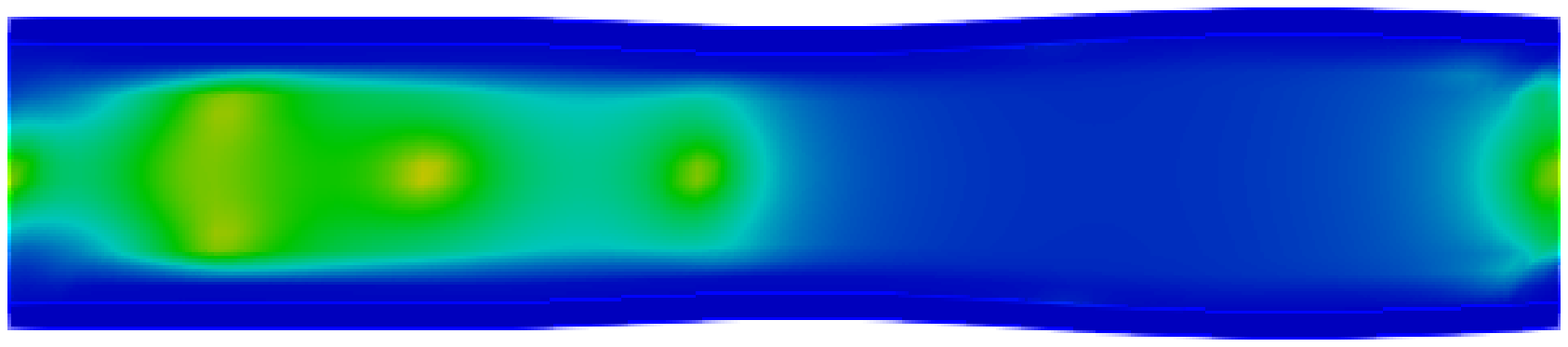}
	\includegraphics[trim=0 0 0 0,scale=0.28]{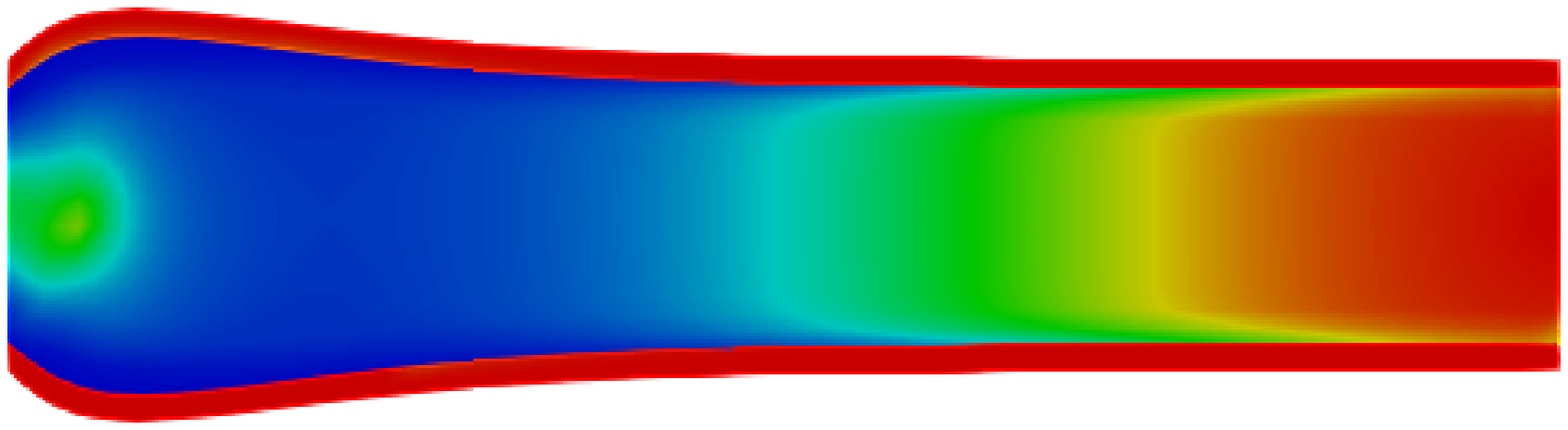}
	\includegraphics[trim=0 0 0 0,scale=0.28]{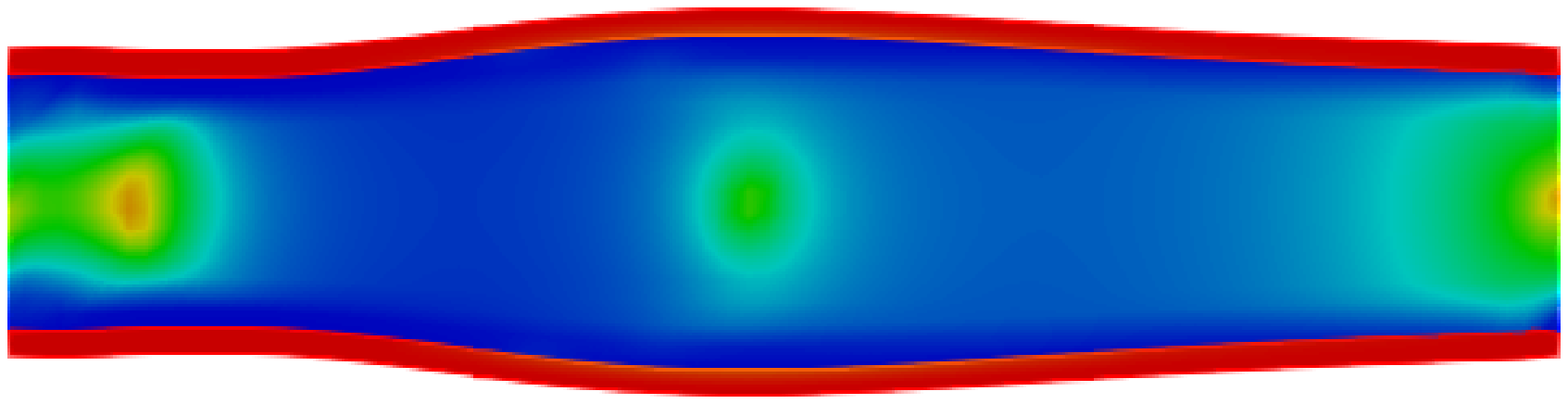}
	\includegraphics[trim=0 0 0 0,scale=0.28]{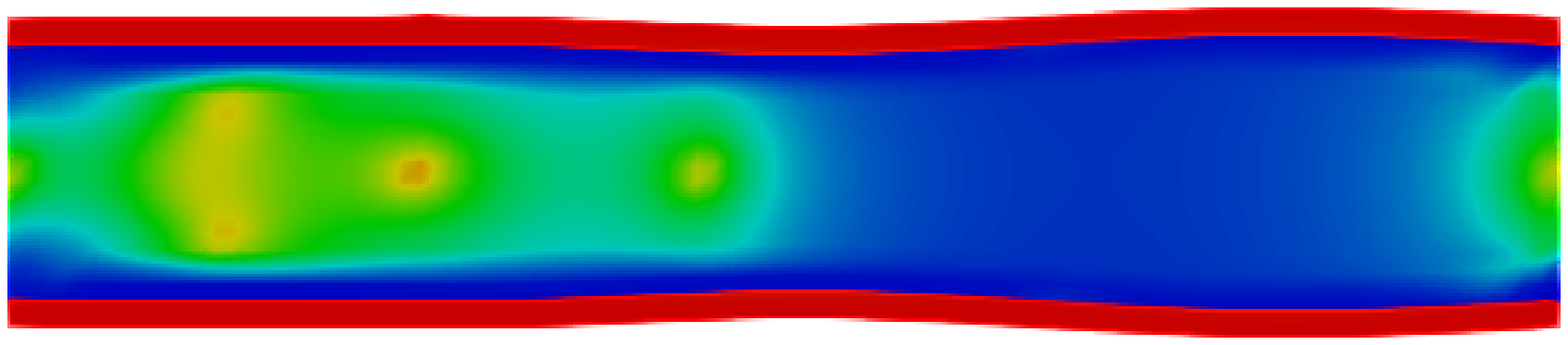}
	\includegraphics[trim=0 10 0 0,scale=0.28]{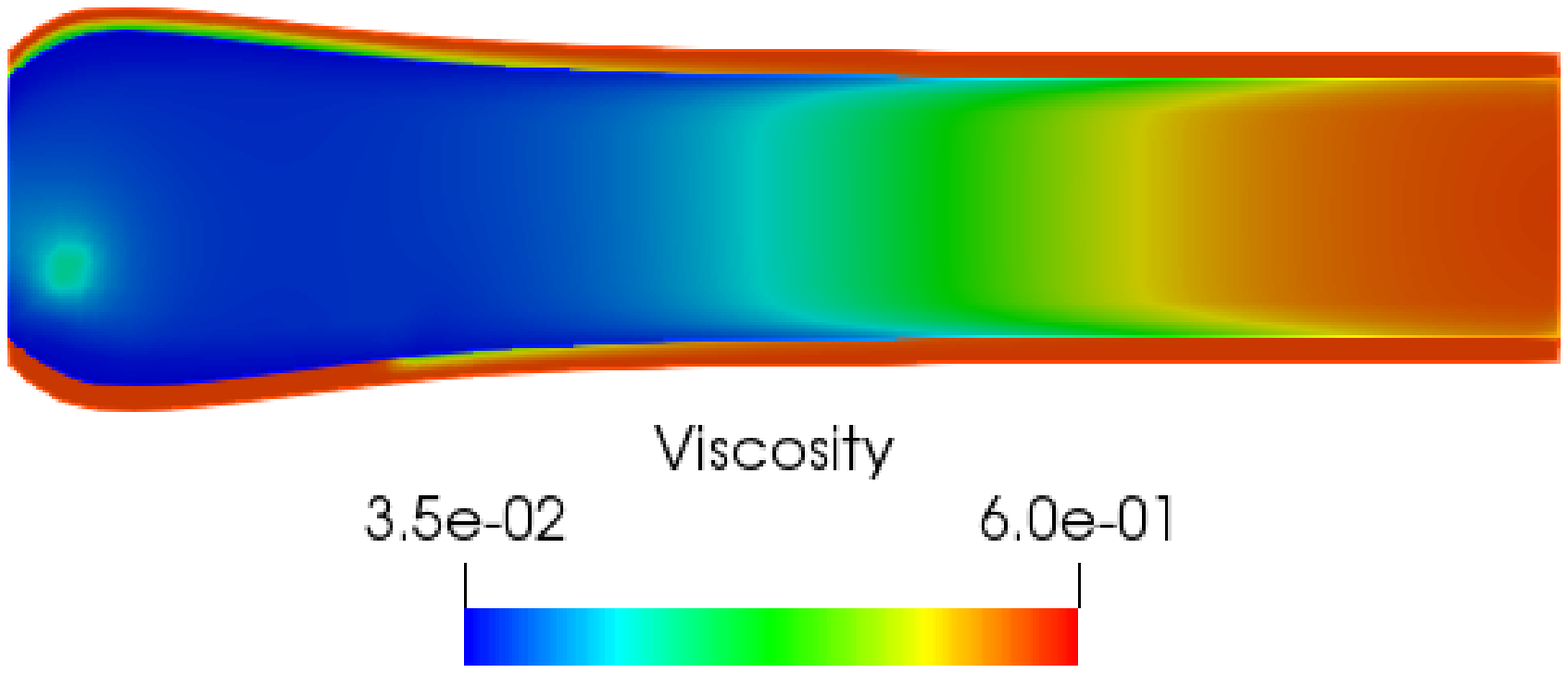}
	\includegraphics[trim=0 10 0 0,scale=0.28]{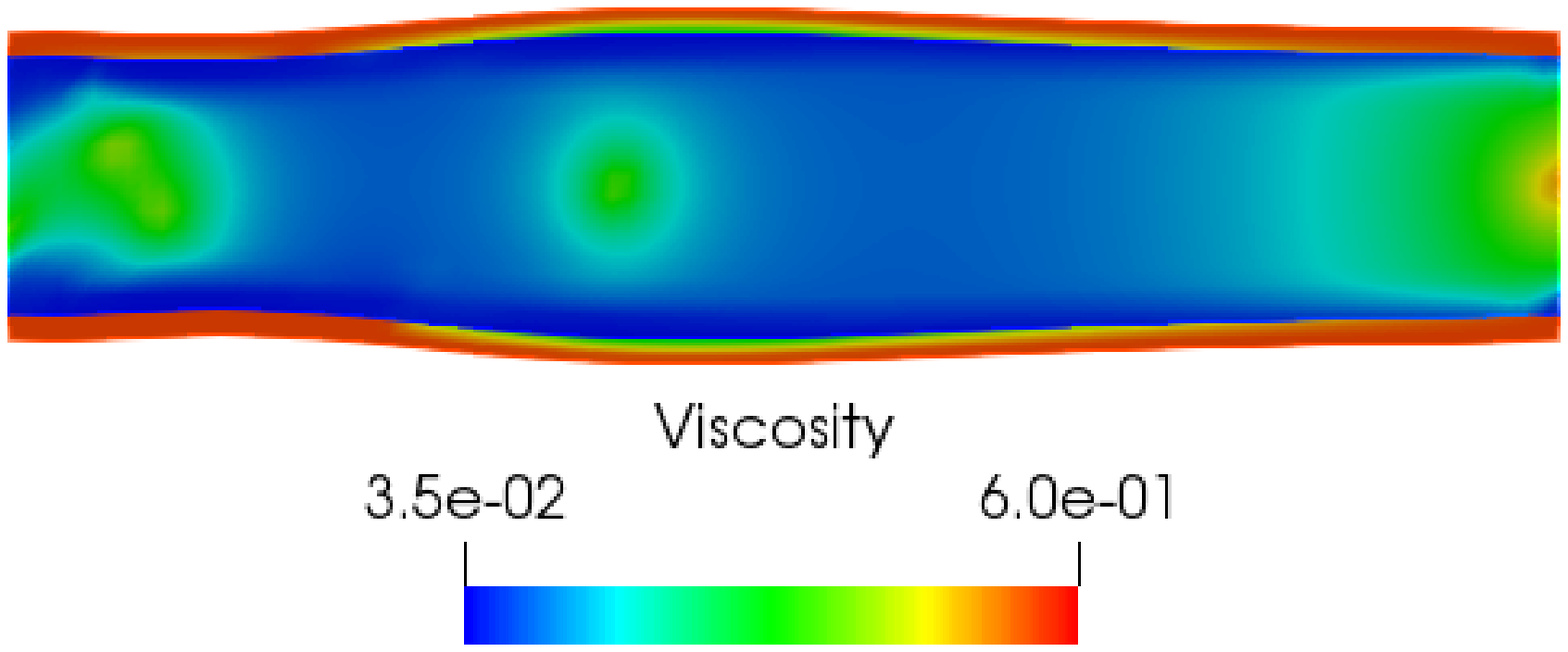}
	\includegraphics[trim=0 10 0 0,scale=0.28]{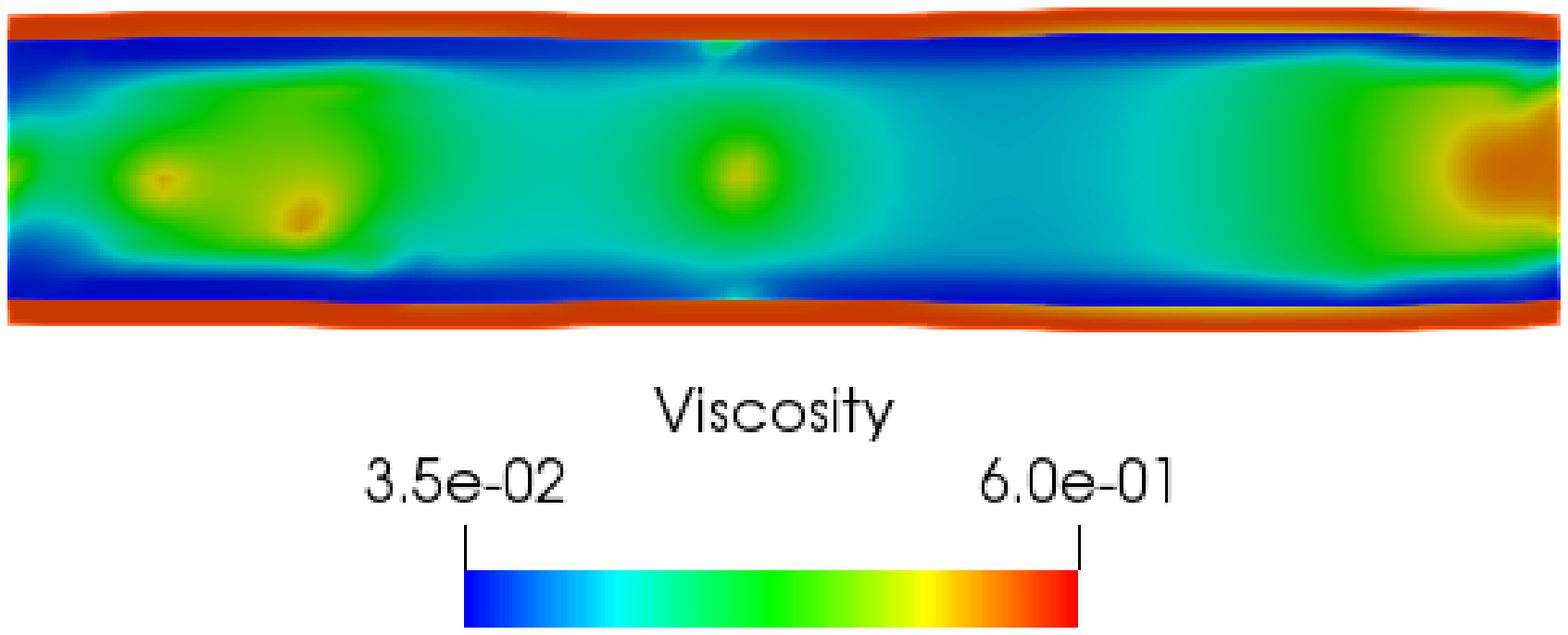}
	\caption{Viscosity at time t=1.8 ms, t=3.6 ms, t=5.4 ms for case 4, case 5, case 6. }
	\label{viscosity}
\end{figure}

In Fig.\ref{velocitycase2}, we present the fluids velocity and the velocity arrow for case 4, case 5 and case 6 at time $t=1.8, 3.6, 5.4$ ms. Note that there are more fluids residuals along the lumen than the Newtonian cases. The velocity in case 6 is overall smaller than the other two cases at the same time as well as the deformation. We also observe the similar phenomena for the pressure waves of case 4, case 5 and case 6 in Fig.\ref{pressurecom2}. 
In Fig.\ref{viscosity}, we present the viscosity of case 4, case 5 and case 6 at time $t=1.8, 3.6, 5.4$ ms. For case 4 with NSE/E model, since there is no Darcy flow in the structure region, we use solid blue instead, and the color in this case has nothing to do with the viscosity value. The viscosity of case 5 is larger than case 6 at the structure regions near the interface and we can detect the shear-thinning phenomena for case 6. At time $t=5.4$ ms, the dynamic viscosity of case 6 in the fluid region is overall larger than the other two cases. 

\begin{figure}[ht!]
	\centering
	\includegraphics[trim=25 60 25 0,scale=0.149]{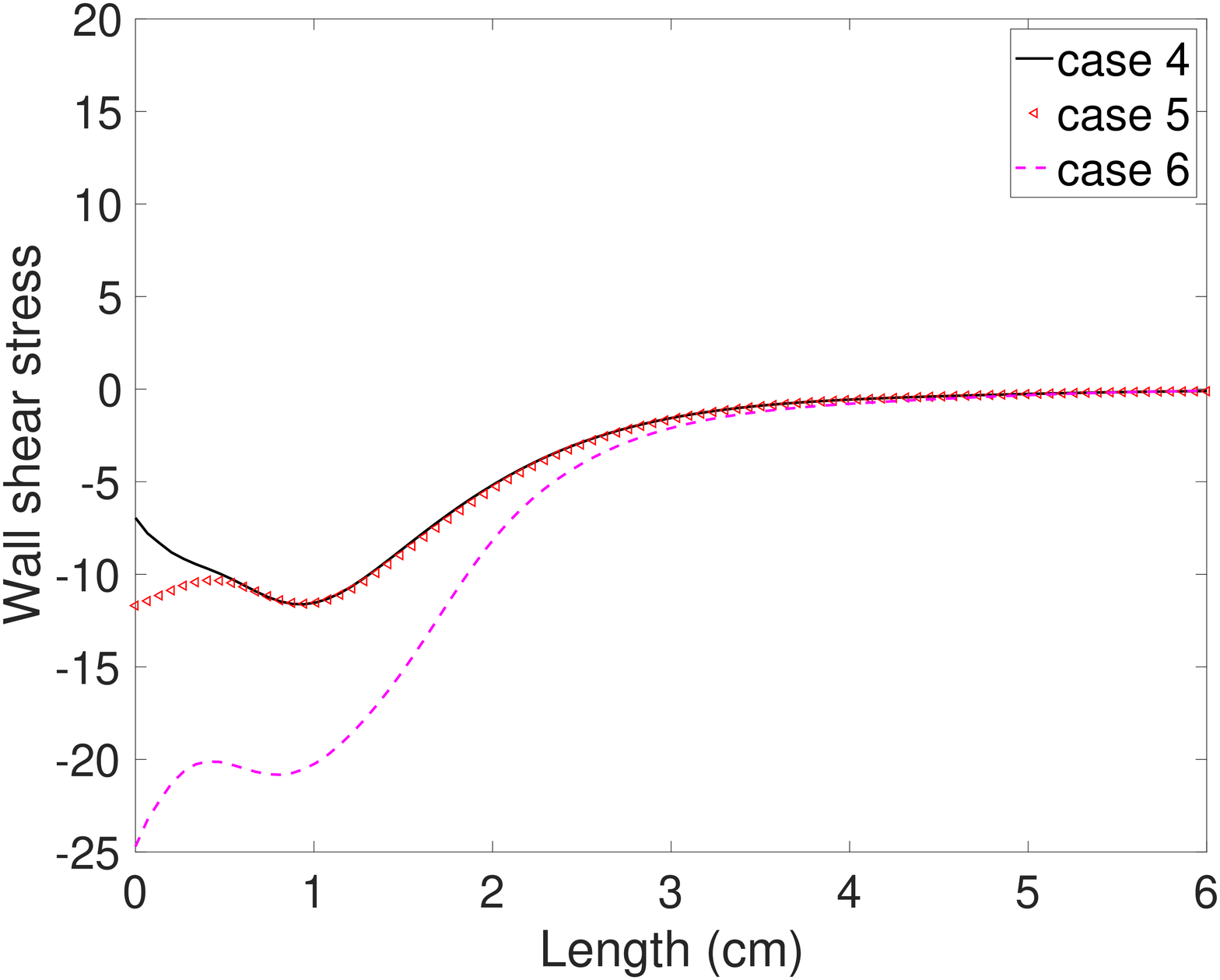}
	\includegraphics[trim=25 60 25 0,scale=0.149]{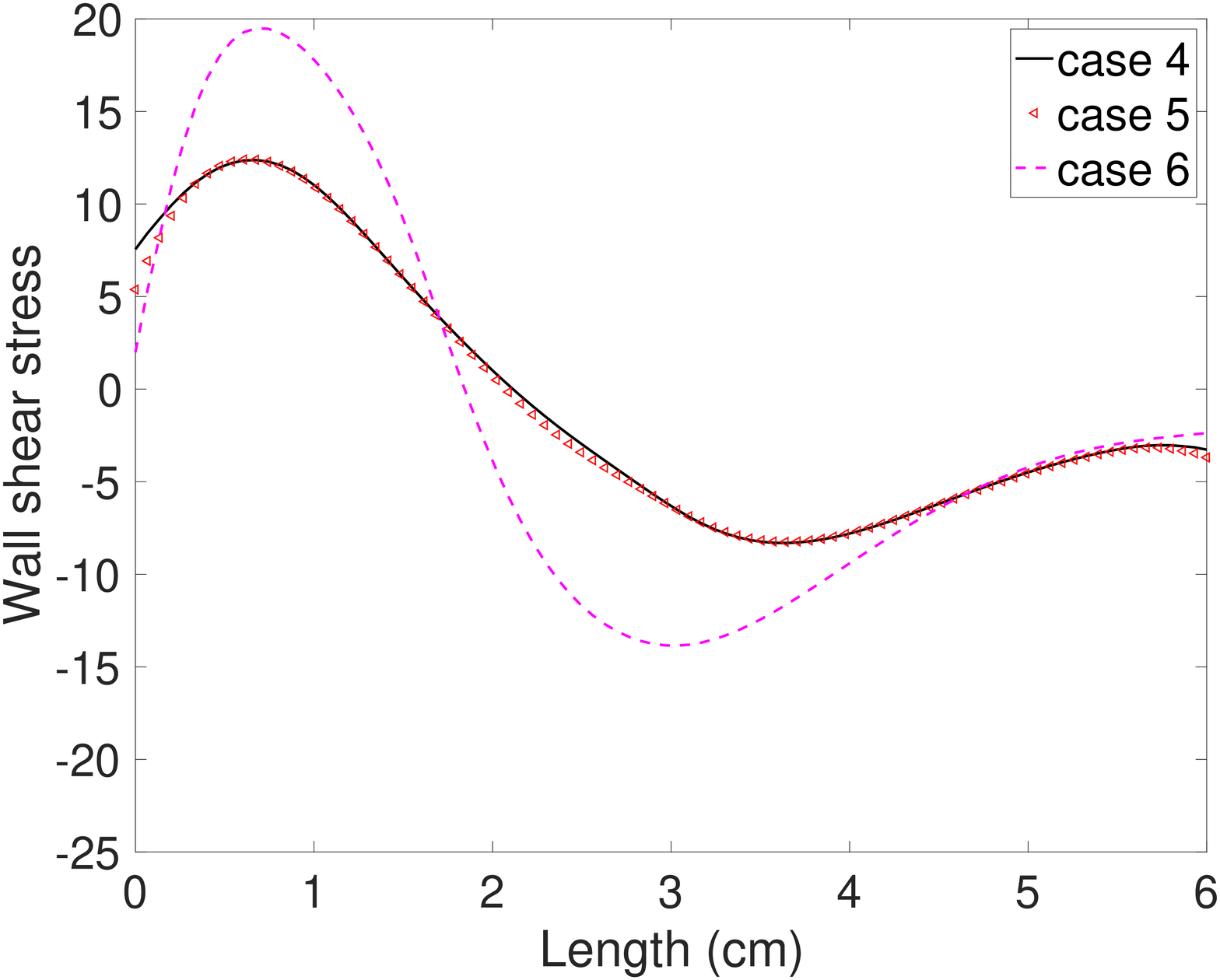}
	\includegraphics[trim=0 60 50 0,scale=0.149]{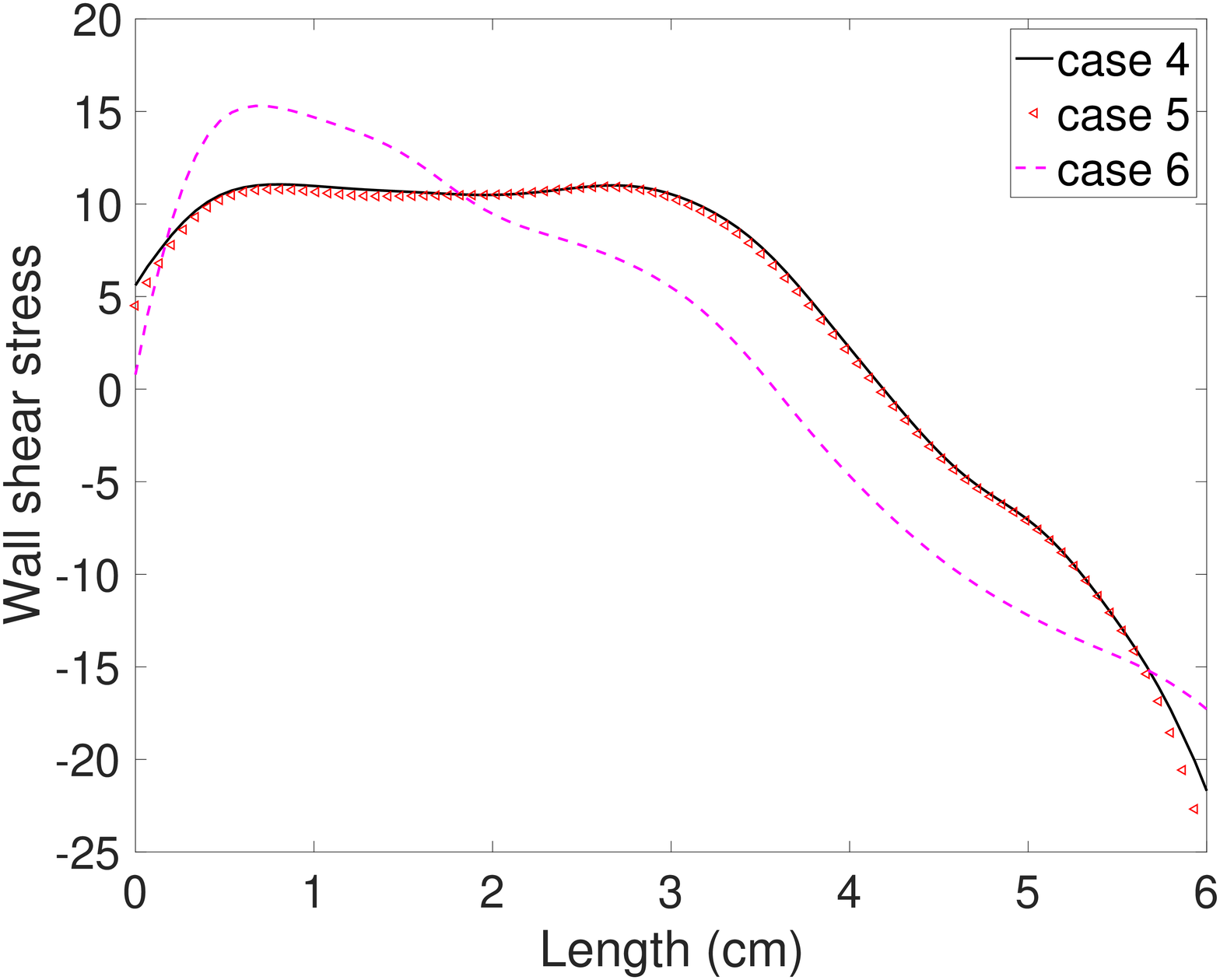}
	\caption{The shear stress $\bsi_f\bn\cdot \bt$ along the top arterial wall at time t=1.8 ms, t=3.6 ms, t=5.4 ms for case 4, case 5, and case 6.}
	\label{shearcompare2}
\end{figure}

\begin{figure}[ht!]
	\centering
	\includegraphics[trim=0 60 0 0,scale=0.149]{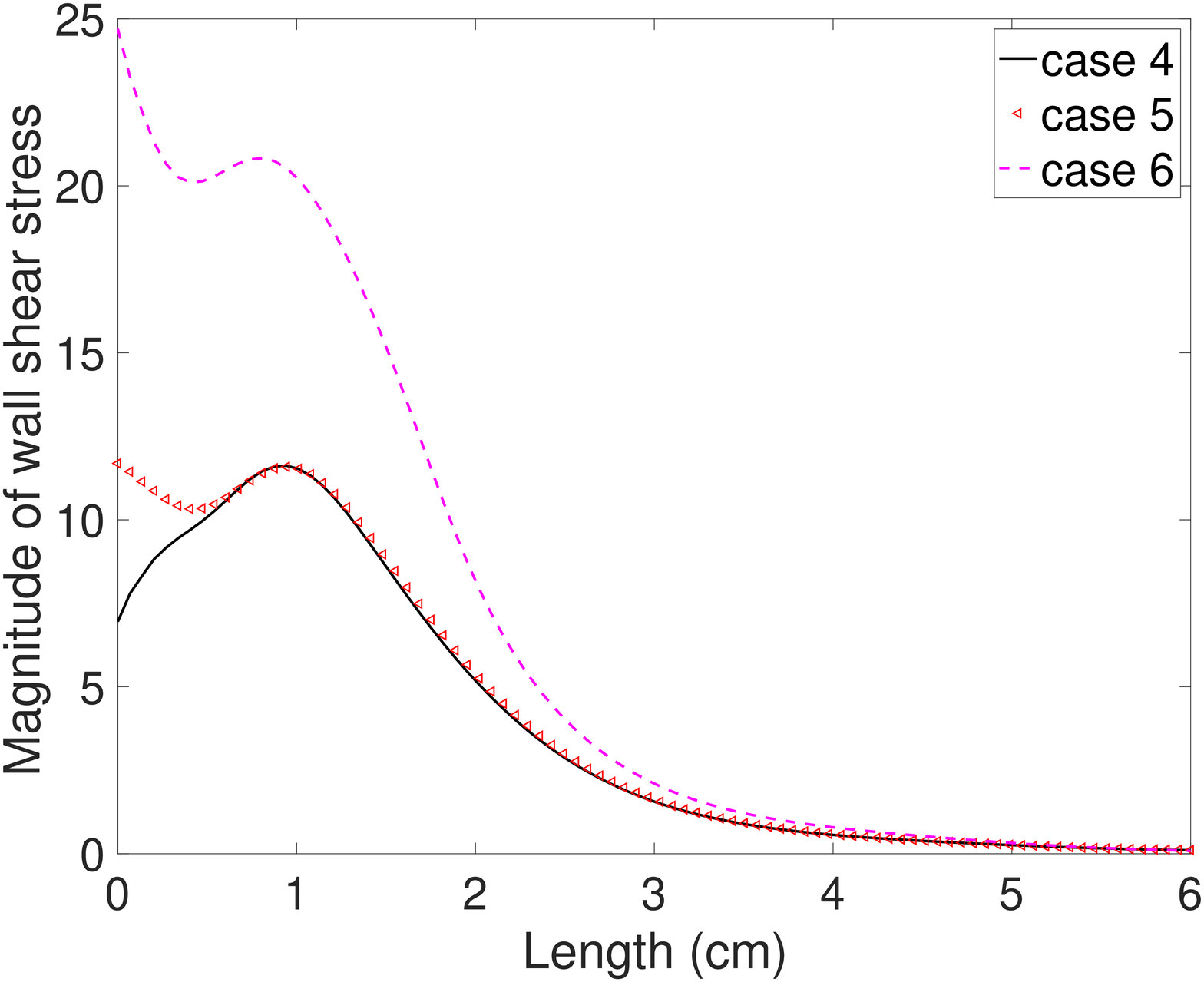}
	\includegraphics[trim=0 60 20 0,scale=0.149]{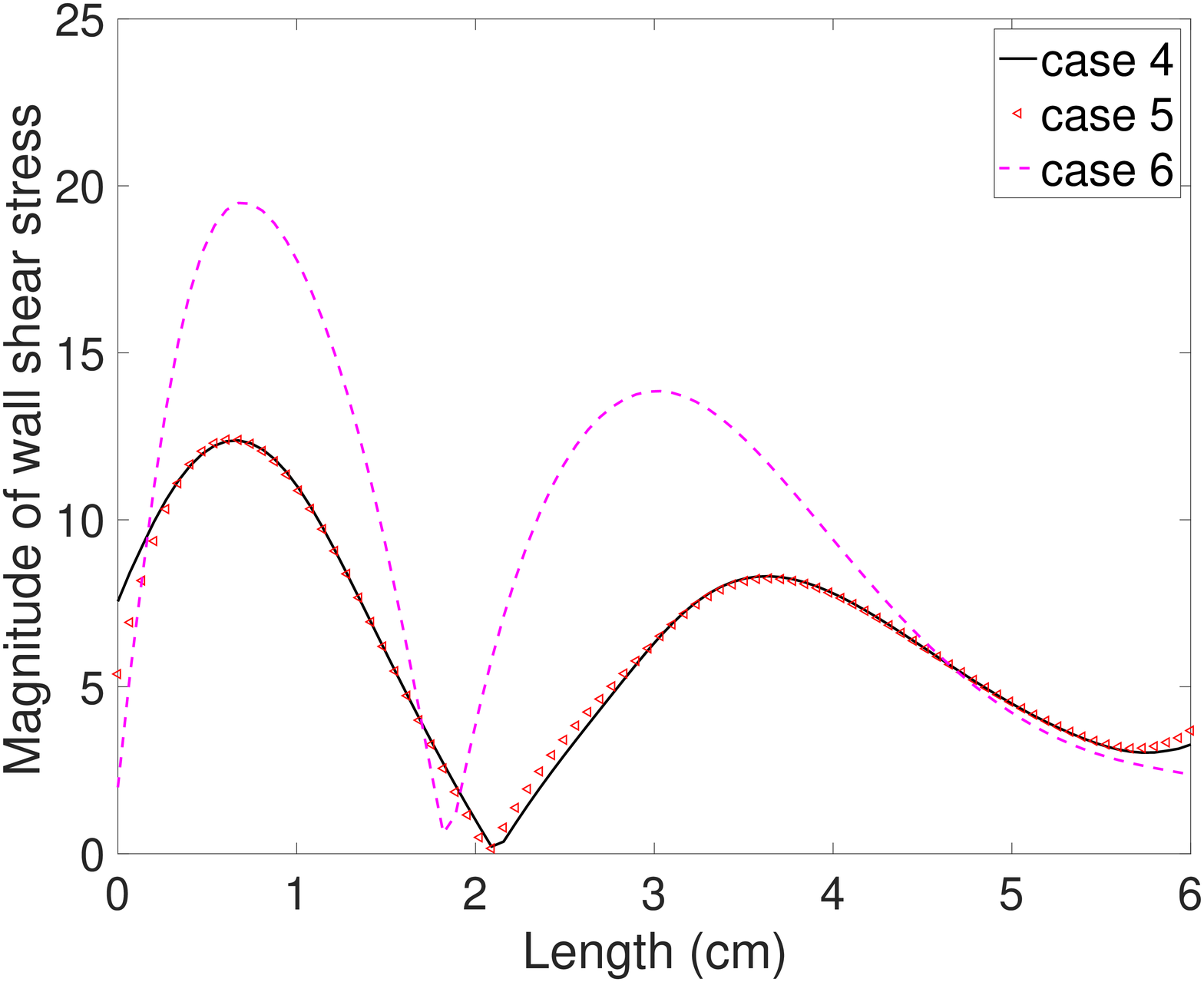}
	\includegraphics[trim=0 60 120 0,scale=0.149]{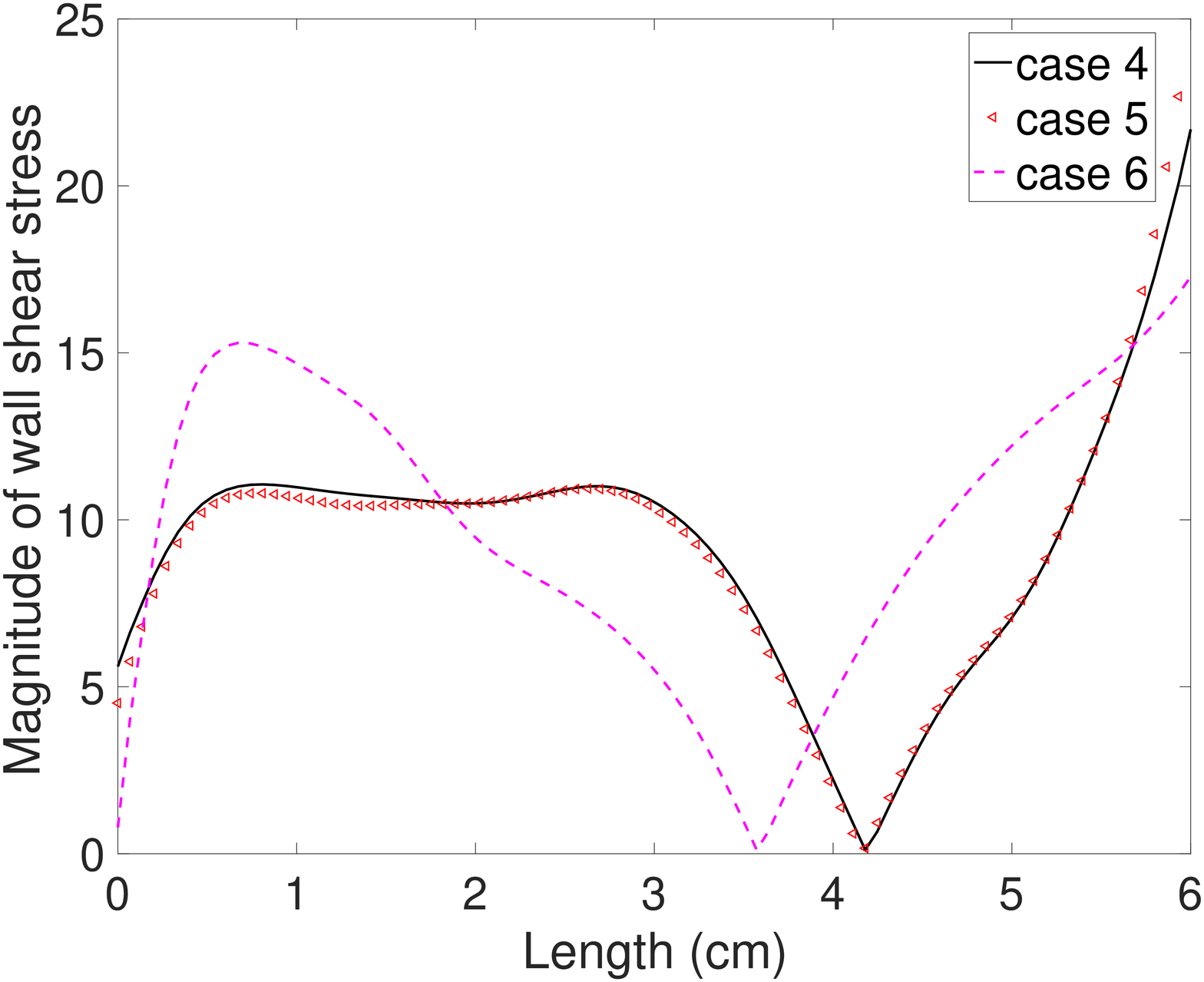}
	\caption{The shear stress $|\bsi_f\bn\cdot \bt|$ along the top arterial wall at time t=1.8 ms, t=3.6 ms, t=5.4 ms for case 4, case 5, and case 6.}
	\label{shearcompare22}
\end{figure}

\begin{figure}[ht!]
	\centering
	\includegraphics[trim=0 60 120 0,scale=0.149]{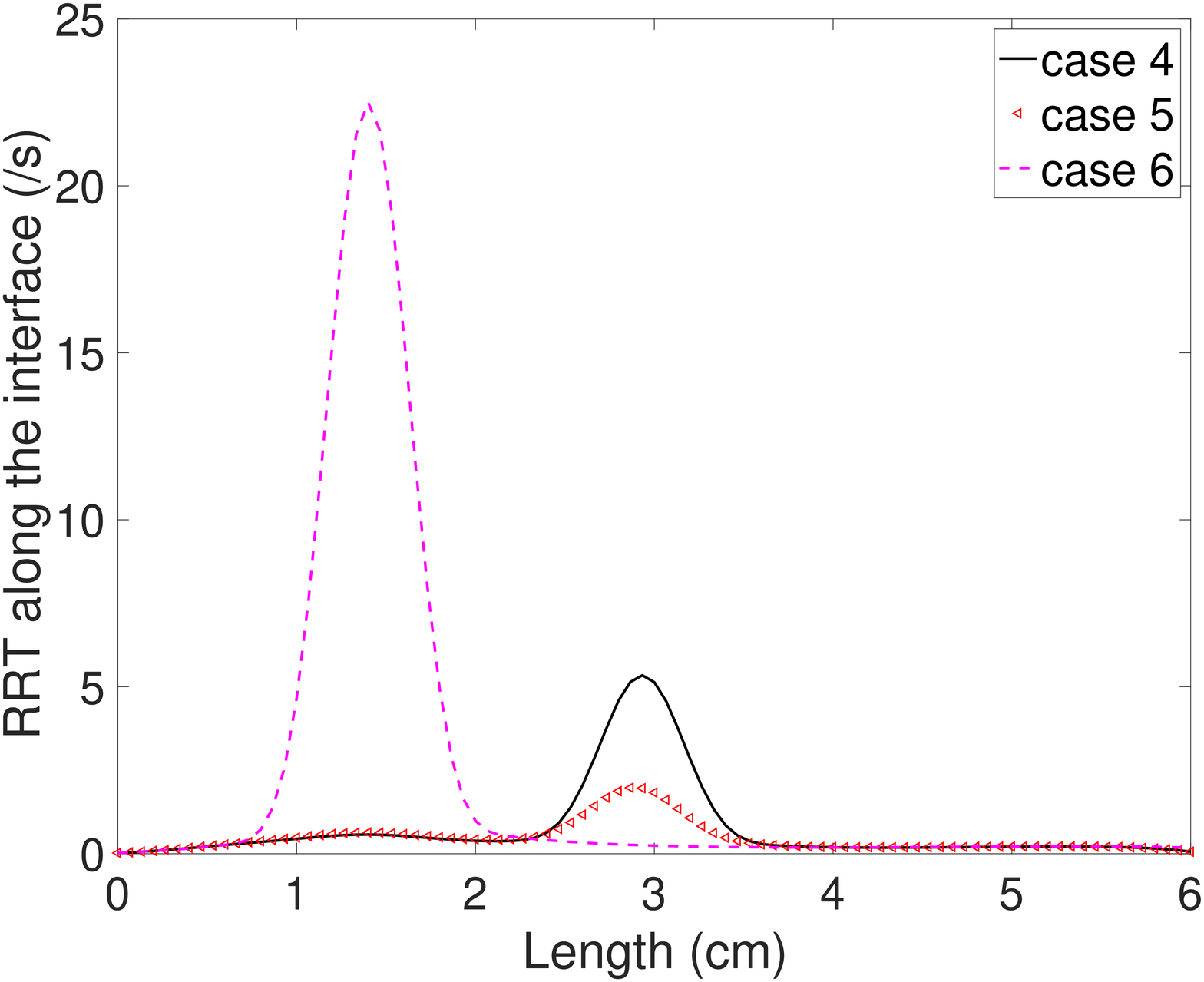}
	\caption{RRT along the top arterial wall  for case 4, case 5 and case 6.}
	\label{RRT2}
\end{figure}

In Fig.\ref{shearcompare2} and Fig.\ref{shearcompare22}, we present the wall shear stress $\bsi_f\bn\cdot \bt$ and its magnitude $\vert \bsi_f\bn\cdot \bt \vert$ respectively for case 4, case 5 and case 6 at time $t=1.8, 3.6, 5.4$ ms along the interface. 
In Fig.\ref{RRT2}, we report the RRT distribution along the lumen for case 4, case 5 and case 6. The peak of case 5 is lower than case 4 even if they show similar dynamic behaviors, such as fluids velocity and pressure. As expected, the peak of case 6 is the highest and is localized ahead of case 4 and case 5. 

\begin{figure}[ht!]
	\includegraphics[trim=0 60 0 0,scale=0.149]{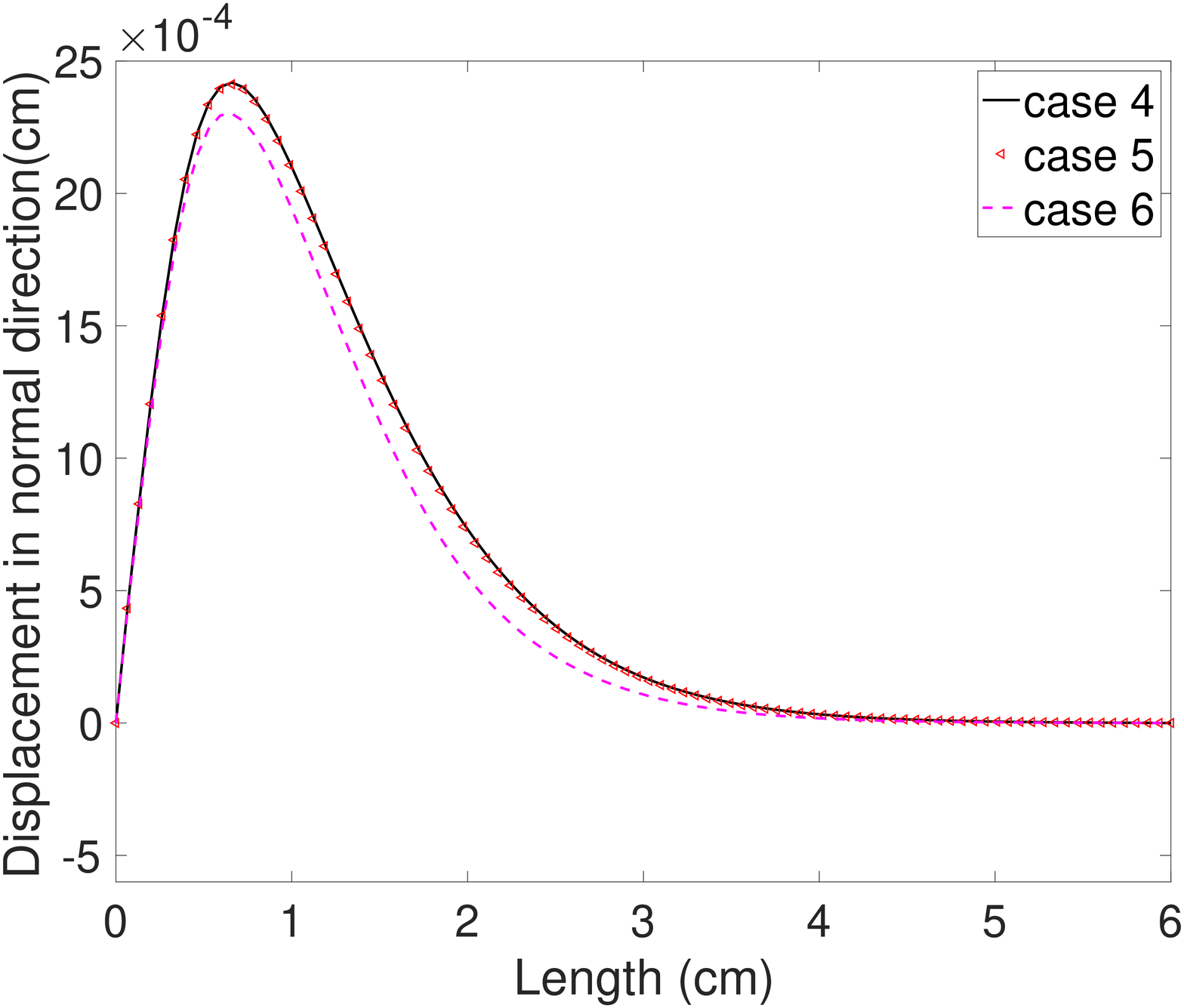}
	\includegraphics[trim=0 60 0 0,scale=0.149]{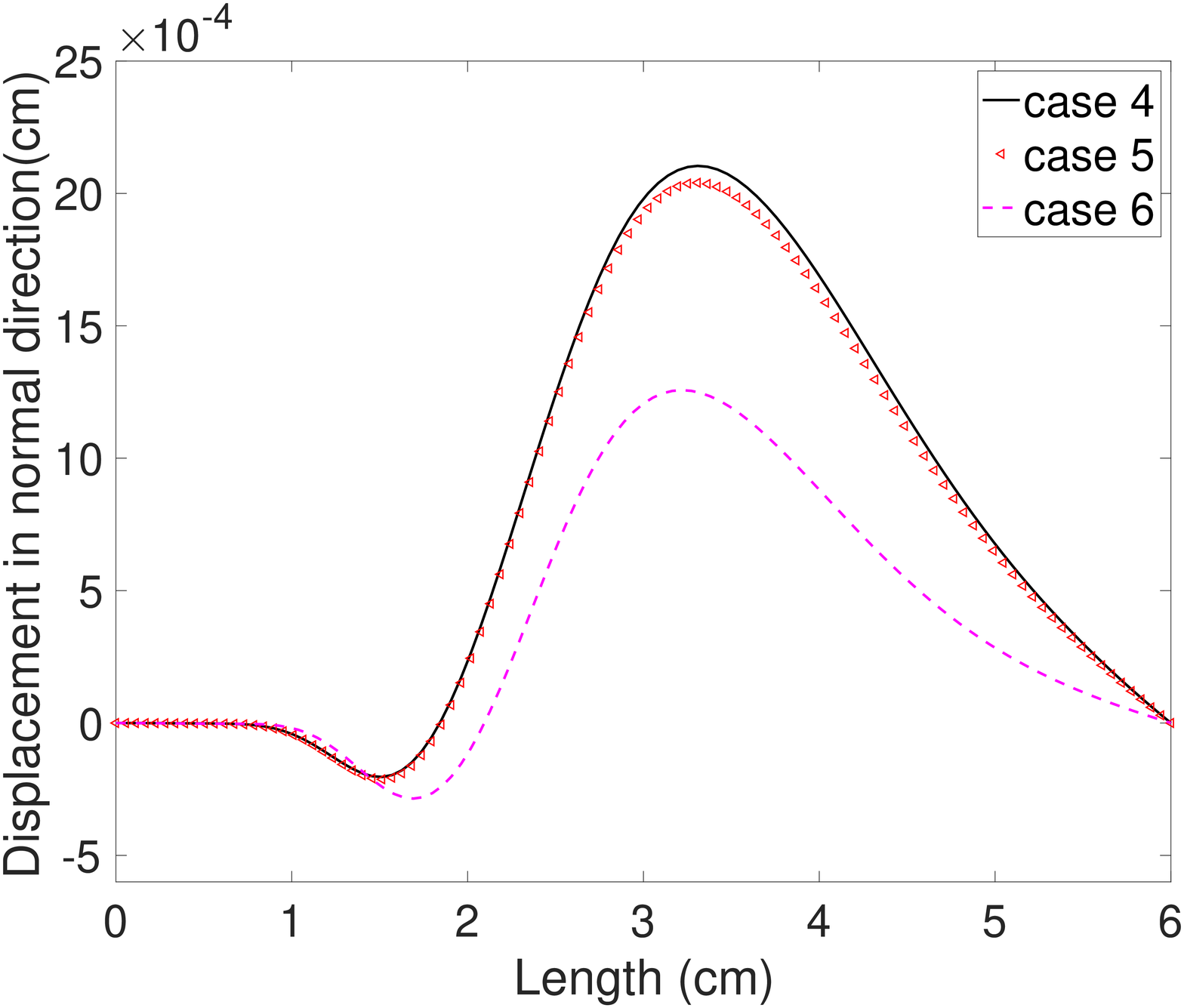}
	\includegraphics[trim=0 60 0 0,scale=0.149]{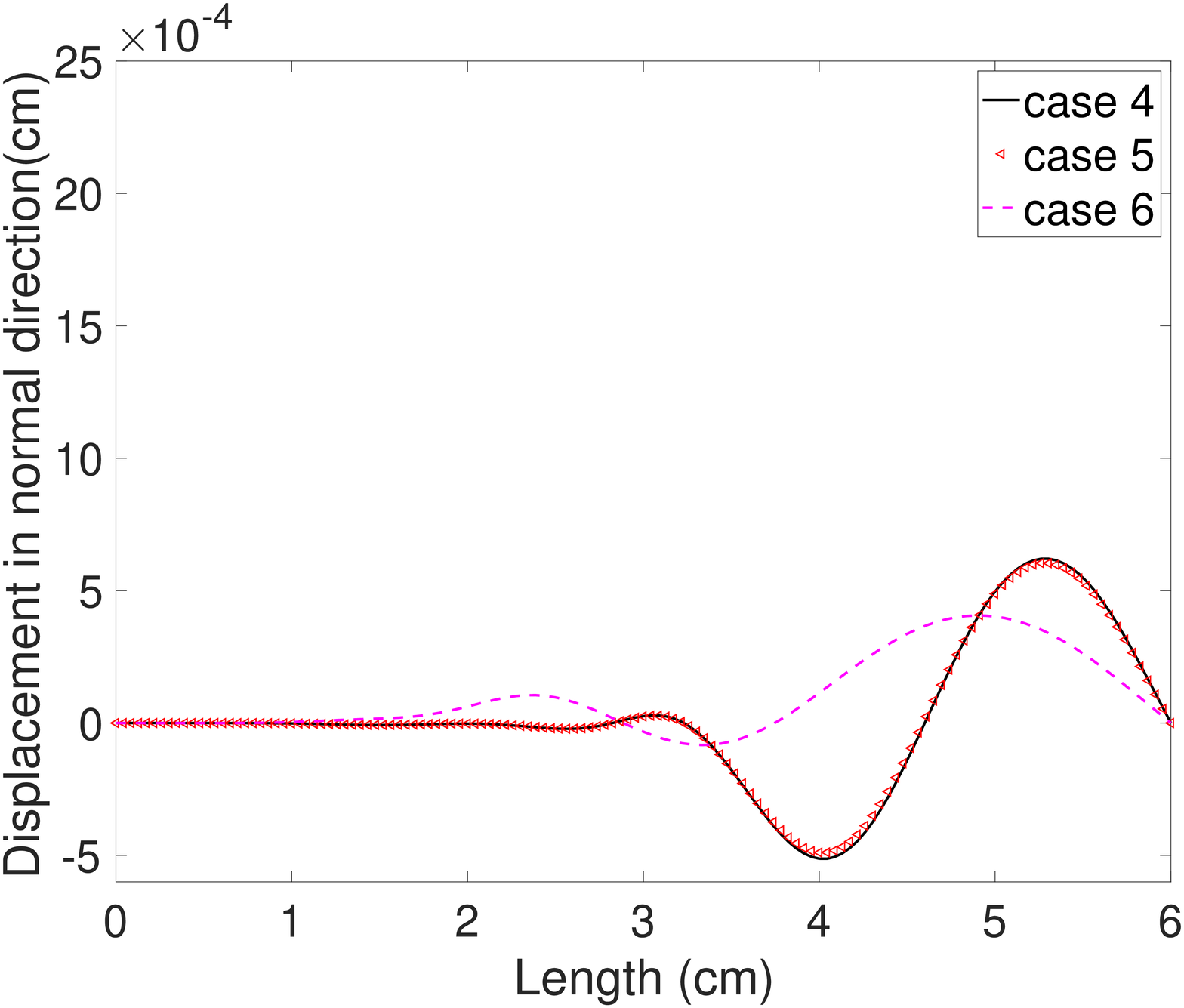}
	\caption{Displacement in the normal direction $\bbeta_{\star}\cdot \bn$ along the top arterial wall at time t=1.8 ms, t=3.6 ms, t=5.4 ms for case 4, case 5, and case 6.}
	\label{discompare2}
\end{figure}

\begin{figure}[ht!]
	\includegraphics[trim=0 60 0 0,scale=0.149]{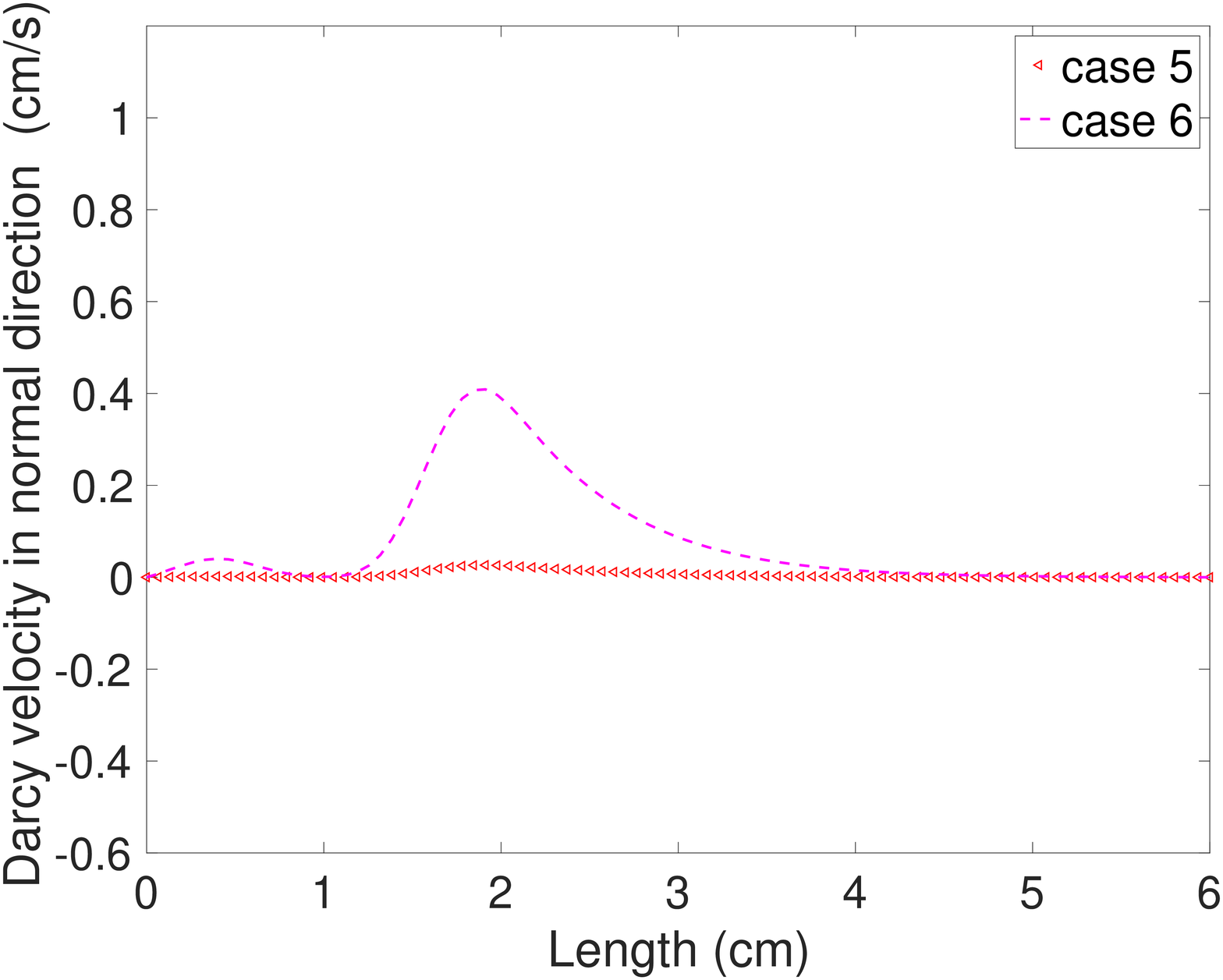}
	\includegraphics[trim=0 60 0 0,scale=0.149]{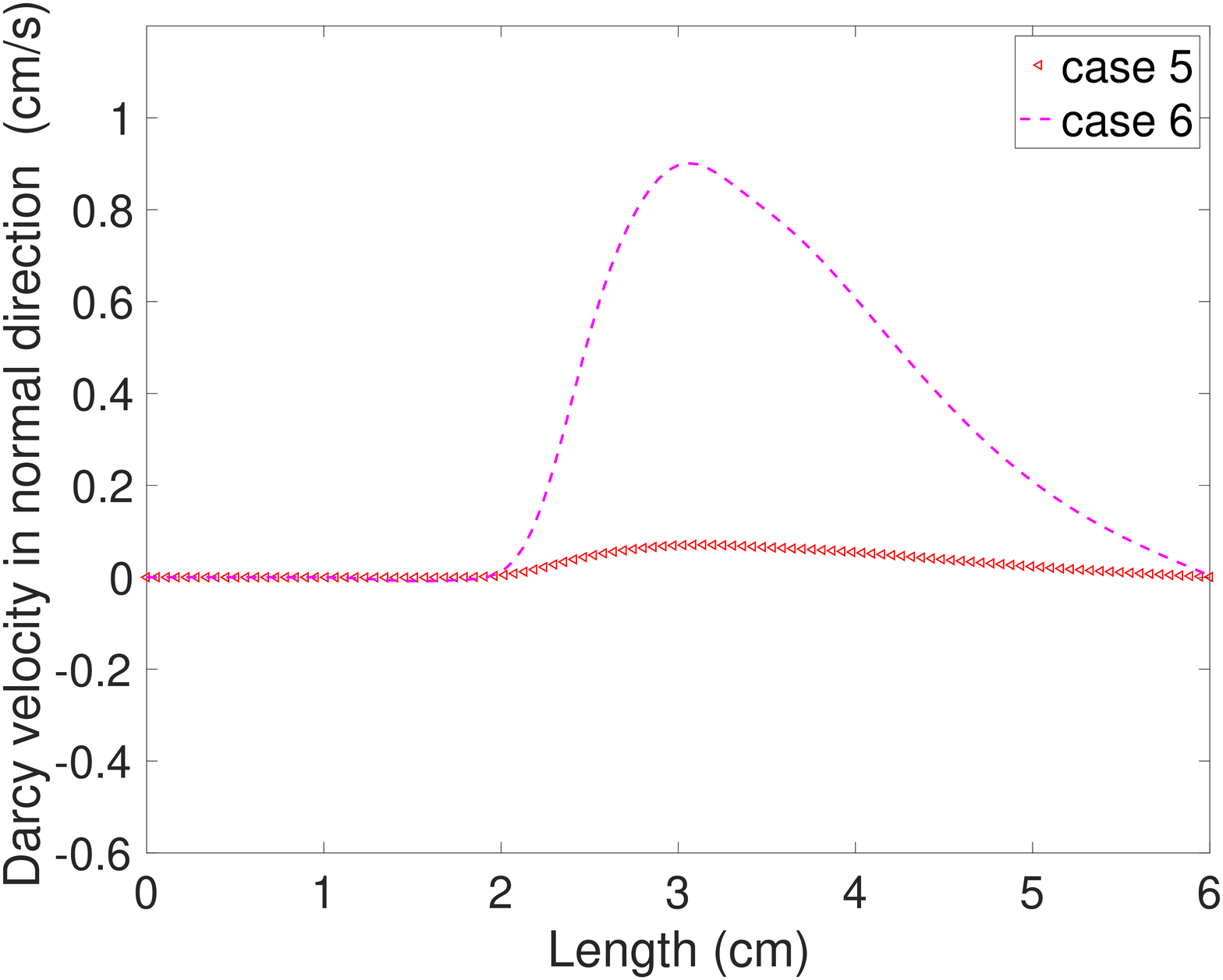}
	\includegraphics[trim=0 60 0 0,scale=0.149]{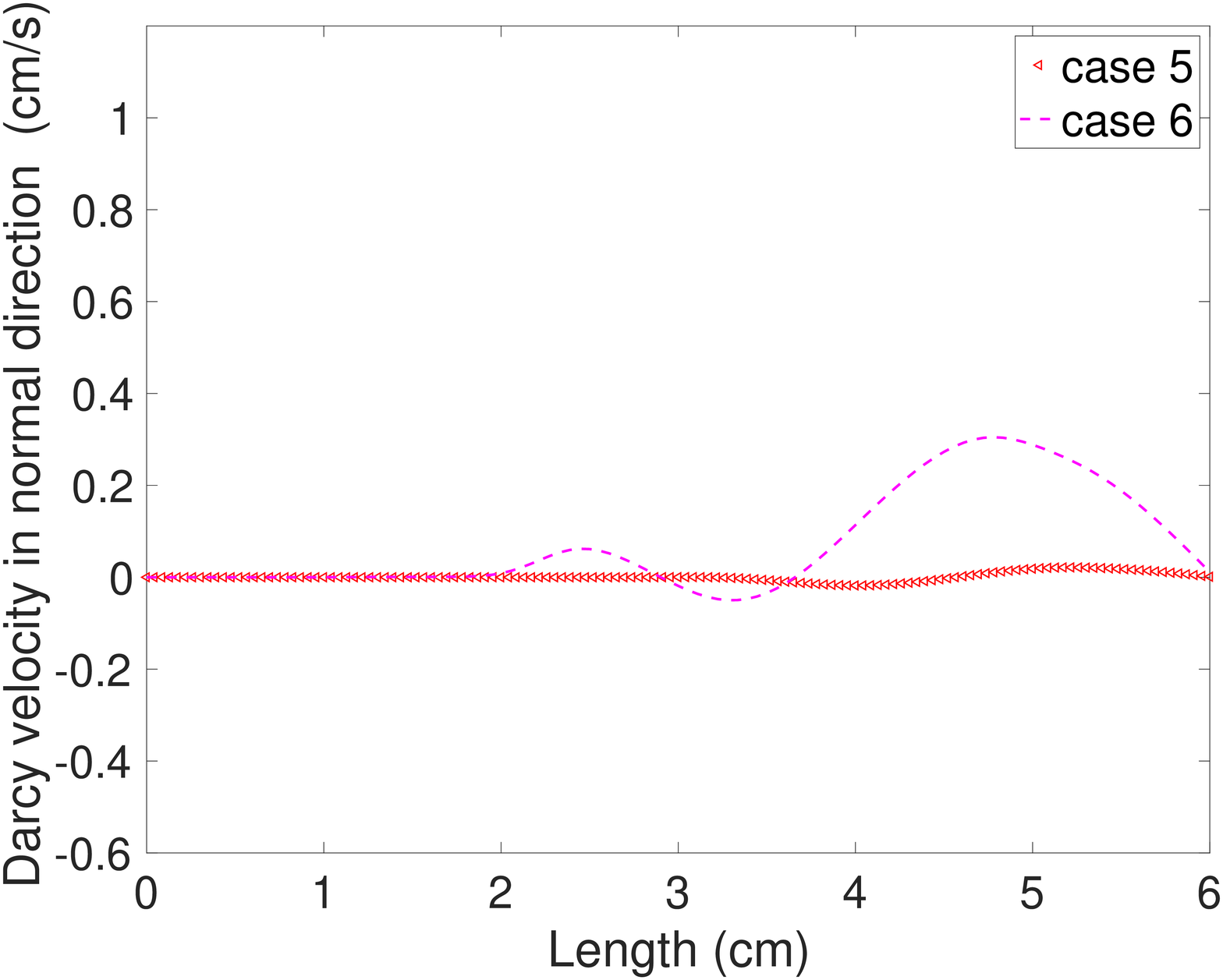}
	\caption{Darcy velocity in the normal direction $\bu_p\cdot \bn$ along the top arterial wall at time t=1.8 ms, t=3.6 ms, t=5.4 ms for case 5, case 6.}
	\label{upncompare2}
\end{figure}

In Fig.\ref{discompare2}, we show the displacement in the normal direction $\bbeta_{\star}\cdot \bn$ along the lumen. In Fig.\ref{upncompare2}, we present the filtration velocity for case 5 and case 6 at time $t=1.8, 3.6, 5.4$ ms. Overall, larger permeability corresponds to larger filtration velocities. 
\begin{figure}[ht!]
	\includegraphics[trim=0 60 0 0,scale=0.149]{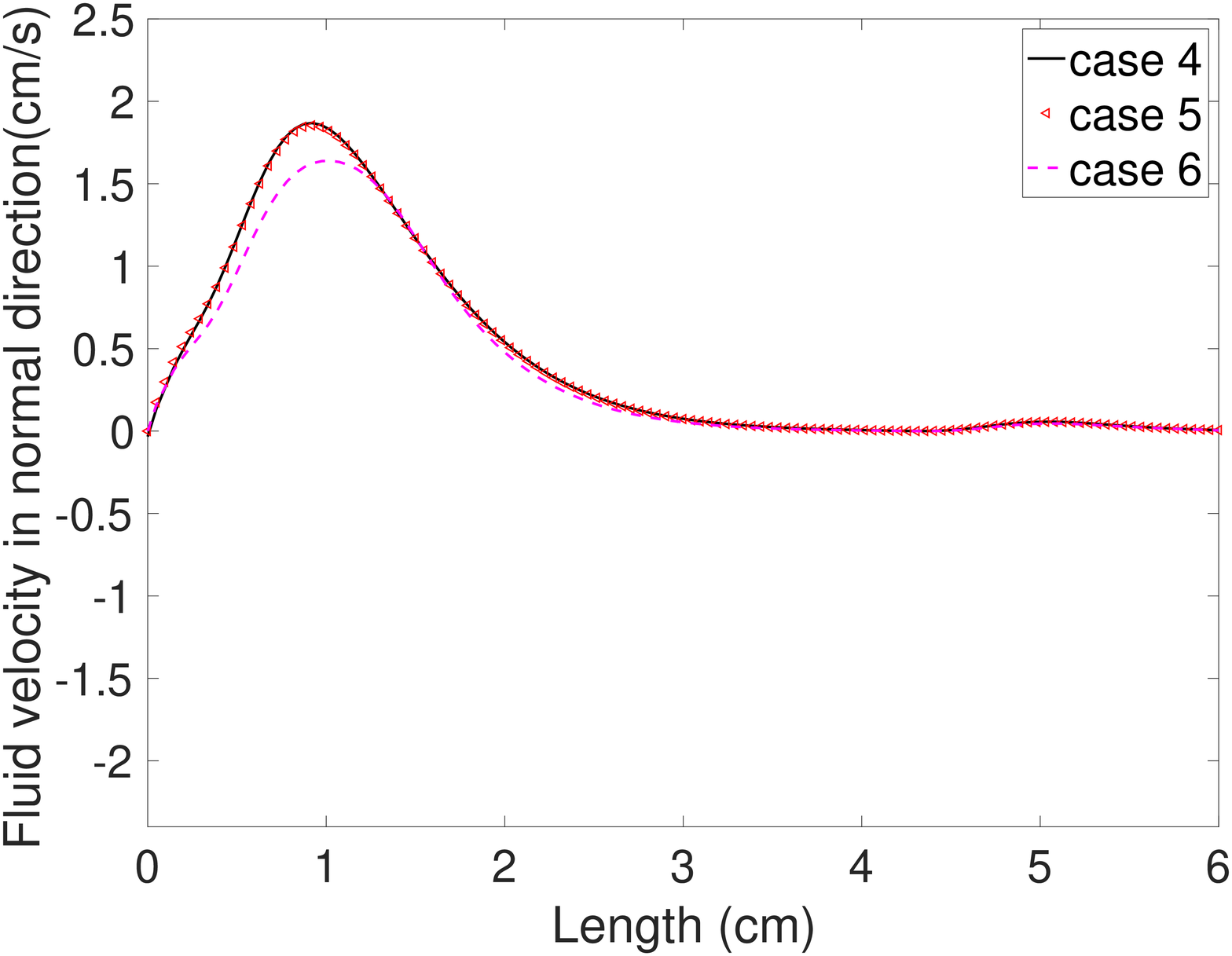}
	\includegraphics[trim=0 60 0 0,scale=0.149]{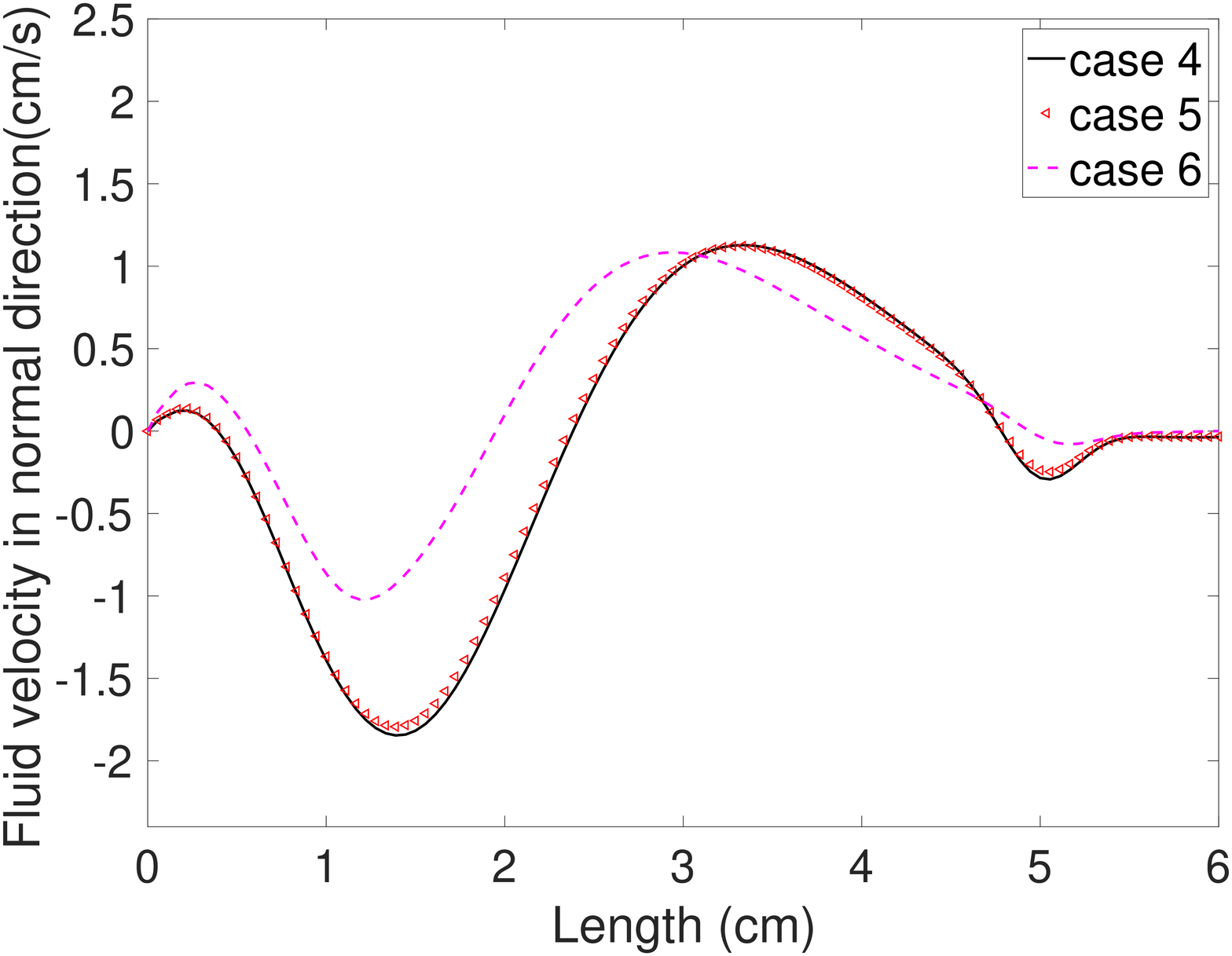}
	\includegraphics[trim=0 60 0 0,scale=0.149]{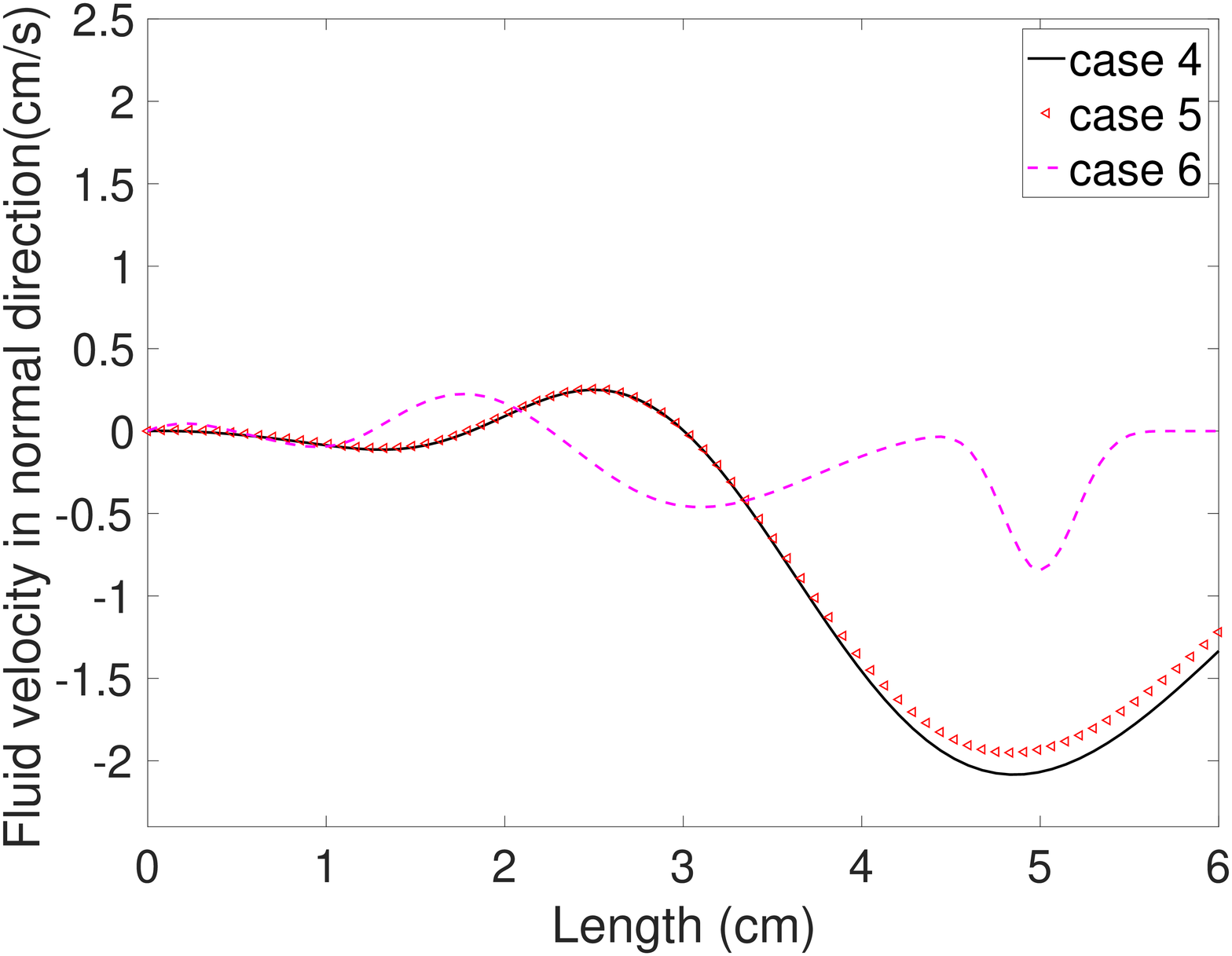}
	\caption{Velocity in the normal direction $\bu_f\cdot \bn$ along the top arterial wall at time t=1.8 ms, t=3.6 ms, t=5.4 ms for case 4, case 5, and case 6.}
	\label{utncompare2}
\end{figure}

\subsection{Comparison between Newtonian and non-Newtonian models}
\begin{figure}[ht!]
	\centering
	\includegraphics[trim=25 60 25 0,scale=0.149]{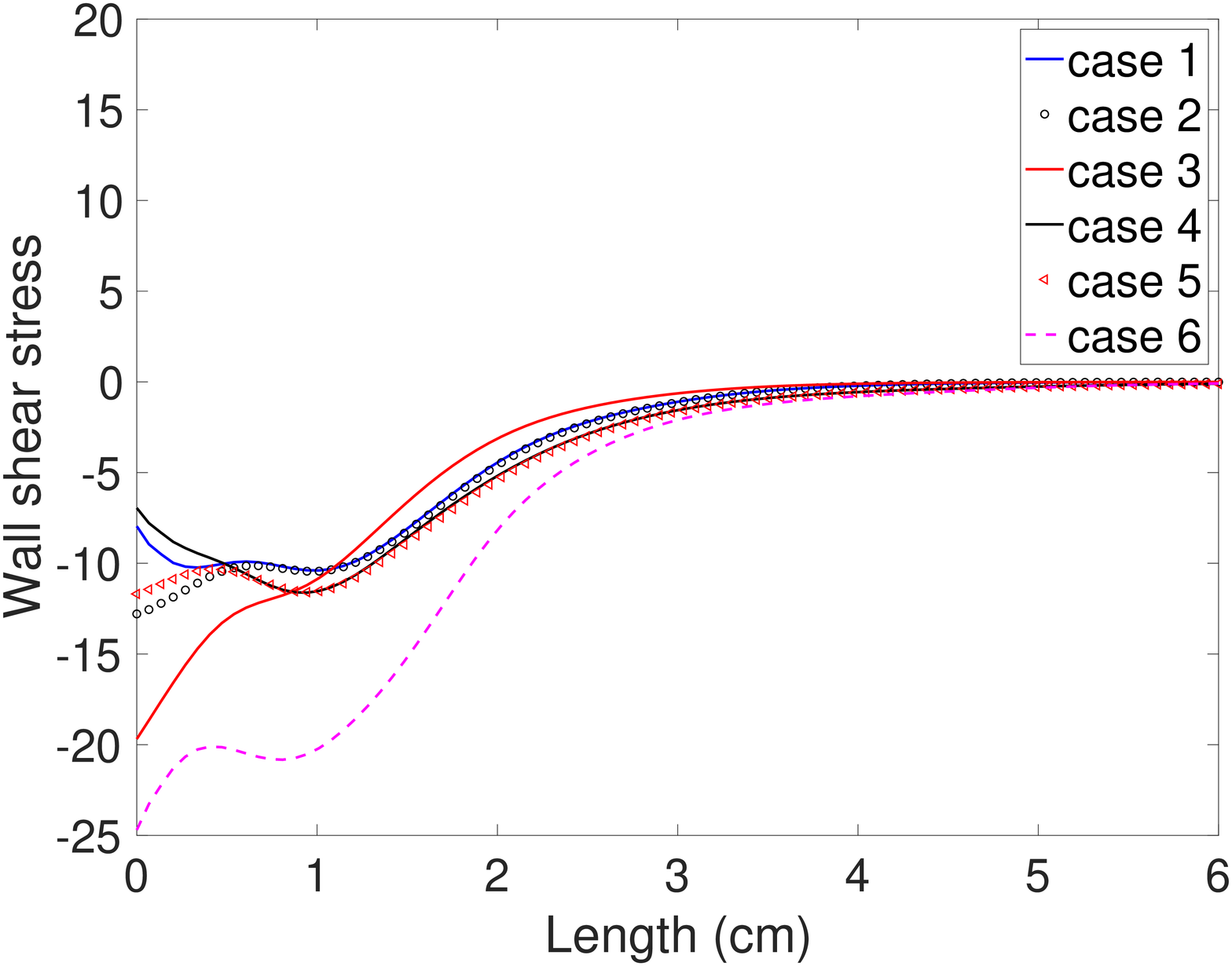}
	\includegraphics[trim=25 60 25 0,scale=0.149]{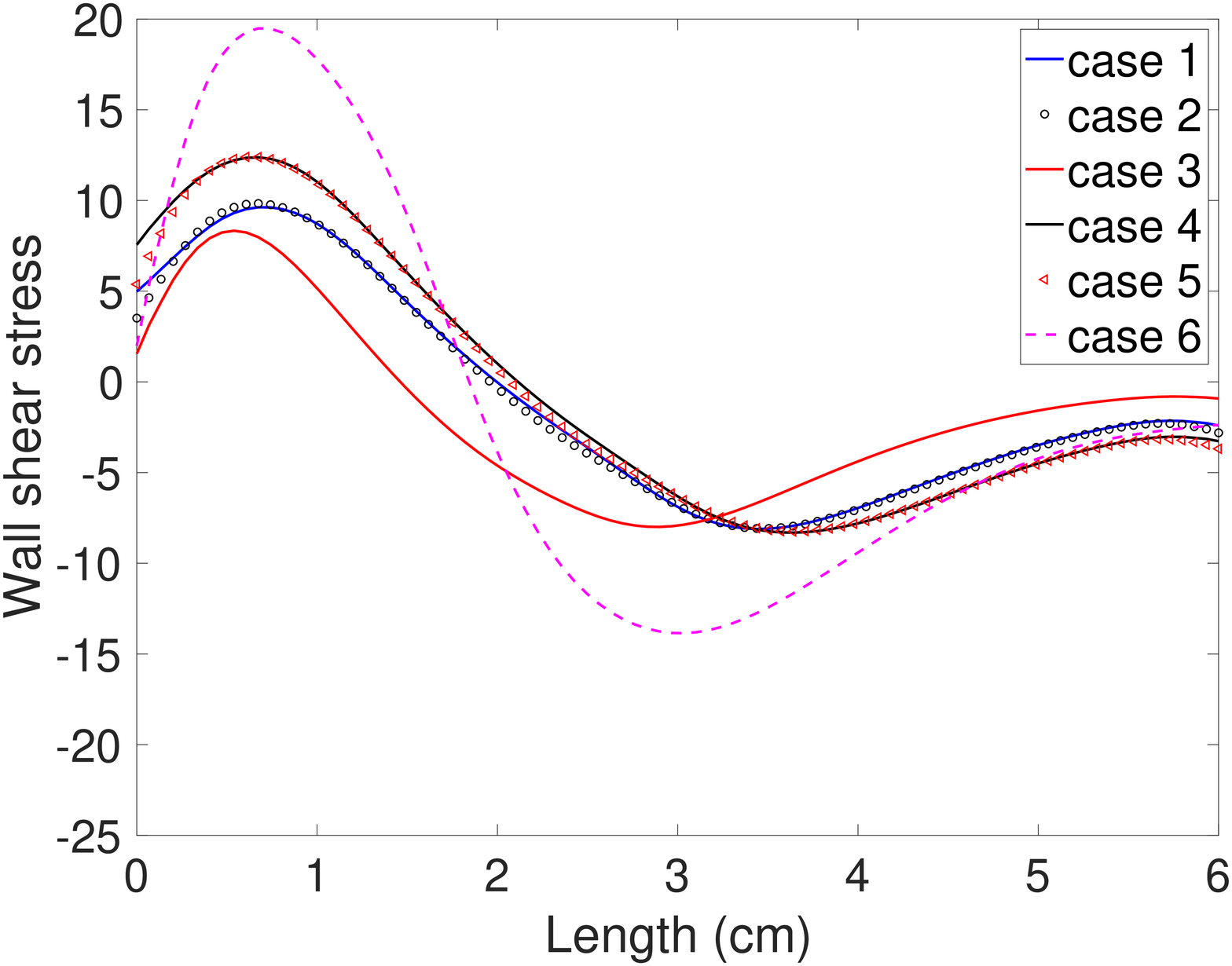}
	\includegraphics[trim=0 60 50 0,scale=0.149]{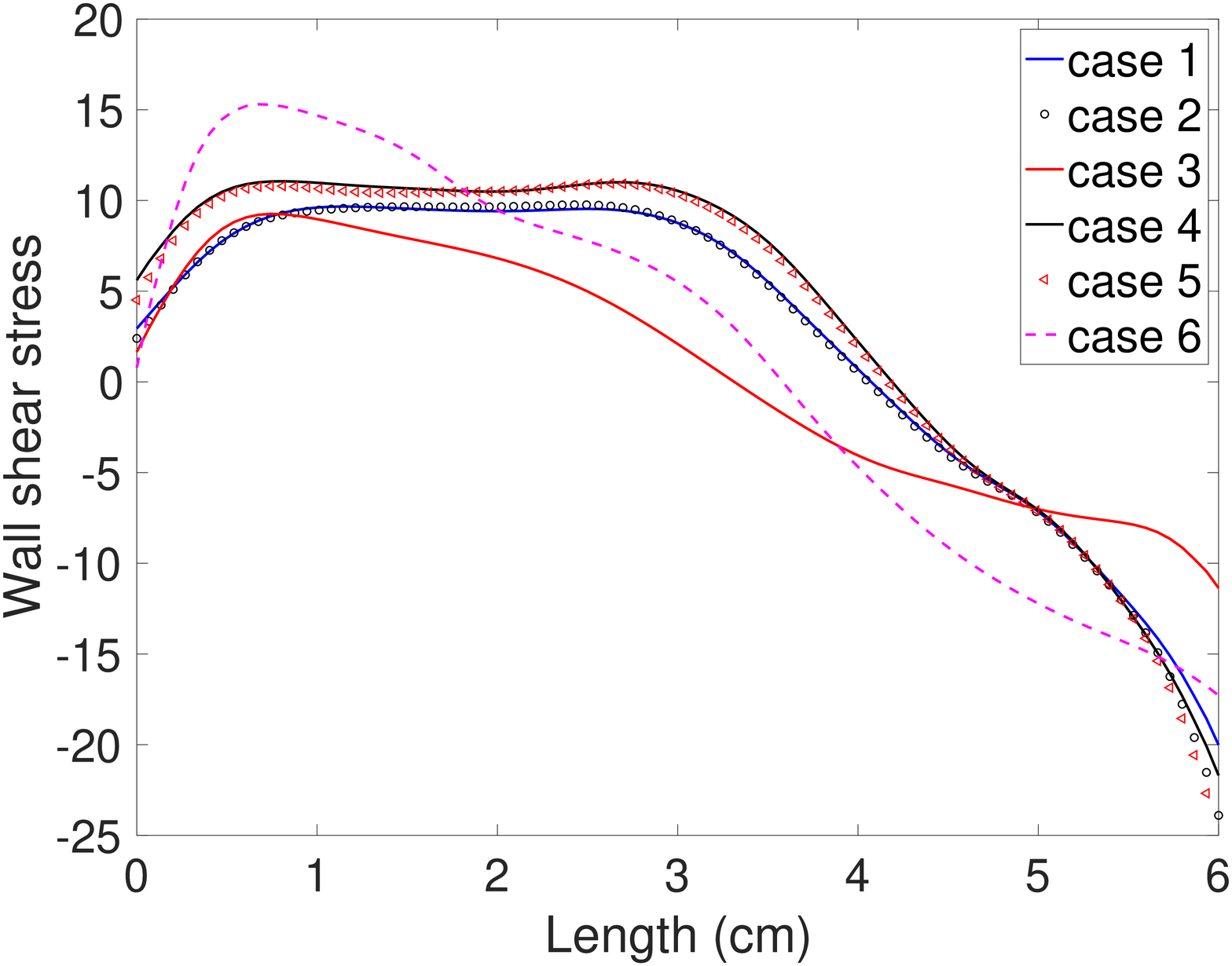}
	\caption{The shear stress $\bsi_f\bn\cdot \bt$ along the top arterial wall at time t=1.8 ms, t=3.6 ms, t=5.4 ms for all six cases.}
	\label{shearcompare3}
\end{figure}

In this section, we will focus on the comprehensive comparison between NSE/E and NSE/P models, and between Newtonian and non-Newtonian models. In Fig.\ref{shearcompare3}, we present the WSS at different time for all six cases above. To understand the effects of poroelasticity, we compare case 1 and case 2, and case 4 and case 5. We do not observe significant differences between these cases, therefore we conclude that for relatively small permeability cases, NSE/E model is sufficient to describe WSS. We then compare case 1 with case 4 and case 2 with case 6 to study non-Newtonian effects, we can draw the conclusion that Newtonian models would typically generate smaller WSS. For Newtonian fluids, larger permability would result in smaller WSS. However for non-Newtonian models, the peak of case 6 with larger permeability is the highest. So for non-Newtonian case with larger permeability, the WSS would also be larger. Finally, comparing case 3 with case 5, the permeability of case 3 is $2\times10^3$ times larger than case 5. However, the difference between case 3 and case 5 is not dramatic. Thus non-Newtonian properties have larger influence on WSS than permeability.
\begin{figure}[ht!]
	\centering
	\includegraphics[trim=25 60 25 0,scale=0.149]{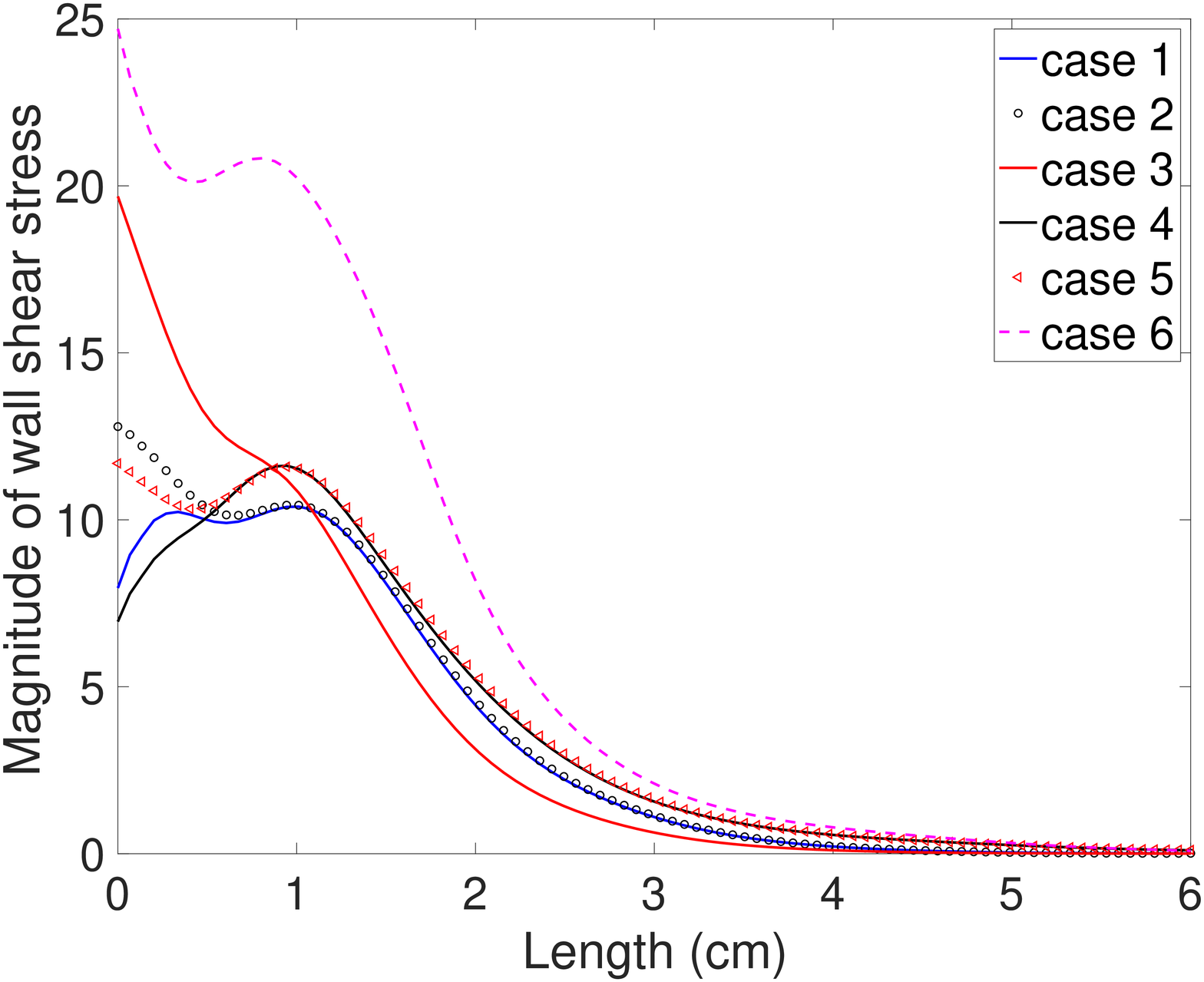}
	\includegraphics[trim=25 60 25 0,scale=0.149]{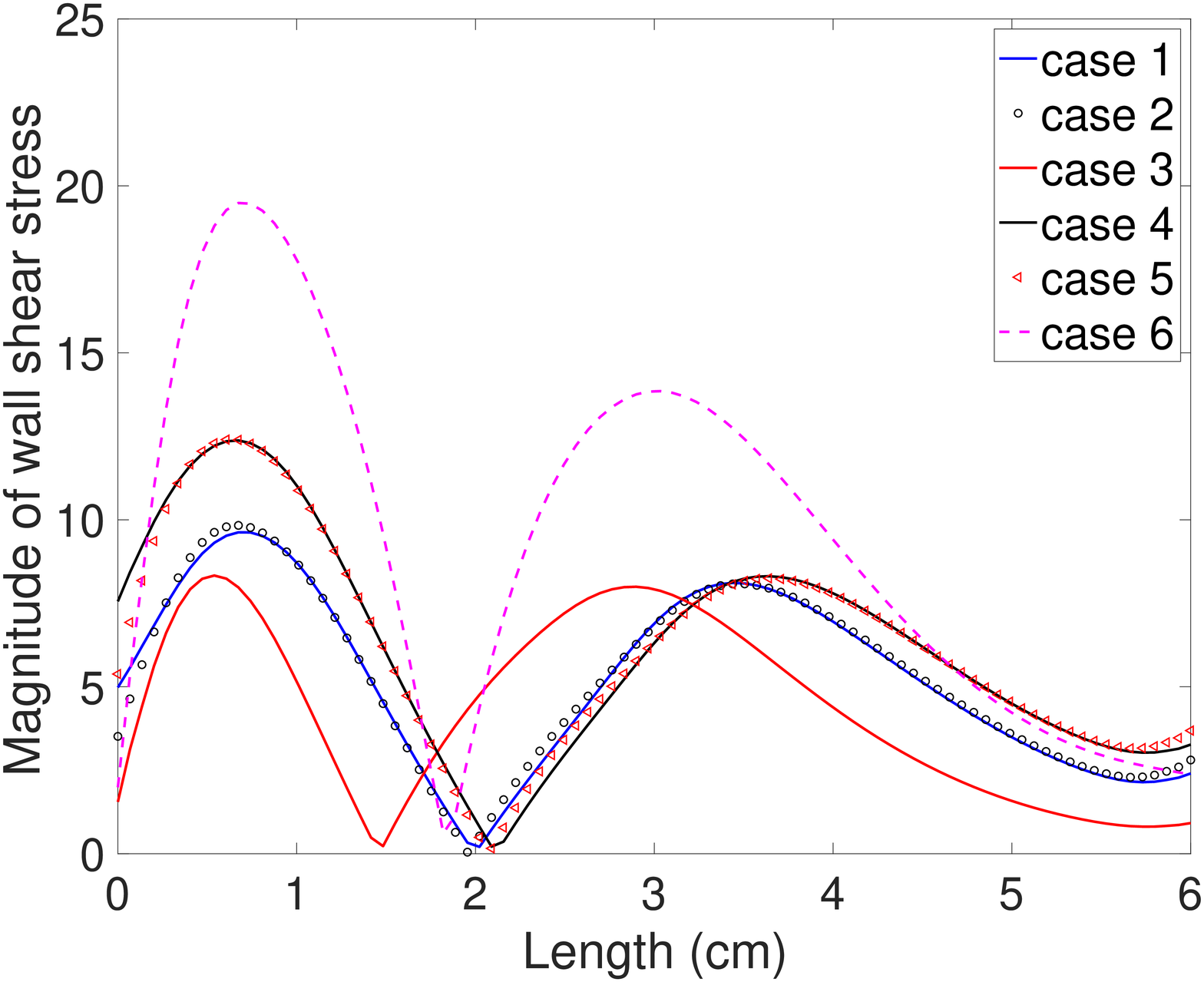}
	\includegraphics[trim=0 60 50 0,scale=0.149]{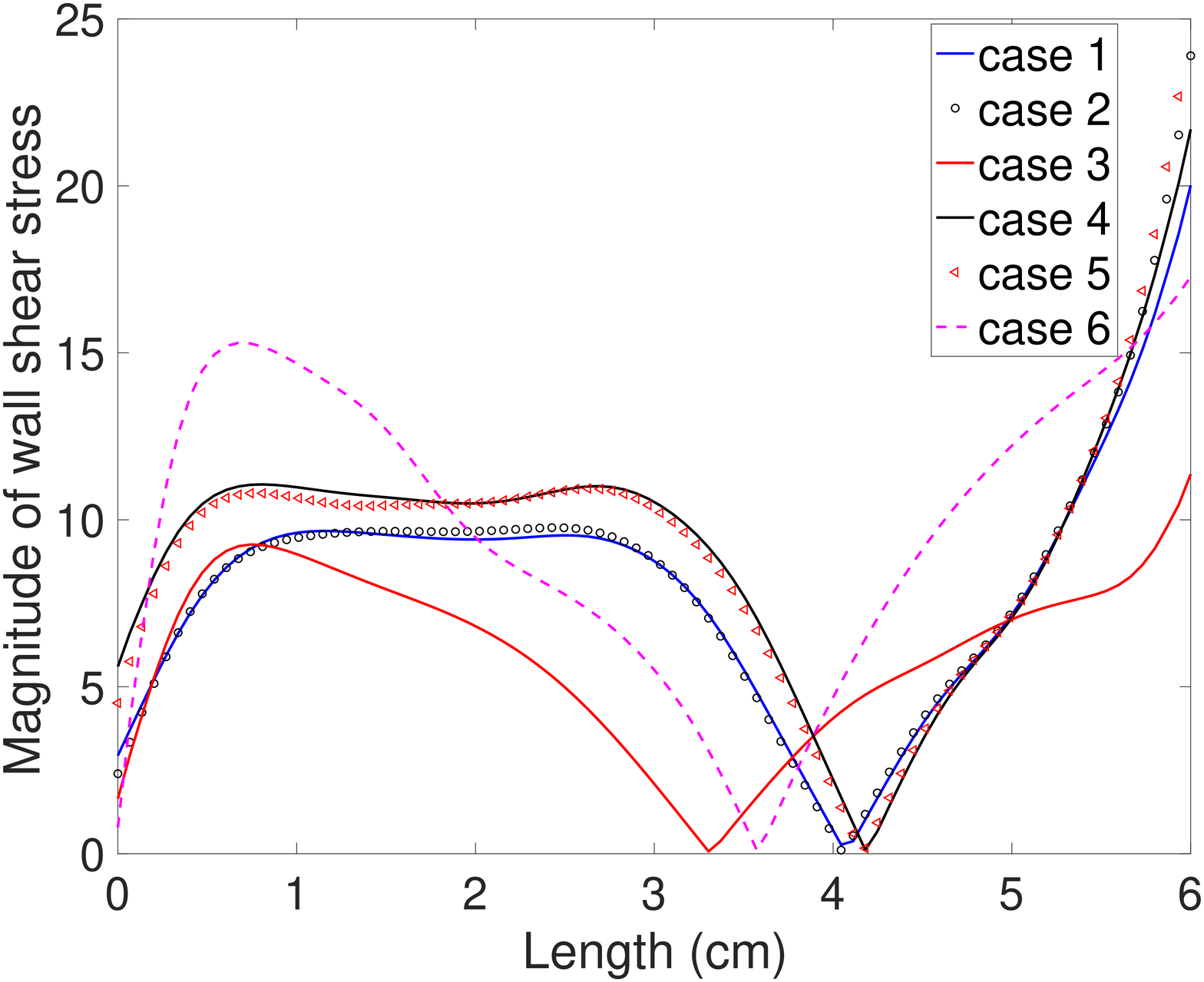}
	\caption{The shear stress $|\bsi_f\bn\cdot \bt|$ along the top arterial wall at time t=1.8 ms, t=3.6 ms, t=5.4 ms for all six cases.}
	\label{shearcompare33}
\end{figure}

\begin{figure}[ht!]
	\centering
	\includegraphics[trim=0 60 110 0,scale=0.149]{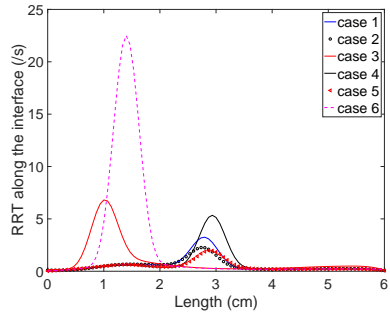}
	\caption{RRT along the top arterial wall  for all six cases.}
	\label{RRT3}
\end{figure}

In Fig.\ref{RRT3}, we present the RRT for all six cases. To study non-Newtonian effects, we compare case 2 and case 6, both of which are NSE/P model with the same permeability. We can see that Newtonian models would typically generate smaller RRT. As for effects of permeability, we can see that in both Newtonian and non-Newtonian fluids, larger permeability $\bK$ corresponds to larger RRT. Overall, we can see from the plots that RRT is greatly affected by different types of models and the permeabilities. 

\begin{figure}[ht!]
	\includegraphics[trim=0 60 0 0,scale=0.149]{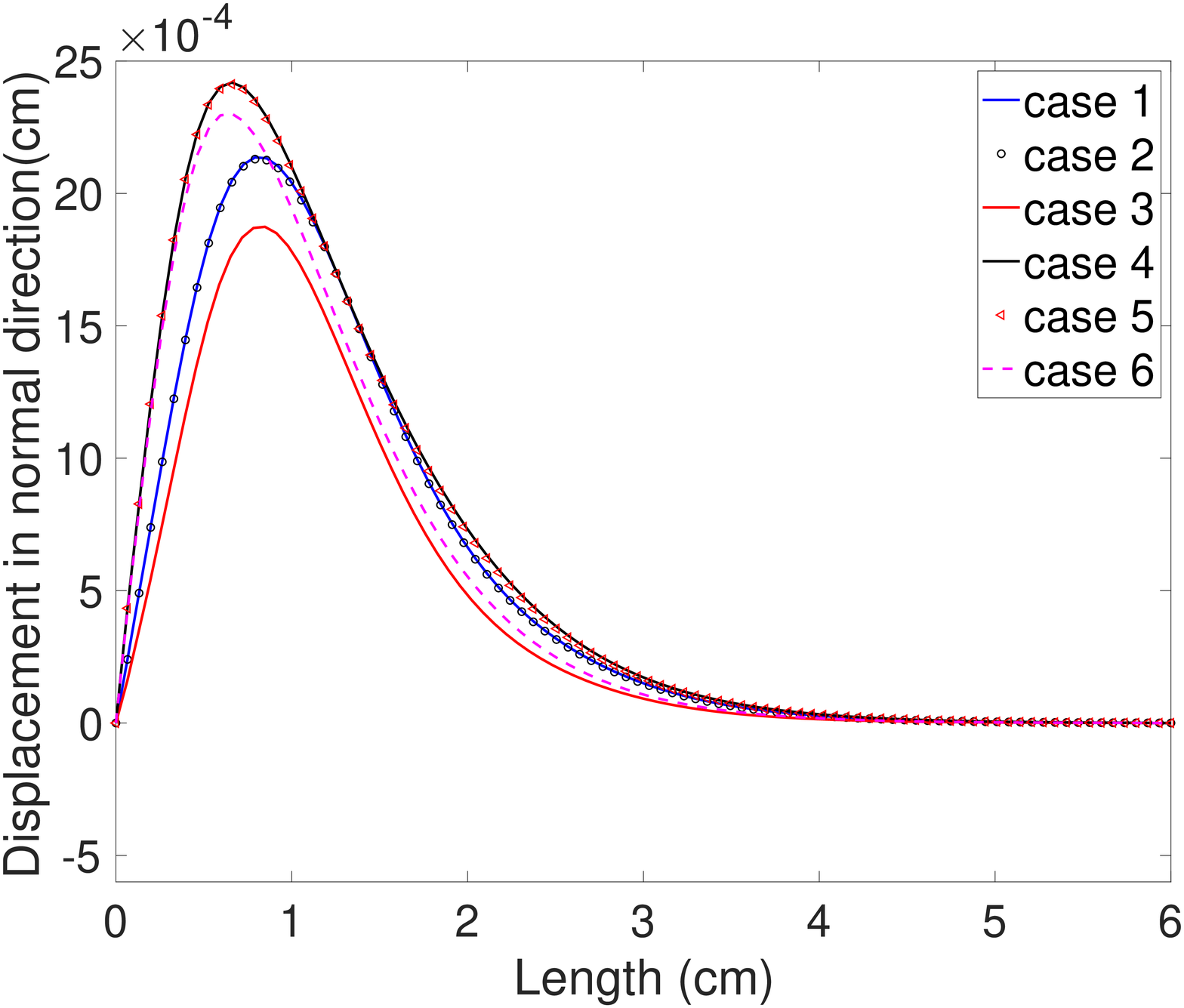}
	\includegraphics[trim=0 60 0 0,scale=0.149]{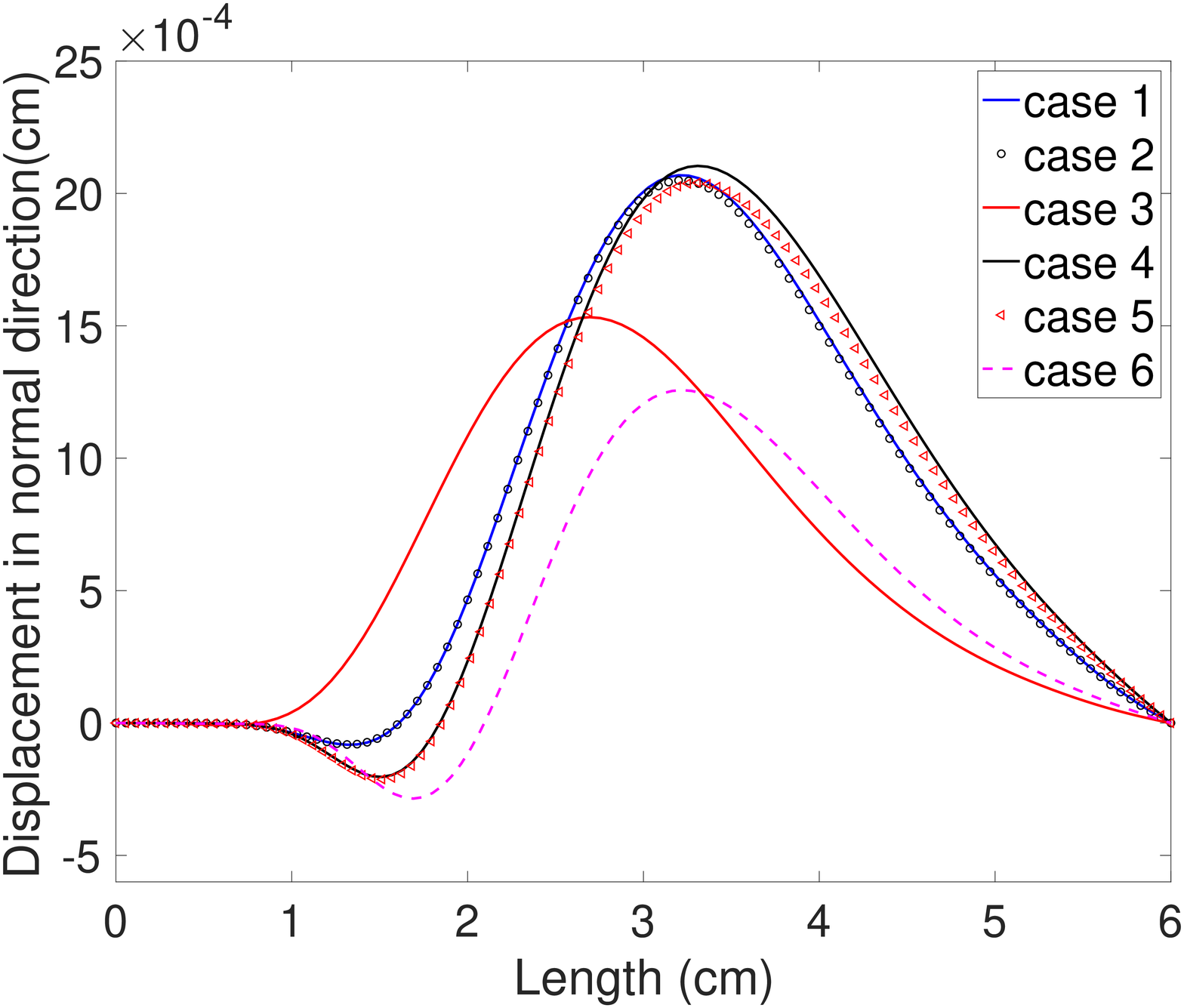}
	\includegraphics[trim=0 60 0 0,scale=0.149]{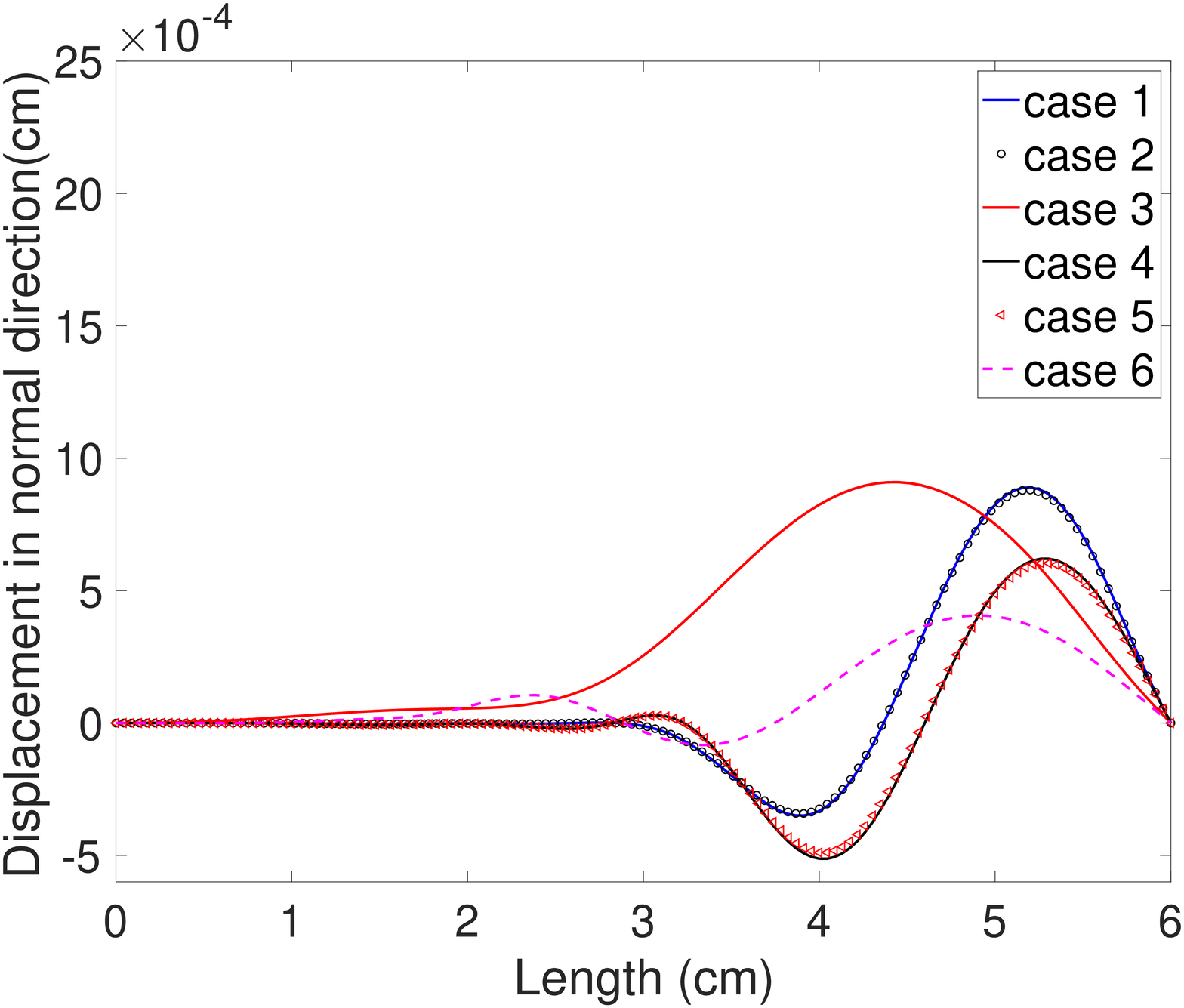}
	\caption{Displacement in the normal direction $\bbeta_{\star}\cdot \bn$ along the top arterial wall at time t=1.8 ms, t=3.6 ms, t=5.4 ms for all six cases.}
	\label{discompare3}
\end{figure}

\begin{figure}[ht!]
	\includegraphics[trim=0 60 0 60,scale=0.149]{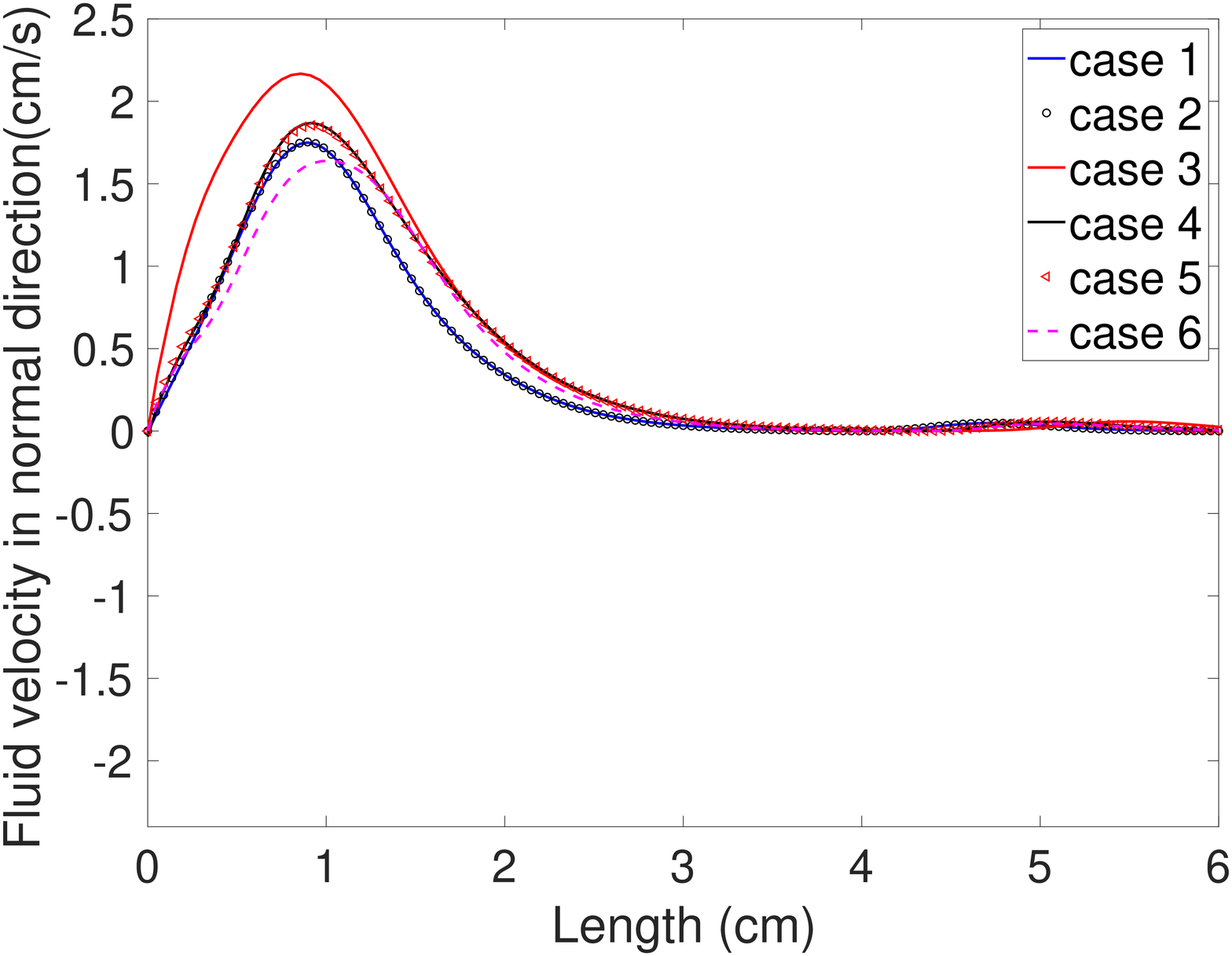}
	\includegraphics[trim=0 60 0 60,scale=0.149]{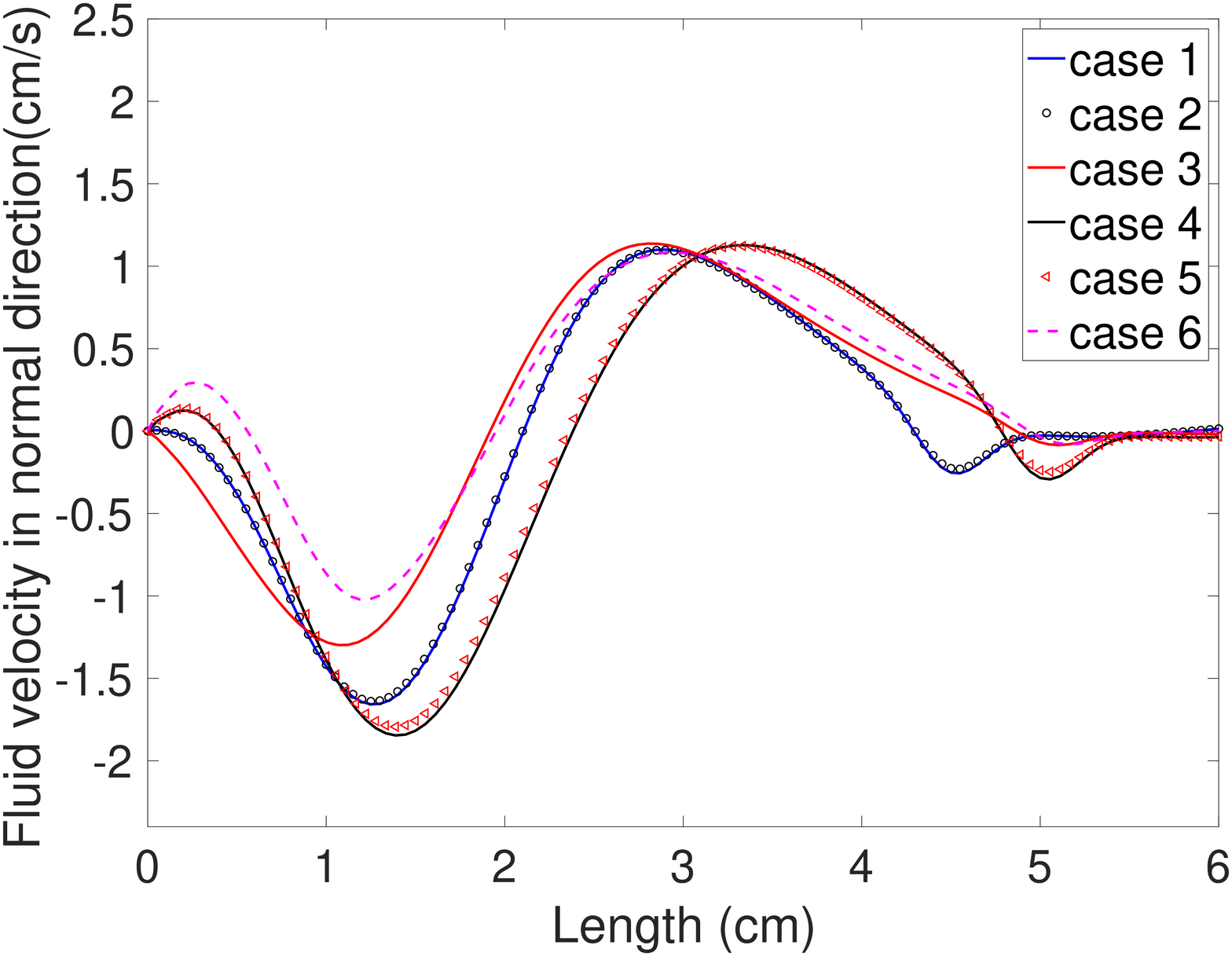}
	\includegraphics[trim=0 60 0 60,scale=0.149]{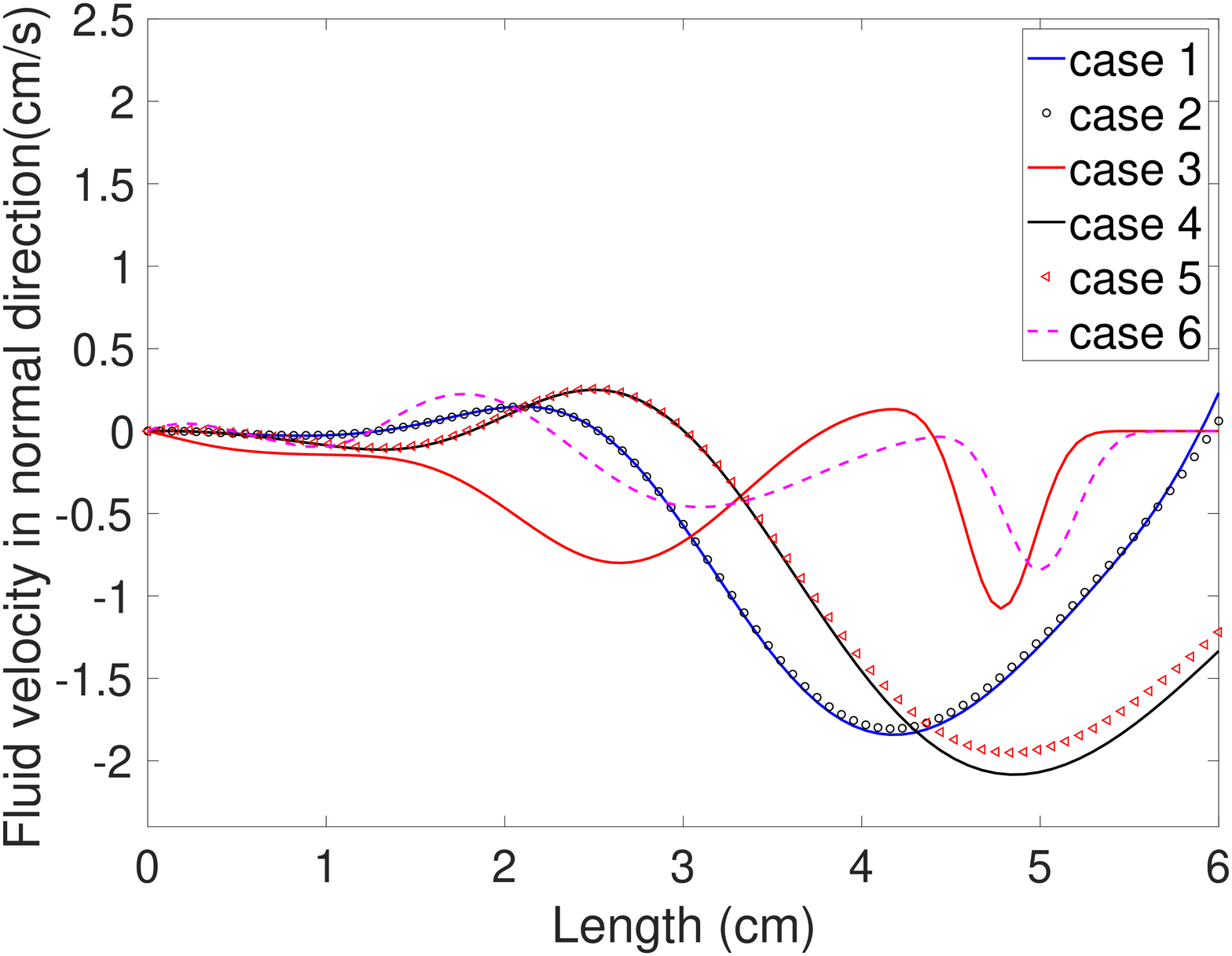}
	\caption{Velocity in the normal direction $\bu_f\cdot \bn$ along the top arterial wall at time t=1.8 ms, t=3.6 ms, t=5.4 ms for all six cases.}
	\label{utncompare3}
\end{figure}

\begin{figure}[ht!]
	\includegraphics[trim=0 60 0 0,scale=0.149]{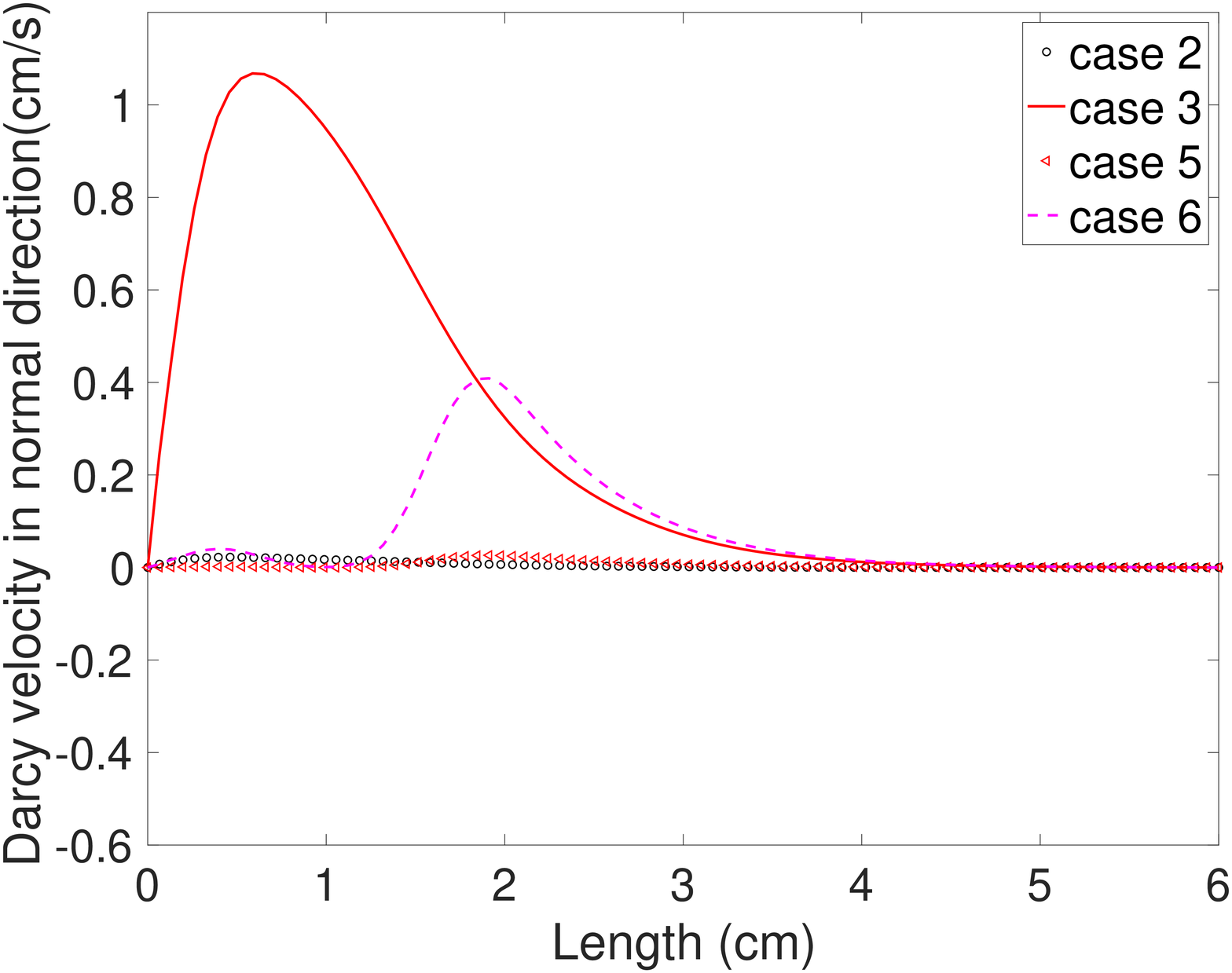}
	\includegraphics[trim=0 60 0 0,scale=0.149]{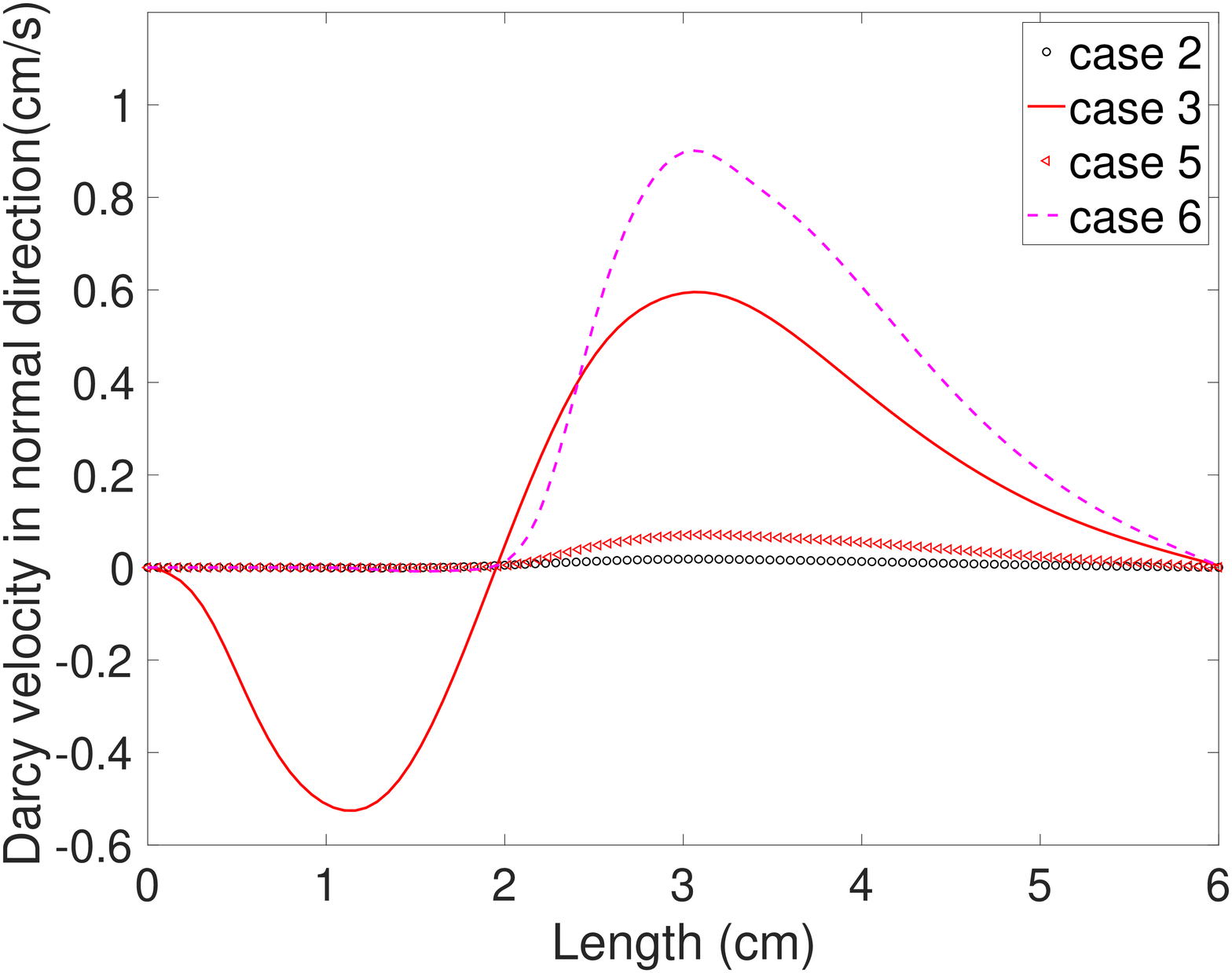}
	\includegraphics[trim=0 60 0 0,scale=0.149]{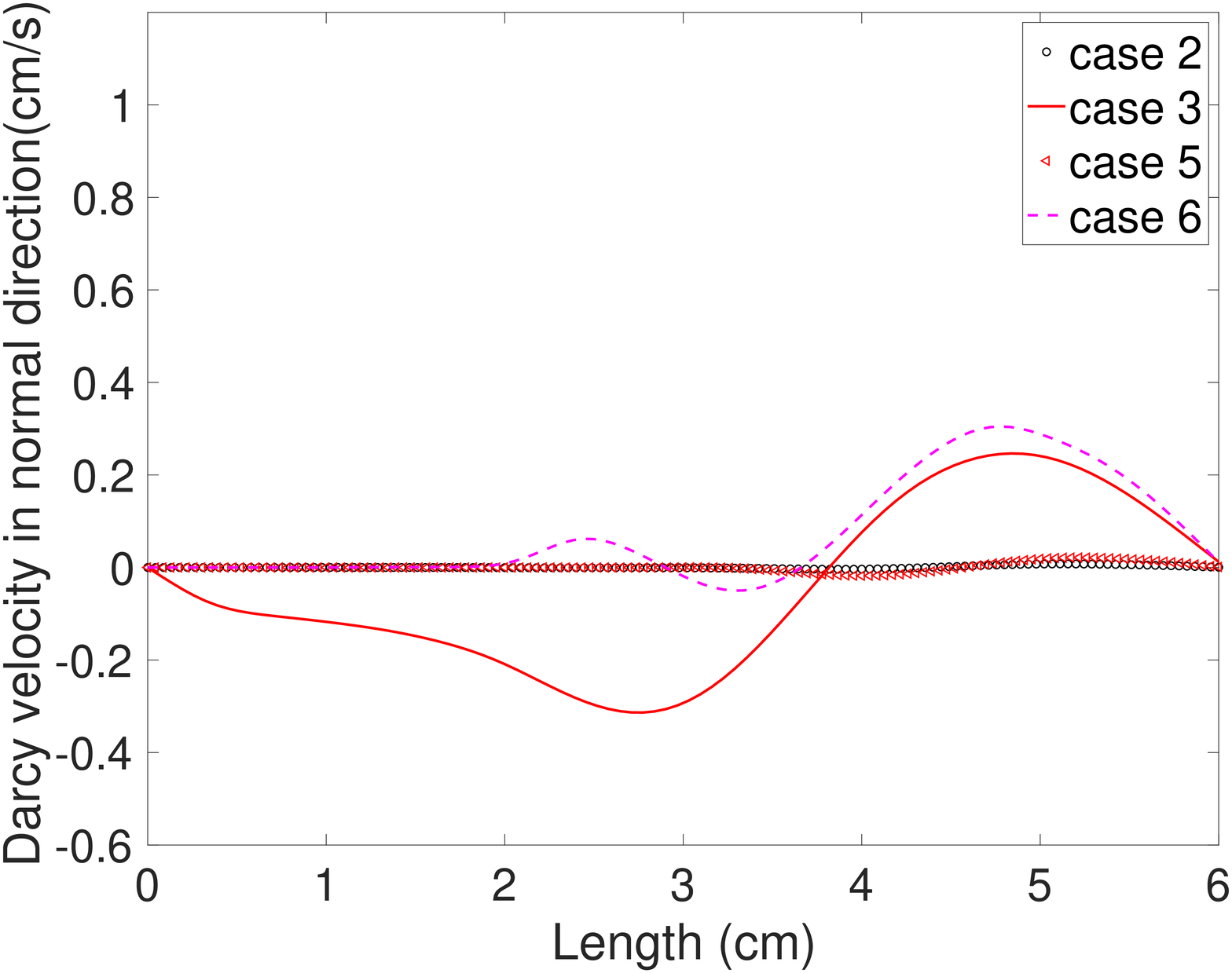}
	\caption{Darcy velocity in the normal direction $\bu_p\cdot \bn$ along the top arterial wall at time t=1.8 ms, t=3.6 ms, t=5.4 ms for case 2, case 3, case 5 and case 6.}
	\label{upncompare3}
\end{figure}

In Fig.\ref{discompare3}, we present the displacement in normal direction $\bbeta_{\star}\cdot \bn$ for all six cases above. We draw the conclusion that larger permeability would result in smaller displacement in the normal direction for both Newtonian and non-Newtonian models. In particular, non-Newtonian cases would generate smaller normal displacement.  
From Fig.\ref{utncompare3}, the non-Newtonian property of fluids and poroelasticity don't affect much the fluids velocity in normal directions along the interface. 
In Fig.\ref{upncompare3}, we conclude that permeability $\bK$ affects the filtration velocity greatly. At time $t=1.8$ ms, the peak of case 3 is the highest since it has the largest permeability $\bK$. At time $t=3.6, 5.4$ ms, even though the permeability of case 3 is $10^2$ times larger than case 6, the filtration velocity are still comparable, which means the non-Newtonian property has a large impact on the filtration velocity. And it's important to include the non-Newtonian characteristic of blood.

\subsubsection{Conclusions}
Our comparisons show that along the interface:
\begin{itemize}
\item For the velocity and pressure fields, NSE/P model with larger permeability present smaller quantities. Similarly, non-Newtonian models tend to generate smaller velocity, pressure and deformations. And the fluids would be more viscous.  
\item Small differences between NSE/E and NSE/P models are found in the quantification of WSS. Great differences are found between non-Newtonian and Newtonian models. Non-Newtonian models tend to generate larger WSS.
\item RRT is affected dramatically by both non-Newtonian behavior of blood flow and poroelasticity of the vessel structures. To accurately compute RRT, we believe it's important to consider not only the non-Newtonian property but also the structure characteristics.
\item No significant differences are found in the fluids velocities in normal direction.
\item Displacement in normal direction is affected greatly by non-Newtonian property, especially for NSE/P models. And non-Newtonian models with large permeability would result in smaller normal displacement.  
\item Both non-Newtonian and poroelasticity make a difference to the filtration velocity. Non-Newtonian models with larger permeability would generate larger filtration velocity.  
\end{itemize}

\subsection{Stenosis model}\label{stenosis}
In this section, we change our computational geometry to an ideal stenosis region shown in Fig.\ref{mesh11} part (b). Stenosis is characterized by local arterial narrowing which is initially due to the deposition of lipid, cholesterol and some other substances on the endothelium \cite{rabby2014pulsatile, razavi2011numerical}. It is important to understand the non-Newtonian fluid dynamical properties of the blood flow using coupled model in the stenotic cases. Abrupt geometrical changes along the arterial wall may cause flow separation and appearance of the recirculation zone at the post-stenotic region \cite{uren1994relation}. For this stenotic model, we use the following cosine function to describe the interfaces: for the top interface,
\begin{equation}
y=
\begin{cases}
0.5,& \,\ \text{for} \,\ 0\leq x\leq2\text{ and }4\leq x\leq6; \\
0.4+0.1\cdot\cos(\pi(x-2)),& \,\ \text{for} \,\ 2\leq x\leq4,
\end{cases}
\end{equation}
for the bottom interface, symmetrically, we use
\begin{equation}
y=
\begin{cases}
-0.5,& \,\ \text{for} \,\ 0\leq x\leq2\text{ and }4\leq x\leq6; \\
-0.4+0.1\cdot\cos(\pi(x-3)),& \,\ \text{for} \,\ 2\leq x\leq4.
\end{cases}
\end{equation}
Note that for the plots along the interface of this section, the x-axis would be $[0,6.048]$ cm.

We still use the same non-linear viscosity in the fluid and structure regions as in equation (\ref{Non-New Viscosity}) as well as the same inflow/outflow boundary conditions shown in section \ref{sec:model-problem}. In this section, we want to investigate how the build-up area affects the non-Newtonian blood rheology. And in the mean time, we will still focus on the difference between NSE/P and NSE/E models. In addition, we add the variation of Lam\'{e} coefficients to fit the stenosis case better. To do so, we introduce the following non-Newtonian cases:
\begin{itemize}
\item case 7: NSE/E model with constant Lam\'{e} coefficients pair shown in Table. \ref{T2} for structures;
\item case 8: NSE/E model with piecewise contant Lam\'{e} coefficients pair;
\item case 9: NSE/P model with constant Lam\'{e} coefficients pair shown in Table. \ref{T2} and constant permeability $\bK=diag(1,1)\times 10^{-9}$ for structures; 
\item case 10: NSE/P mode with piecewise constant Lam\'{e} coefficients pair and permeability.
\end{itemize}

We start by describing the piecewise constant Lam\'{e} coefficients in details. For the Lam\'{e} coefficients in case 8 and case 10, we still adopt $\lambda_\star=4.28\times 10^{6}$ and $\mu_\star=1.07\times 10^{6}$ in the non-stenosis area, namely when $x\in [0,6], y\in[0.5,0.6]$ or $y\in [-0.6,-0.5]$. While for the stenosis area, we keep the same Poisson's ratio $\tilde{\nu}$ but decrease the Young's modulus $E$ from $2.996\times10^{6}$ into $2.996\times10^{4}$. For case 8 and case 10, we use the following Lam\'{e} coefficients for the stenosis area only:
\begin{equation}
	\lambda_\star(x,y)=\begin{cases}
	4.28\times 10^{4},& \,\ \text{for} \,\ 2\leq x\leq4 \text{ and } 0.4+0.1\cdot\cos(\pi(x-2))\leq y\leq 0.5; \\
	4.28\times 10^{4},& \,\ \text{for} \,\ 2\leq x\leq4 \text{ and } -0.5\leq y\leq-0.4+0.1\cdot\cos(\pi(x-3));
	\end{cases}
\end{equation}
\begin{equation}
\mu_\star(x,y)=\begin{cases}
1.07\times 10^{4},& \,\ \text{for} \,\ 2\leq x\leq4 \text{ and } 0.4+0.1\cdot\cos(\pi(x-2))\leq y\leq 0.5; \\
1.07\times 10^{4},& \,\ \text{for} \,\ 2\leq x\leq4 \text{ and } -0.5\leq y\leq-0.4+0.1\cdot\cos(\pi(x-3)); 
\end{cases}
\end{equation}
Normally the build-up stenosis area would be not only softer but also more permeable. We will use the following piecewise constant function for case 10 for the stenotic area:
\begin{equation}\label{K1}
\bK(x,y)=\begin{cases}
diag(1,1)\times 10^{-7},& \,\ \text{for} \,\ 2\leq x\leq4 \text{ and } 0.4+0.1\cdot\cos(\pi(x-2))\leq y\leq 0.5; \\
diag(1,1)\times 10^{-7},& \,\ \text{for} \,\ 2\leq x\leq4 \text{ and } -0.5\leq y\leq-0.4+0.1\cdot\cos(\pi(x-3)); 
\end{cases}
\end{equation}
And for the remainning structure regions in case 10, we still use $\bK=diag(1,1)\times 10^{-9}$.
\begin{figure}[ht!]
	\includegraphics[trim=0 0 0 30, scale=0.28]{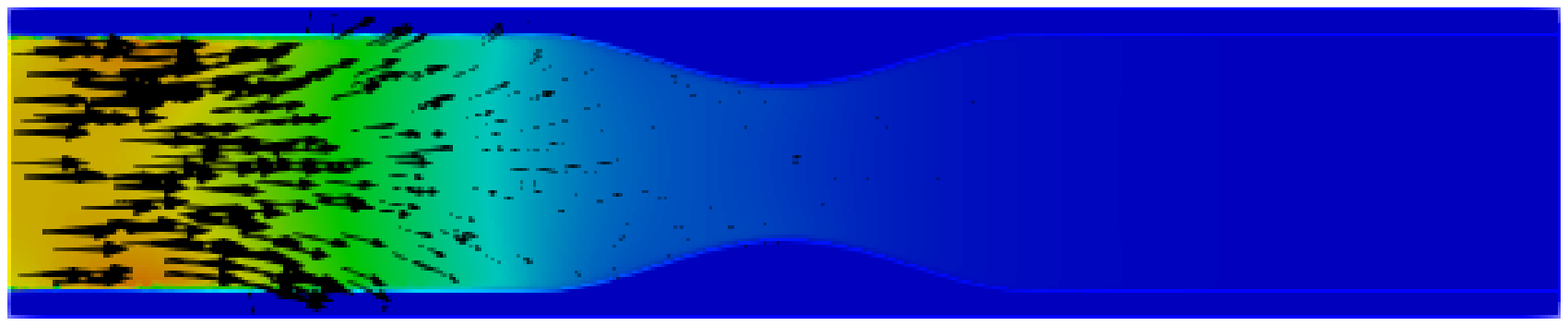}
	\includegraphics[trim=0 0 0 30, scale=0.28]{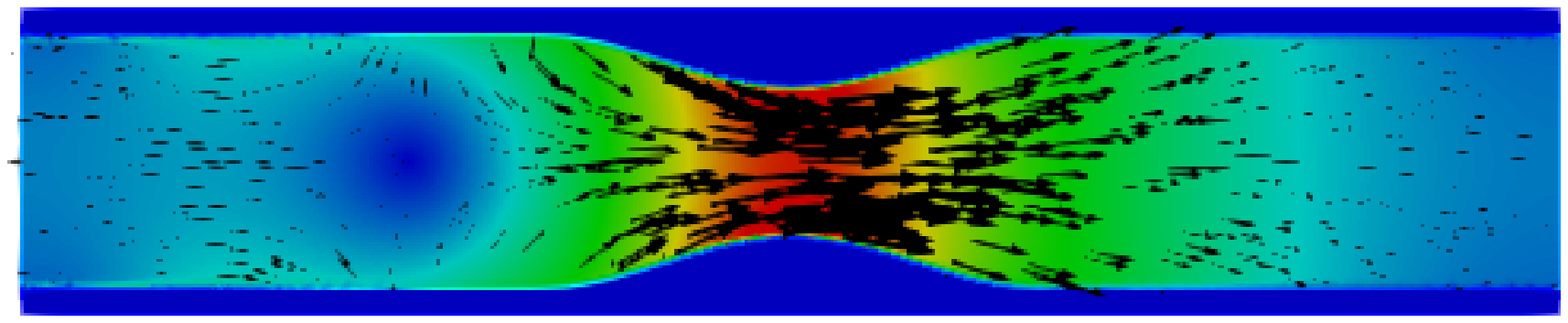}
	\includegraphics[trim=0 0 0 30, scale=0.28]{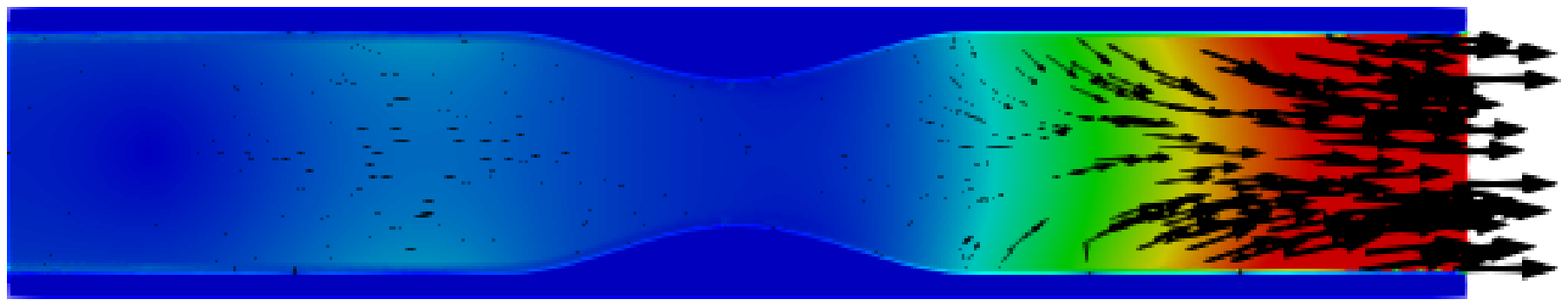}
	\includegraphics[trim=0 0 0 0,scale=0.28]{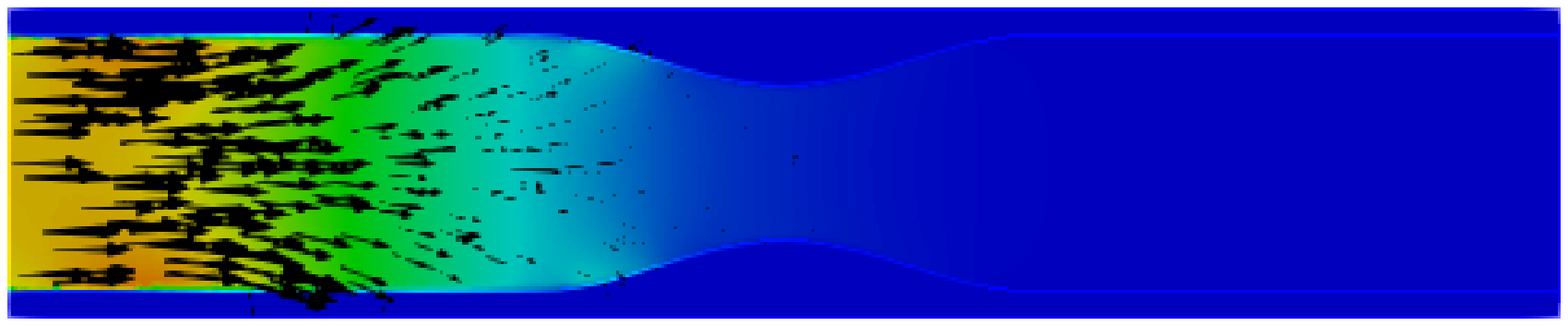}
	\includegraphics[trim=0 0 0 0,scale=0.28]{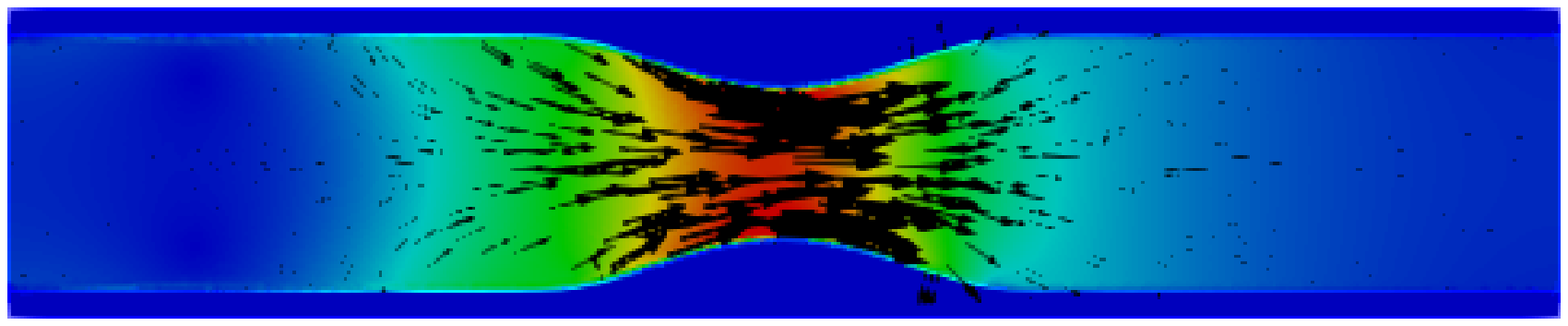}
	\includegraphics[trim=0 0 0 0,scale=0.28]{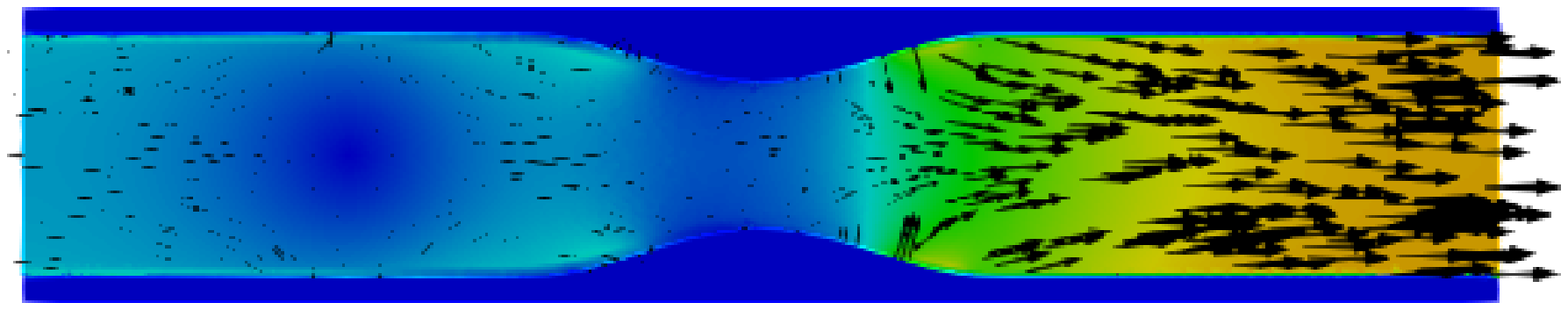}
	\includegraphics[trim=0 0 0 0,scale=0.28]{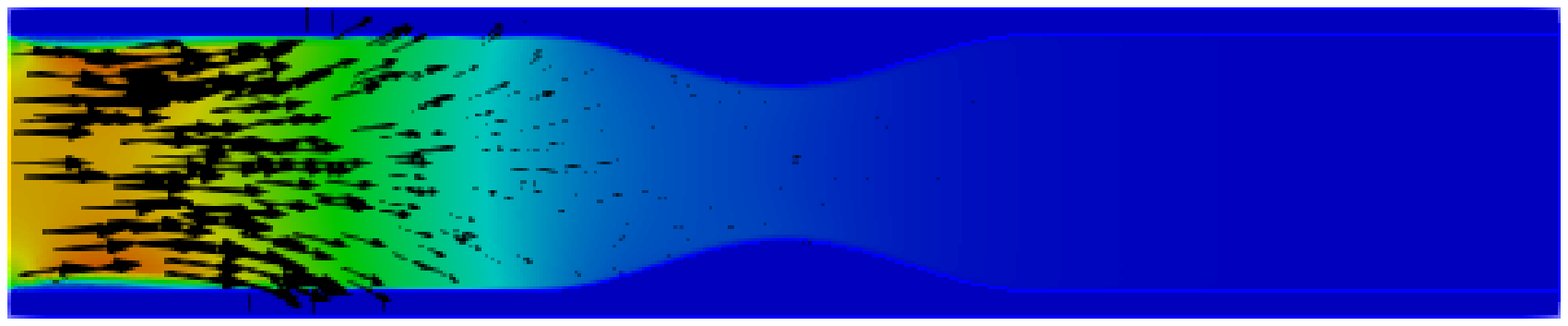}
	\includegraphics[trim=0 0 0 0,scale=0.28]{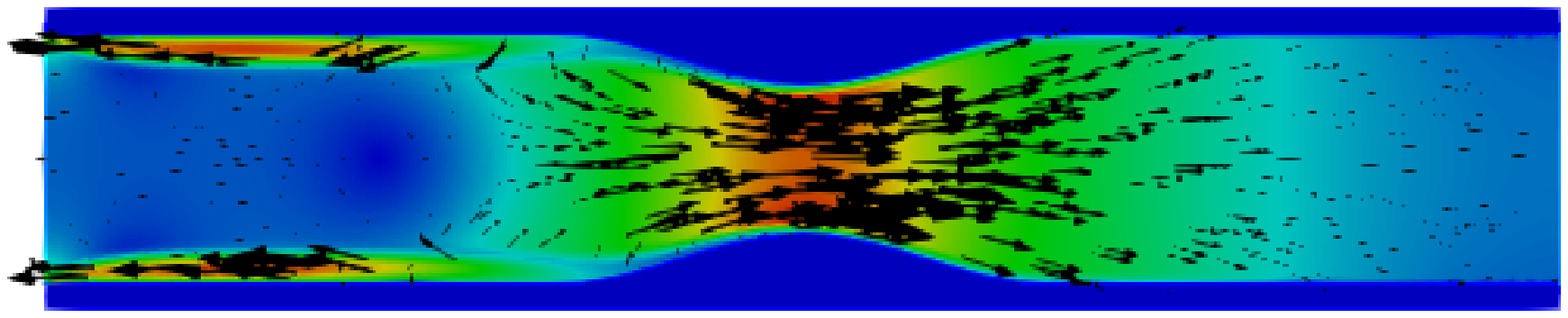}
	\includegraphics[trim=0 0 0 0,scale=0.28]{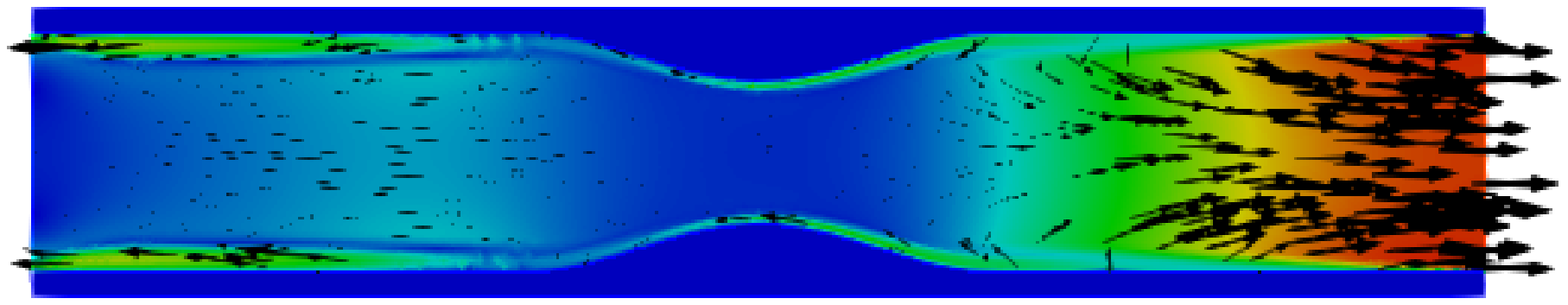}
	\includegraphics[trim=0 10 0 0,scale=0.28]{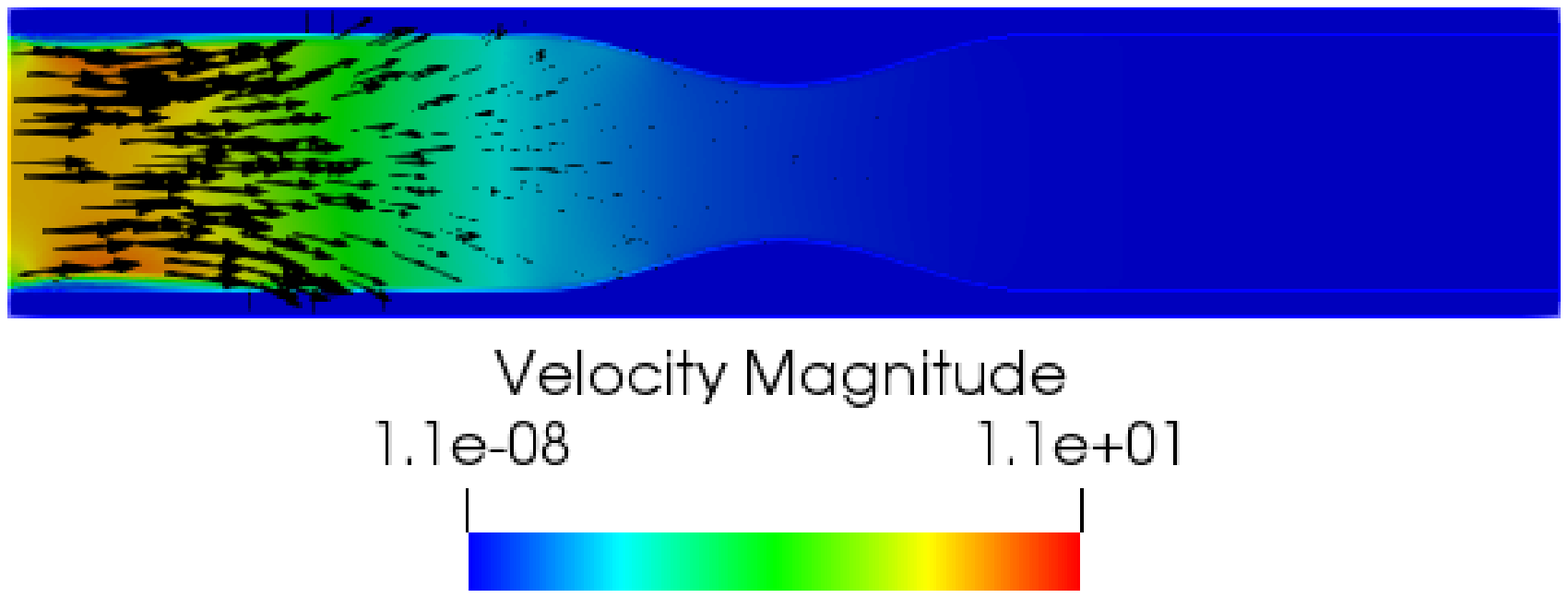}
	\includegraphics[trim=0 10 0 0,scale=0.28]{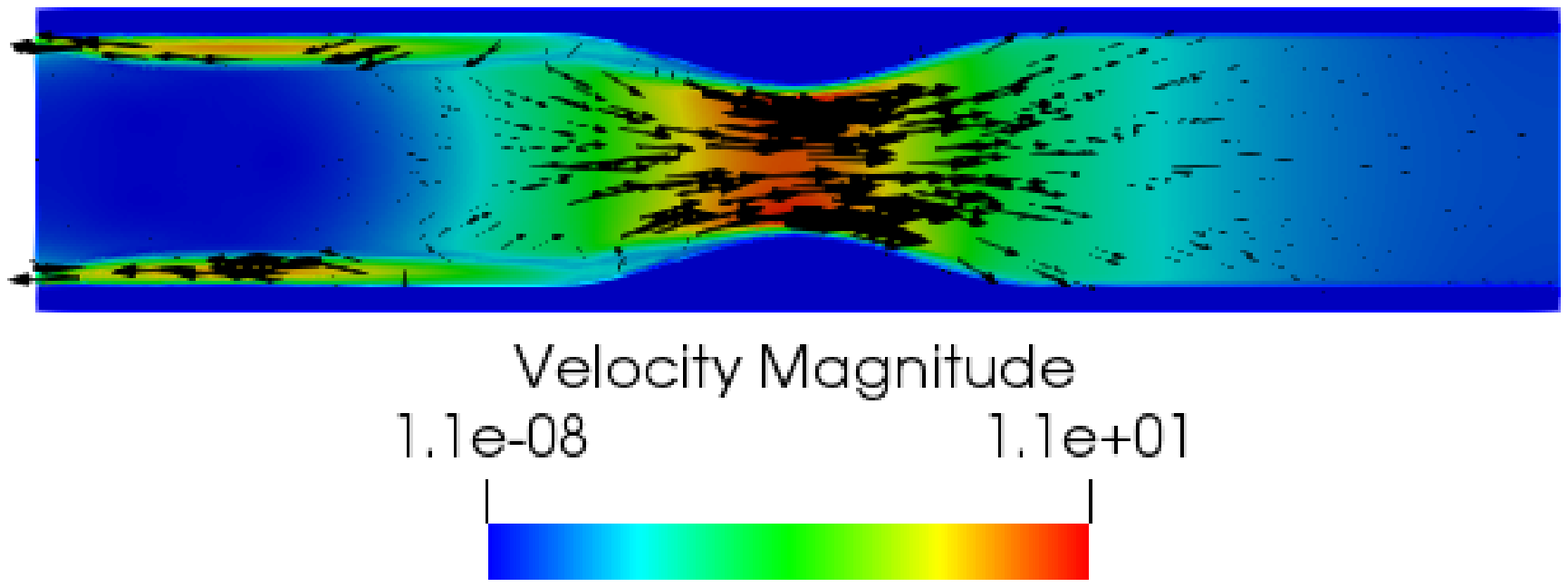}
	\includegraphics[trim=0 10 0 0,scale=0.28]{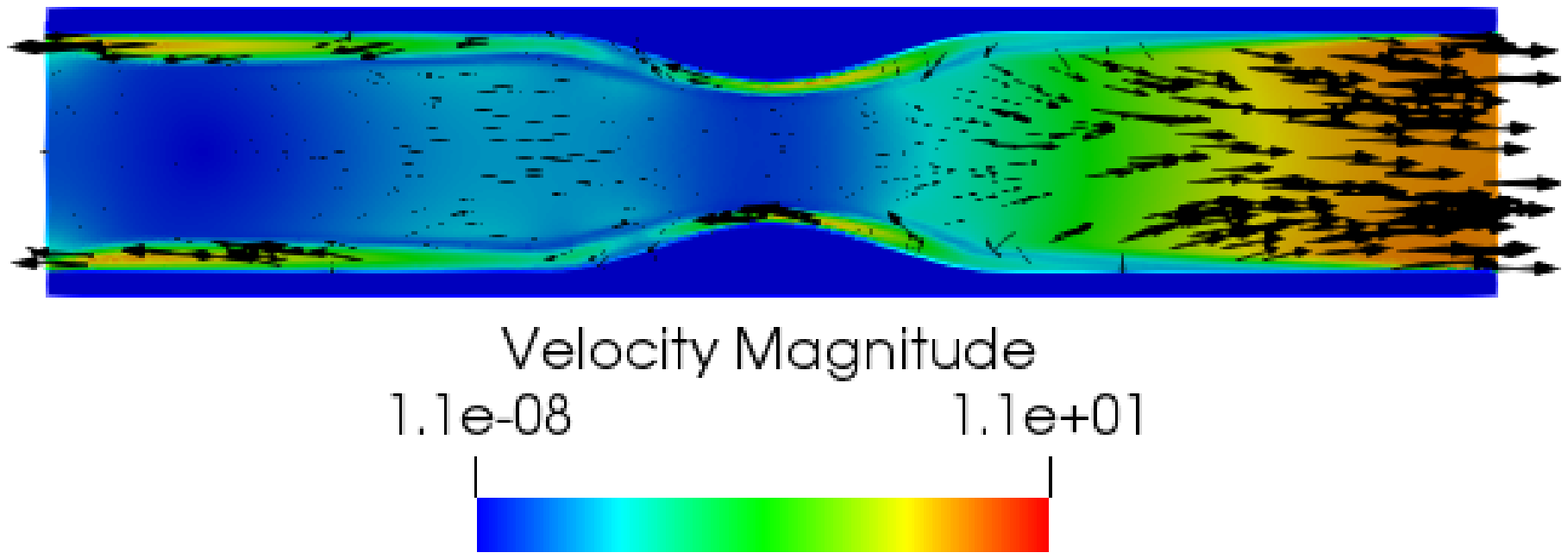}
	\caption{Fluid velocity magnitude together with scaled velocity arrows at time t=1.8 ms, t=3.6 ms, t=5.4 ms for case 7, case 8, case 9 and case 10.  }
	\label{velocitycase3}
\end{figure}

\begin{figure}[ht!]
\includegraphics[trim=0 0 0 30,scale=0.28]{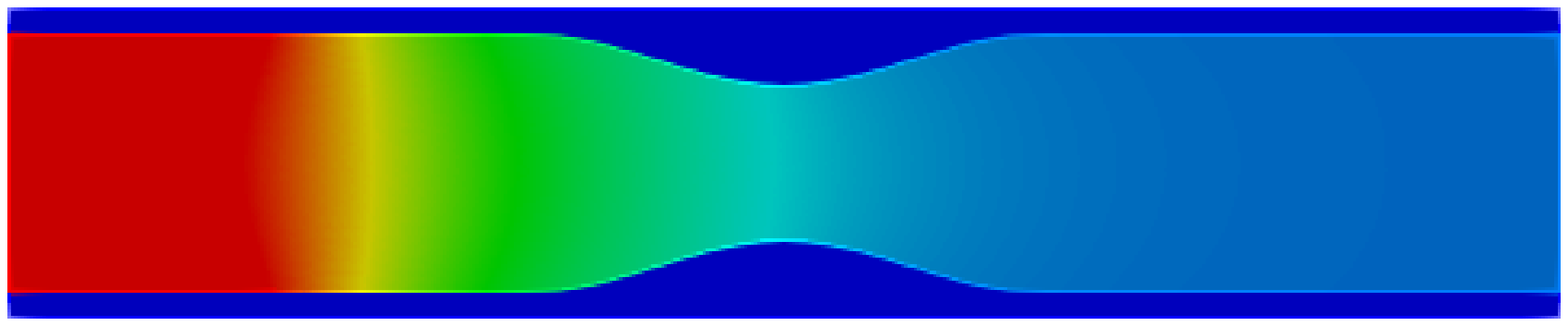}
\includegraphics[trim=0 0 0 30,scale=0.28]{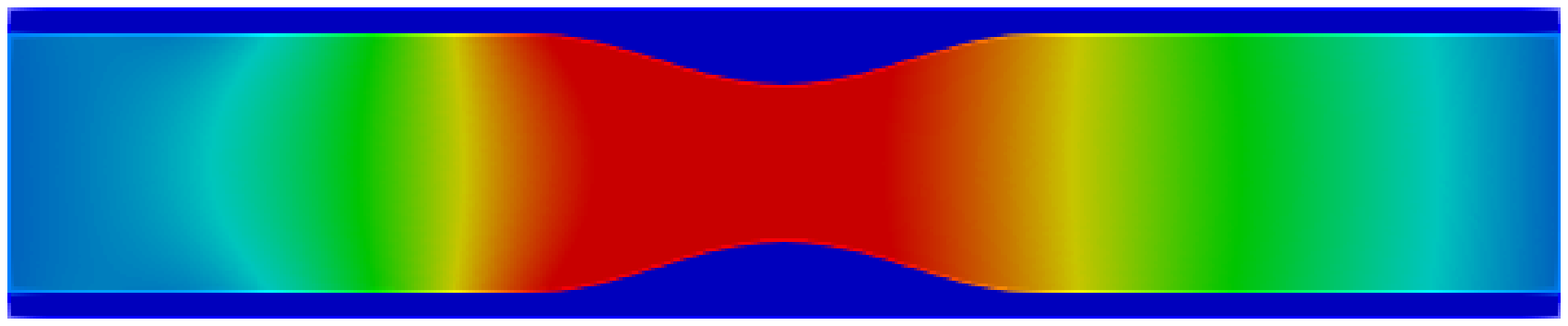}
\includegraphics[trim=0 0 0 30,scale=0.28]{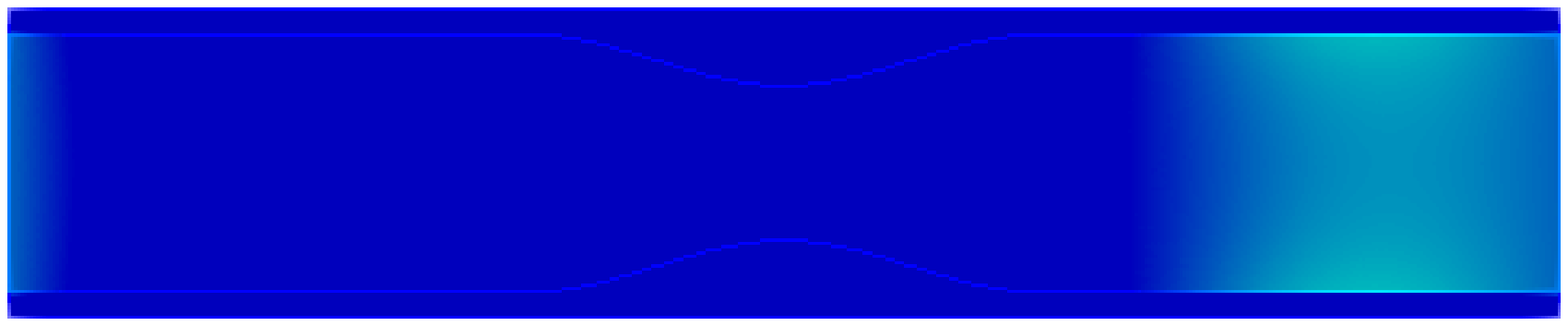}
\includegraphics[trim=0 0 0 0,scale=0.28]{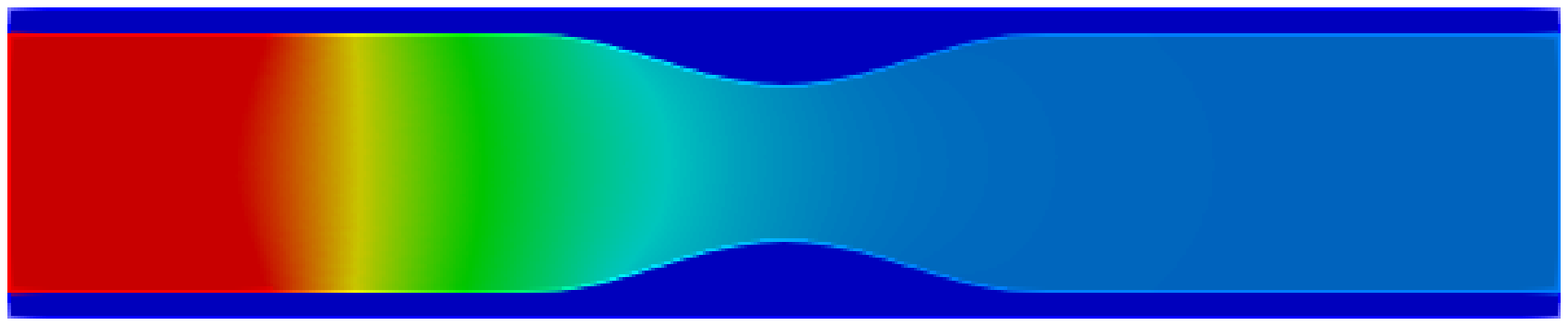}
\includegraphics[trim=0 0 0 0,scale=0.28]{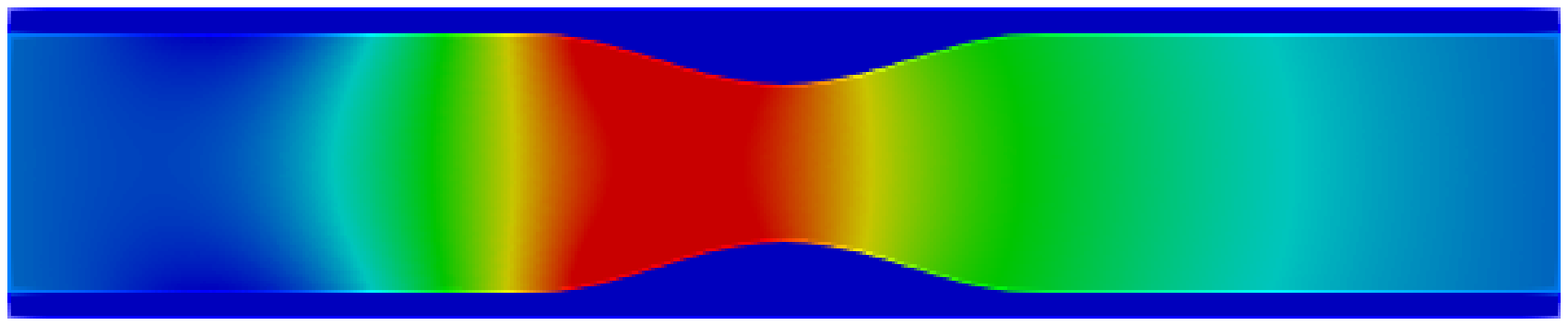}
\includegraphics[trim=0 0 0  0,scale=0.28]{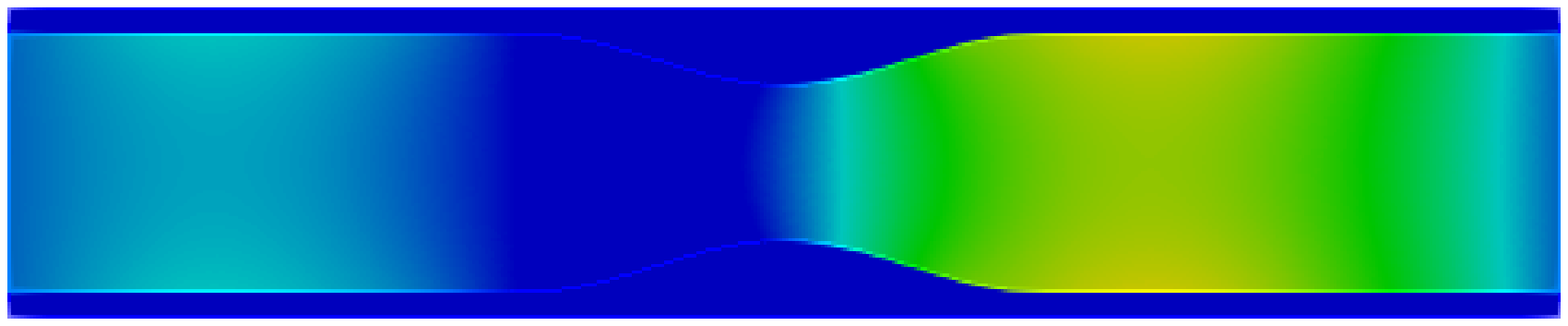}
\includegraphics[trim=0 0 0 0,scale=0.28]{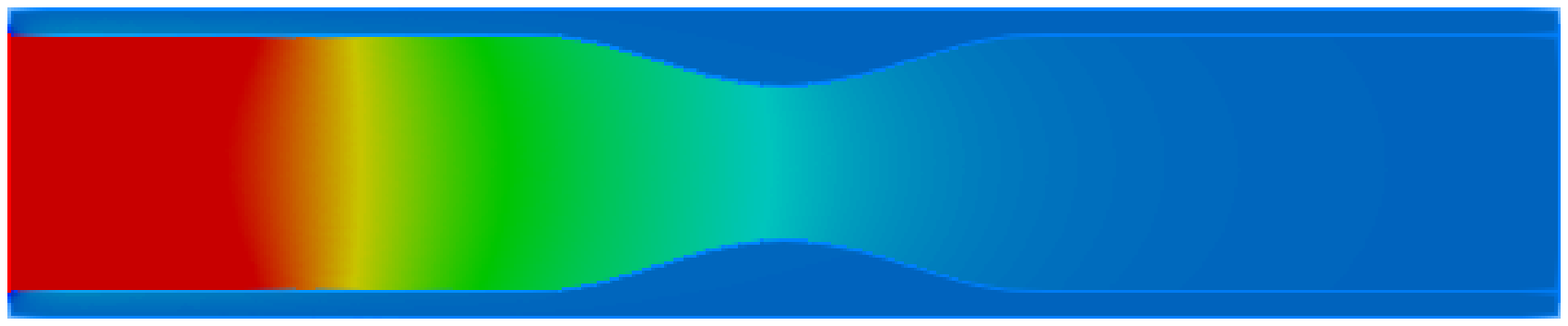}
\includegraphics[trim=0 0 0 0,scale=0.28]{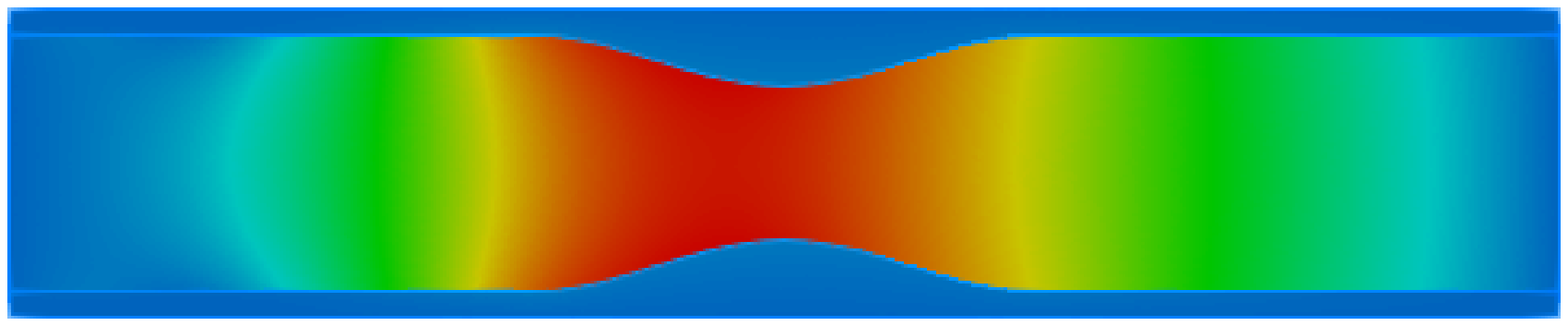}
\includegraphics[trim=0 0 0 0,scale=0.28]{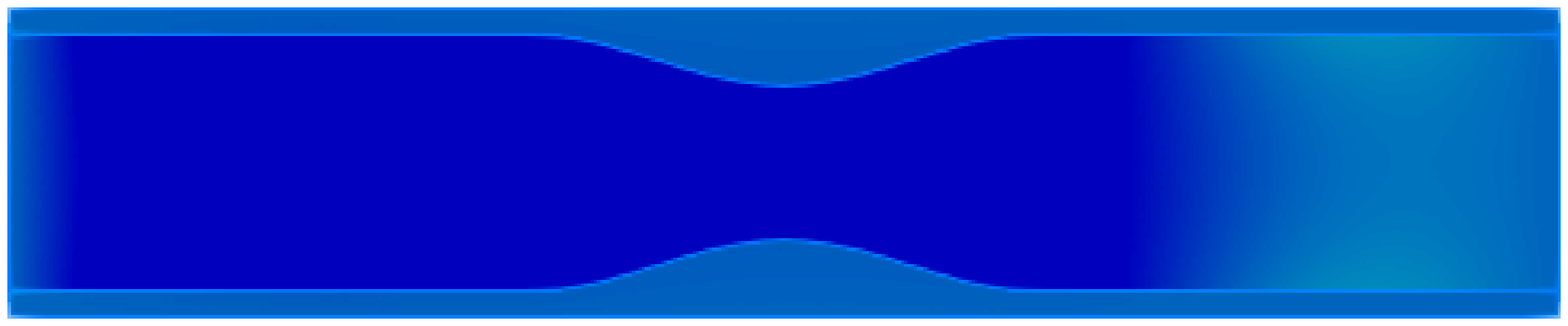}
\includegraphics[trim=0 10 0 0,scale=0.28]{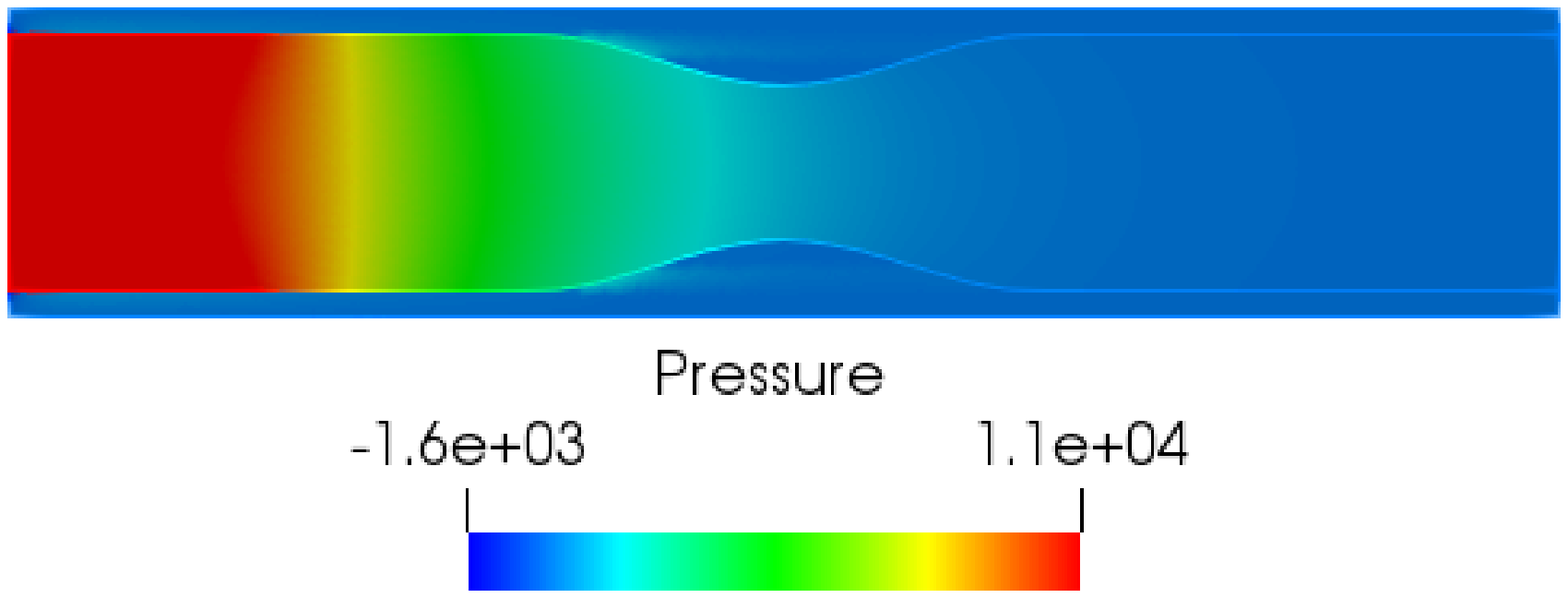}
\includegraphics[trim=0 10 0 0,scale=0.28]{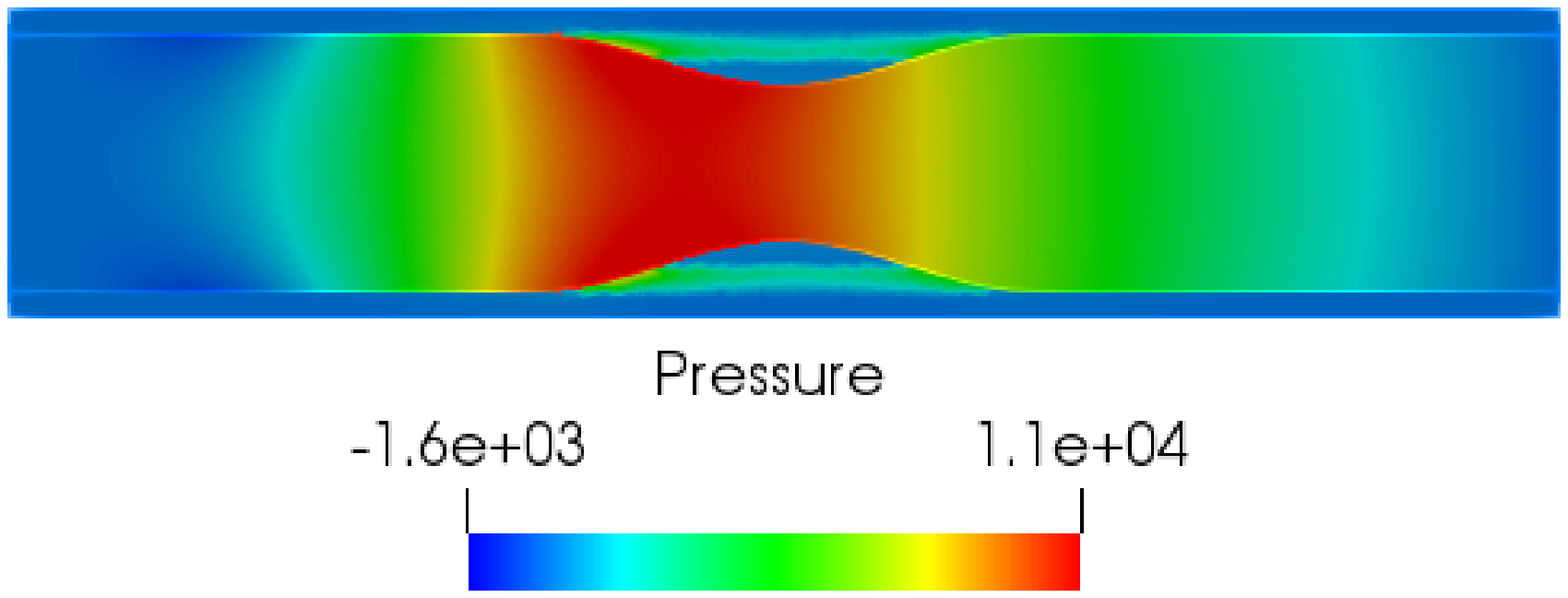}
\includegraphics[trim=0 10 0 0,scale=0.28]{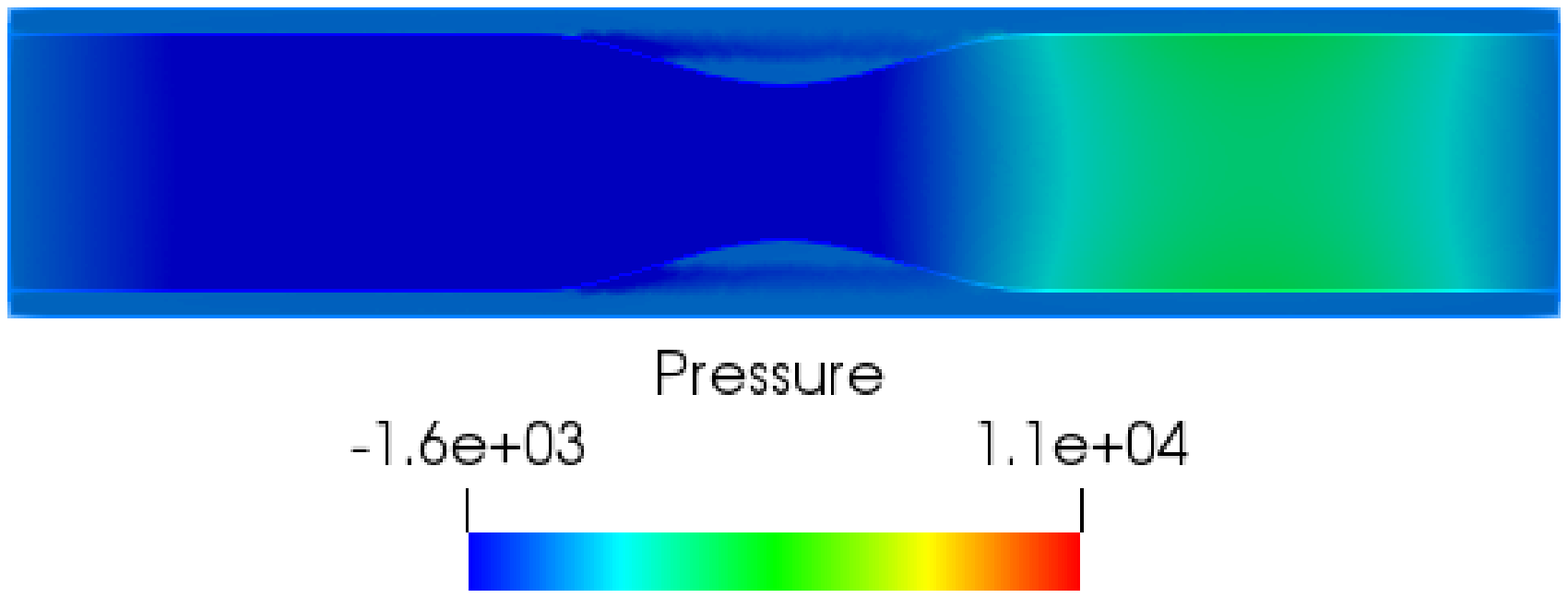}
\caption{Pressure waves along the vessel at time t=1.8 ms, t=3.6 ms, t=5.4 ms for case 7, case 8, case 9 and case 10. }
	\label{pressurecase3}
\end{figure}

In Fig.\ref{velocitycase3} and Fig,\ref{pressurecase3}, we present the fluids velocity and pressure waves along the vessel at time $t=1.8, 3.6, 5.4$ ms. Note that we still use moving mesh in the simulation for this section. However, we don't magnify the deformation as in the former sections to avoid confusion with geometries. From the plots, we note that the velocity waves are strongly influenced by the stenosis. In Fig. \ref{pressurecase3}, we can observe further differences on pressure fields among four stenotic cases. In particular, we can detect clear increasements of $p_p$ in the stenotic areas in case 10. Since we use a piecewiese permeability $K$ for the stenosis in case 10, more fluids are allowed in the stenosis areas, which would result in higher pressure bands in the structure areas. 
\begin{figure}[ht!]
	\includegraphics[trim=0 0 0 0,scale=0.28]{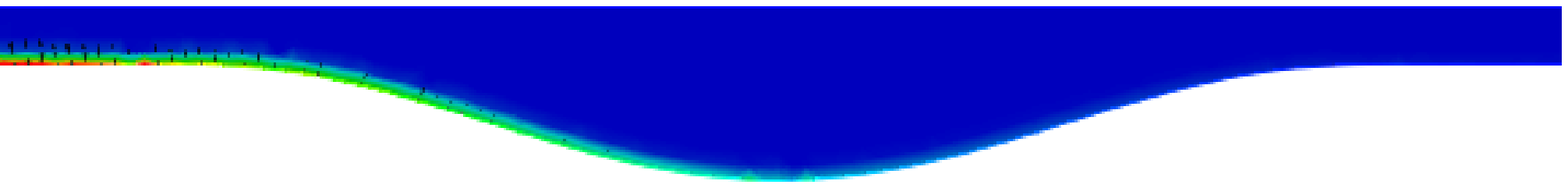}
	\includegraphics[trim=0 0 0 0,scale=0.28]{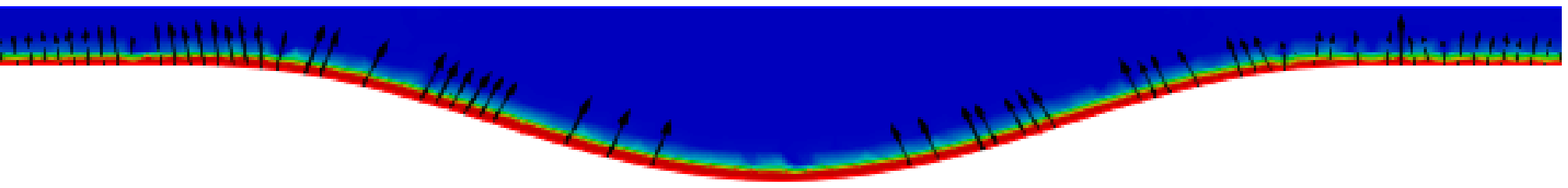}
	\includegraphics[trim=0 0 0 0,scale=0.28]{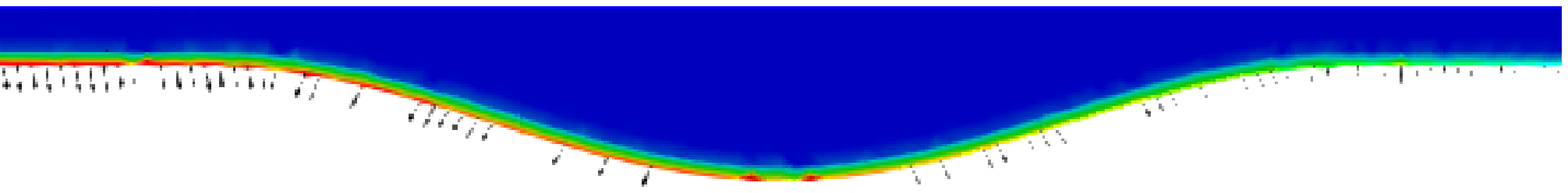}
	\includegraphics[trim=0 10 0 0,scale=0.28]{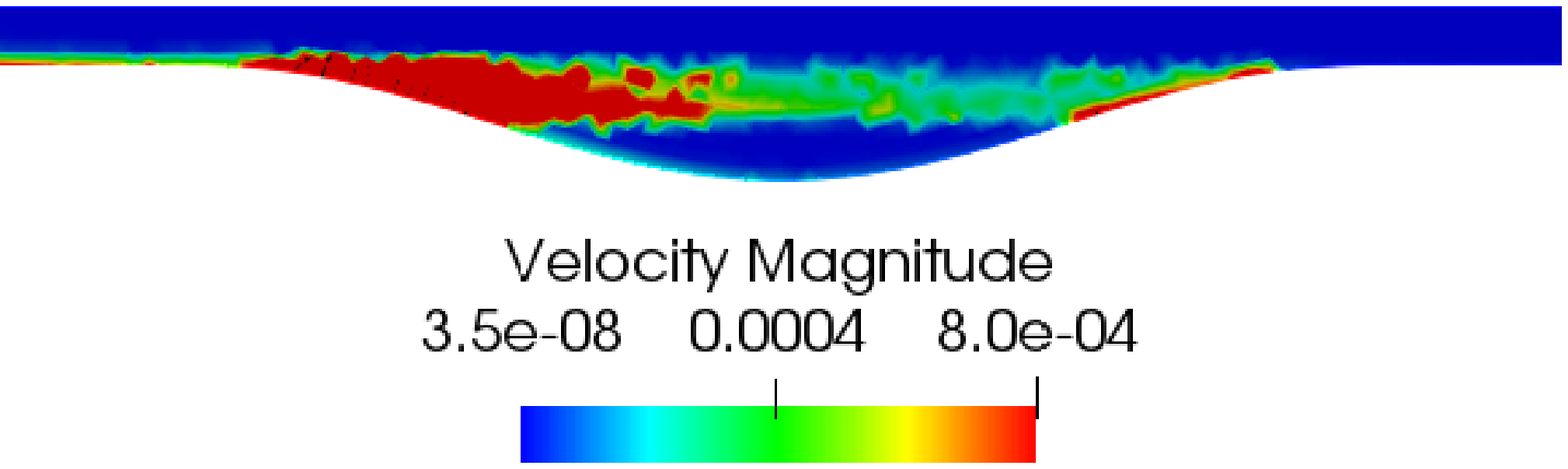}
	\includegraphics[trim=0 10 0 0,scale=0.28]{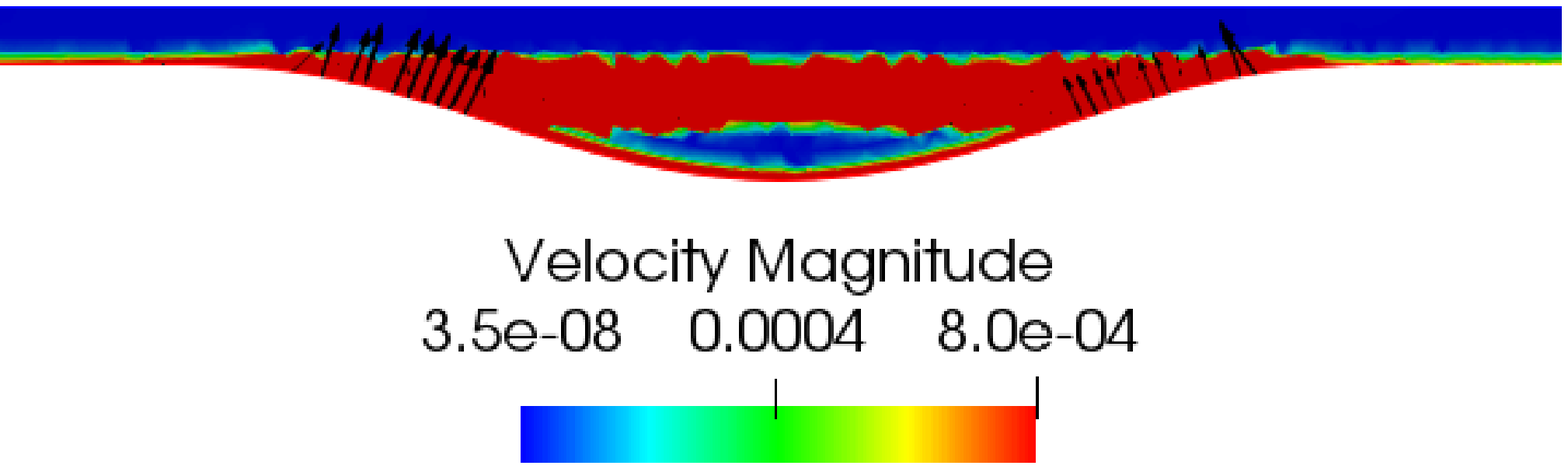}
	\includegraphics[trim=0 10 0 0,scale=0.28]{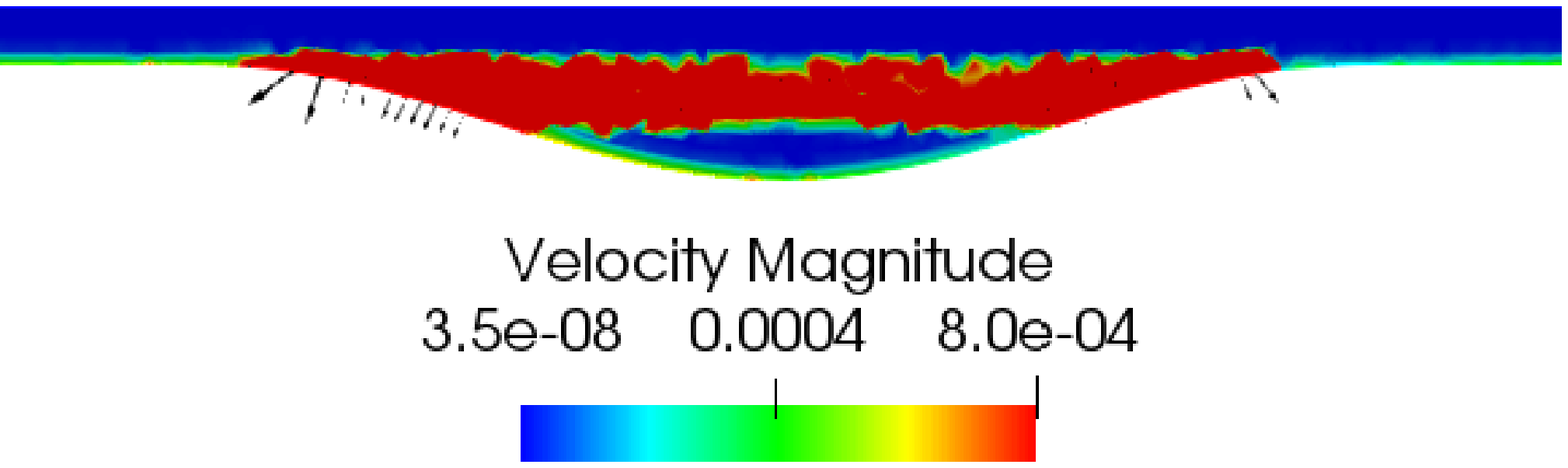}
	\caption{Darcy velocity magnitude together with the scaled velocity arrow in the stenotic area for the top structure at time t=1.8 ms, t=3.6 ms, t=5.4 ms for case 9 and case 10. Note that for better observation, the length of velocity arrow in case 9 are magnified for 40 times while we only magnify case 10 for 5 times. }
	\label{bucase3}
\end{figure}
\begin{figure}[ht!]
	\includegraphics[trim=0 0 0 0,scale=0.28]{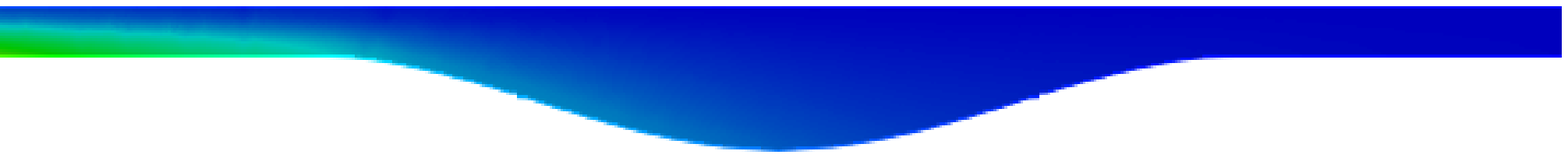}
	\includegraphics[trim=0 0 0 0,scale=0.28]{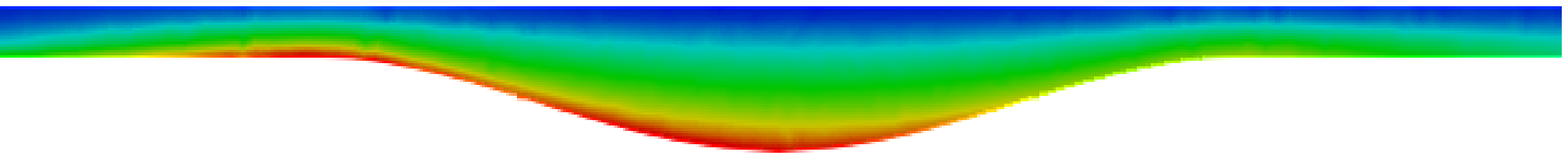}
	\includegraphics[trim=0 0 0 0,scale=0.28]{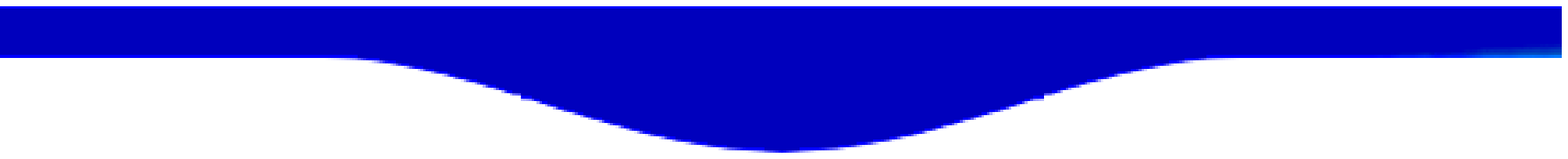}
	\includegraphics[trim=0 10 0 10,scale=0.28]{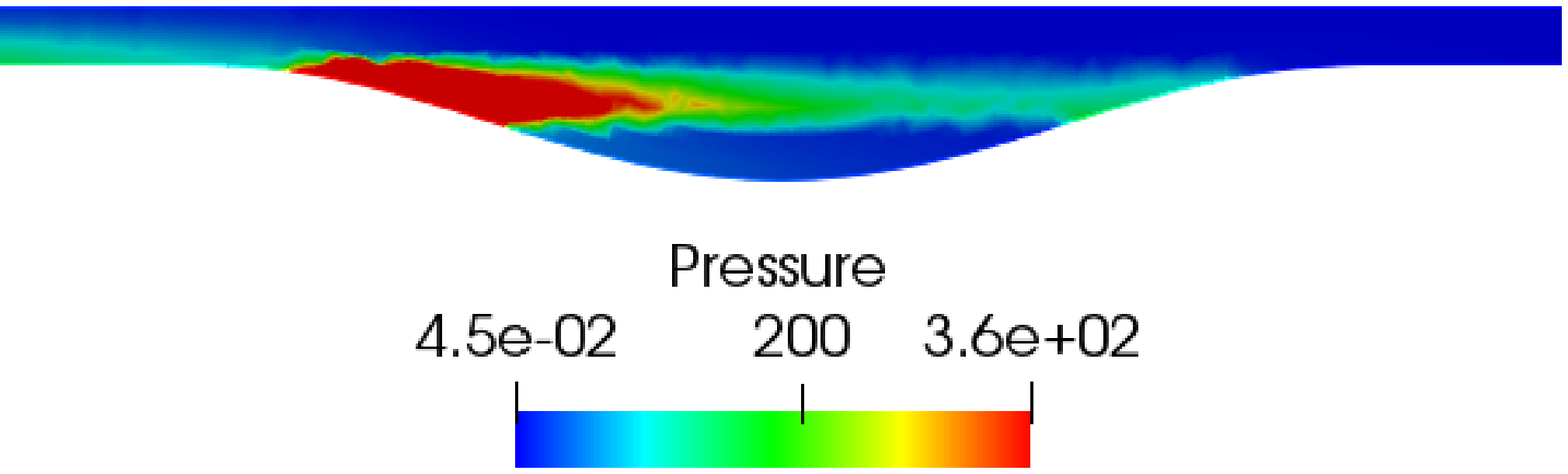}
	\includegraphics[trim=0 10 0 10,scale=0.28]{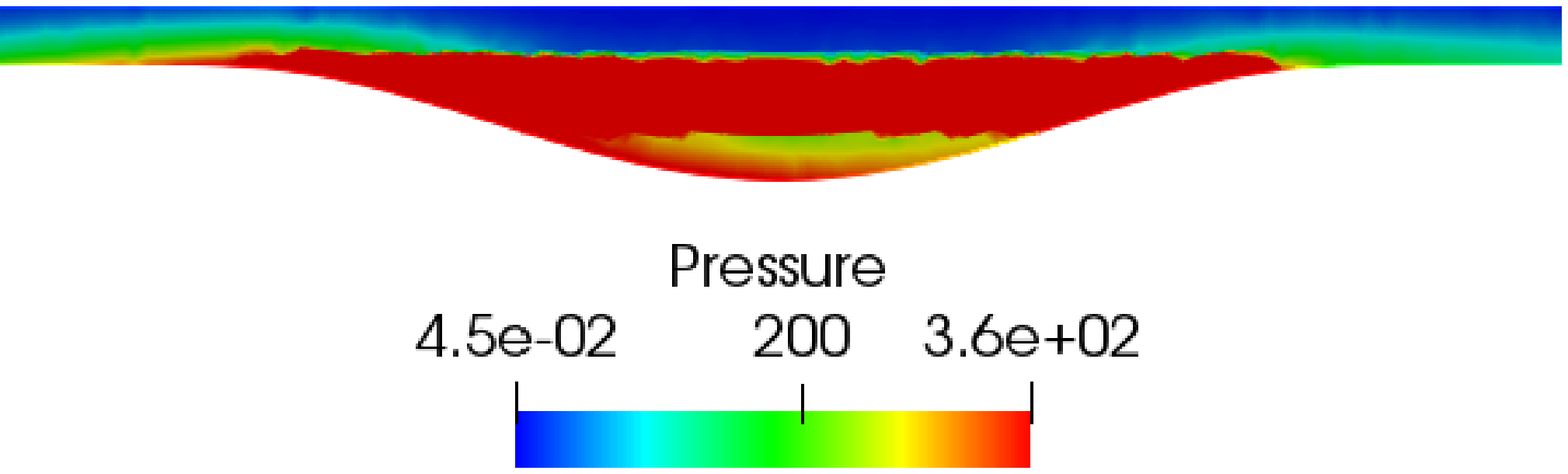}
	\includegraphics[trim=0 10 0 10,scale=0.28]{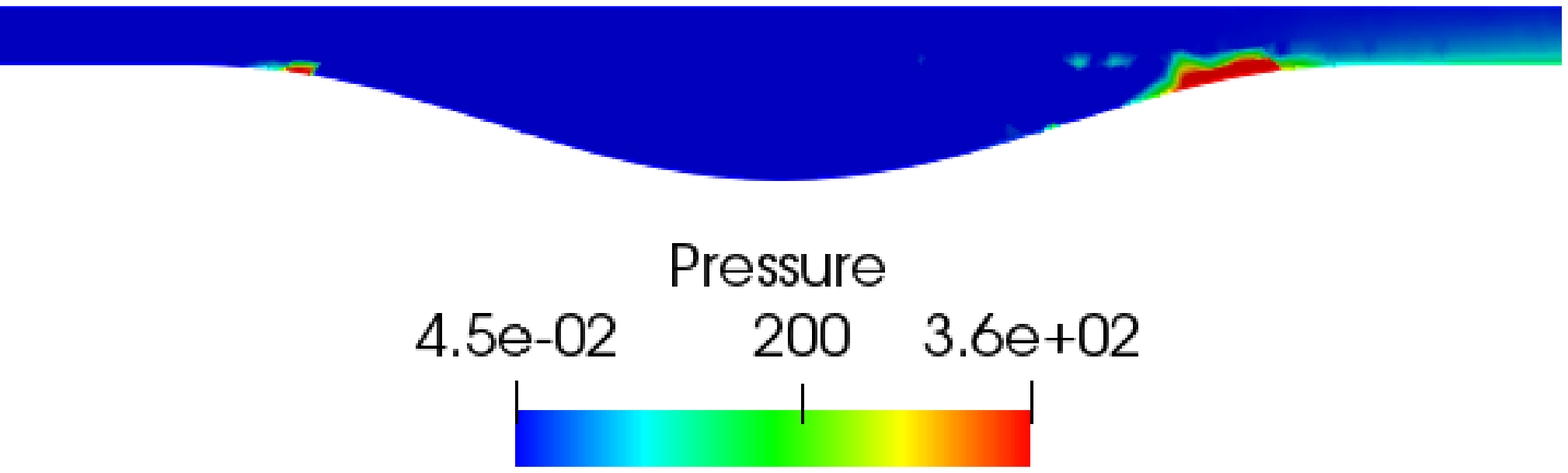}
	\caption{Pressure waves in the stenotic area for the top structure at time t=1.8 ms, t=3.6 ms, t=5.4 ms for case 9 and case 10.}
	\label{bpcase3}
\end{figure}

In Fig.\ref{bucase3} and Fig.\ref{bpcase3}, we also display Darcy velocity and pressure in the top stenotic area for case 9 and case 10 seperately. From Fig. \ref{bucase3}, we can see the fluids would concentrate along the interface for case 9. Less filtration would be found in case 9 even for the stenotic area. But for case 10, fluids can actually penetrate through the stenotic area. From Fig. \ref{bpcase3}, we can detect a clear discontinuity of pressure fields of case 10, comparing with case 9. 

\begin{figure}[ht!]
	\includegraphics[trim=0 0 0 30,scale=0.28]{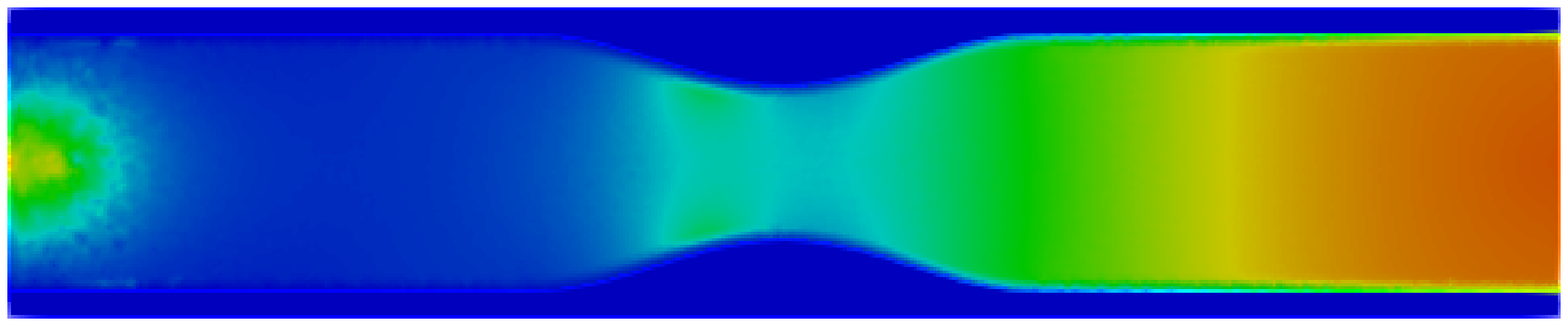}
   \includegraphics[trim=0 0 0 30,scale=0.28]{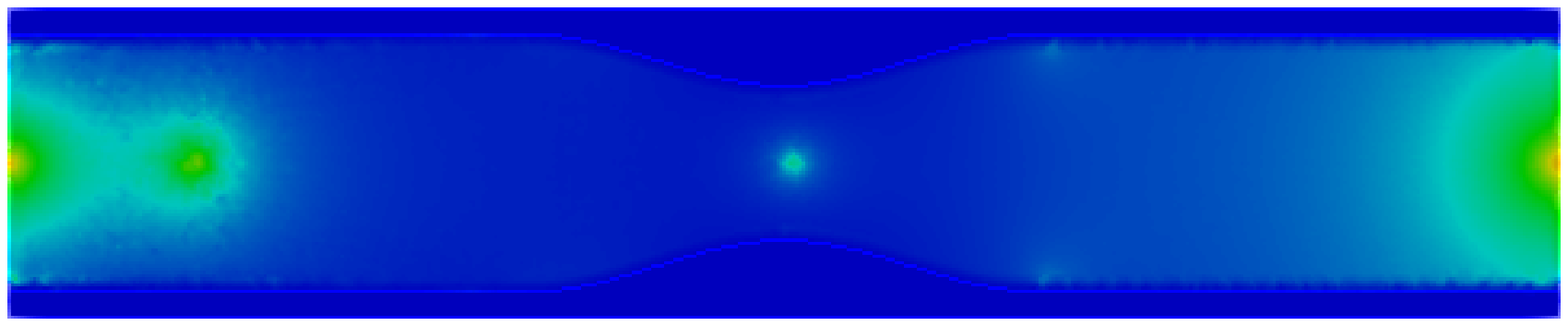}
   \includegraphics[trim=0 0 0 30,scale=0.28]{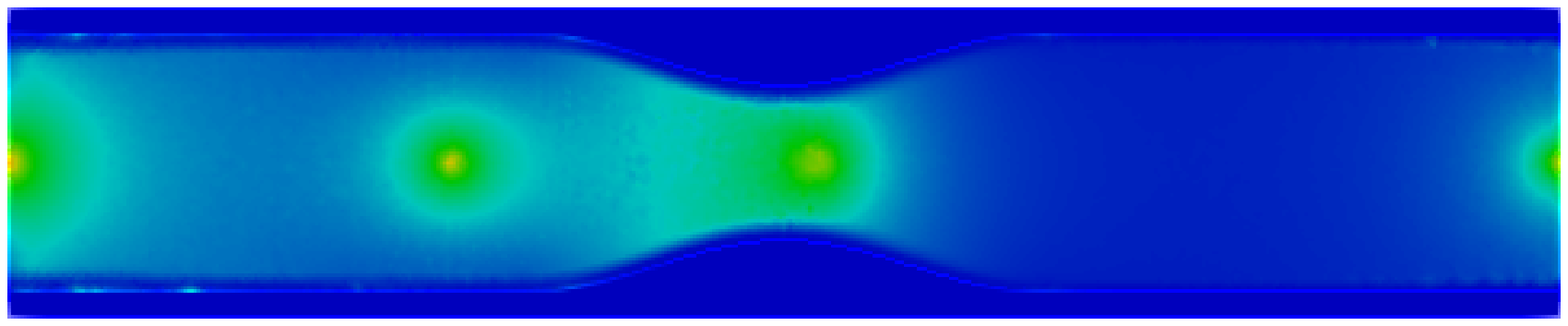}
	\includegraphics[trim=0 0 0 0,scale=0.28]{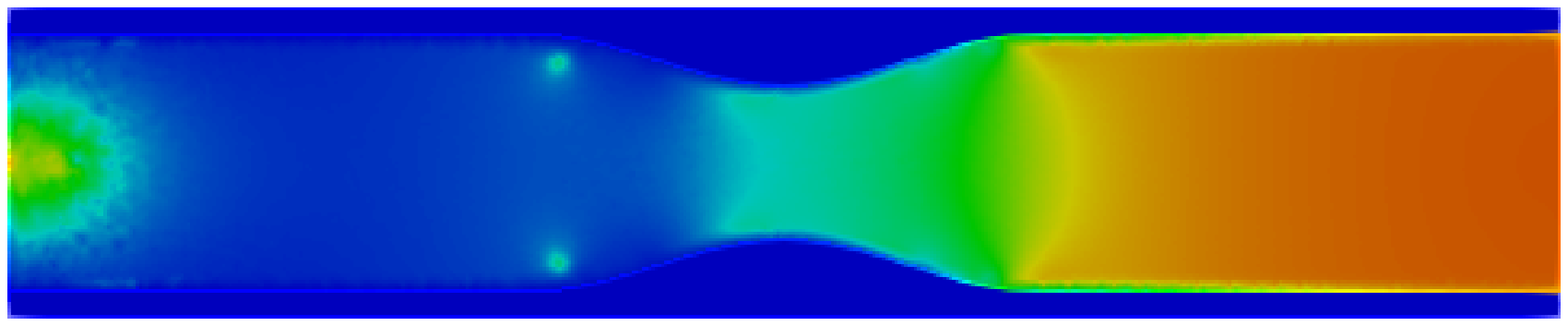}
	\includegraphics[trim=0 0 0 0,scale=0.28]{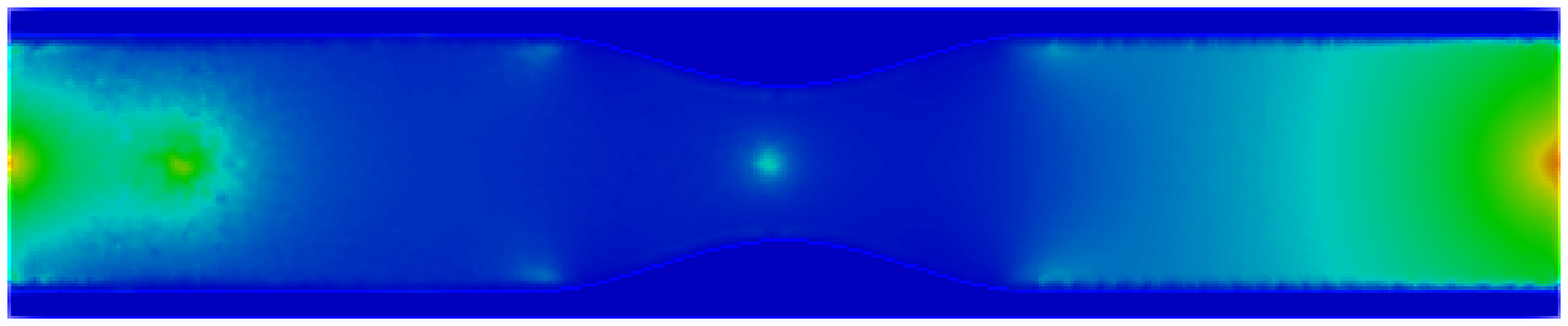}
	\includegraphics[trim=0 0 0 0,scale=0.28]{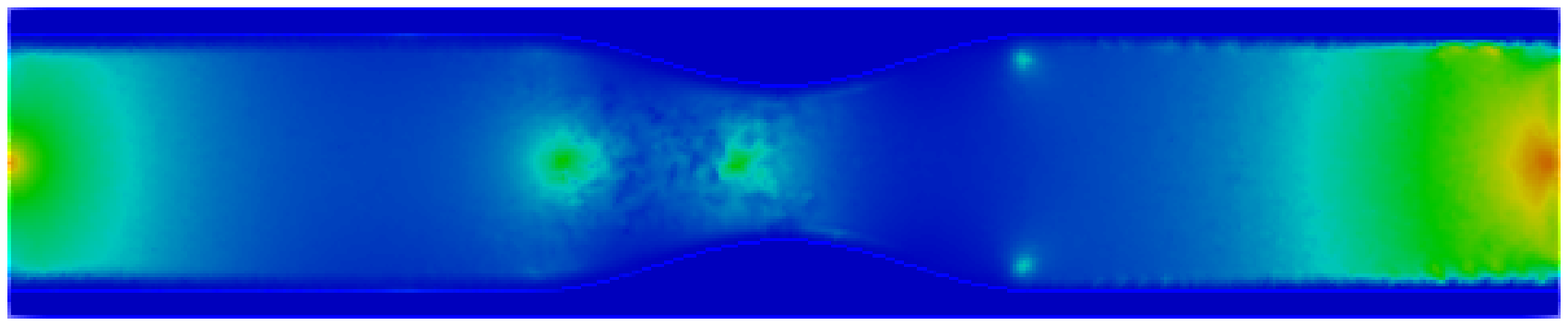}
		\includegraphics[trim=0 0 0 0,scale=0.28]{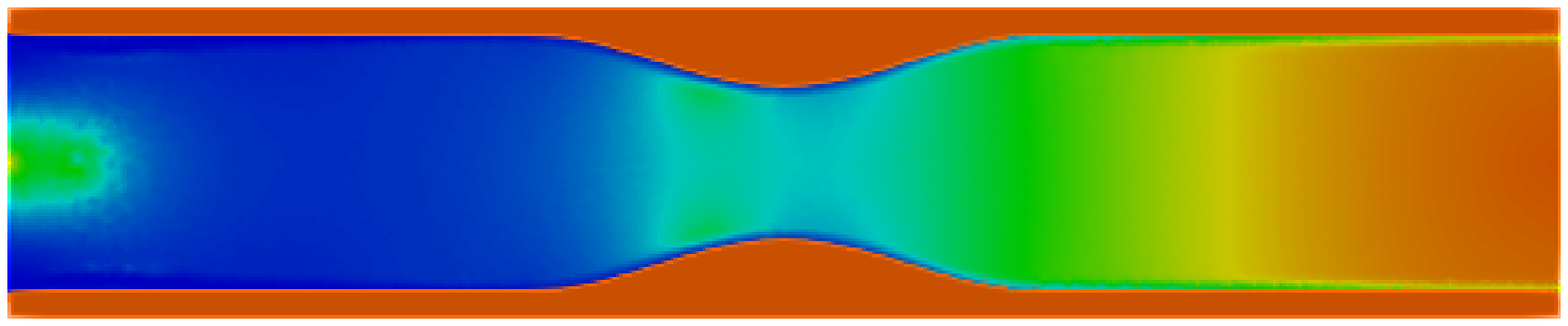}
	\includegraphics[trim=0 0 0 0,scale=0.28]{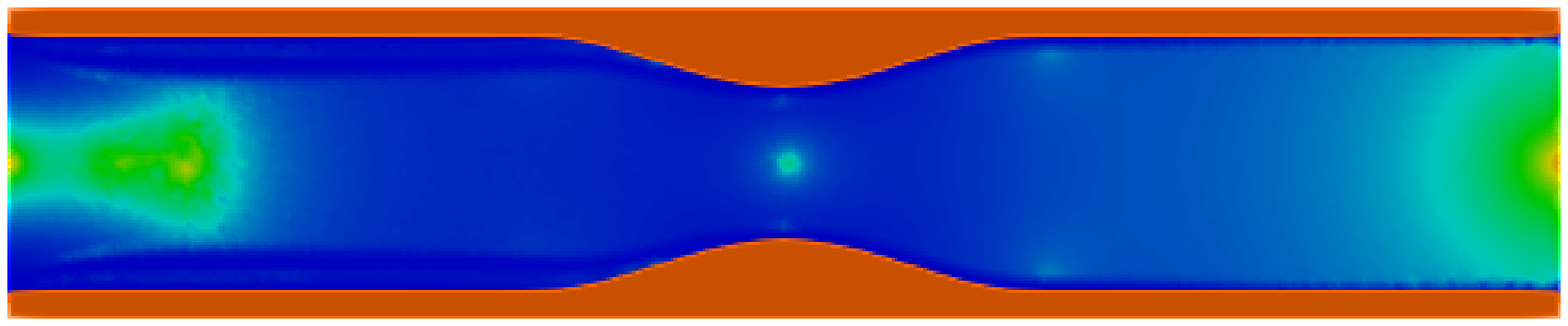}
	\includegraphics[trim=0 0 0 0,scale=0.28]{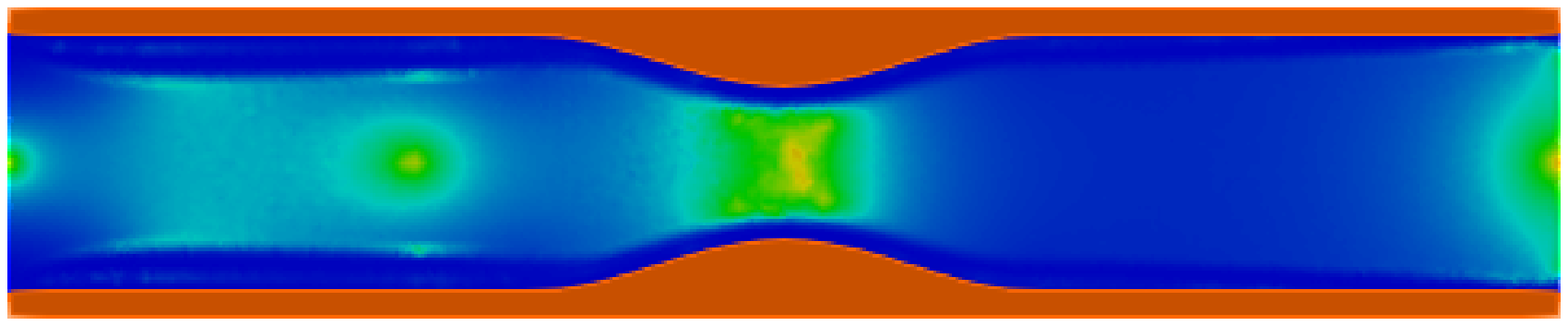}
	\includegraphics[trim=0 10 0 0,scale=0.28]{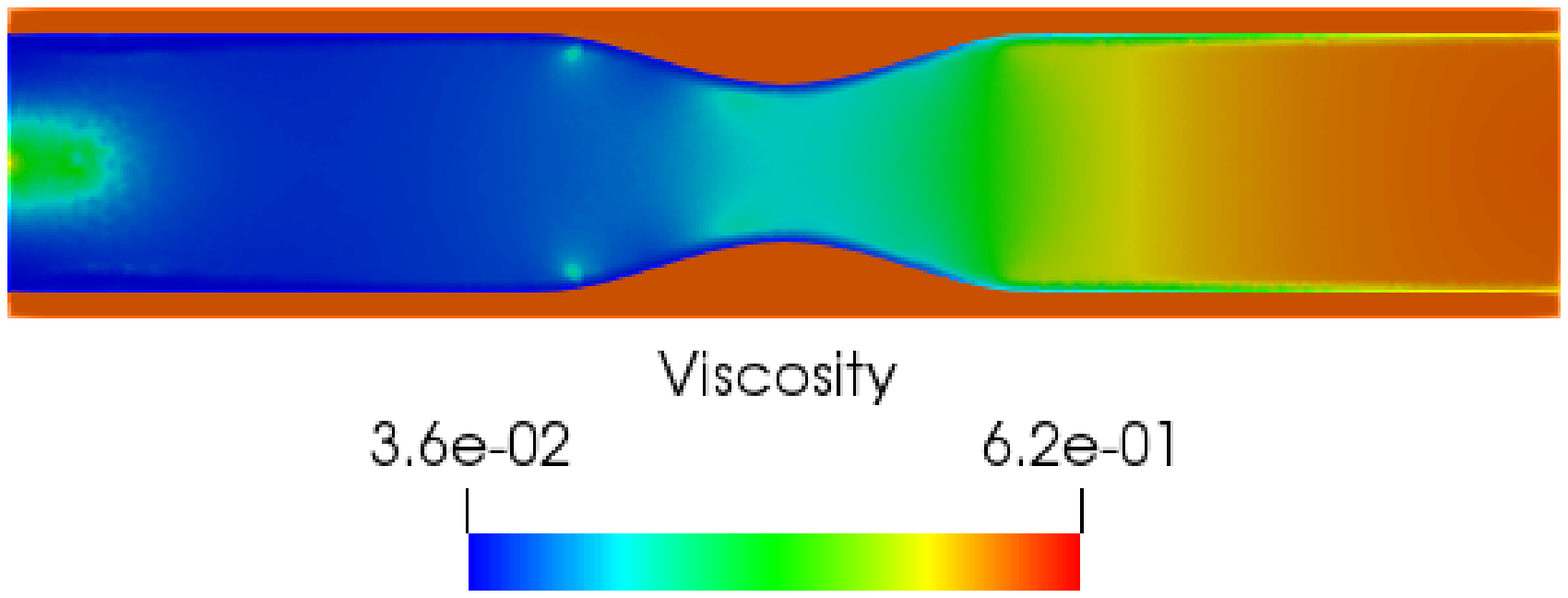}
	\includegraphics[trim=0 10 0 0,scale=0.28]{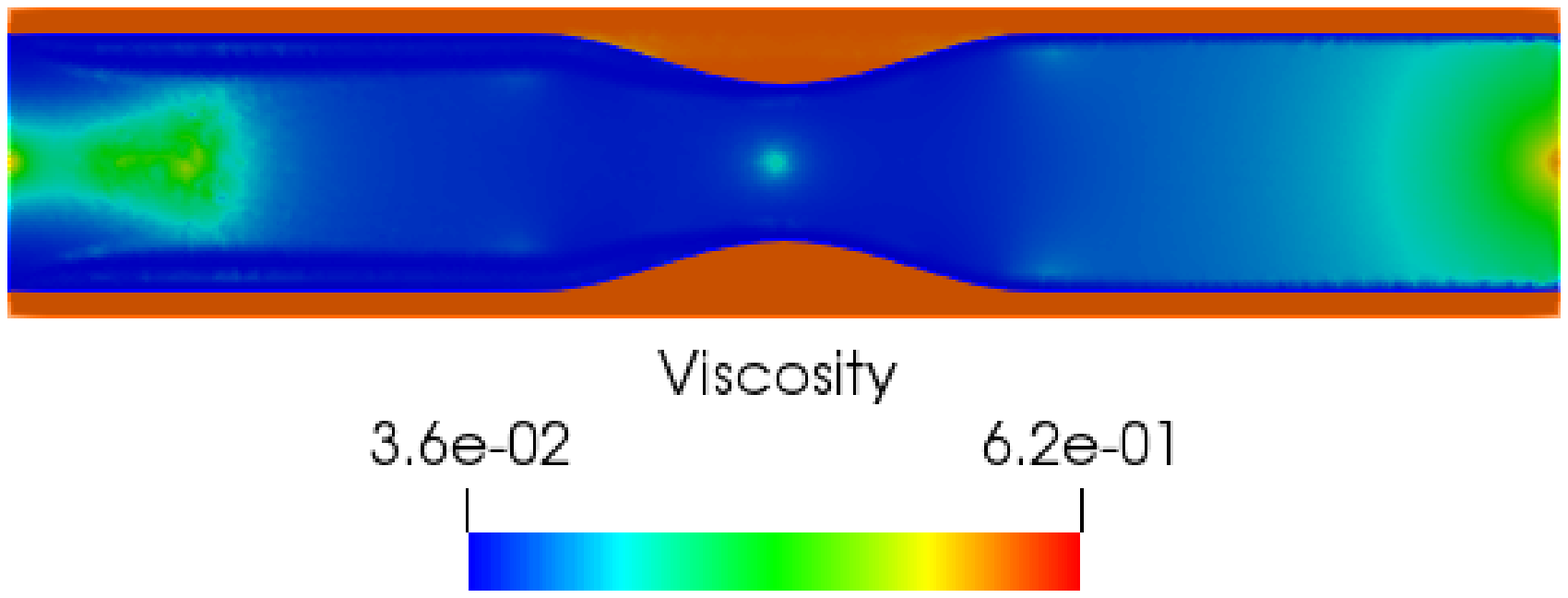}
	\includegraphics[trim=0 10 0 0,scale=0.28]{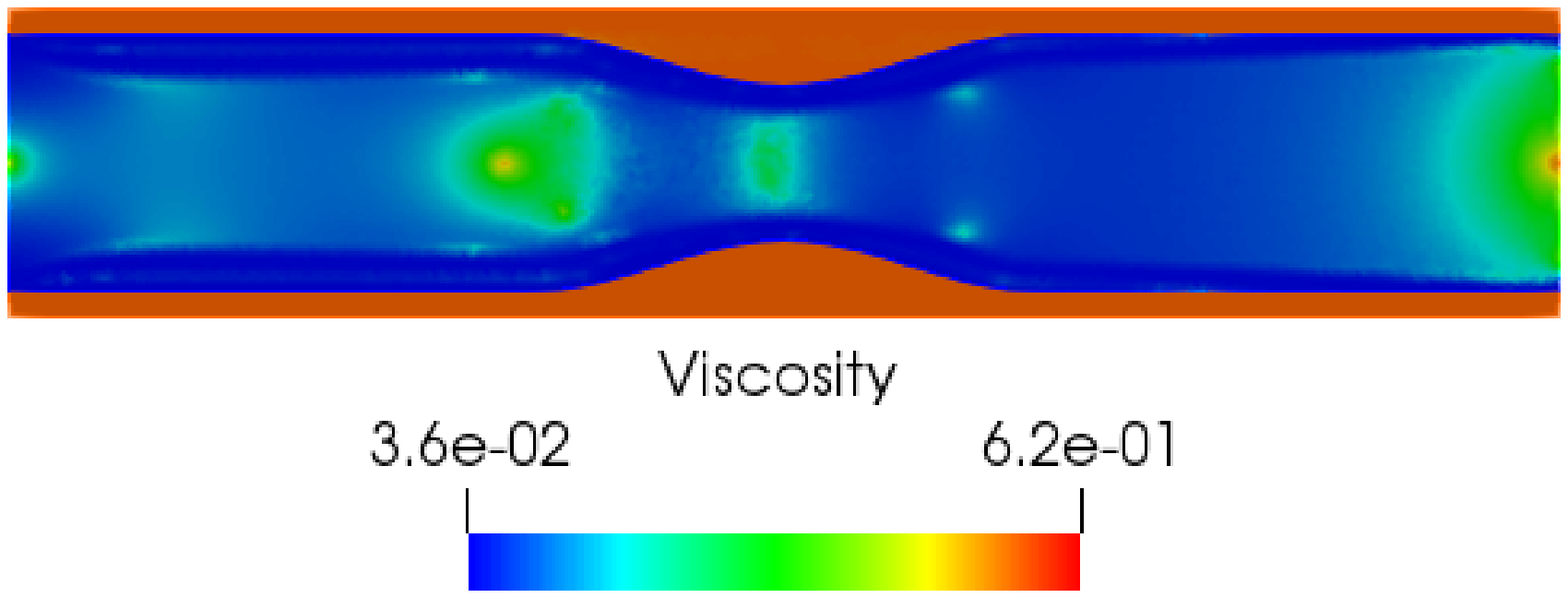}
	\caption{Viscosity in the vessel at time t=1.8 ms, t=3.6 ms, t=5.4 ms for case 7, case 8, case 9 and case 10. }
	\label{viscase3}
\end{figure}

\begin{figure}[ht!]
	\includegraphics[trim=0 0 0 0,scale=0.28]{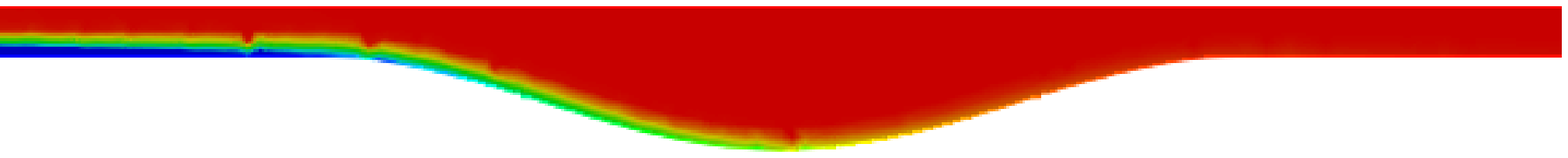}
	\includegraphics[trim=0 0 0 0,scale=0.28]{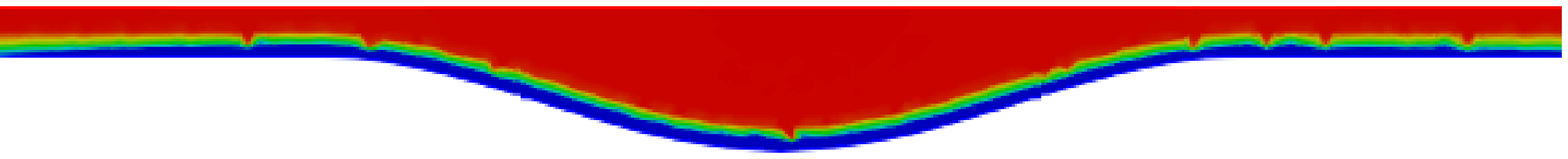}
	\includegraphics[trim=0 0 0 0,scale=0.28]{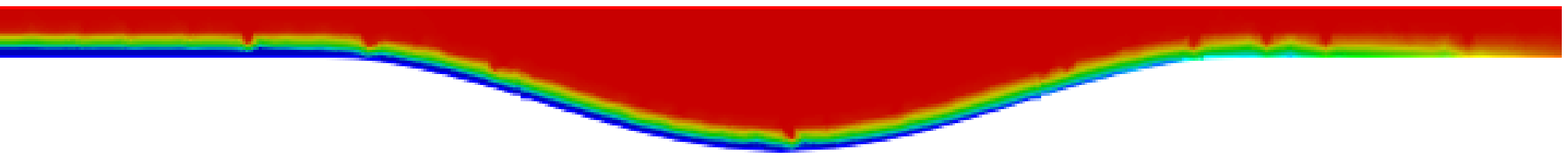}	
	\includegraphics[trim=0 10 0 0,scale=0.28]{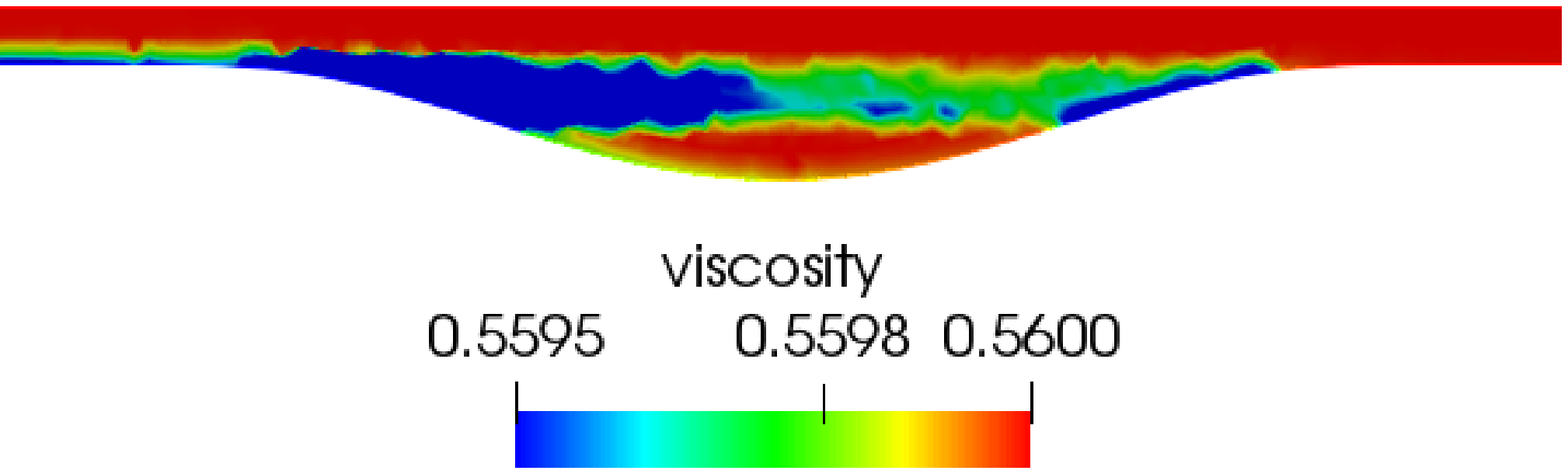}
	\includegraphics[trim=0 10 0 0,scale=0.28]{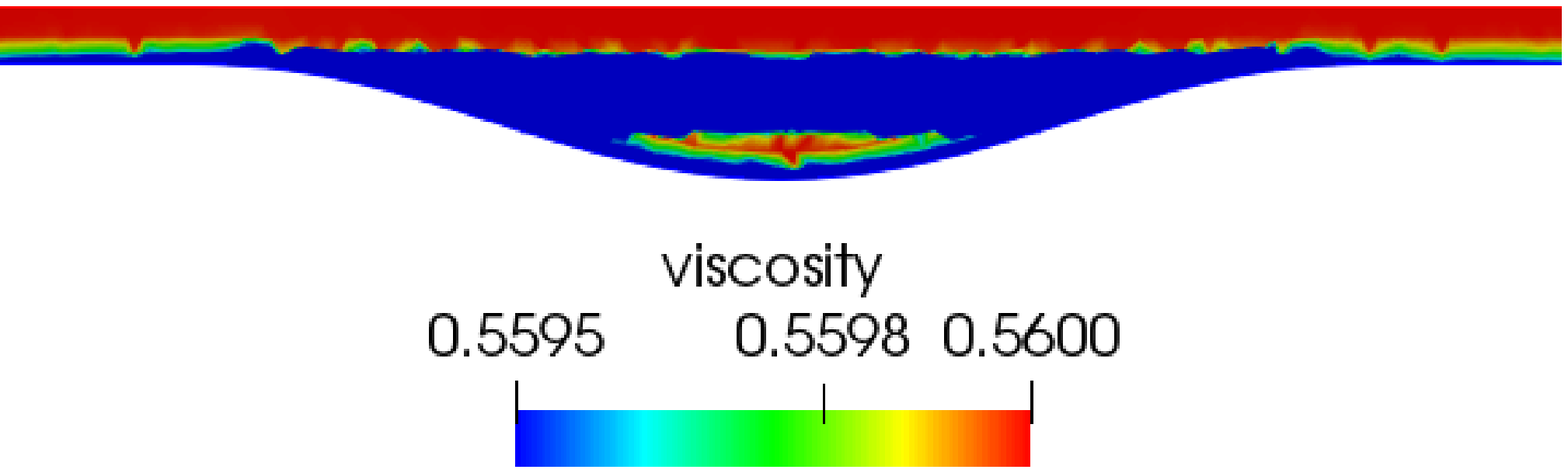}
	\includegraphics[trim=0 10 0 0,scale=0.28]{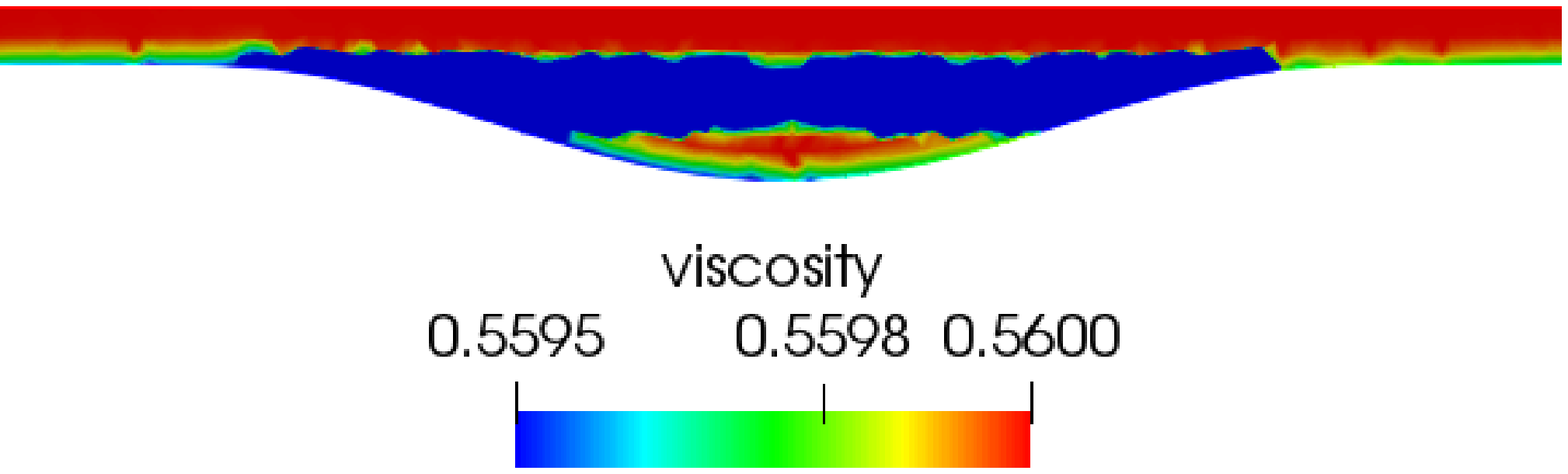}
	\caption{Viscosity in the stenotic area for the top structure at time t=1.8 ms, t=3.6 ms, t=5.4 ms for case 9 and case 10.}
	\label{bviscase3}
\end{figure}

In Fig.\ref{viscase3}, we present the fluid viscosity in the fluid and structure regions. 
More differences are observed for the top build-up structure areas shown in Fig. \ref{bviscase3}. Being consistent with Fig. \ref{bucase3}, the decrease of viscosity would only show up along the interface of case 9. We can see a clear decrease of viscosity of case 10 through the structure. 

\begin{figure}[ht!]
	\includegraphics[trim=0 60 0 0,scale=0.149]{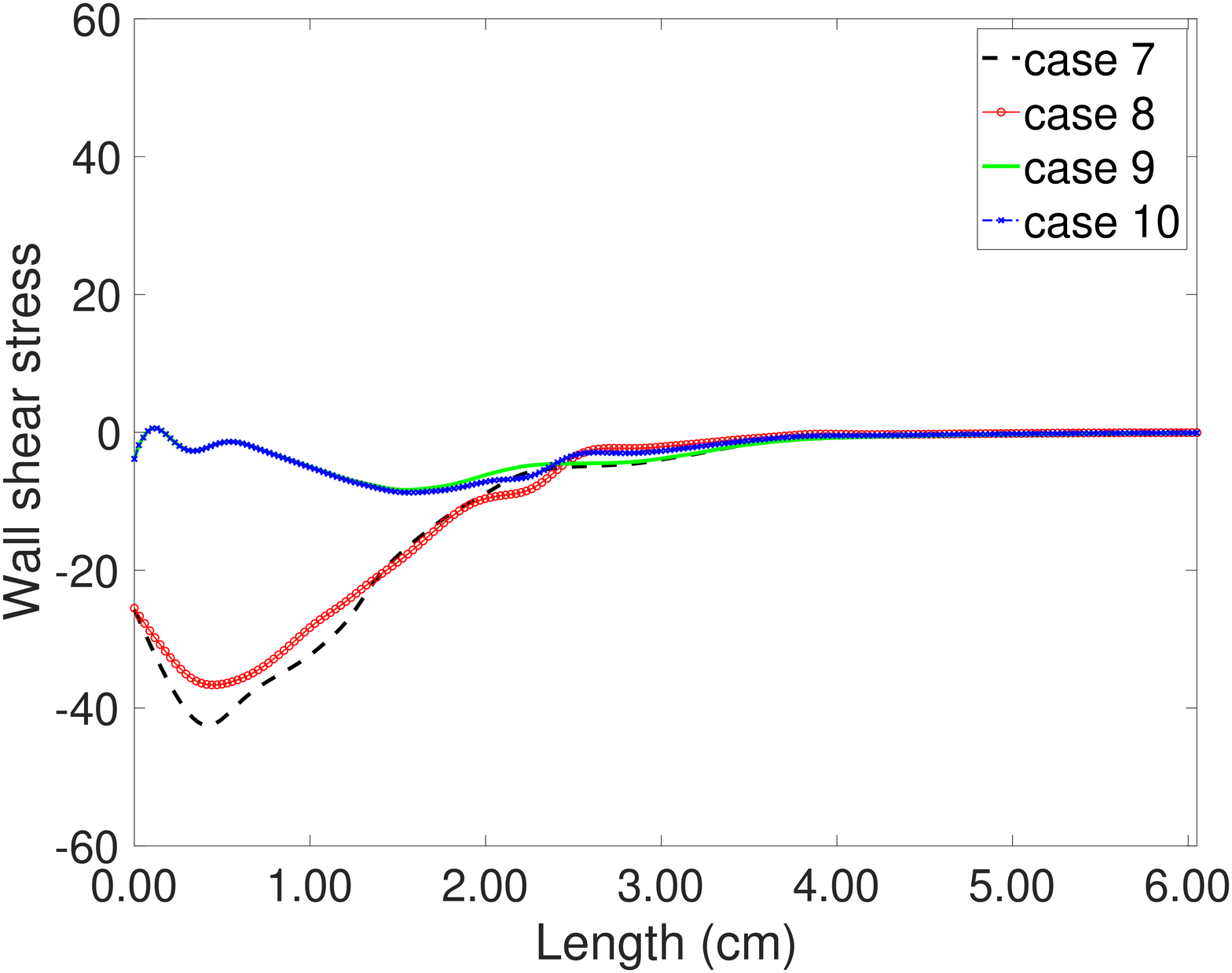}
	\includegraphics[trim=0 60 0 0,scale=0.149]{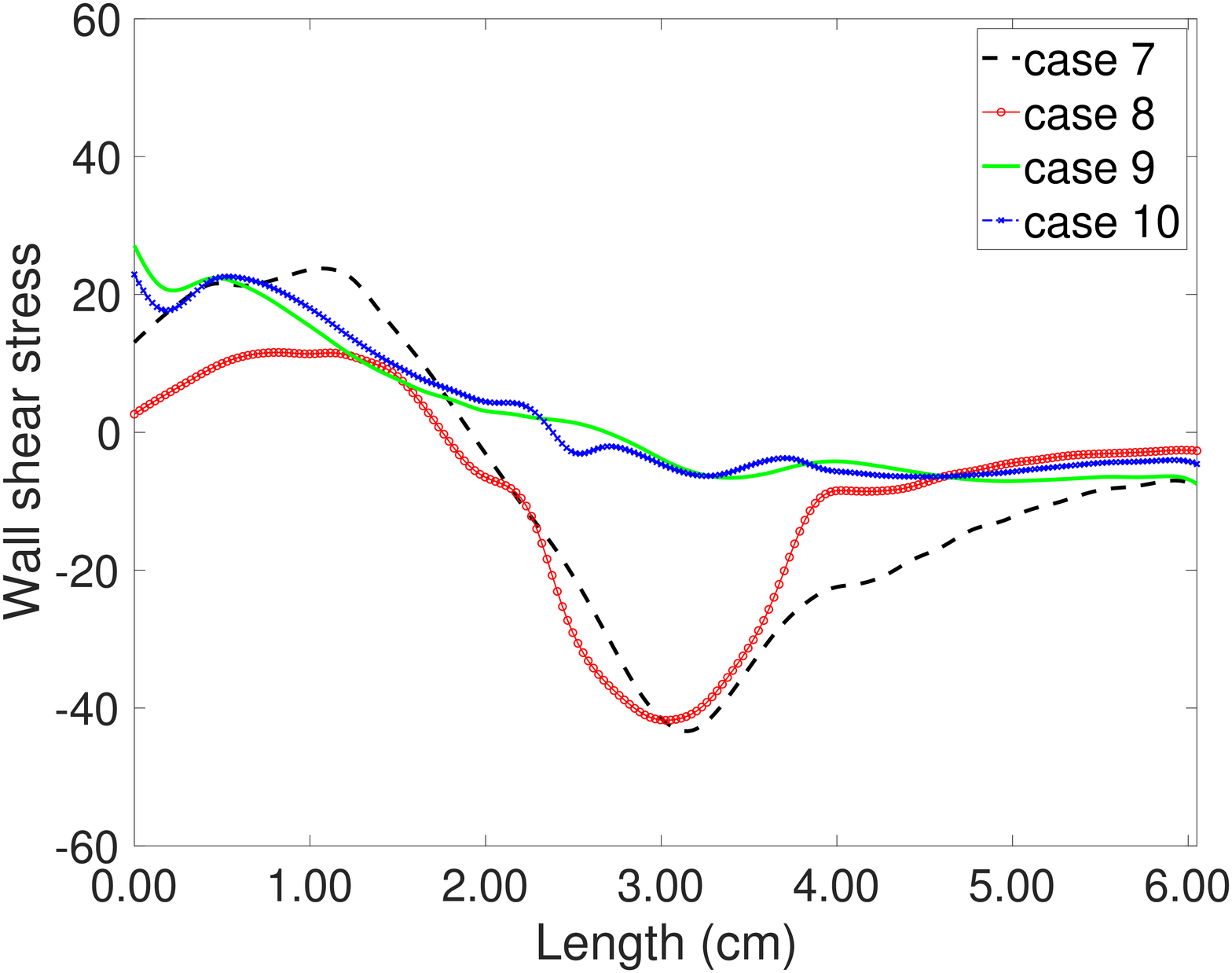}
	\includegraphics[trim=0 60 0 0,scale=0.149]{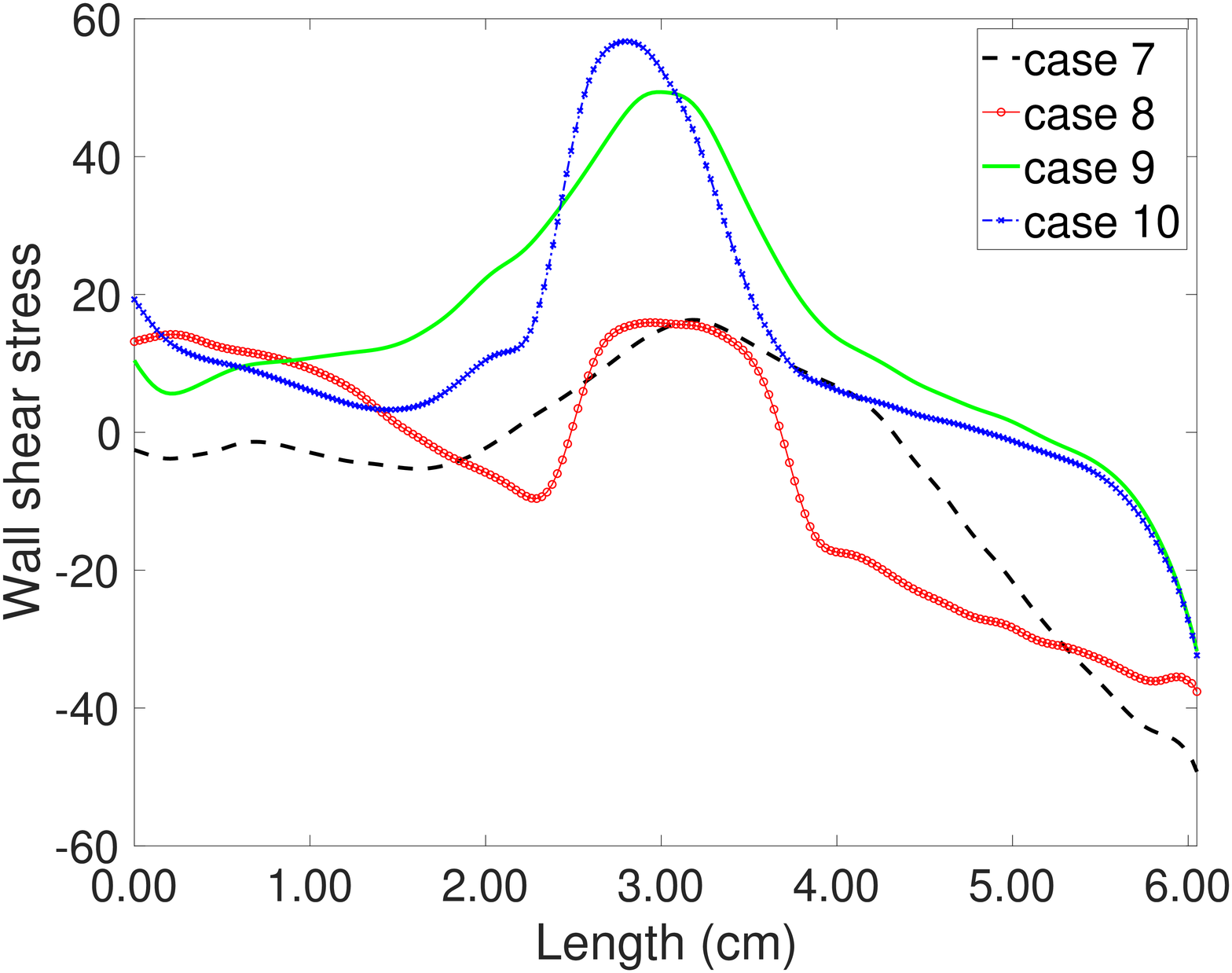}
	\caption{The wall shear stress $\bsi_f\bn\cdot \bt$ along the top arterial wall at time t=1.8 ms, t=3.6 ms, t=5.4 ms for case 7, case 8, case 9 and case 10.}
	\label{swss}
\end{figure}

\begin{figure}[ht!]
	\includegraphics[trim=25 60 25 0,scale=0.149]{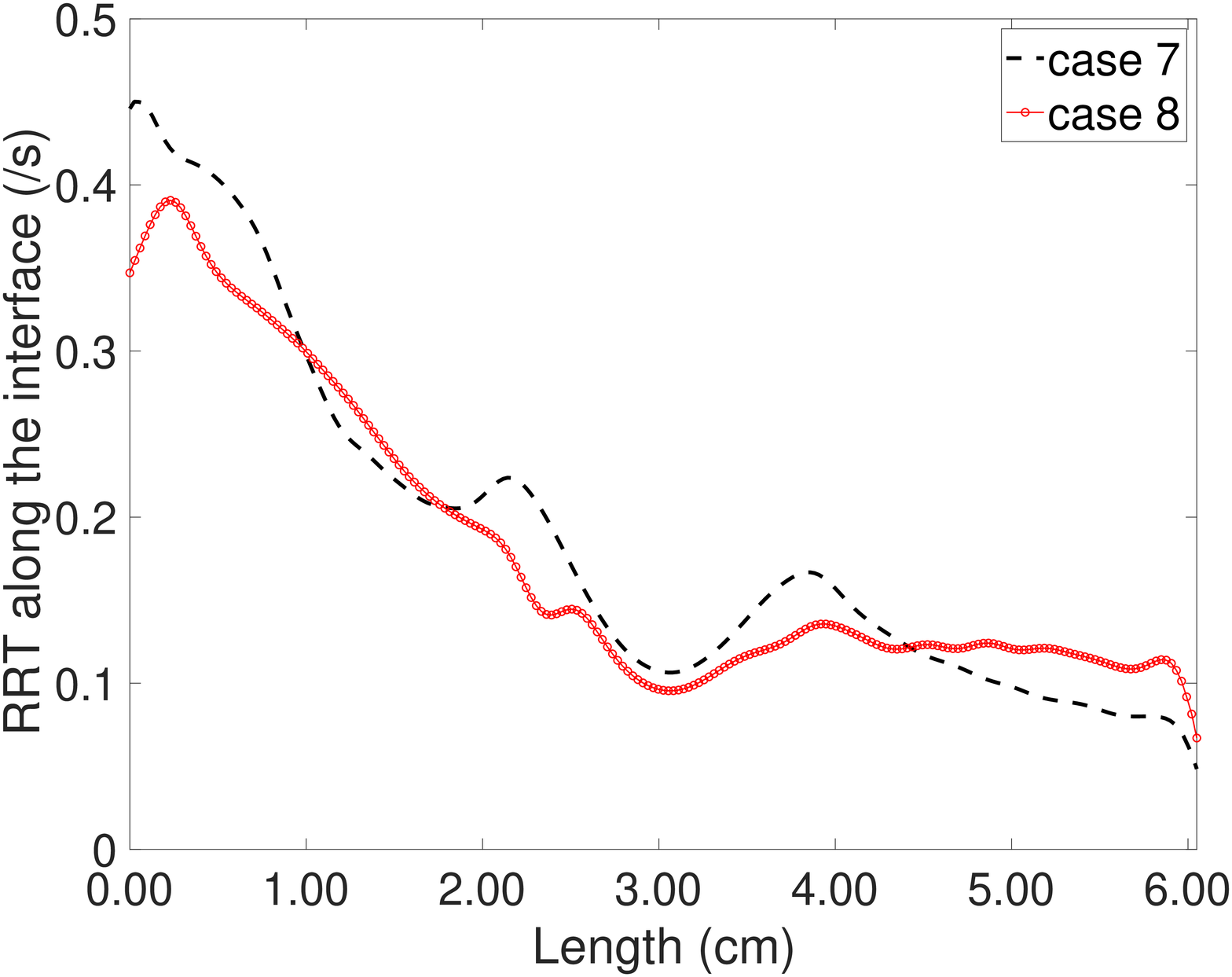}
		\includegraphics[trim=25 60 25 0,scale=0.149]{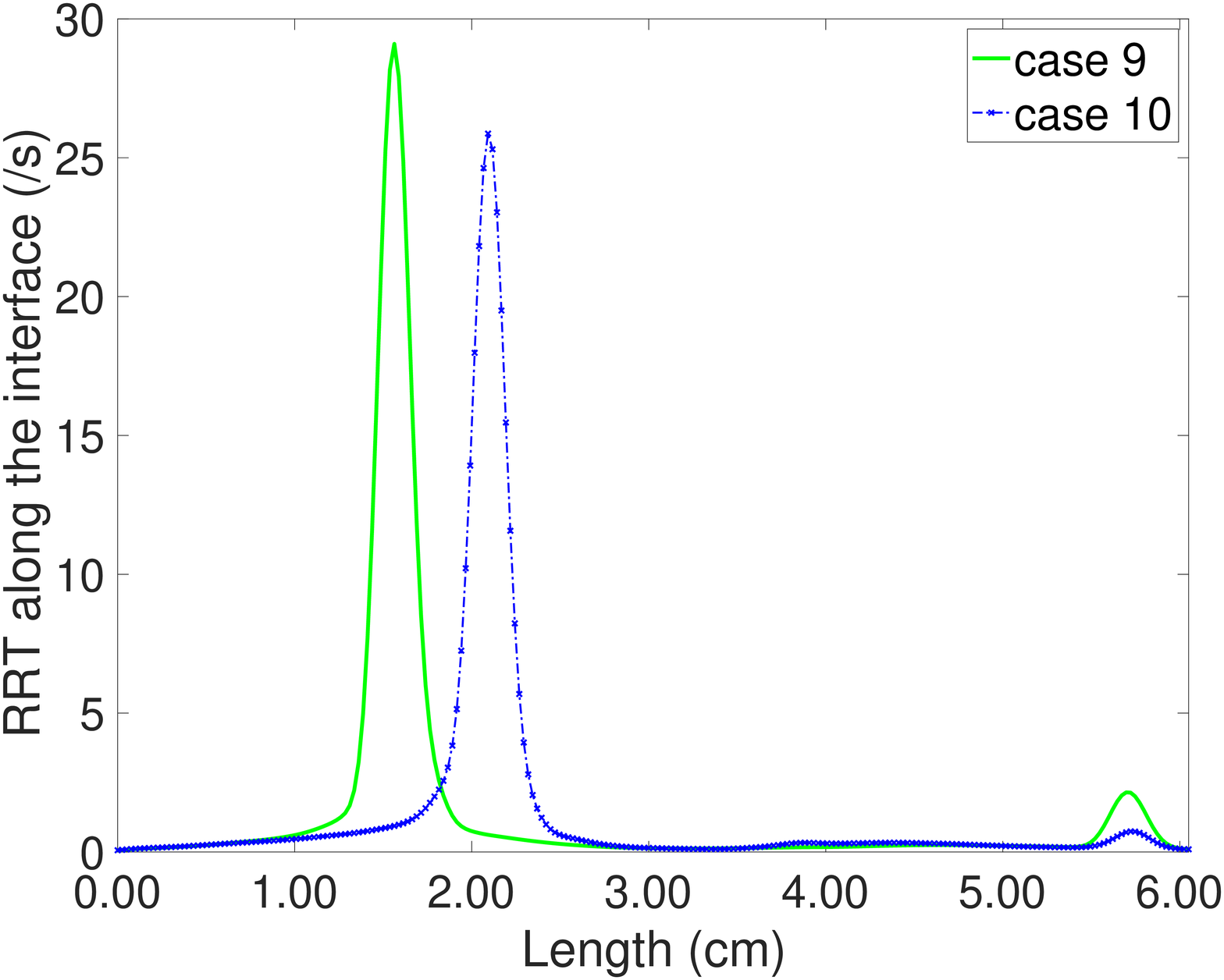}
		\centering
	\caption{RRT along the top arterial wall. The left plot is for case 7, case 8. The right plot is for case 9 and case 10. We use different plots since the scales of RRT for four cases are at difference. We will lose some information of case 7 and case 8 if we put them together. The same pattern is also applied for some of the following plots. }
	\label{SRRT}
\end{figure}

In Fig.\ref{swss}, the WSS for case 7, case 8, case 9 and case 10 are shown respectively. As expected, at three different time, the WSS of case 7 and case 8 are close with each other, while case 9 would be similar with case 10. 
Comparing case 9 and case 10, we can make a conclusion that the discontinuities of permeability and Lam\'{e} don't affect the WSS much. As a conclusion, it is the poroelasticity that makes a difference on WSS. In Fig. \ref{SRRT}, we present the RRT for four stenosis cases. There are huge differences between NSE/E and NSE/P models. The peaks of NSE/P models would be much higher than that in NSE/E models. The peaks of case 7 and case 8 are only attaining $0.45$ $s^{-1}$. While for case 9, the peak could reach to approximately 30 $s^{-1}$. In addition, for NSE/P models, the peaks of RRT would be typically showing up before the stenosis part.                                                
\begin{figure}[ht!]
	\includegraphics[trim=0 60 0 0,scale=0.149]{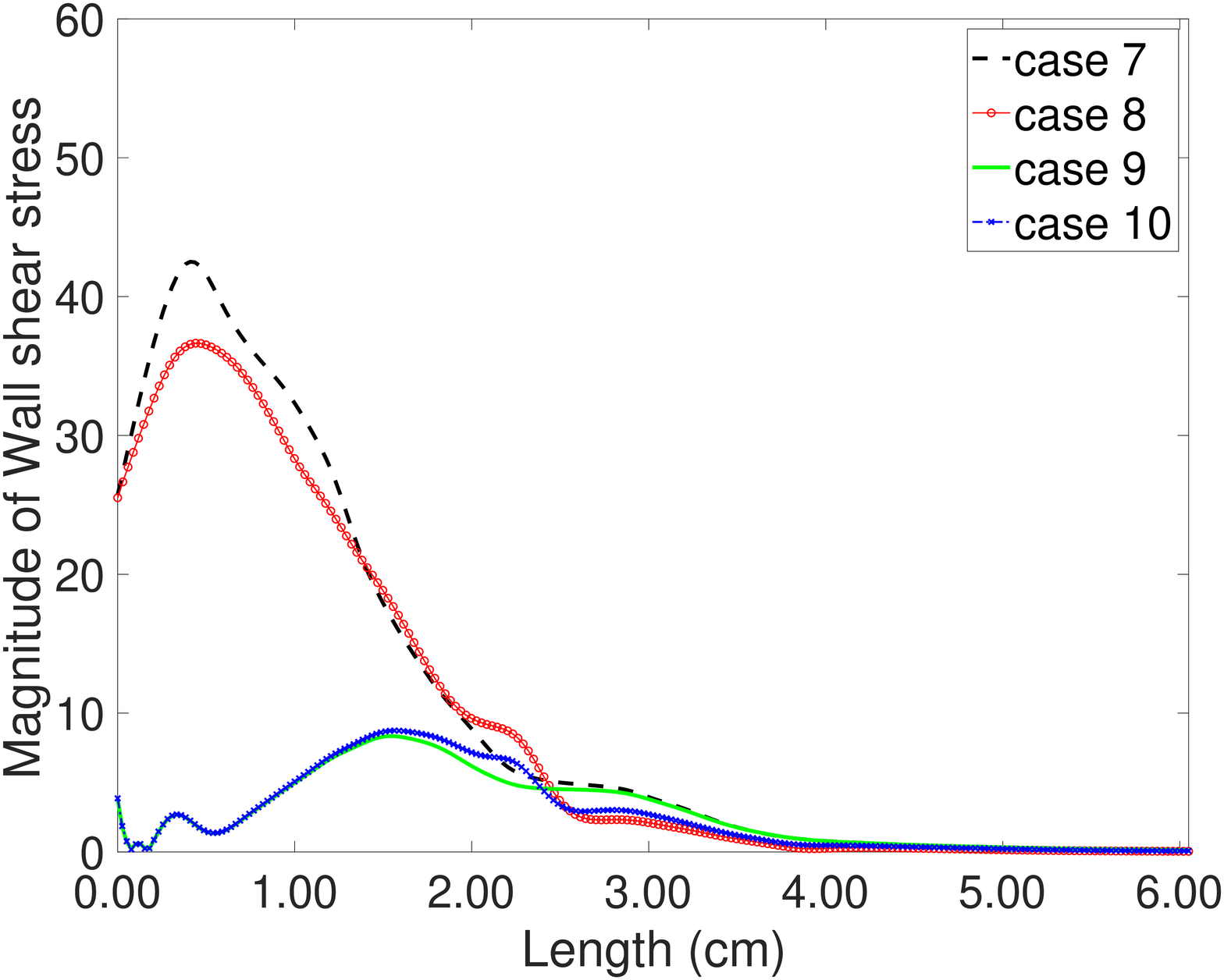}
	\includegraphics[trim=0 60 0 0,scale=0.149]{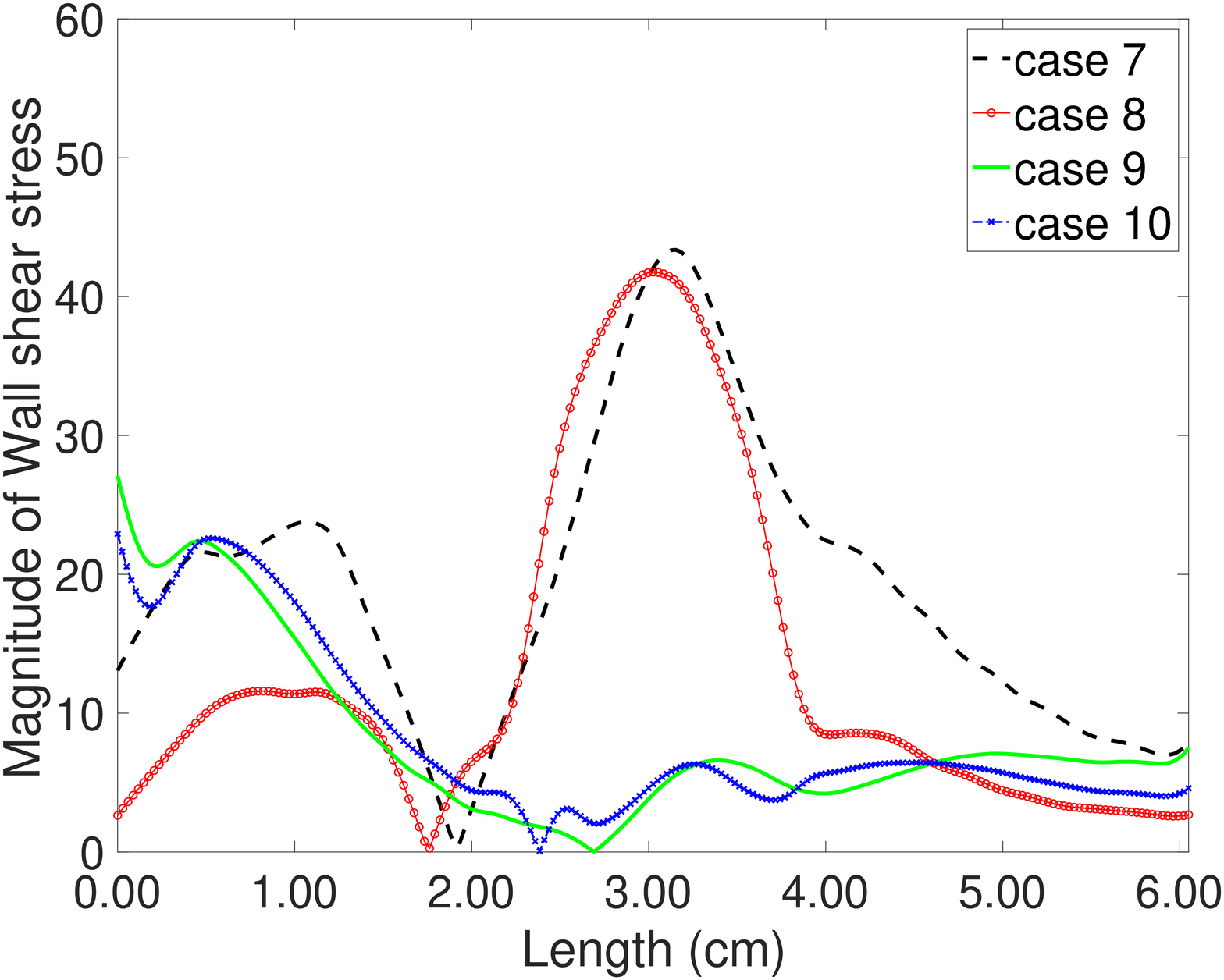}
	\includegraphics[trim=0 60 0 0,scale=0.149]{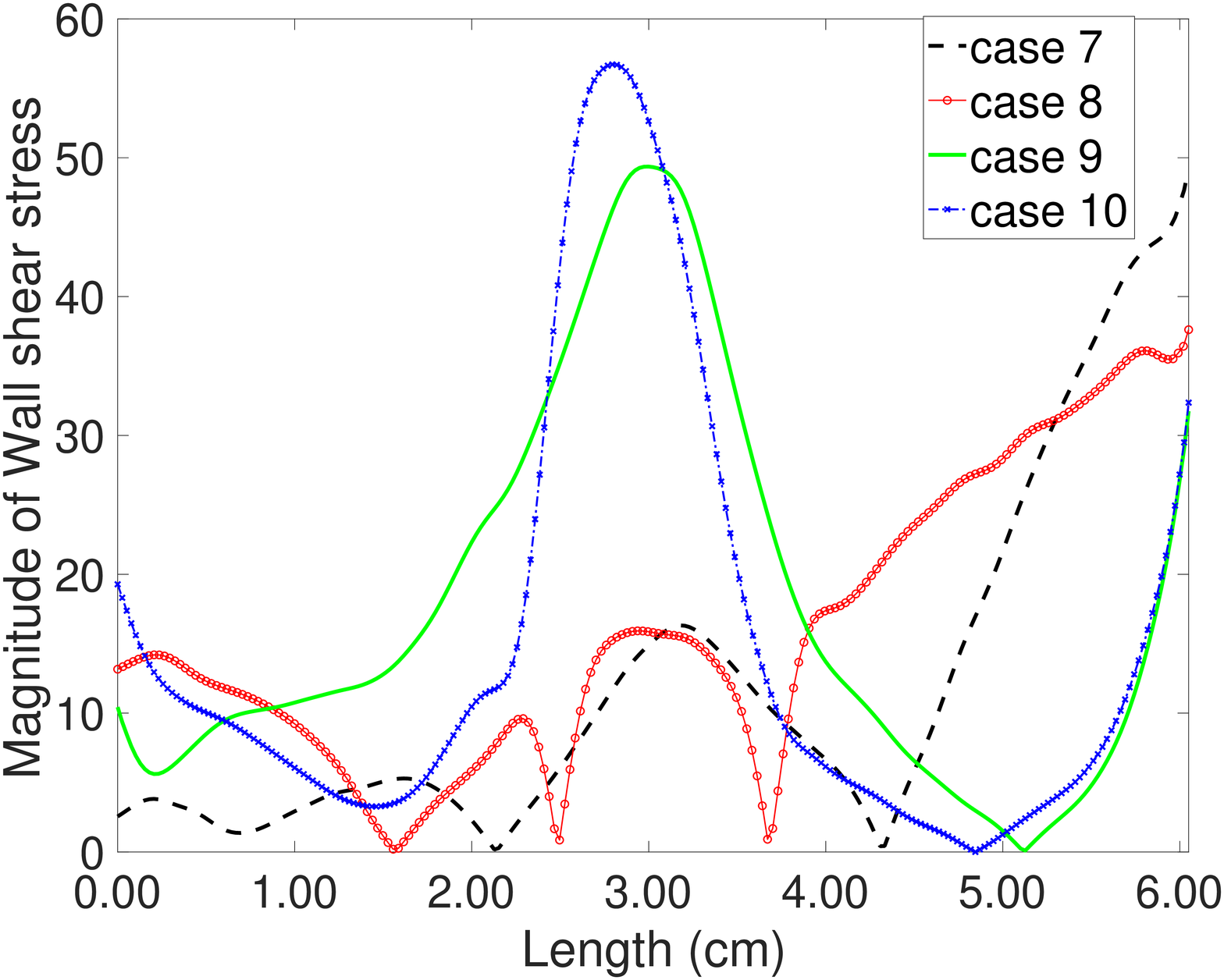}
	\caption{The magnitude of the wall shear stress $|\bsi_f\bn\cdot \bt|$ along the top arterial wall at time t=1.8 ms, t=3.6 ms, t=5.4 ms for case 7, case 8, case 9 and case 10.}
	\label{sss}
\end{figure}

\begin{figure}[ht!]
	\includegraphics[trim=0 60 0 0,scale=0.149]{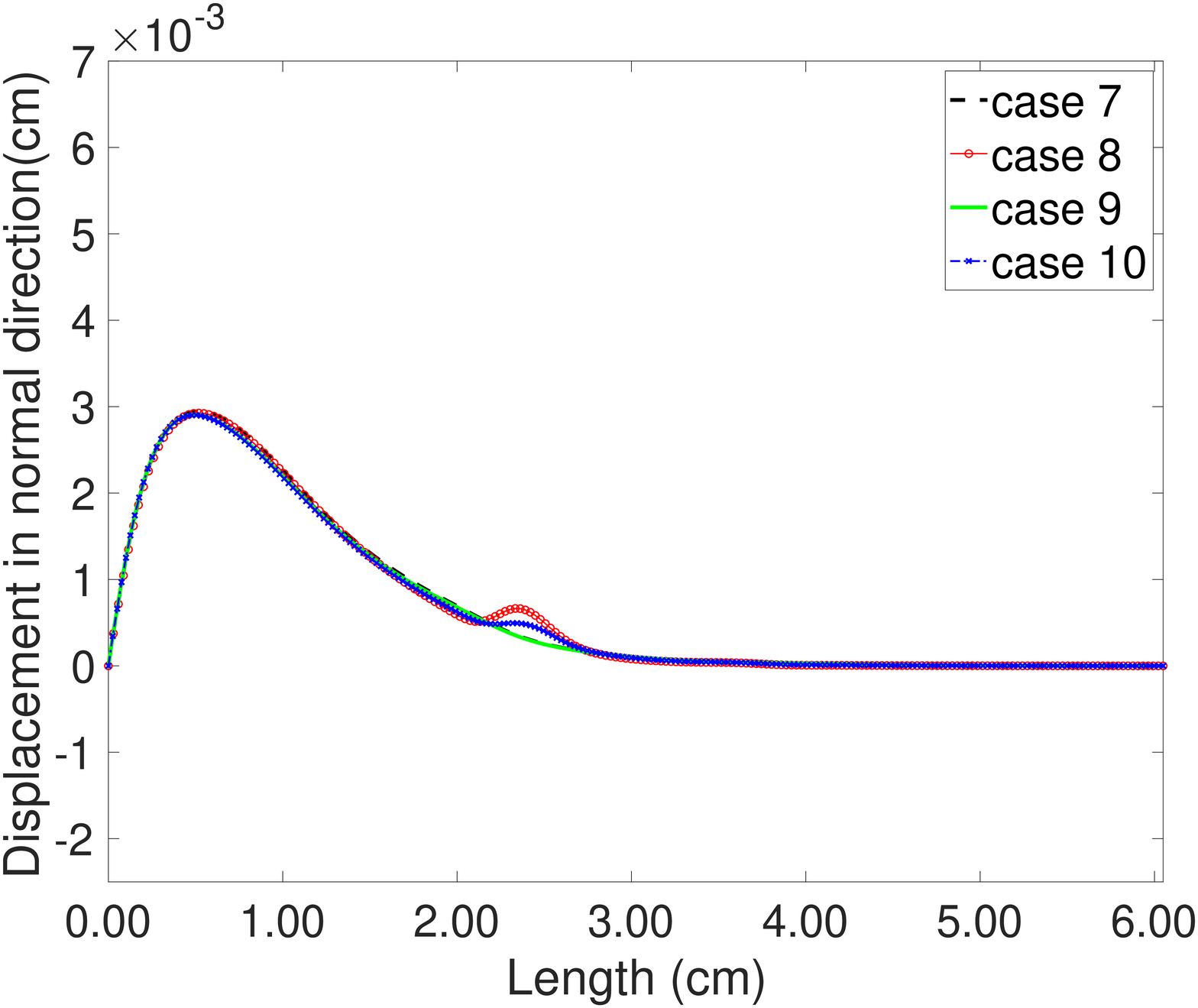}
	\includegraphics[trim=0 60 0 0,scale=0.149]{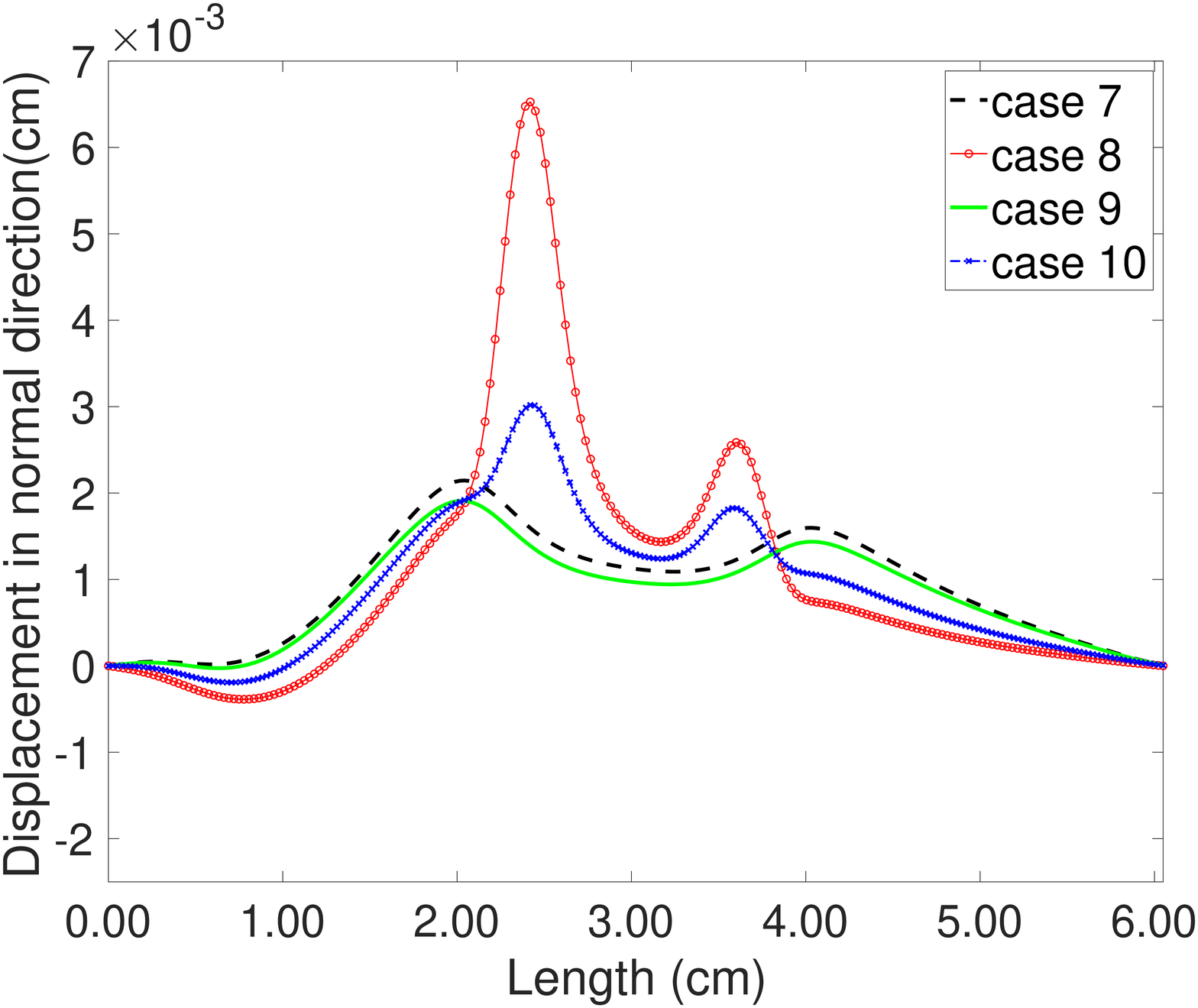}
	\includegraphics[trim=0 60 0 0,scale=0.149]{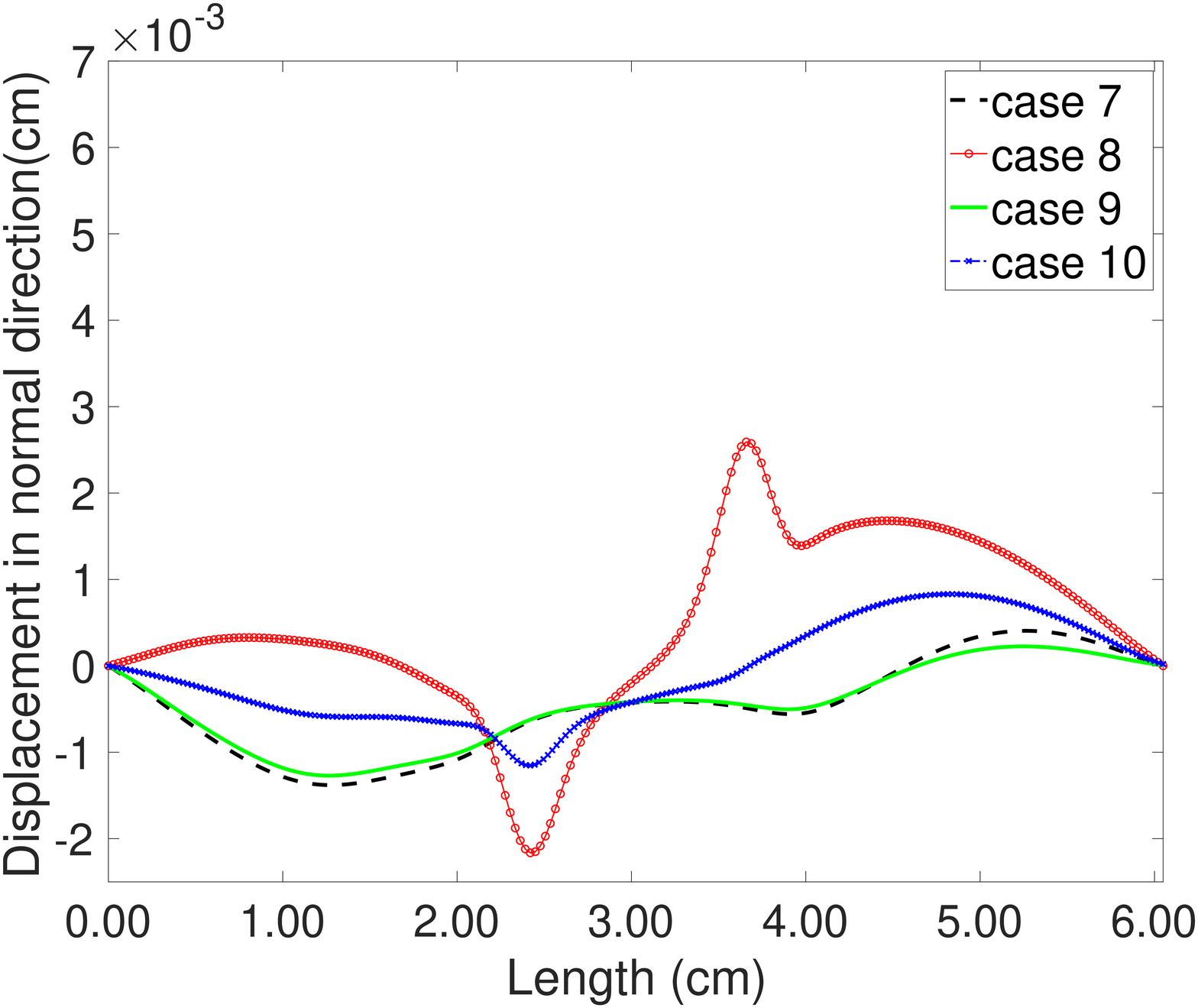}
	\caption{Displacement in the normal direction $\bbeta_{\star}\cdot \bn$ along the top arterial wall at time t=1.8 ms, t=3.6 ms, t=5.4 ms for case 7, case 8, case 9 and case 10.}
	\label{sdis}
\end{figure}

In Fig.\ref{sdis}, we present the displacement in the normal direction $\bbeta_{\star}\cdot \bn$ along the arterial wall at time $t=1.8, 3.6, 5.4$ ms. Comparing case 7 with case 8, we note that smaller Lam\'{e} coefficients would contribute to larger displacement in normal direction. While when we compare case 8 with case 10, NSE/P model would result in smaller normal displacement. We conclude that both Lam\'{e} coefficients and permeability $K$ affect the normal displacement. Another interesting phenomena is that at time $t=3.6$ ms, unlike the previous cases with healthy structures, the peaks of $\bbeta_{\star}\cdot \bn$ of all four stenotic cases, showing up at the beginning part of stenosis, don't conincide with the peaks of pressure $p_f$. 

\begin{figure}[ht!]
	\includegraphics[trim=0 60 0 0,scale=0.149]{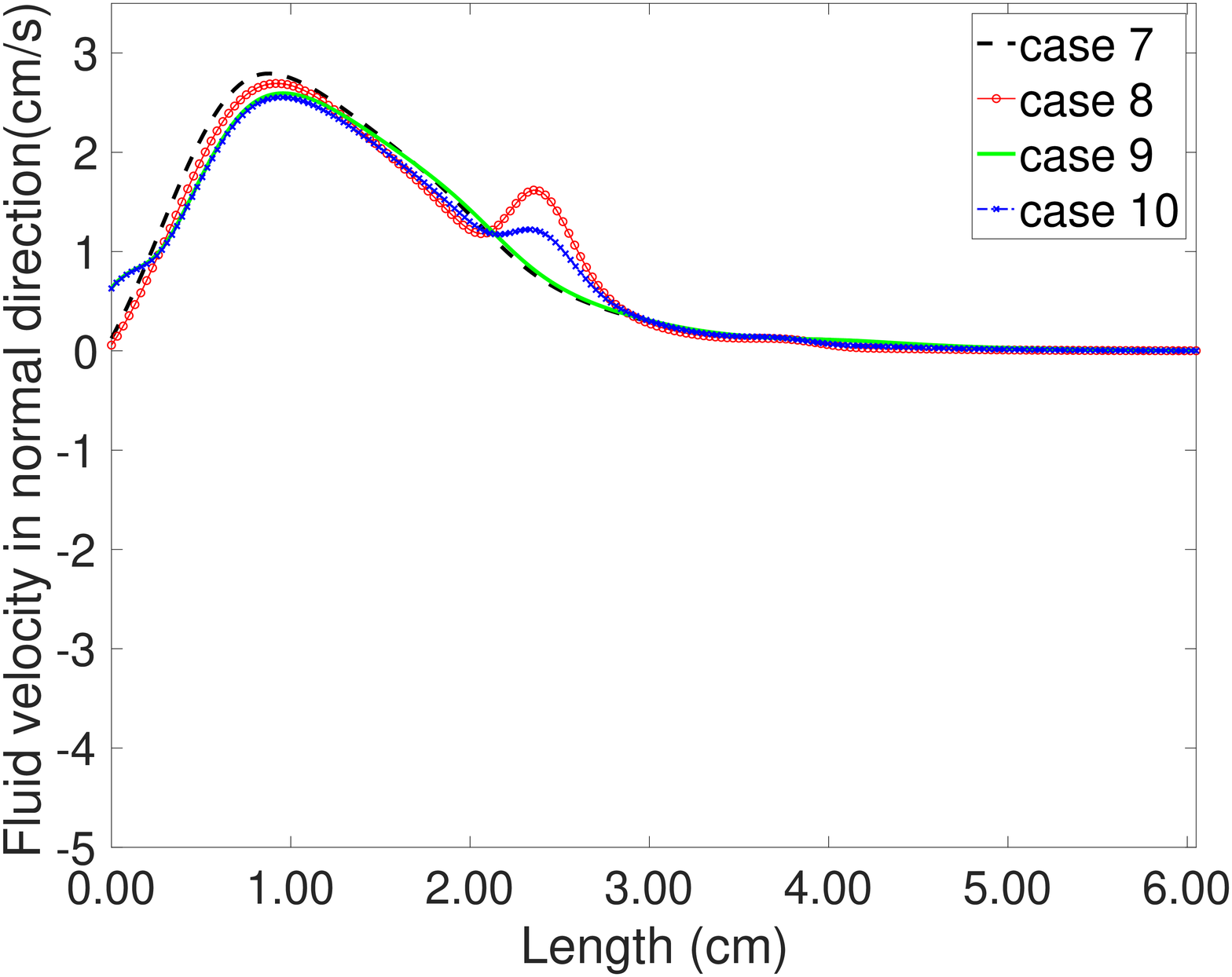}
	\includegraphics[trim=0 60 0 0,scale=0.149]{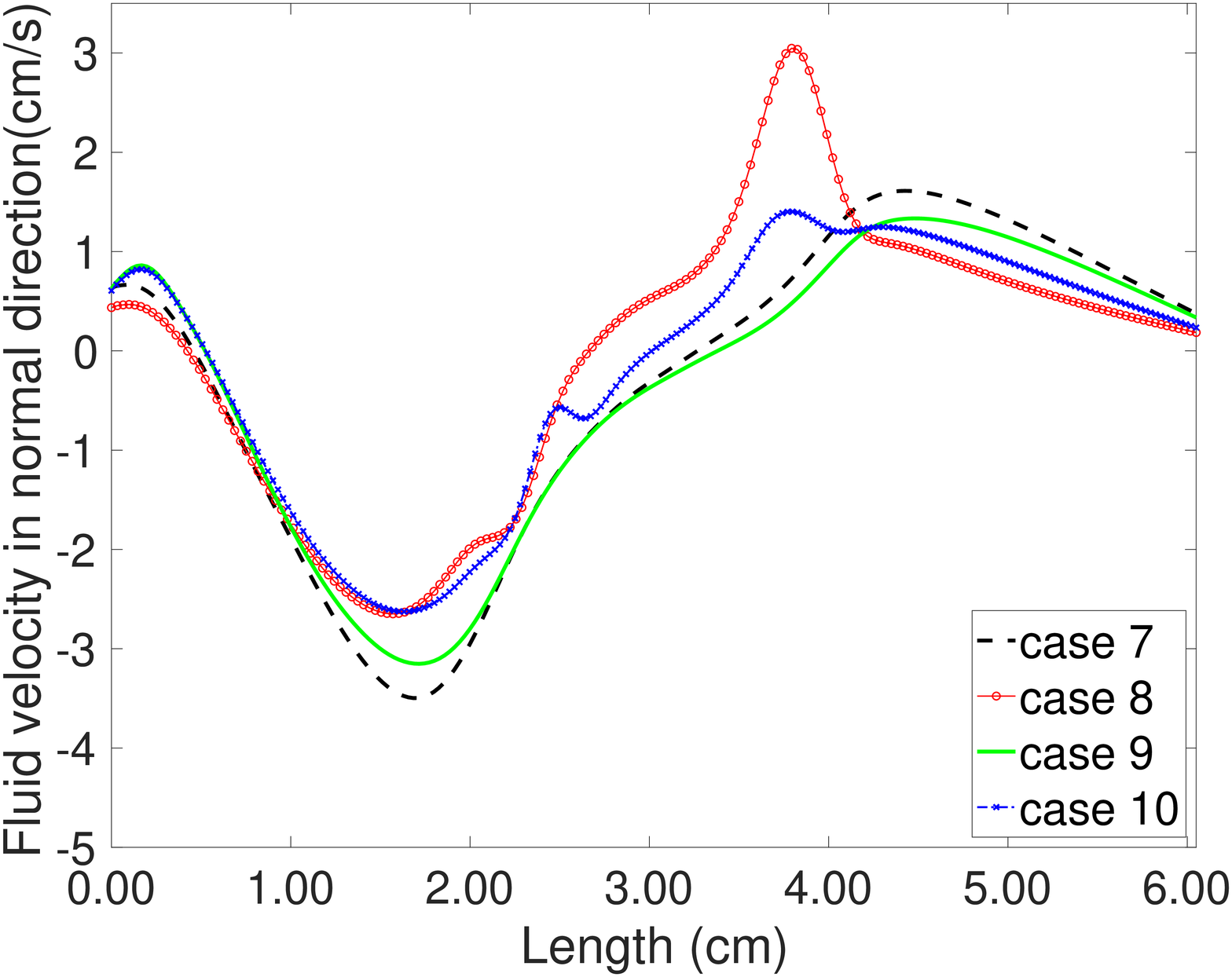}
	\includegraphics[trim=0 60 0 0,scale=0.149]{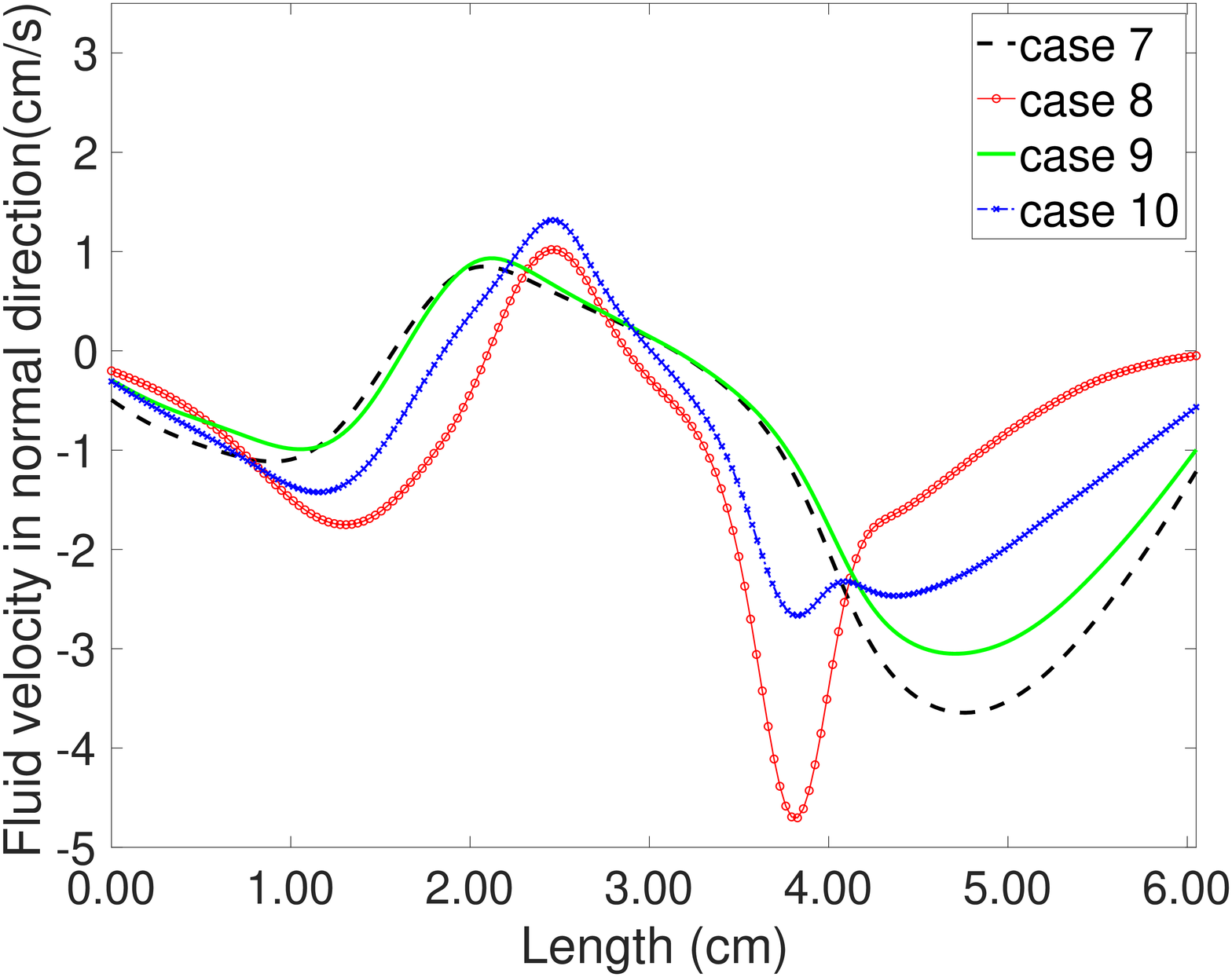}
	\caption{Fluid velocity in the normal direction $\bu_{f}\cdot \bn$ along the top arterial wall at time t=1.8 ms, t=3.6 ms, t=5.4 ms for case 7, case 8, case 9 and case 10. }
		\label{sufn}
\end{figure}

In fig.\ref{sufn}, we present the fluid velocity in normal direction $\bu_{f}\cdot \bn$ along the arterial wall at different time for case 7, case 8, case 9 and case 10, respectively. 

\begin{figure}[ht!]
	\includegraphics[trim=0 0 0 0,scale=0.149]{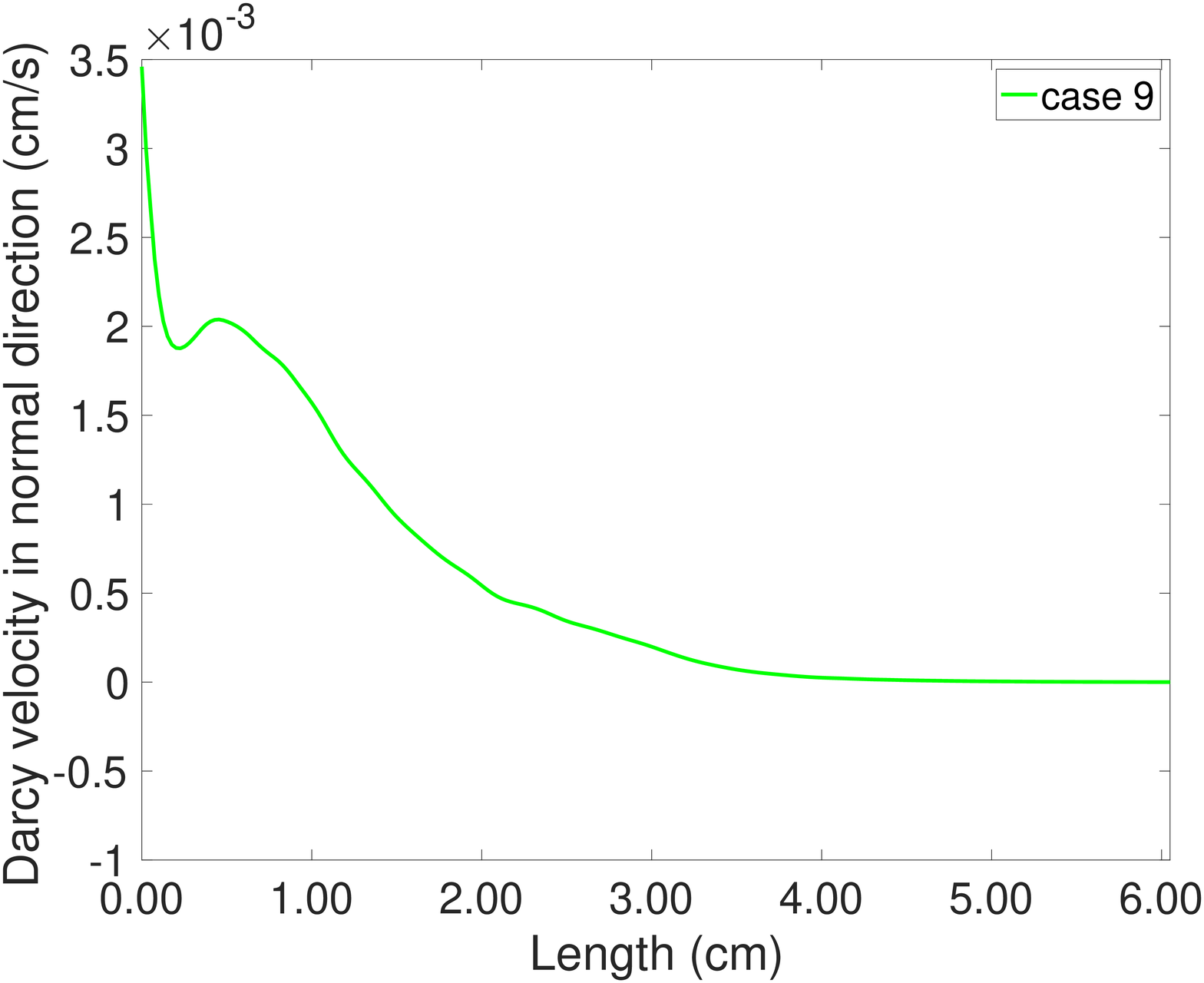}
	\includegraphics[trim=0 0 0 0,scale=0.149]{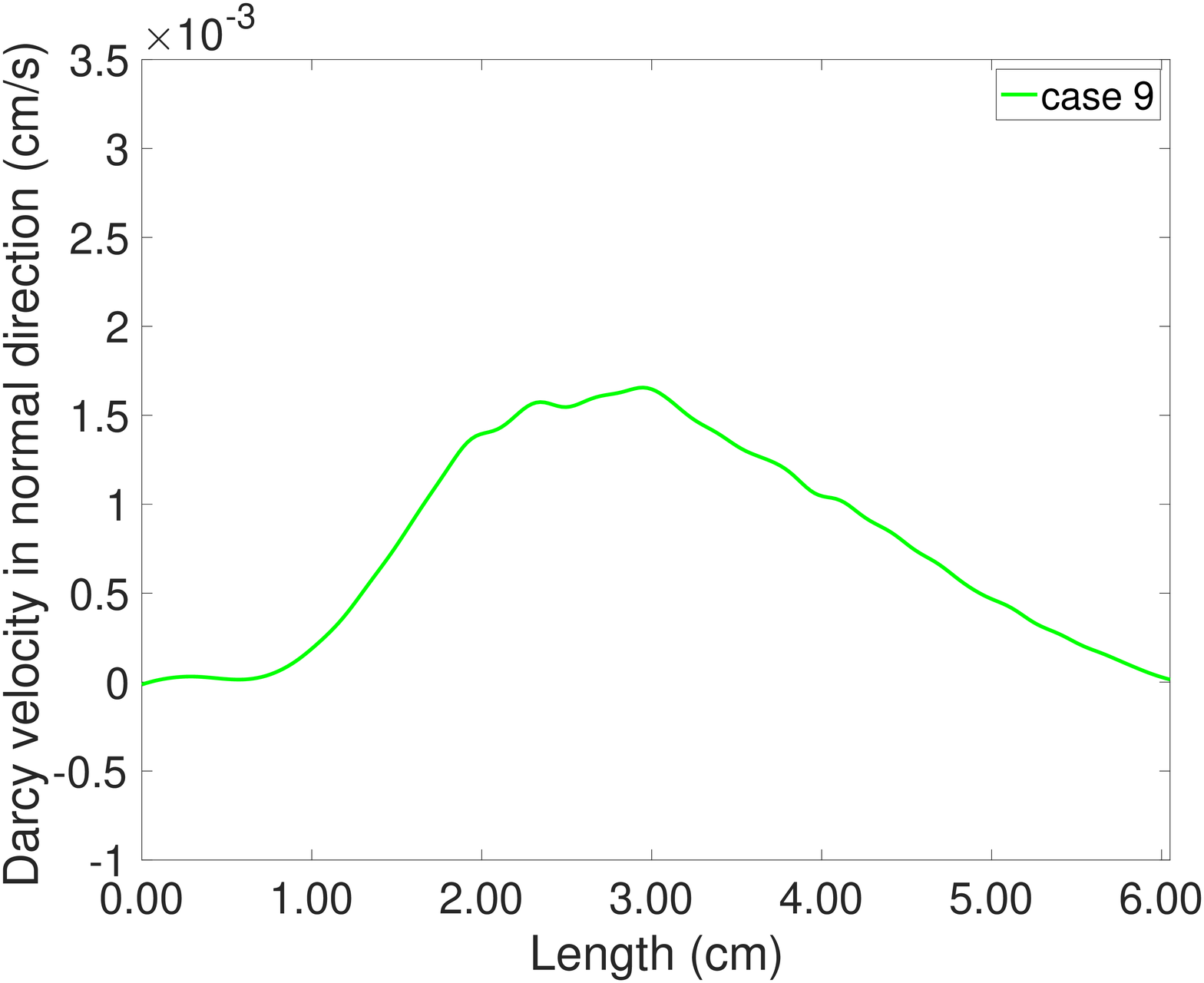}
	\includegraphics[trim=0 0 0 0,scale=0.149]{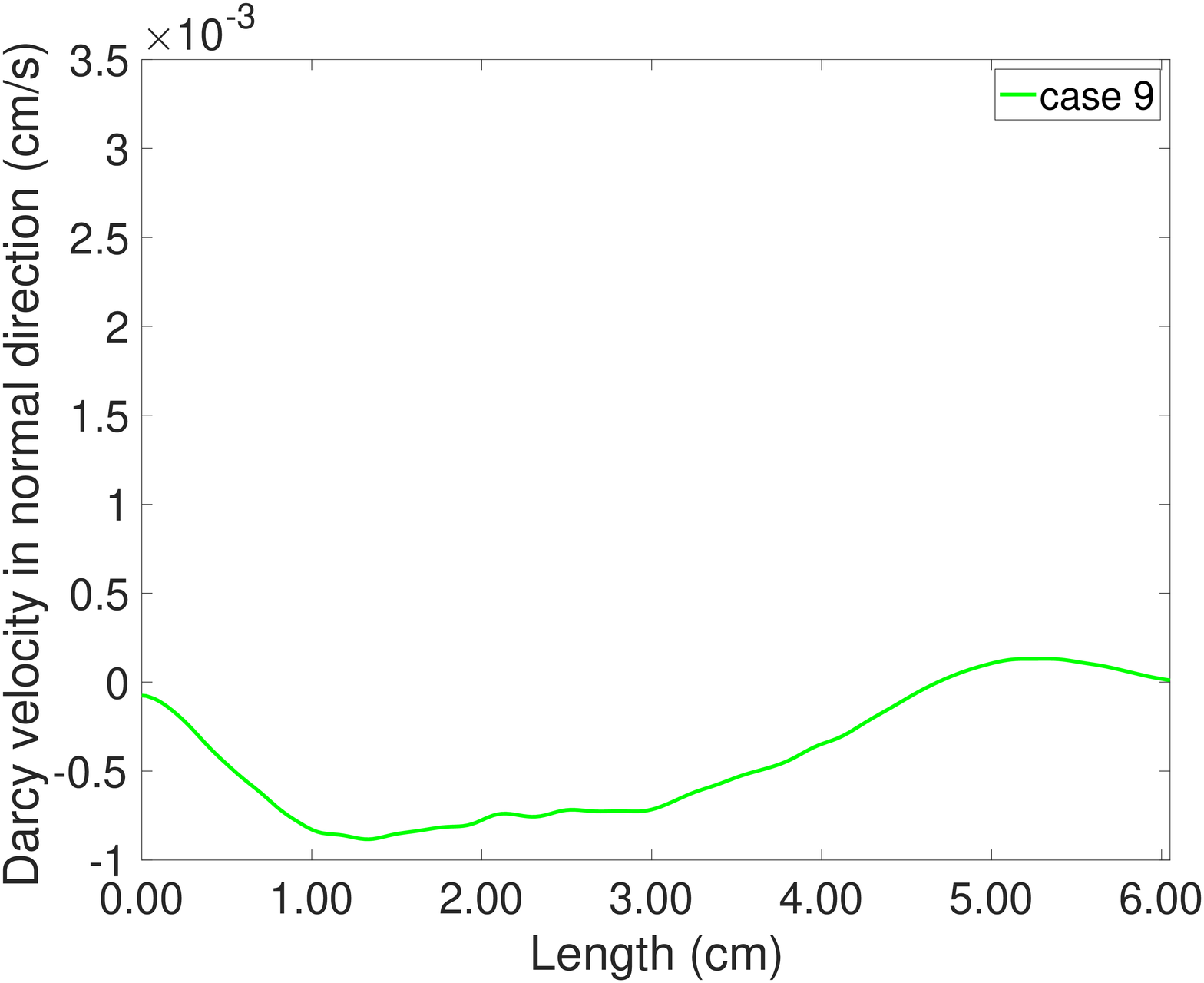}
	\includegraphics[trim=0 60 0 0,scale=0.149]{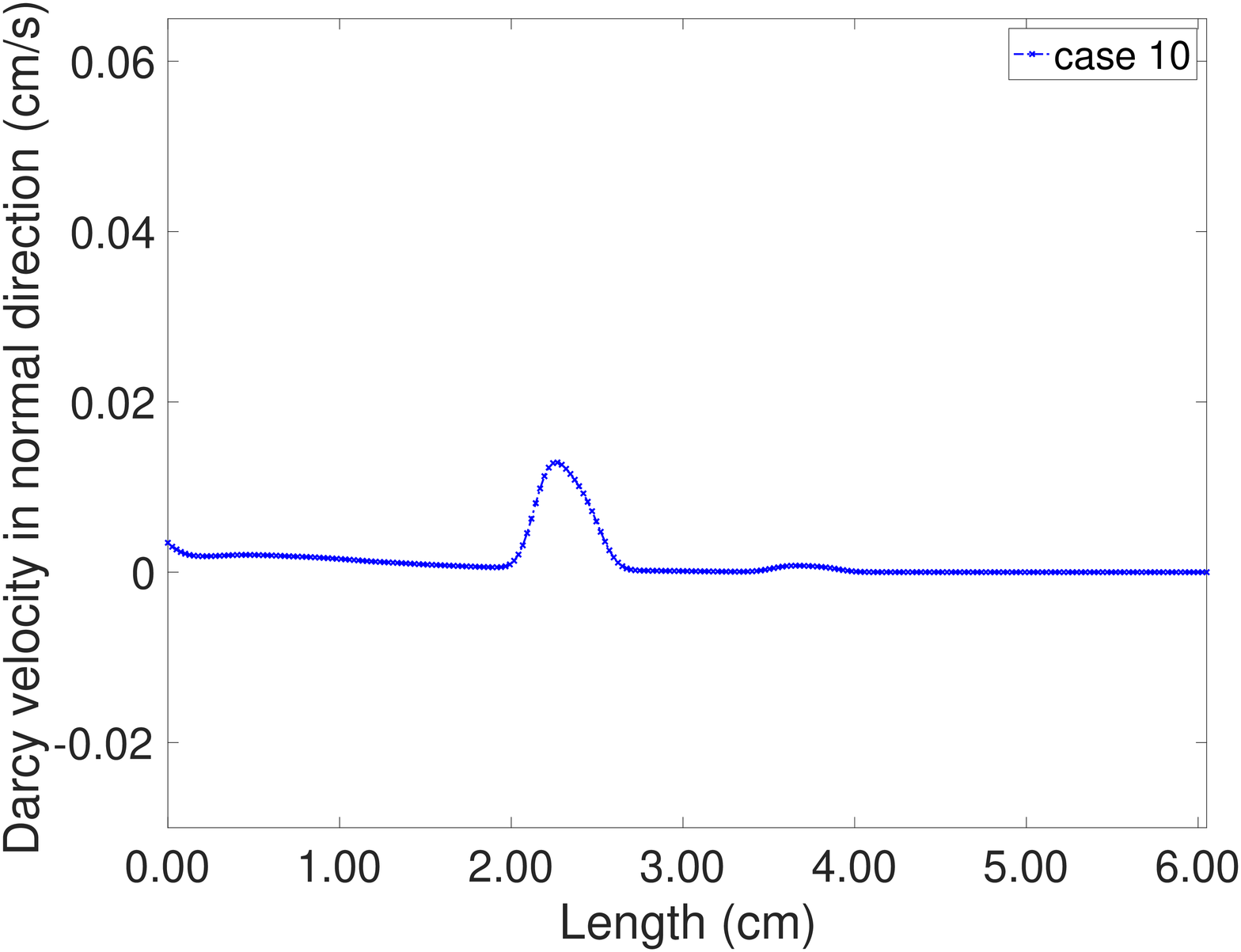}
	\includegraphics[trim=0 60 0 0,scale=0.149]{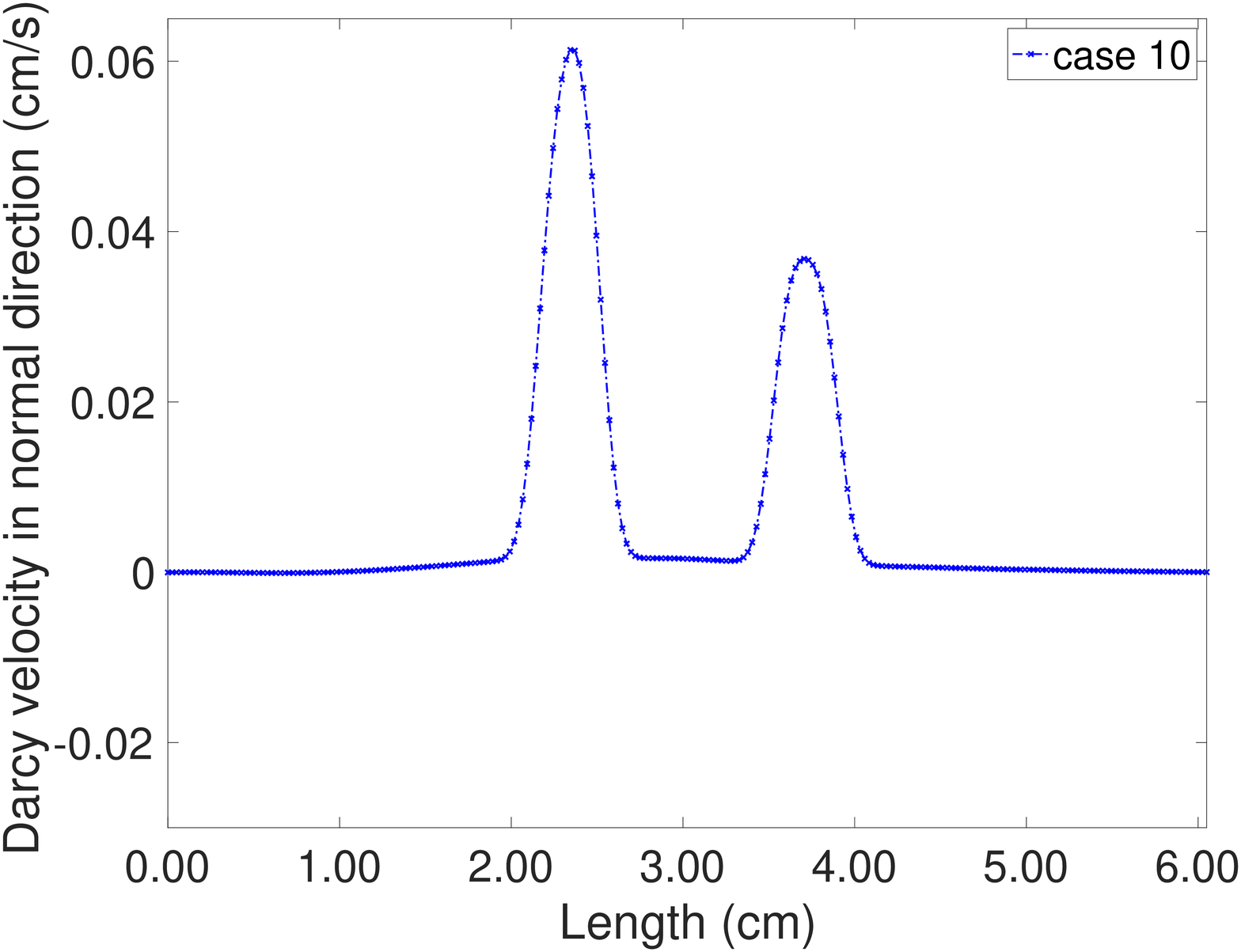}
	\includegraphics[trim=0 60 0 0,scale=0.149]{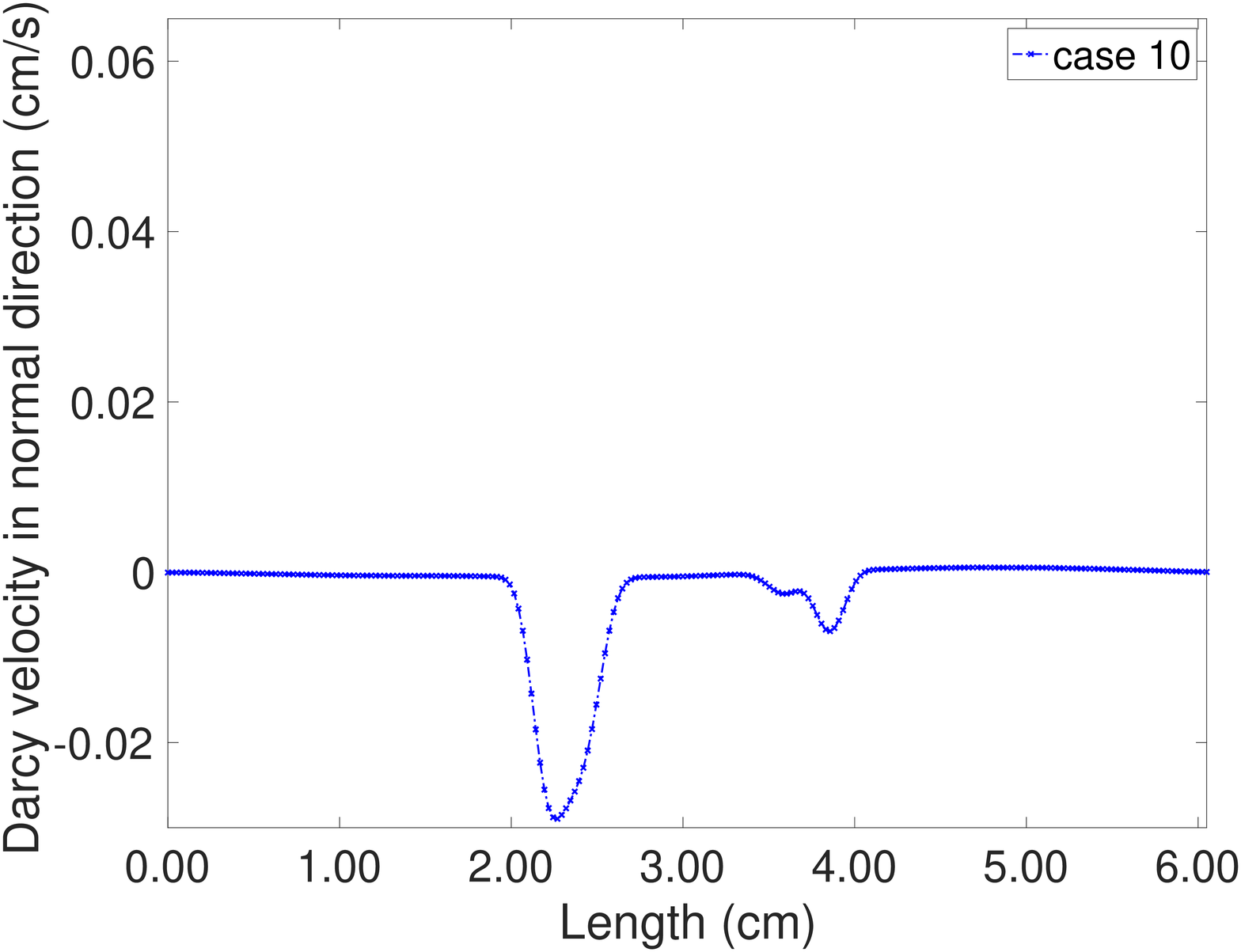}
	\caption{Darcy velocity in the normal direction $\bu_{p}\cdot \bn$ along the top arterial wall at time t=1.8 ms, t=3.6 ms, t=5.4 ms for case 9 and case 10. The top panel of this figure is for case 9, and the bottom panel is case 10. }
		\label{supn}
\end{figure}

\begin{figure}[ht!]
	\includegraphics[trim=0 0 0 0,scale=0.149]{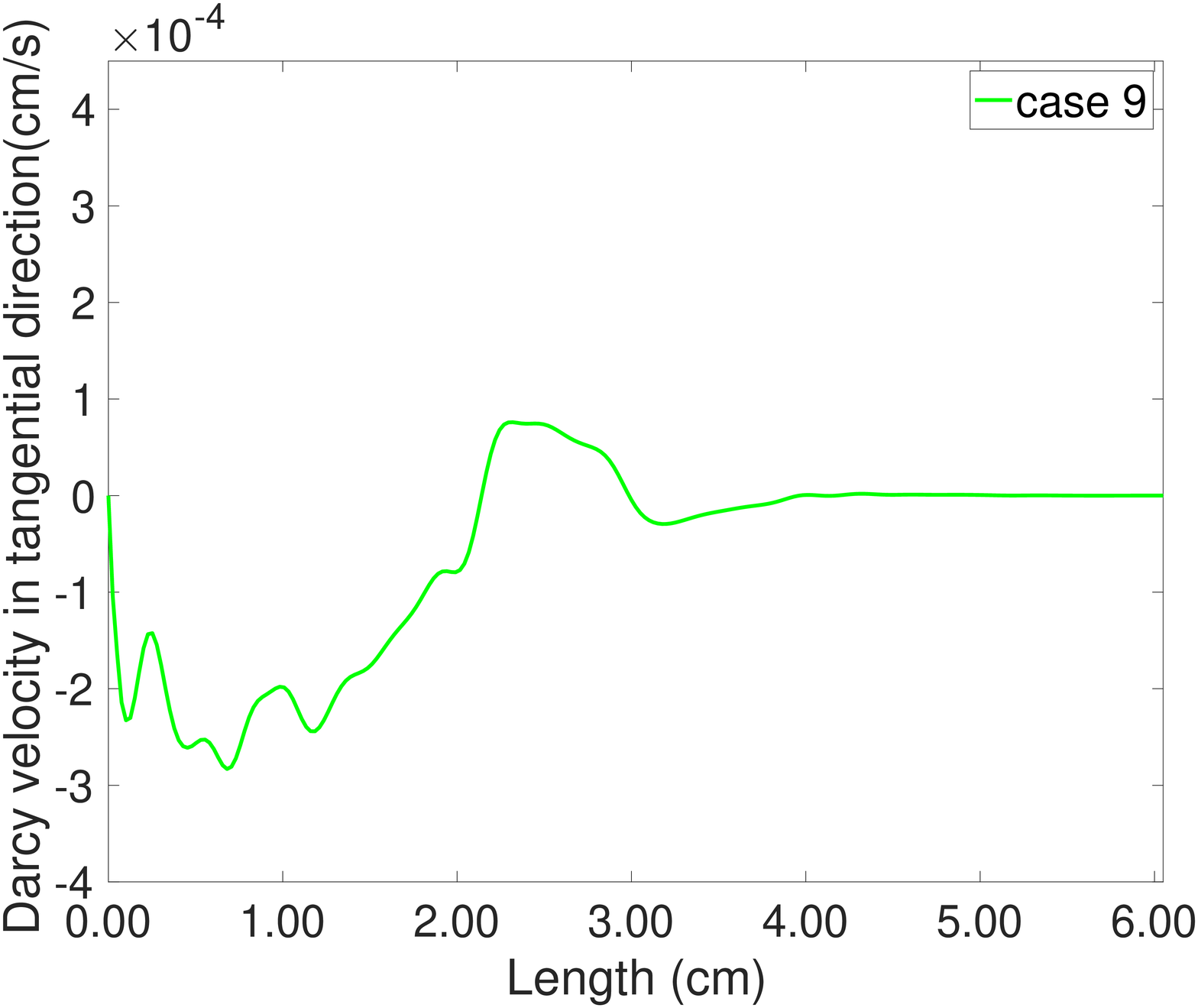}
	\includegraphics[trim=0 0 0 0,scale=0.149]{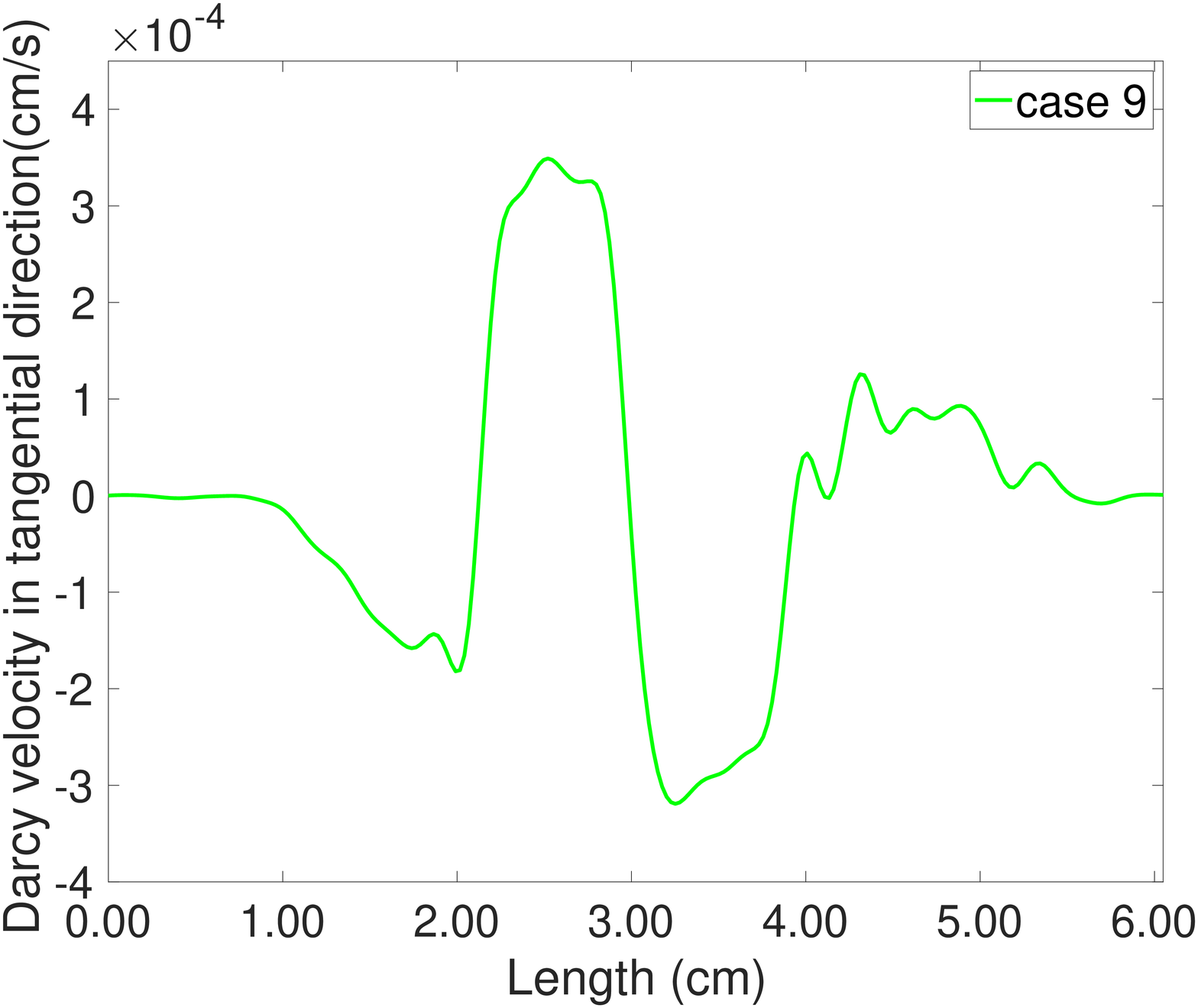}
	\includegraphics[trim=0 0 0 0,scale=0.149]{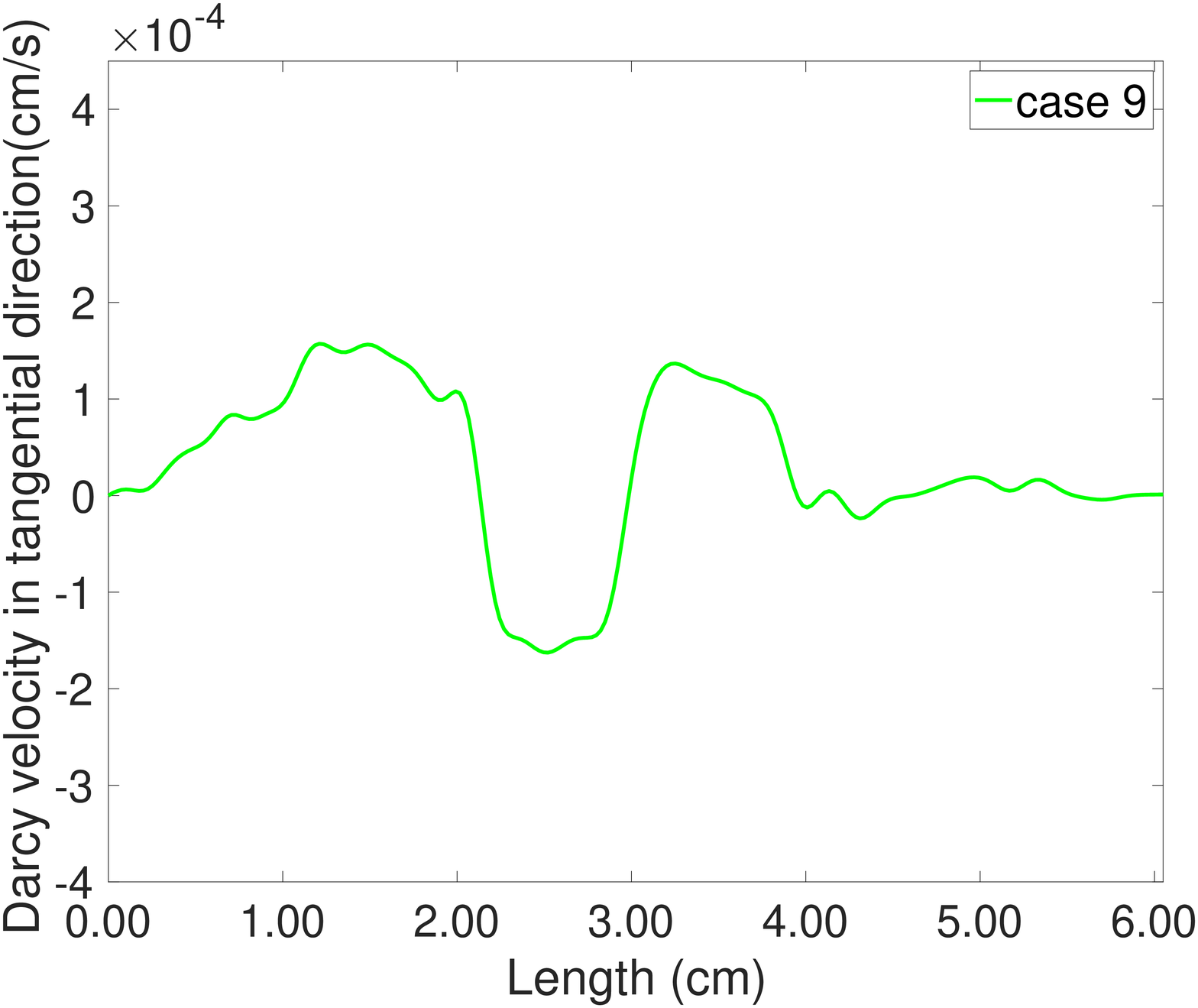}
	\includegraphics[trim=0 60 0 0,scale=0.149]{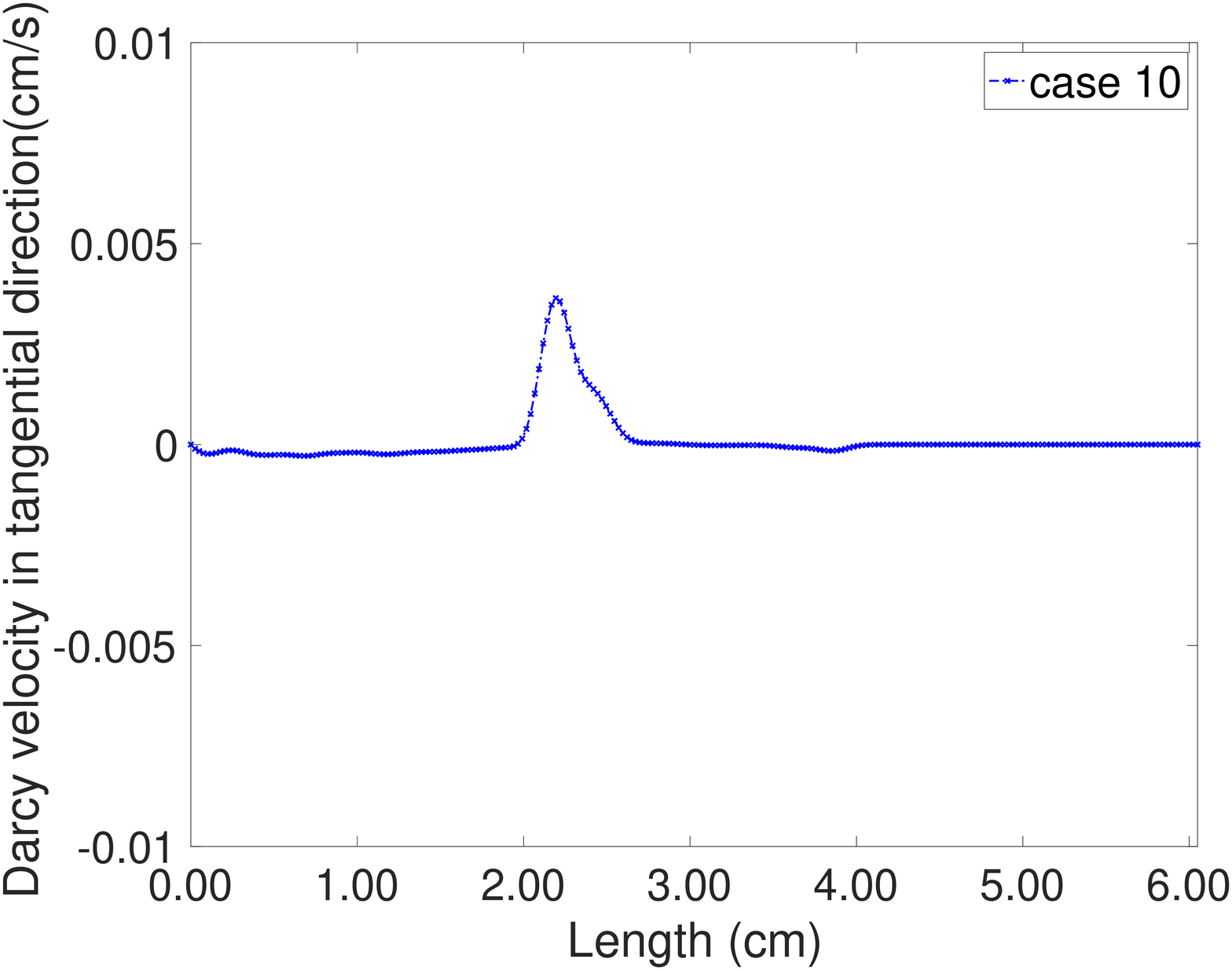}
	\includegraphics[trim=0 60 0 0,scale=0.149]{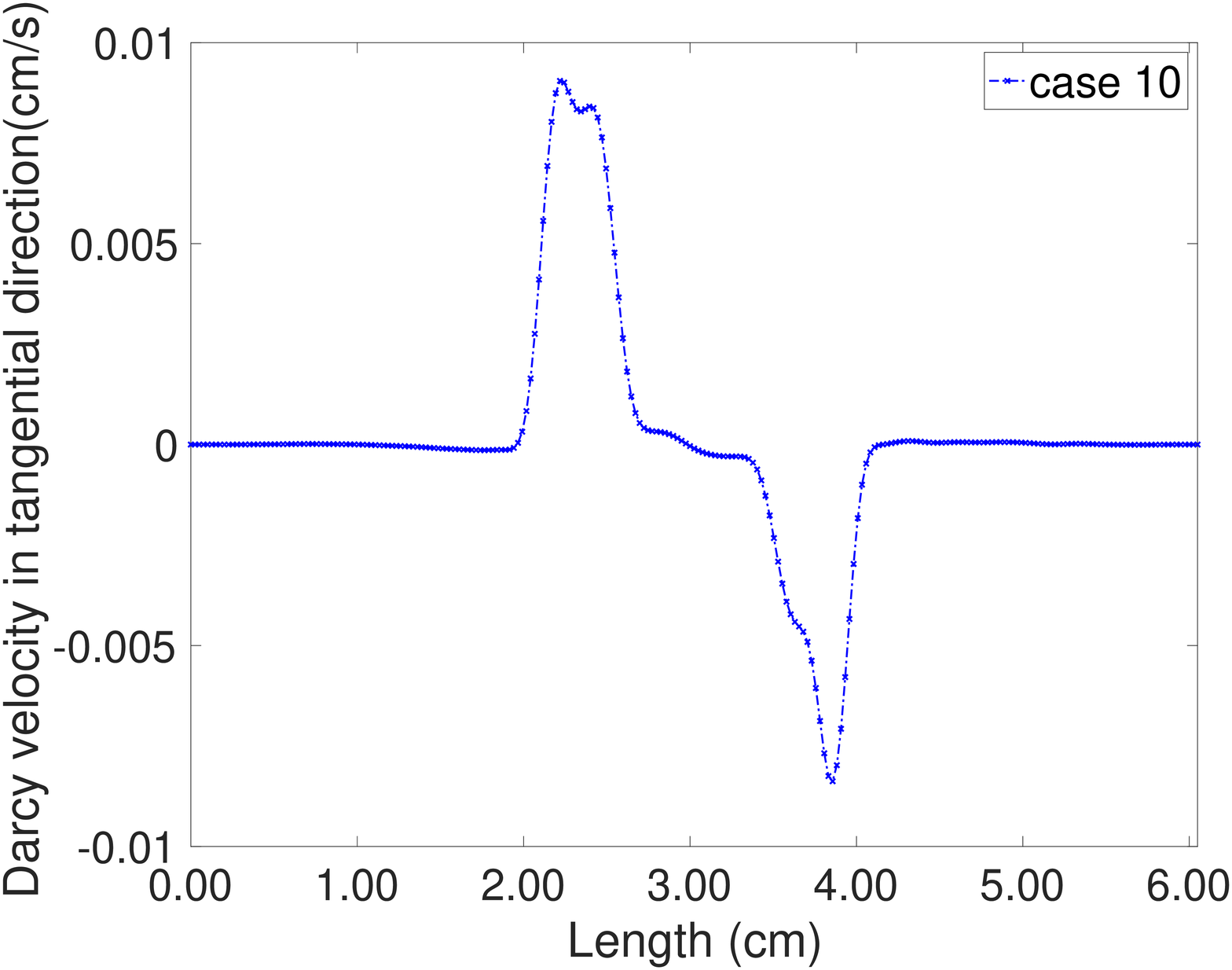}
	\includegraphics[trim=0 60 0 0,scale=0.149]{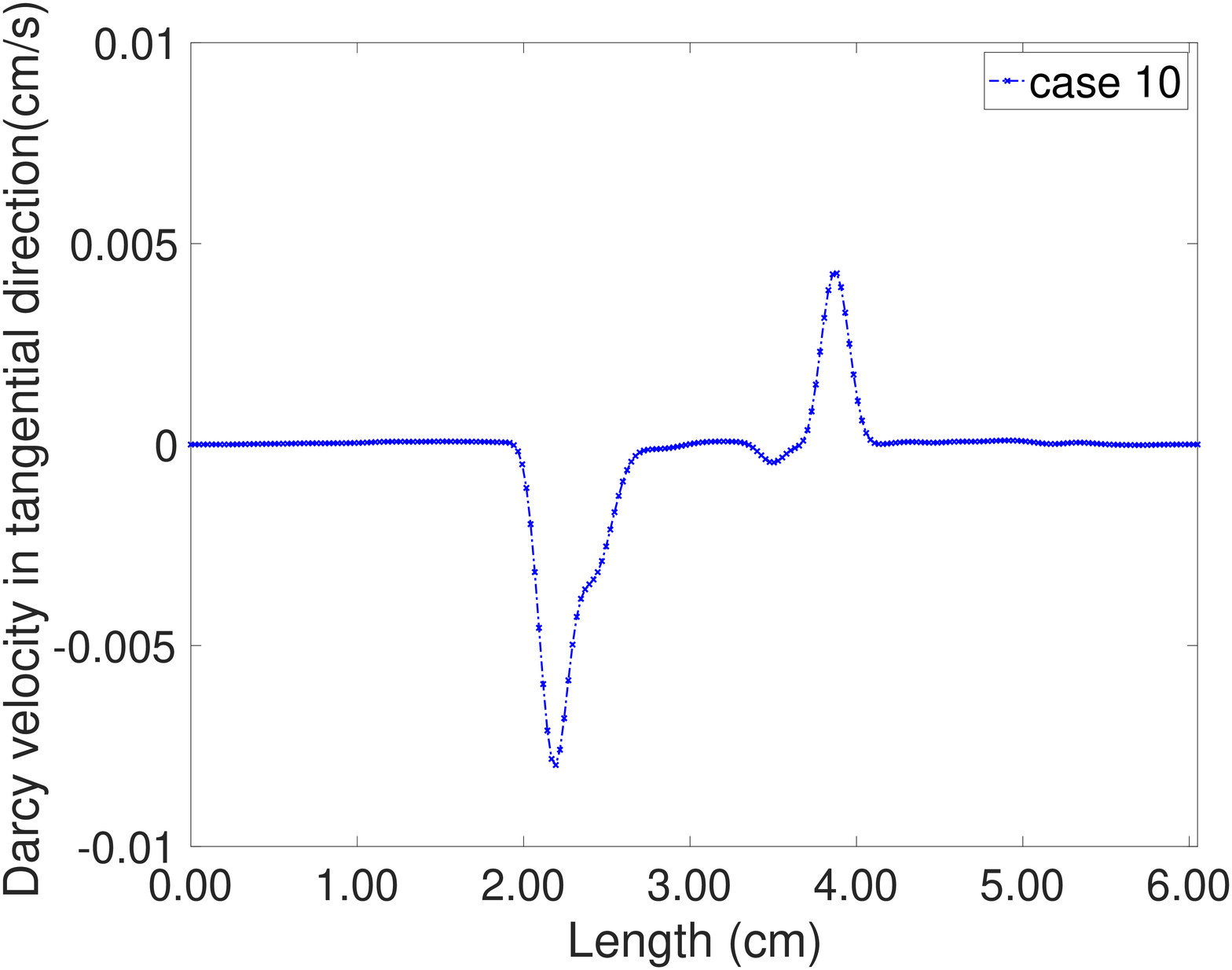}
	\caption{Darcy velocity in the tangential direction $\bu_{p}\cdot \bt$  along the top arterial wall at time t=1.8 ms, t=3.6 ms, t=5.4 ms for case 9 and case 10. The top panel is for case 9, and the bottom panel is case 10. }
		\label{supt}
\end{figure}

In Fig.\ref{supn} and Fig.\ref{supt}, we present the Darcy velocities in the normal $\bu_{p}\cdot \bn$ and tangential $\bu_{p}\cdot \bt$ directions along the interface for case 9 and case 10, respectively. As expected, there are remarkable differences in the filtration velocities $\bu_{p}\cdot \bn$ for these NSE/P cases. For case 9 with constant permeability, the magnitude of filtration velocity $\bu_{p}\cdot \bn$ is much smaller than that of case 10, and so is $\bu_{p}\cdot \bt$. In addition, when we compare case 9 with case 6, which have differences only on the geometry, Darcy velocities of case 9 in both directions are much smaller than case 6. As conclusion, on one hand, similar with the previous healthy structure cases, larger permeability would result in larger Darcy velocities in both directions; on the other hand, since the change of geometry, the appearance of stenosis would result in smaller Darcy velocity in both directions. 
\subsubsection{Conclusions}
From the stenosis models, we draw the following conclusions:
\begin{itemize}
	\item Comparing with the healthy vessel structure, the stenotic geometry of arteries has a vital effect on blood flow patterns. Flow variables (velocities and pressure fields) are affected dramatically. The general flow rates are large and the pressure loss are slight even for larger permeability. 
	\item Poroelasticity affects the WSS dramatically. WSS of stenotic models is in general larger than the healthy structure models. 
	\item Great differences between NSE/E and NSE/P models are found in the quantification of RRT for the stenotic geometry. For NSE/P models, the peak of RRT typically appears at the prior-stenotic area. It is the combination of permeability and Lam\'{e} coefficients that make a difference. There would be smaller risks of vessel lesions for softer and more permeable build-ups. NSE/E model would typically generate smaller RRT. We believe that the poriferous property of vessel should not be neglected to calculate RRT accurately. Models without considering the structure would typically underestimate the arterial lesion. 
	\item For stenosis models, the displacement in the normal direction is not affected much by the permeability $K$, but by the Lam\'{e} coefficients. The peak of normal displacement would generally occur at the prior-stenosis regions and it would not coincide with the the peak of pressure fields. 
	\item For fluids velocity in normal direction along the arterial wall, no significant differences are detected.
	\item For Darcy velocities in normal and tangential directions along the lumen, they are greatly influenced by permeability $K$. NSE/P models with larger permeability would result in larger Darcy velocities in both normal and tangential directions. Comparing healthy structure, the stenotic geometry tends to impede the penetration along the interface. 
\end{itemize}

Limitations of this work are the abscence of 3D patient-specific stenotic vessels and of turbulence models in the carotids.  

\bibliographystyle{abbrv}
\bibliography{main}

\end{document}